\documentclass[longauth]{aa} 
\usepackage{natbib}
\usepackage{graphicx}
\usepackage{lscape}
\usepackage{longtable}
\usepackage{txfonts}
\usepackage{color}                                                             
\newcounter{subtable}

\begin{document}

   \title{DUst Around NEarby Stars. The survey observational results
    \thanks{Herschel is an ESA space observatory with science instruments 
provided 
            by European-led Principal Investigator consortia and with important 
            participation from NASA.
     } \fnmsep
    \thanks{Tables are only available in the electronic version of the paper or
    at the CDS via anonymous ftp to cdsarc.u-strasbg.fr (130.79.128.5)
    or via http://cdsweb.u-strasbg.fr/cgi-bin/qcat?J/A+A/
    }}

   \subtitle{}

   \author{C. Eiroa 
          \inst{1}
          \and  J.P. Marshall 
          \inst{1}
          \and A. Mora 
          \inst{2}
          \and B. Montesinos 
          \inst{3}
            \and O. Absil\inst{4}
          \and J. Ch. Augereau
          \inst{5}
          \and A. Bayo\inst{6,7} 
         \and G. Bryden\inst{8} 
          \and W. Danchi\inst{9}
         \and C. del Burgo\inst{10}
         \and S. Ertel\inst{5}
          \and M. Fridlund\inst{11} 
          \and A. M. Heras\inst{11}
          \and A. V. Krivov\inst{12}  
           \and R. Launhardt\inst{7}
           \and R. Liseau\inst{13}
           \and  T. L\"ohne\inst{12}
           \and J. Maldonado\inst{1} 
           \and G.L. Pilbratt\inst{11}
            \and A. Roberge\inst{9} 
           \and J. Rodmann\inst{14}           
            \and J. Sanz-Forcada\inst{3}
            \and E. Solano\inst{3}
            \and K. Stapelfeldt\inst{9}
            \and P. Th\'ebault\inst{15} 
            \and S. Wolf\inst{16}  
             \and D. Ardila\inst{17}
            \and M. Ar\'evalo\inst{3,33}
            \and C. Beichmann\inst{18}
            \and V. Faramaz\inst{5}
            \and B.M. Gonz\'alez-Garc\'\i a\inst{19}
            \and R. Guti\'errez\inst{3}
            \and J. Lebreton\inst{5}
            \and R. Mart\'\i nez-Arn\'aiz\inst{20} 
            \and G. Meeus\inst{1}
            \and  D. Montes\inst{20}
             \and G. Olofsson\inst{21}
             \and K.Y.L.  Su\inst{22}
             \and G. J. White\inst{23,24}
             \and D. Barrado\inst{3,25}
             \and M. Fukagawa\inst{26}
             \and E. Gr\"un\inst{27}
             \and I. Kamp\inst{28}
             \and R. Lorente\inst{29}
              \and A. Morbidelli\inst{30}
             \and S. M\"uller\inst{12}
             \and  H. Mutschke\inst{12}
             \and T. Nakagawa\inst{31}
              \and I. Ribas\inst{32}
              \and H. Walker\inst{24}
          }
\institute{Universidad Aut\'onoma de Madrid, Dpto. F\'isica Te\'orica, 
                   M\'odulo 15, Facultad de Ciencias, Campus de Cantoblanco, 
                   E-28049 Madrid, Spain
 \and ESA-ESAC Gaia SOC. P.O. Box 78
 E-28691 Villanueva de la Ca{\~n}ada, Madrid, Spain
\and Centro de Astrobiolog\'\i a (INTA-CSIC), ESAC Campus,  P.O.Box 78, E-28691 Villanueva de la Ca{\~n}ada, Madrid, Spain
 \and Institut d'Astrophysique et de Géophysique, Université de Liège, 17 Allée du Six Août, B-4000 Sart Tilman, Belgium
\and UJF-Grenoble 1 / CNRS-INSU, Institut de Plan\'etology et d'Astrophysique de Grenoble (IPAG), UMR 5274, Grenoble, F-38041, France
 \and European Space Observatory, Alonso de Cordova 3107, Vitacura
Casilla 19001, Santiago 19, Chile 
\and Max-Planck Institut f\"ur Astronomie, K\"onigstuhl, 69117, Germany
\and Jet Propulsion Laboratory, California Institute of Technology, 4800 Oak Grove Drive, Pasadena, CA 91109, USA
 \and NASA Goddard Space Flight Center, Exoplanets and Stellar Astrophysics, Code 667, Greenbelt, MD 20771.USA
 \and Instituto Nacional de Astrof\'isica, Optica y Electr\'onica, Luis Enrique Erro 1, Sta. Ma. Tonantzintla, Puebla, M\'exico
 \and ESA Astrophysics \& Fundamental Physics Missions Division, ESTEC/SRE-SA,  Keplerlaan 1, NL-2201 AZ Noordwijk,
 The Netherlands
  \and   Astrophysikalisches Institut und Universit{\"a}tssternwarte, Friedrich-Schiller-Universit{\"a}t,
 Schillerg{\"a}{\ss}chen 2-3, 07745 Jena, Germany
 \and Department of Earth and Space Sciences, Chalmers University of Technology, Onsala Space Observatory, Se-439 92 Onsala, Sweden
 \and  Georg-August-Universit\"at G\"ottingen, Institut f\"ur Astrophysik, Friedrich-Hund-Platz 1, 37077 G\"ottingen, Germany
 \and LESIA, Observatoire de Paris, 92195 Meudon France  
\and Christian-Albrechts-Universit\"at zu Kiel, Institut f\"ur Theoretische 
Physik und Astrophysik, Leibnizstr. 15, 24098 Kiel, Germany
\and NASA Herschel Science Center, California Institute of Technology, 1200 E. California Blvd., Pasadena, CA 91125, USA
\and NASA ExoPlanet Science Institute California Institute of Technology, 1200 E. California Blvd., Pasadena, CA 91125, USA 
\and ISDEFE at ESAC, E-28691 Villanueva de la Ca{\~na}da, Madrid, Spain
  \and Universidad Complutense de Madrid, Facultad de Ciencias F\'\i sicas, Dpt. Astrof\'\i sica, av. Complutense s/n. 28040 Madrid, Spain
 \and Department of Astronomy, Stockholm University, AlbaNova University Center, Roslagstullsbacken 21, SE-106 91 Stockholm, Sweden
 \and Steward Observatory, University of Arizona, 933 North Cherry Avenue, Tucson, AZ 85721, USA    
 \and Rutherford Appleton Laboratory, Chilton OX11 0QX, UK  
 \and  Department of Physics and Astrophysics, Open University, Walton Hall, Milton Keynes MK7 6AA, UK
 \and Calar Alto Observatory, Centro Astron\'onomico Hispano-Alem\'an, c/ Jes\'us Durb\'an Rem\'on, 2-2, 04004, Almer\'ia, Spain
\and Nagoya University, Japan. 
\and Max-Planck-Institut f\"ur Kernphysik, Saupfercheckweg 1, D-69117 Heidelberg, Germany 
  \and Kapteyn Astronomical Institute, Postbus 800, 9700 AV Groningen, The Netherlands
 \and Herschel Science Center, ESAC/ESA, P.O. BOX 78, 28691 Villanueva de la 
Ca\~nada, Madrid, Spain
 \and Observatoire de la C\^ote d'Azur, Boulevard de l'Observatoire, B.P. 4229, 06304 Nice Cedex 4, France. 
 \and Institute of Space and Astronautical Science (ISAS), Japan Aerospace Exploration Agency (JAXA), 3-1-1, Yoshinodai, Chuo-ku, Sagamihara, Kanagawa, 252-5210, Japan 
 \and Institut de Ci\`encies de l'Espai (CSIC-IEEC), Campus UAB,
Facultat de Ci\`encies, Torre C5, parell, 2a pl., E-08193 Bellaterra, Barcelona, Spain     
\and ESA-ESAC, Science Archives Team, P-O- Box 78, E-28691 Villanueva de la Ca\~nada, Madrid, Spain         
         }

   \offprints{C. Eiroa \\  \email{carlos.eiroa@uam.es}}
   \date{Received , ; Accepted  }

 
\abstract
   {Debris discs  are a  consequence  of  the planet
formation  process and constitute  the fingerprints  of planetesimal
systems.  Their solar system's counterparts  are the asteroid and 
Edgeworth-Kuiper belts.
   }
   {The DUNES survey aims at detecting extra-solar analogues to the 
Edgeworth-Kuiper belt around solar-type stars, putting 
in this way the solar system into context. The survey allows us to address 
some  questions related to the prevalence
and properties of planetesimal systems.  
   }
   {We used  {\it Herschel}/PACS  to observe a sample of 
nearby FGK  stars. Data at 100 and 160
$\mu$m were obtained, complemented in some cases with observations at 70
$\mu$m, and at 250, 350 and 500 $\mu$m 
using  SPIRE. The observing strategy was to integrate as
deep as possible at 100 $\mu$m to detect the stellar photosphere.
     }
{ Debris discs have been detected at a fractional luminosity level down to 
several times that of the Edgeworth-Kuiper belt. The incidence rate of discs 
around the DUNES stars is increased from a rate of $\sim$ 12.1\% $\pm$ 5\% before 
\emph{Herschel} to  $\sim$ 20.2\% $\pm$ 2\%. A significant
fraction ($\sim$ 52\%) of the discs  are resolved, which represents an enormous step  
ahead from  the previously known resolved discs. Some stars are associated 
with faint far-IR excesses attributed to a new class of cold  discs. 
Although it cannot be excluded that these excesses are produced by coincidental
alignment of background galaxies, statistical arguments  suggest 
that at least some of them are true debris discs. Some  discs 
display peculiar SEDs with spectral indexes in the 70--160 $\mu$m  range  
 steeper than the Rayleigh-Jeans one. An analysis of the debris disc
parameters   suggests that a decrease might exist of the mean black body radius
from the F-type to the K-type stars. In addition, a weak trend is
suggested for a correlation of disc sizes and an anticorrelation
of disc temperatures with the stellar age.
}
   {}

   \keywords{stars: circumstellar matter -- stars: planetary systems -- infrared: stars}

   \maketitle

\section{Introduction}

Circumstellar discs  are formed around stars as  a by-product required
by  angular momentum  conservation.  In  their earliest  phases, stars
accrete a  significant part of their  masses from gas and  dust in the
discs.  Meanwhile,  those circumstellar accretion discs  evolve from a
gas-dominated  protoplanetary phase  to a  gas-poor  debris-disc phase
where  large planetesimals  and  full-sized planets  may have  formed,
after  primordial  submicron-sized  dust  grains settle  in  the  disc
midplane and  coagulate to form  dust aggregates, pebbles,  and larger
rocky bodies.  Most  likely this is the formation  pathway followed by
the currently  known exoplanets  (close to one  thousand) and  our own
solar system to form. However, it  is not known which are the ultimate
circumstances  and  physical  conditions  that make  planet  formation
possible, and  whether planet formation is nearly  as universal during
disc evolution  as is  the formation of  discs during  star formation.
This issue is central to understand the incidence of planetary systems
in general, and consequently the formation of Earth-like planets.

In the solar system, the  planets together with asteroids, comets, the
zodiacal material  and   the  Edgeworth-Kuiper   Belt  (EKB)   are  the
fingerprints of  such dynamical processes.   Planet formation resulted
in  a nearly total depletion  of  planetesimals
inside  the orbit  of Neptune,  with the  remarkable exception  of the
asteroid belt.   Leftover planetesimals not  incorporated into planets
arranged to  form the  EKB, beyond the  orbit of  Neptune, dynamically
sculpted and excited by  the giant planets.  Mutual collisions between
EKB  objects and  erosion  by interstellar  dust  grains release  dust
particles that spread over  the EKB region \citep{jewitt2009}.  If the
EKB could  be observed from  afar, it would  appear as  an extended 
($\sim$50 AU) and very faint ($L_\mathrm{d}/L_\odot \sim 10^{-7}$) 
emission with a temperature of 70--100 K  
\citep{backman1995,vitense2010,vitense2012}, with  a huge central hole
caused  by the massive  planets \citep{moromartin2005}.   

The discovery of IR excesses in main-sequence stars such as Vega,
Fomalhaut or $\beta$ Pic was one  of the most significant accomplishements of the
IRAS  satellite  \citep{aumann1984}.  The  observed excess  was  attributed  to
thermal  emission  from  solid  particles around  the  stars.  Optical
imaging of $\beta$ Pic convincingly demonstrated that the dust was located
in  a  flattened  circumstellar disc  \citep{smithterrile1984}.  Since
lifetimes   of    dust   grains   against    radiative/wind   removal,
Poynting-Robertson  drag and collisional  disruption are  much shorter
than the age of the stars, one must conclude that these dust particles
are  not  remnants of the primordial discs, instead they are  
the  result  of  ongoing processes.  Nearly all
modelling  efforts  explain "debris discs"  dust
production as a  result of   collisions  of   larger  bodies
\cite[][and references   therein]{wyatt2008,krivov2010}.  Given that
debris discs  survive over  billions of years,  there must be  a large
reservoir  of leftover  planetesimals and  solid bodies that collide and are intimately
related to  the dust particles. Furthermore, dust particles respond in
different ways to the gravity  of planetary perturbers depending on their 
size distribution and can be used as a tracer of  planets
\citep{augereau2001,quillen2002,moromartin2007,mustill2009,thebault2012}. 
Consequently,
observations of   debris discs  sheds  light onto  the  processes related  to
planet and planetesimal formation.

Much observational as well as modelling progress has occurred in
the last two decades primarily from infrared (IR)
and  (sub)-millimetre  facilities. The first  debris discs were discovered by the InfraRed Astronomical Satellite (IRAS), mainly around A stars  due to sensitivity limitations.  The Infrared Space Observatory (ISO) extended
the  study of debris discs  and  added important  information on the age
distribution of debris discs \citep{habing2001}.  More  recently, {\it
  Spitzer}  added  a  wealth  of  new  information  in  a  variety  of
aspects. For example, the  incidence rate was found  to be larger  for A stars
and then it decreased   with  later spectral   types   up  to   M   stars
\citep{su2006,gautier2007}. An incident rate  of $\sim$ 16\% was found
around solar-type FGK stars \citep{trilling2008}, not dependent on the
stellar  metallicity \citep{beichman2006},  although a  marginal trend
might  exist,  as  recently  suggested  by  \cite{maldonado2012}.  The
presence of  exoplanets is  not necessarily a sign  for a higher  incidence of
debris  discs  \citep{kospal2009}, although \cite{wyatt2012} have recently claimed 
that the debris incidence rate is higher around stars with low mass planets, 
and there may be trends  
 between  some debris  discs  and  planet  properties when  both
simultaneously exist  \citep{maldonado2012}. {\it Spitzer}  also found
that typical  debris discs around solar-type stars  emit much stronger
at 70 $\mu$m than at 24  $\mu$m, with the detection rate for hot discs
being  very  low.  Spectral  energy  distributions  (SEDs) imply the dust
is located at several tens of AU and dust temperatures $\sim$ 50--150 K
\citep{trilling2008,moor2011}.  However,  the distance, dust  mass and
optical  properties are  degenerate  with the  (unknown) particle  size
distribution. 

In spite  of its remarkable  contribution {\it Spitzer}  suffered from
two severe  constraints.  Firstly, its  moderate sensitivity, $L_{\rm
  d}/L_\star$ $\sim$ $10^{-5}$ \citep{trilling2008}, i.e., about  two  orders  of   magnitude  above  the  EKB  luminosity, and its wavelength coverage, in practice up to 70 $\mu$m, limited its ability
to detect cold dust.  Secondly, its moderate spatial resolution
prevented detailed  studies of  the spatial  structure  in debris
discs since it resolved only a few discs.   Significantly higher spatial resolution is 
required in order to determine
the  location of  the dust  and its
spatial distribution, which traces  rings, warps, cavities, or asymmetries,  and which 
can be used to  infer the  potential presence  of planets
\citep{mouillet1997,lagrange2010}. 
The   ESA  {\it  Herschel}  space
telescope \citep{pilbratt2010}  overcomes these limitations  thanks to
its  larger  mirror and  instruments  PACS \citep{poglitsch2010}  and
SPIRE  \citep{griffin2010},  which  allow  for a  better  sensitivity,
wavelength coverage and higher spatial resolution.

In this paper  we summarize the observational results  obtained in the
frame  of the {\it  Herschel} Open  Time Key  Programme DUNES\footnote{http://www.mpia-hd.mpg.de/DUNES/},  {\it DU}st
around {\it NE}arby {\it S}tars, (KPOT\_ceiroa\_1 and SDP\_ceiroa\_3).
This  programme   aims  at  detecting  EKB   analogues  around  nearby
solar-type  stars;  putting  in  this  manner the  solar  system  into
context. The  content of this paper addresses  the DUNES observational
results  presented  as  a  whole.   Detailed analysis  or  studies  of
individual sources  or groups of objects  are out of the  scope of this
work.  For such  more detailed  and deeper  studies we  refer  to some
already                     published                    observational
\citep{liseau2010,eiroa2010,marshall2011,eiroa2011},   and   modelling
papers \citep{ertel2012,loehne2012},  as well as  to forthcoming ones.
The    current    paper    is    organized   as    follows:    section
\ref{section:objectives}  describes the  scientific  rationale and
the  observing strategy.   Section  \ref{section:sample} presents  the
sample   of    stars.    Section   \ref{section:herschel_observations}
describes  the {\it  Herschel} PACS  and SPIRE  observations  and data
reduction,  while  the  treatment of PACS noise  and  the  results  are
presented      in     sections      \ref{section:noise_method}     and
\ref{section:results}, respectively.  The analysis of the  data of the
non-excess sources with the  upper limits of the fractional luminosity
of  the dust  are in  Section \ref{section:non-excess}.   The detected
debris discs  are presented in Section  \ref{section:excess}, where the
background   contamination    and   some   general    properties   and
characteristics    of    the    discs    are    described.     Section
\ref{section:discussion} presents a  discussion of disc properties and
some stellar  parameters. Finally, section  \ref{section:summary} contains a
summary  and our conclusions.    In  addition,  several  appendixes  give
some fundamental  parameters  of  the   stars,  the  method  used  for  the
photospheric fits, and a short description of some spurious objects. 

\section{DUNES Scientific objectives: Survey rationale}
\label{section:objectives}

The primary  scientific objective of  DUNES is the  identification and
characterization of  faint exosolar analogues to the  solar system EKB
in an  unbiased sample of nearby solar-type  stars. Strictly speaking,
the detection of  such faint discs is a direct  proof of the incidence
of  planetesimal   systems  and  an  indirect  one   of  the  presence
planets. The  survey design allows  us to additionally  address several
fundamental, specific  questions that help to  evaluate the prevalence
and  properties of  such  planetesimal and  planetary systems.   These
are: i) the fraction of  solar-type stars  with faint,
EKB-like  discs;  ii)  the collisional  and  dynamical  evolution  of  EKB
analogues;  iii) the dust  properties  and  size  distribution;  and  iv)
the incidence of EKB-like discs versus the presence of planets.


 According to  the recent  EKB model  of  \cite{vitense2012}, the
  predicted  infrared excess  peaks at  $\sim$50 $\mu$m  and  the flux
  levels in  the PACS bands  would be between  0.1 and 0.4  mJy.  This
  flux  is  about  an  order  of magnitude  lower  than  the  expected
  photospheric  fluxes  from  nearby  solar-type stars  (see  appendix
  \ref{appendix:predictions}), and few  times lower than the predicted
  pre-launch sensitivity of PACS (PACS observer's manual, version 1.3,
  04/July/2007).  Therefore,  the   challenge  of  detecting  a  faint
  infrared excess,  which could be considered as  an exo-EKB analogue,
  is the the detection of a  faint far-IR signal from a debris disc on
  top of  a weak photospheric signal  which is few  times the expected
  measurement uncertainties.

\begin{figure}[ht!]
\centering
\includegraphics[angle=270,scale=0.33]{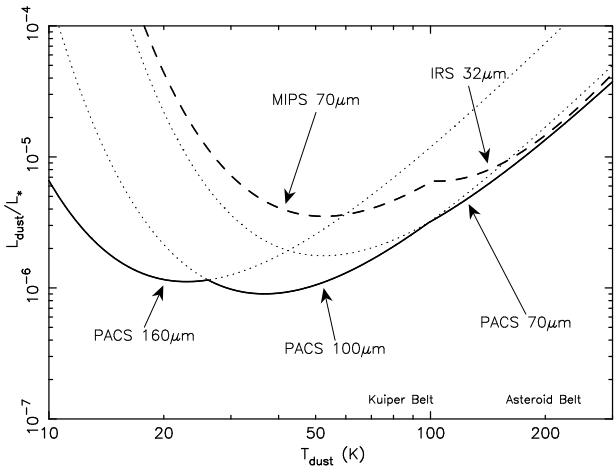}
\caption{Detection  limits  for a  G2V  star at  10  pc  for the  {\it
    Herschel}  70, 100,  and 160  $\mu$m  bands compared  to the  {\it
    Spitzer} instruments MIPS at 70 $\mu$m and IRS at 32 $\mu$m.}
\label{fig:sensitivity_limits}
\end{figure}


The observing strategy is also  modulated by the choice of the optimal
wavelength.  The  equilibrium temperature of  a dust grain  depends on
the  stellar luminosity,  the radial  distance to  the star,  and dust
properties (size, chemical composition, mineralogy).  For distances of
$\sim$30-100 AU, grains  of about 10 $\mu$m in  size have temperatures
in  the  range   $\sim$30-50  K  (Krivov  et  al.    2008).  At  these
temperatures, the bulk  of the thermal re-emission is  radiated in the
far-IR covered by the PACS  photometric bands centered at 70, 100, and
160 $\mu$m.   Figure \ref{fig:sensitivity_limits}  highlights the
  unique {\it Herschel} PACS discovery space compared to {\it Spitzer}
  MIPS and IRS.  The limits in Figure \ref{fig:sensitivity_limits} are
  calculated assuming  PACS 1$\sigma$ accuracies of 1.5,  1.5, and 3.5
  mJy  at  70,  100,   and  160  $\mu$m,  respectively.  A  systematic
  uncertainty    of   5\%   is    also   included    for   calibration
  uncertainty. Note that these  accuracies are larger than the typical
  uncertainties     found    in     this     survey    (e.g.     Table
  \ref{table:nonexcess_sources}), and  that the systematic uncertainty
  is   larger  than  that   reported  in   the  PACS   technical  note
  PICC-ME-TN-037\footnote{Technical               note              in
    http://herschel.esac.esa.int}. Spitzer/MIPS limits are based on an
  assumed  photometry  accuracy  of   3  mJy  and  a  10\%  systematic
  contribution \citep[e.g.][]{bryden2009}. Spitzer/IRS is limited to a
  2  \%  uncertainty at  32  $\mu$m  \citep{lawler2009}.  The  assumed
  photospheric uncertainty for  both PACS and MIPS is  2\%.  The plot
shows in  particular that PACS  100 $\mu$m provides the  most suitable
range to  detect very faint discs  for dust temperatures  in the range
 from $\sim$20 to $\sim$100  K. Further, with a detection limit of
  $L_{\rm  d}/L_\star$ few times  $10^{-7}$, PACS  100 $\mu$m  has the
  ability to reveal dust discs  with emission levels close to the EKB.
  We  note that although  the PACS  70 $\mu$m  band has  a sensitivity
  similar to PACS 100 $\mu$m for EKB temperatures around 100 K, and is
  more  competitive  in terms  of  background  confusion and  stellar
photospheric detection,  100 $\mu$m  provides a better  contrast ratio
between the emission of cold  dust and the stellar photosphere, and is
in  fact much  more sensitive  than PACS  70 $\mu$m  for  probing very
faint, cold discs.

Given  the above considerations  concerning flux  levels from  the EKB
analogues  and the  stars together  with the  optimal  wavelength, the
choice  to fulfil the  DUNES objectives  was to  integrate as  deep as
needed  to  achieve the  estimated  photospheric  flux  levels at  100
$\mu$m.
\begin{table}[!ht]
\caption{Summary of spectral types in  the DUNES sample and the shared
  sources observed by DEBRIS.}
\label{table:sample_spectraltypes}
\begin{tabular}{lllll}
\hline\noalign{\smallskip}
Sample                                        & F stars &G stars  & K stars & Total   \\
\hline
Solar-type stars observed by                  &         &         &         &         \\
DUNES (the DUNES sample)                      &  27     &    52   &   54    &  133     \\    
20 pc DUNES subsample                         &  20     &    50   &   54    &  124    \\
Shared solar-type stars                       &         &         &         &         \\
observed by Debris                            & 51      &    24   &    8    &   83  \\
Shared 20 pc subsample                        &  32     &    16   &    8    &   56      \\
\hline
\end{tabular}
\end{table}


\begin{table*}
\centering
\caption{The   DUNES  stellar  sample.   Columns  correspond   to  the
  following: {\it  Hipparcos} and HD  numbers as well as  usual stars'
  names; spectral types and ranges (see text); equatorial and galactic
  coordinates; parallaxes  with errors and stars'  distances. Only the
  first 5 lines  of the table are presented here.  The full version is
  available as online material.}
\begin{small}
\begin{tabular}{llllllllll}
\hline\noalign{\smallskip}
HIP &  HD    &Name              &SpT    & SpT range &               ICRS (2000)   &      Galactic    &$\pi$(mas) &d(pc)     \\
\hline
171    &224930 &HR 9088             &G3V       &G2V - G5V        &00 02 10.156 +27 04 56.13&109.6056 -34.5113 &   82.17$\pm$ 2.23& 12.17  \\ 
544    &166    &V439 And            &K0V       &G8V - K0V        &00 06 36.785 +29 01 17.40&111.2636 -32.8326 &   73.15$\pm$ 0.56& 13.67  \\
910    &693    &6 Cet               &F5V       &F5V - F8V        &00 11 15.858 -15 28 04.73&082.2269 -75.0650 &   53.34$\pm$ 0.64& 18.75  \\
2941   &3443   &HR 159              &K1V+...   &G7V - G8V        &00 37 20.720 -24 46 02.18&068.8453 -86.0493 &   64.93$\pm$ 1.85& 15.40  \\
3093   &3651   & 54 Psc             &K0V       &K0V - K2V        &00 39 21.806 +21 15 01.71&119.1726 -41.5331 &   90.42$\pm$ 0.32& 11.06  \\
\hline\noalign{\smallskip}
\end{tabular}
\end{small}
\label{table:stars_sample}
\end{table*}

\section{The stellar sample} 
\label{section:sample}

\begin{figure*}
\centering
\includegraphics[angle=270,scale=0.33]{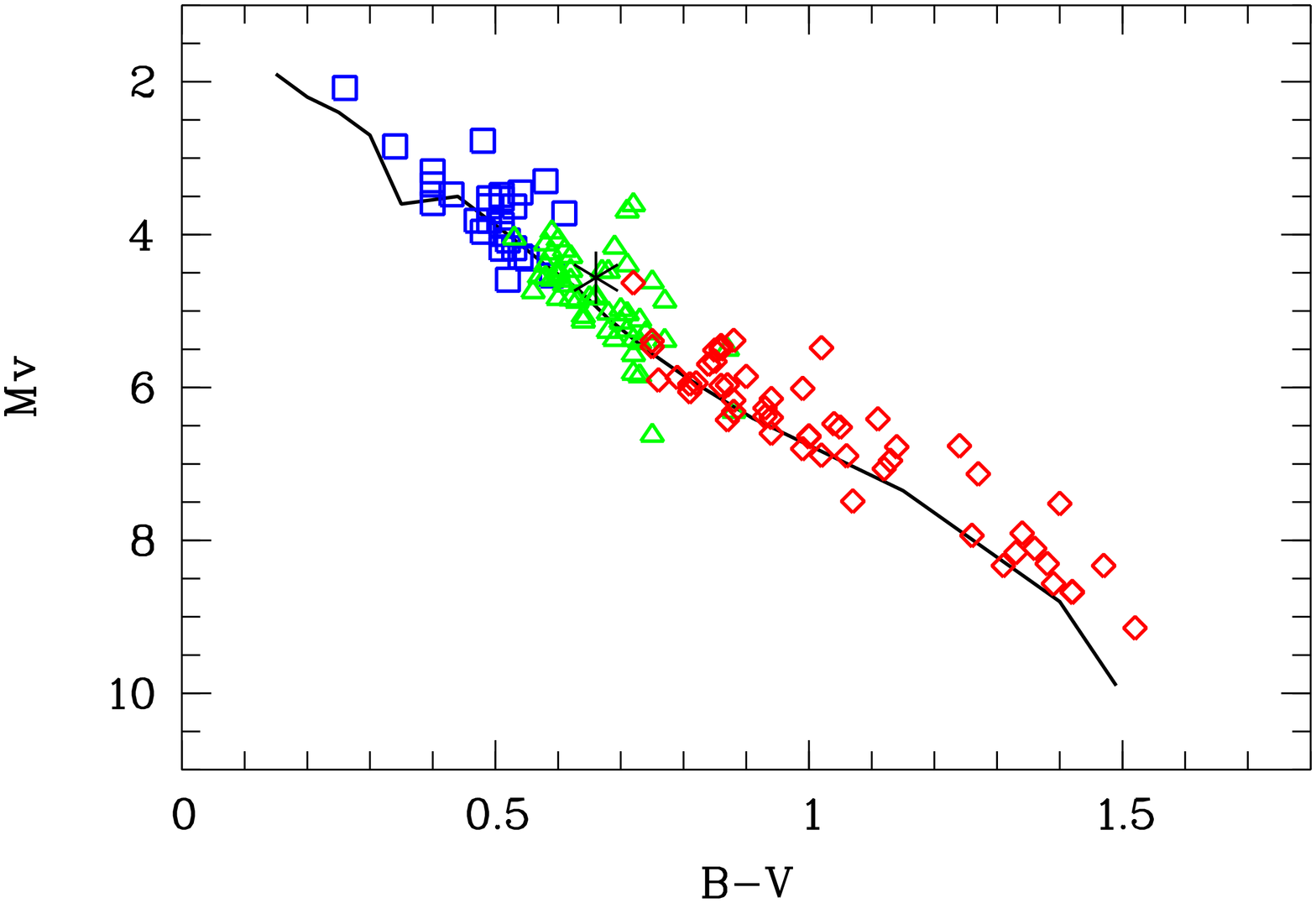}
\includegraphics[angle=270,scale=0.33]{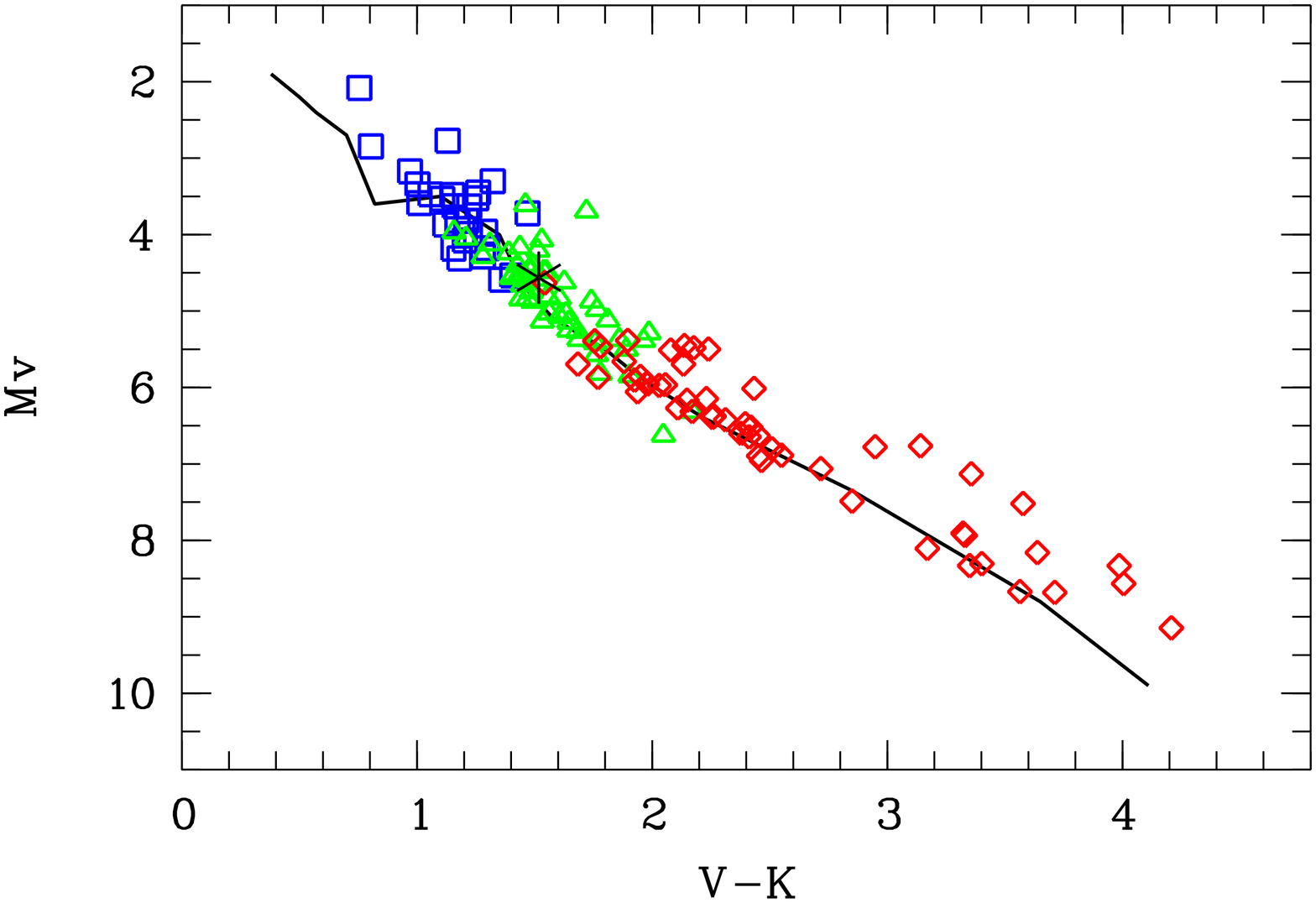}
\caption{Colour-absolute  magnitude  diagrams  of the  DUNES  sources.
  Spectral  types as  in Table  1 are  distinguished by  symbols: blue
  squares  (F-type),   green  triangles  (G-type)   and  red  diamonds
  (K-type).   The   solid  line   in  both  diagrams   represents  the
  main-sequence while  the star symbol  indicates the position  of the
  Sun \citep{cox2000}.}
\label{HR_diagram}
\end{figure*}

\begin{table*}
\setcounter{subtable}{1}
\centering
\caption{Photometric magnitudes  and fluxes  of the DUNES  stars. Only
  the first  5 lines  with the optical  (Johnson and  Str\"omgrem) and
  2MASS     photometry    are     shown    here     (see    appendixes
  \ref{appendix:stellar_properties}   and   \ref{appendix:predictions}
  ). The  full version  of the table  including further  near-IR data,
  AKARI,  WISE, IRAS  and {\it  Spitzer} MIPS  is available  as online
  material.}
\begin{tabular}{lcccccccccc}
\hline\noalign{\smallskip}
HIP      &  $V$      & $B-V$ & $V-I$ &$b-y $& $m_1$& $c_1$ &        $J$        &       $H$  & $K_{\rm s}$      &  Q \\  
\hline        
171      & 5.80    &0.69  & 0.82 & 0.432& 0.184& 0.218 & 4.702$\pm$0.214 & 4.179$\pm$ 0.198 & 4.068$\pm$ 0.236& CCD\\    
544      & 6.07    &0.75  & 0.80 & 0.460& 0.290& 0.311 & 4.733$\pm$0.019 & 4.629$\pm$ 0.144 & 4.314$\pm$ 0.042& EBE\\
910      & 4.89    &0.49  & 0.59 & 0.328& 0.130& 0.405 & 4.153$\pm$0.268 & 3.800$\pm$ 0.208 & 3.821$\pm$ 0.218& DCD\\
2941     & 5.57    &0.72  & 0.78 & 0.435& 0.254& 0.287 & 4.437$\pm$0.266 & 3.976$\pm$ 0.224 & 4.027$\pm$ 0.210& DDC\\
3093     & 5.88    &0.85  & 0.83 & 0.507& 0.384& 0.335 & 4.549$\pm$0.206 & 4.064$\pm$ 0.240 & 3.999$\pm$ 0.036& CDE\\
\hline\noalign{\smallskip}
\end{tabular}
\noindent 
\label{table:stellar_fluxes}
\end{table*} 

\begin{table*}[!ht]
\centering
\caption{Fundamental  stellar parameters  and some  properties  of the
  DUNES  sources   (see  appendix  \ref{appendix:stellar_properties}).
  Only the first  5 lines are shown. The full version  of the table is
  available as online material.}
\begin{tabular}{llrrrrrrrrr}
\hline\noalign{\smallskip}
HIP   &SpT     &T$_{\rm eff}$&log g &[Fe/H] & $v \sin i$ &Lbol &  Lx/Lbol & AgeX &$\log R'_{\rm HK}$ & Age(Ca {\sc ii})   \\
      &        & (K)        &(cm/s$^2$)&(dex)&(km/s)   &(L$_\odot$)& (log)  & (Gyr)&       &(Gyr)  \\
\hline                                                                                              
171   &G3V     &5681    & 4.86  & -0.52 & 1.8  & 0.614&  -5.9  &  3.12     &-4.851 & 3.96  \\
544   &K0V     &5577    & 4.58  &  0.12 & 3.4  & 0.616&  -4.4  &  0.32     &-4.328 & 0.17  \\
910   &F5V     &6160    & 4.01  & -0.38 & 3.8  & 3.151&  -7.6  &  12.53    &-4.788 & 3.04  \\
2941  &K1V+... &5509    & 4.23  & -0.14 & 1.6: & 1.258&        &           &-4.903 & 4.83  \\
3093  &K0V     &5204    & 4.45  &  0.16 & 1.15 & 0.529&  -6.0  &  4.53     &-4.991 & 6.43  \\
\hline\noalign{\smallskip}
\end{tabular}
\noindent 
\label{table:stellar_parameters}
\end{table*} 

The  preliminary stellar sample  was chosen  from the  {\it Hipparcos}
catalogue  \citep{1997ESASP1200}  following   the  sole  criterion  of
selecting main-sequence, luminosity  classes V-IV/V, stars closer than
25  pc  without  any  further  bias concerning  any  property  of  the
stars. Since  the {\em Herschel} observations were  designed to detect
the photosphere,  the only restriction  to build the final  sample was
that the  stars could  effectively be detected  by PACS at  100 $\mu$m
with a $S/N  \geq 5$, i.e., the expected  100 $\mu$m photospheric flux
should  be  significantly  higher  than  the  expected  background  as
estimated by the {\em Herschel} HSPOT tool at that wavelength.  Taking
into account the total observing  time finally allocated for the DUNES
survey  (140 hours)  as  well  as the  complementarity  with the  {\em
  Herschel}  OTKP   DEBRIS  \citep{2010A&A...518L.135M},  the  stellar
sample  for this  study was  reduced to  main-sequence  FGK solar-type
stars located at distances smaller  than 20 pc.  In addition, from the
original sample  we retained  FGK stars between  20 and 25  pc hosting
exoplanets  (3 stars,  1  F-type and  2  G-type, at  the  time of  the
proposal writing)  and previously known debris discs,  mainly from the
{\it Spitzer} space telescope (6  stars, all F-type).  Thus, the final
sample of stars directly observed  by DUNES, formally called the DUNES
sample in  this paper,  is formed by  133 stars,  27 out of  which are
F-type, 52 G-type, and 54 K-type  stars. The 20 pc subsample is formed
by  124   stars  -  20  F-type,   50  G-type  and   54  K-type.  Table
\ref{table:sample_spectraltypes}   summarizes    the   spectral   type
distribution of the samples.

The OTKP  DEBRIS project was  defined as a  volume limited study  of A
through M stars selected from the ``UNS'' survey \citep{phillips2010},
observing  each  star   to  an  uniform  depth,  i.e.,   DEBRIS  is  a
flux-limited survey.   In order to  optimize the results  according to
the  DUNES and DEBRIS  scientific goals,  the complementarity  of both
surveys was  achieved by  dividing the common  stars of  both original
samples considering if the  stellar photosphere could be detected with
the DEBRIS uniform integration time.   Those stars were assigned to be
observed by DEBRIS. In that  way, the DUNES observational objective of
detecting the  stellar photosphere was  satisfied. The few  A-type and
M-type stars common in both surveys were also assigned to DEBRIS.

The net result of this exercise  was that 106 stars observed by DEBRIS
satisfy the DUNES photospheric detection condition and are, therefore,
shared  targets.  Specifically,  this sample  comprises  83 FGK
stars  - 51  F-type, 24  G-type and  8 K-type  (the rest  are A  and M
stars). Since the assignment to one of the teams was made on the basis
of  both DUNES  and  DEBRIS  original samples,  the  number of  shared
targets located closer  than 20 pc, i.e., the  revised DUNES distance,
are  less:  56  FGK -  32  F-type,  16  G-type,  and 8  K-type  stars 
(see Table \ref{table:sample_spectraltypes}).
Considering  {\it {\it  Hipparcos}} completeness,  the total  sample -
DUNES  stars plus  the shared  stars observed  by DEBRIS  -  should be
fairly complete  (with the constraint that the  photosphere is detected
with a $S/N \geq 5$ at 100 $\mu$m) up to the distance of 20 pc for the
F and G stars, while it is most likely incomplete for distances larger
than around 15 pc for the  K-type stars, particularly for the latest K
spectral types. We point out  that because of the imposed condition of
a photospheric detection over the background with S/N $>$ 5 the number
of ``rejected sources'' sources according to the {\it {\it Hipparcos}}
catalogue are 10 F-type, 43 G-type, and 213 K-type stars.

Table \ref{table:stars_sample} provides  some basic information on the
133 stars in  the DUNES sample.  Columns 1 and  2 give {\it Hipparcos}
and HD numbers, respectively, while column 3 gives the stars' names as
provided  by SIMBAD.   {\it  Hipparcos} spectral  types  are given  in
column 4; in order to check the consistency of these spectral types we
have     explored     VIZIER     using     the     DUNES     discovery
tool\footnote{http://sdc.cab.inta-csic.es/dunes/}             (Appendix
\ref{DUNES_VO}).  Results of this exploration are summarized in column
5 which gives the spectral type  range of each star taken into account
SIMBAD,    \cite{gray2003,gray2006},    \cite{wright2003}   and    the
compilation made  by \cite{skiff}. Typical spectral type  range is 2-3
subtypes.  Columns  6 and 7 give equatorial  and galactic coordinates,
respectively. Finally, columns 8 and 9 give parallaxes with errors and
distances, respectively.  These two  latter columns are taken from the
recent                compilation               given               by
\cite{2007A&A...474..653V,2008yCat.1311....0V}.   Parallax  errors are
typically less  than 1 mas, although  there are few  stars with errors
larger than  2 mas; those  stars are either spectroscopic  binaries or
are listed in  the Catalogue of the Components  of Double and Multiple
Stars (CCDM) \citep{ccdm} as orbit/astrometric binaries.  There are 10
stars in Table  1 with distances between 20 and 25  pc.  Those are the
previously mentioned stars with  known exoplanets (HIP 3497, HIP 25110
and HIP 109378),  and with identified {\it Spitzer}  debris discs (HIP
14954, HIP  51502, HIP 72603, HIP  73100, HIP 103389  and HIP 114948).
In addition, the distance to HIP 36439 is 20.24 pc ($\pi$ = 49.41 mas)
after     the    revised     {\it    {\it     Hipparcos}}    catalogue
\citep{2008yCat.1311....0V} but  19.90 pc (  $\pi$ = 50.25  mas) after
the  original  one  \citep{1997ESASP1200}.   We  also  note  that  the
distance  to  HIP  73100 is  25.11  pc  (  $\pi$  = 39.83  mas)  after
\cite{2008yCat.1311....0V},  but 24.84  pc ($\pi$  = 40.25  mas) after
\cite{1997ESASP1200}.

Tables \ref{table:stellar_fluxes}  (a, b, c  and d) give  the optical,
near-IR,  AKARI, WISE,  IRAS  and {\it  Spitzer}  MIPS magnitudes  and
fluxes of the  DUNES stars, while Table \ref{table:stellar_parameters}
gives        various        stellar       parameters.         Appendix
\ref{appendix:stellar_properties}  gives  some   details  on  how  the
stellar properties were  collected.  Figure \ref{HR_diagram} shows the
($B-V$,  $M_{\rm  v}$)   and  ($V-K$,  $M_{\rm  v}$)  colour-magnitude
diagrams of the  sources where one can see how  they spread across the
stellar  main-sequence.  The  K-type  star located  within the  G-type
locus is  HIP 2941.  This is  likely a misclassification  of {\it {\it
    Hipparcos}}; in fact, the  range of spectral types in \cite{skiff}
indicates an  earlier type, G5V -  G9V. This is also  supported by the
high  effective  temperature,  T$_{\rm  eff}$  $\sim$  5500  K  (Table
\ref{table:stellar_parameters}),  too high  for a  K1 star.   The main
stellar parameters (T$_{\rm  eff}$, $\log$ g and [Fe/H])  were used to
compute  a set of  synthetic spectra  from the  PHOENIX code  for GAIA
\citep{brott2005}, which were later normalized to the observed SEDs of
the stars  in order  to estimate the  photospheric fluxes at  the {\it
  Herschel}  bands.  The  whole procedure  is described  in  detail in
Appendix \ref{appendix:predictions}.

\section{Herschel observations and data reduction} 
\label{section:herschel_observations}

\subsection{PACS observations}
\label{section:PACS}

PACS scan  map observations of  all 133 DUNES targets  (comprising 130
individual  fields, due  to  close binaries  allowing  doubling up  of
sources in  the cases  of HIP 71382/4,  HIP 71681/3 and  HIP 104214/7)
were taken  with the  100/160 channel combination.   Additional 70/160
observations  were carried  out  for 47  stars,  some of  them with  a
\textit{Spitzer}  MIPS  70~$\mu$m  excess. Following  the  recommended
parameters  laid  out  in  the  scan  map  release  note\footnote{see:
  PICC-ME-TN-036 for details.}  each scan  map consisted of 10 legs of
3$\arcmin$ length, with a 4$\arcsec$ separation between legs, scanning
at the  medium slew speed  (20$\arcsec$ per second).  Each  target was
observed   at   two  array   orientation   angles  (70$^{\circ}$   and
110$^{\circ}$)  to improve  noise  suppression and  to  assist in  the
removal  of low  frequency (1/$f$)  noise, instrumental  artifacts and
glitches from  the images. A summary  of the PACS  observations can be
found  in  Table ~\ref{table:dunes_pacs}  where  the  PACS bands,  the
observation  identification number  of  each scan,  and the  on-source
integration time are given.

\begin{table}
\centering
\caption{Summary of all DUNES PACS observations, including the 100/160
  and  70/160 channel combinations.  The Obs  Ids of  both cross-scans
  (columns 3rd and 4th) and on-source integration time are given. Only
  the first 5 lines of the  table are presented here; the full version
  is available as online material.}
\label{table:dunes_pacs}
\begin{small}
\begin{tabular}{lcccc}
\hline\noalign{\smallskip}
HIP & PACS & Scan & X-Scan & On-source time [s] \\
\hline
171 & 100/160 & 1342212800 & 1342212801 &      900 \\
544 & 100/160 & 1342213512 & 1342213513 &     1440 \\
910 & 100/160 & 1342199875 & 1342199876 &      360 \\
2941 & 100/160 & 1342212844 & 1342212845 &     540 \\
3093 &  70/160 & 1342213242 & 1342213243 &     180 \\
\hline
\end{tabular}
\end{small}
\end{table}

\subsection{SPIRE observations}
\label{section:SPIRE}

SPIRE small map  observations were taken of 20  DUNES targets selected
because they were known as excess stars or as follow-up to the results
of  the PACS  observations.  Each SPIRE  observation  was composed  of
either two or five repeats  (equivalent on-source time of either 74 or
185~s)     of     the     small    scan     map     mode\footnote{see:
  http://herschel.esac.esa.int/Docs/SPIRE/html/spire\_om.pdf        for
  details.},  producing   a  fully  sampled  map   covering  a  region
4$\arcmin$ around  the target.  A  summary of the  SPIRE observations,
observation   identification  and   on-source  integration   time,  is
presented in Table ~\ref{table:dunes_spire}.


\begin{table}
\centering
\caption{Summary of DUNES SPIRE observations. Obs Ids and observing time are 
given.}
\label{table:dunes_spire}
\begin{tabular}{llr}
\hline\noalign{\smallskip}
HIP & Obs Id & Time [s] \\
\hline
544 & 1342213493 & 74 \\
7978 & 1342195666 & 185 \\
13402 & 1342213481 & 74 \\
15371 & 1342198448 & 185 \\
17439 & 1342214553 & 74 \\
22263 & 1342203629 & 185 \\
32480 & 1342204066 & 185 \\
40843 & 1342219959 & 74 \\
51502 & 1342214703 & 74 \\
72603 & 1342213475 & 74 \\
83389 & 1342198192 & 185 \\
84862 & 1342203593 & 185 \\
85235 & 1342213451 & 74 \\
85295 & 1342203588 & 185 \\
92043 & 1342204948 & 185 \\
101997 & 1342206205 & 185 \\
105312 & 1342209303 & 185 \\
106696 & 1342206206 & 185 \\
107649 & 1342209300 & 185 \\
108870 & 1342206207 & 185 \\
\hline
\end{tabular}
\end{table}

\subsection{Data reduction}
\label{section:reduction}

The  PACS  and SPIRE  observations  were  reduced  using the  Herschel
Interactive Processing Environment, HIPE \citep{ott2010}, user release
version 7.2, PACS calibration version 32 and SPIRE calibration version
8.1. The individual PACS scans  were processed with a high pass filter
to remove  background structure,  using high pass  filter radii  of 15
frames  at  70~$\mu$m,  20  frames  at 100~$\mu$m  and  25  frames  at
160~$\mu$m, suppressing structure larger than 62$\arcsec$, 82$\arcsec$
and 102$\arcsec$ in the  final images, respectively. For the filtering
process,  regions of  the map  where the  pixel brightness  exceeded a
threshold defined as twice the standard deviation of the non-zero flux
elements in the map were masked from inclusion in the high pass filter
calculation.   Deglitching  was carried  out  using  the second  level
spatial deglitching  task, following issues  with the clipping  of the
cores of  bright sources  using the MMT  deglitching method.   The two
individual PACS scans were mosaicked  to reduce sky noise and suppress
1/$f$  stripping effects from  the scanning.   Final image  scales were
1$\arcsec$ per pixel at 70  and 100~$\mu$m and 2$\arcsec$ per pixel at
160~$\mu$m compared to native instrument pixel sizes of 3$\farcs$2 and
6$\farcs$4. For  the SPIRE observations,  the small maps  were created
using the  standard pipeline routine  in HIPE, using the  naive mapper
option.  Image  scales of 6$\arcsec$, 10$\arcsec$  and 14$\arcsec$ per
pixel   were  used  at   250  $\mu$m,   350  $\mu$m   and  500~$\mu$m,
respectively.

\section{Noise analysis of the DUNES PACS images}
\label{section:noise_method}

The DUNES  sample is mostly composed  of faint targets  in the far-IR.
Their  fluxes  are  negligible   compared  to  the  telescope  thermal
emission,  which is  the  main contributor  in  the form  of a  large
background.   Confusion noise  is also  a concern  for some  very deep
observations, particularly  for the 160 $\mu$m band.   The optimum S/N
ratio is affected by the choice of the aperture to estimate the source
flux  and  the background.   Poisson  statistics  describe the  energy
collected from both noise sources: thermal emission and confusion.

The map noise  properties can be studied using  two different metrics:
i)  $\sigma_{\rm  pix}$  is  the  dispersion of  the  background  flux
measured  on   regions  sufficiently  large  to   avoid  small  number
statistics, and sufficiently small to avoid the effects of large scale
sky  inhomogeneities,  e.g.    cirrus.   $\sigma_{\rm  pix}$  is  best
estimated  taking  the median  value  of  several  such areas  in  the
image. ii)  $\sigma_{\rm sky}$ is  the standard deviation of  the flux
collected by  several apertures placed  in clear areas in  the central
portion of the image.

In  an ideal  scenario with  purely  random high  Poisson noise,  both
parameters would be related by:

\begin{equation}
  \sigma_{\rm  sky} = \sigma_{\rm pix} \alpha_{\rm  corr}
                     \sqrt{N_{\rm pix}^{\rm  circ}}
\label{eq:sigmas}
\end{equation}

\noindent where $N_{\rm pix}^{\rm circ}$ is the total number of pixels
in  a   circular  aperture  and  $\alpha_{\rm  corr}$   is  the  noise
correlation  factor.   However,  the  real  far-IR  sky  is  far  from
homogeneous, specially  for wavelengths  around 160 $\mu$m.   In addition,
the reduction  procedure is not  perfect and some  residual artificial
structure  appears superimposed.   This  ``corrugated'' noise  usually
makes  $\sigma_{\rm  sky}$ be  larger  than  the  expected value  from
Eq.~\ref{eq:sigmas}.

Noise correlation is a feature  of PACS scan maps that appears because
the  signal in  a given  output pixel  partially  depends  on the
values  recorded in  the  neighborhood.  Correlations  appear due  to
three main  reasons.  First, the  scan procedure entangles  the output
pixel counts via  the signal recorded by the  discrete bolometers at a
given time. Second, the output  maps have pixels much smaller than the
real  pixel size  of the  bolometers, which  is done  with the  aim of
providing better spatial resolution. Third, the $1/f$ noise introduced
by small instabilities in the array temperature and electronics.  

\subsection{Signal to noise ratio and optimal aperture}

Aperture photometry  estimates the flux  of a source integrating  in a
circle centered  on it  and containing a  significant fraction  of the
flux. The flux is given by:

\begin{equation}
  {\rm Signal} = F_\star {\rm EEF}(r)
\end{equation}

\noindent  where $F_\star$  is the  flux of  the point  source  in the
circle  with radius  $r$, and  ${\rm EEF}(r)$  is the  enclosed energy
fraction in  the circular aperture.  The radius is chosen  to maximize
the signal to  noise ratio.  The noise has  two main contributions. The
uncertainty in  the flux inside  the aperture, Noise$_\star$,  and the
uncertainty in the background,  Noise$_{\rm back}$. There are two ways
to estimate  the noise,  based on the  metrics $\sigma_{\rm  pix}$ and
$\sigma_{\rm sky}$.

In terms of $\sigma_{\rm  pix}$, the  aperture noise is given by:

\begin{equation}
  {\rm Noise_\star} = \sigma_{\rm pix} \alpha_{\rm corr} \sqrt{N_{\rm pix}^{\rm circ}}
                   = \sigma_{\rm pix} \alpha_{\rm corr} \sqrt{\pi} r_{\rm pix}
\end{equation}

The background flux is typically  determined using an annulus of inner
$r_i$ and  $r_o$ outer  radii (pixel units). The  flux coming  from the
point source at the location of the annulus due to the large extension
of the  PSF is assumed negligible  compared to the  noise, because the
DUNES sources are typically  faint.  The background noise contribution
can be estimated as:

\begin{small}
\begin{equation}
  {\rm Noise_{\rm back}}
     = \sigma_{\rm pix} \alpha_{\rm corr} N_{\rm pix}^{\rm circ} / \sqrt{N_{\rm pix}^{\rm annulus}}
     = \sigma_{\rm pix} \alpha_{\rm corr} \sqrt{\pi} r_{\rm pix}^2 / \sqrt{ r_o^2 - r_i^2 }
\end{equation}
\end{small}

The total  noise is the  quadratic sum of  both the aperture and background 
contributions:

\begin{equation}
  \rm Noise = \sqrt{ \rm Noise_\star^2 + Noise_{back}^2 }
\end{equation}

Alternatively, in  terms of $\sigma_{\rm sky}$ the  sky background and
the associated  uncertainty can be estimated measuring  the total flux
in $n_{\rm  sky}$ apertures  with the same  size used for  the source.
The apertures are  located in clean fields, in  order to avoid biasing
the statistics,  and as close as  possible to the source,  in order to
get uniform exposure times.  In this case, the noise is given by:

\begin{equation}
  {\rm Noise} = \sigma_{\rm sky} \sqrt{1 + \frac{1}{n_{\rm sky}} }
\end{equation}

The $1/n_{\rm sky}$  factor comes from the finite  number of apertures
used and quickly  goes to zero.  This approach  has the advantage that
no  correlated  noise  factor   is  required  for  sufficiently  large
apertures.   However,  it  provides  a conservative  estimate  if  the
background  is  variable, due  to  sky  inhomogeneities  or $1/f$  noise
filtering residuals, as it is the case for the DUNES observations.

\begin{table}
\caption{Gaussian noise propagation in the absence of noise
correlation. The RMS dispersion of the sky flux $\sigma_{\rm sky}$
in different windows is consistent with propagating the single pixel
uncertainty $\sigma_{\rm pix}$ according to the window size in
pixels $N_{\rm pix}$.} 
\label{tab:noCorrelatedNoise}
\begin{center}
\begin{tabular}{cccc}
\hline\noalign{\smallskip}
Box size (pix) &
$\sigma_{\rm pix}$ &
$\sigma_{\rm sky}$ &
$\sigma_{\rm pix} \sqrt{ N_{\rm  pix} }$ \\
\hline
 7 & 100 &  610 &  698 \\
15 & 101 & 1310 & 1520 \\
22 & 100 & 2290 & 2200 \\
\hline
\end{tabular}
\end{center}
\end{table}

In  order  to  validate  the  consistency  of  both  noise  estimation
procedures we have carried out several tests using both survey reduced
images  and  synthetic  noise  frames.  The  theoretical  relationship
between    $\sigma_{\rm    sky}$    and    $\sigma_{\rm    pix}$    in
Eq. ~\ref{eq:sigmas} has been tested for small to moderately large
box sizes, which is a way to verify the error propagation scheme under
large Poisson  noise conditions.  For  the synthetic noise  frames, we
have built  an image of 200x200  pixels with an  arbitrarily large sky
level of  10000 photons and gaussian  noise of 100  photons, since the
Poisson distribution can  be well approximated by a  gaussian for high
fluxes.  This  image simulates the  noise introduced by  the telescope
emission, which is  the dominant factor for DUNES  - faint sources and
broad band  photometry. Multiple regions  (25+) have been  selected in
the  image with square  box sizes  of 7,  15 and  22 pixels  per side.
$\sigma_{\rm  pix}$ and  $\sigma_{\rm  sky}$ have  been estimated  for
these boxes, and the latter  values have been compared to $\sigma_{\rm
  pix} \sqrt{ N_{\rm pix} }$ (Table~\ref{tab:noCorrelatedNoise}).  The
differences  are  below  15\%,  consistent  with  Poisson  propagation
noise. It has thus been verified that noise propagation works well for
images not affected by correlated  noise. In addition, small boxes can
be used to provide reliable estimates.

Further, a  comparison of  both methods by  the HSC team  (B. Altieri,
private communication) showed that the multiple apertures $\sigma_{\rm
  sky}$ method provides in general larger uncertainties than the error
propagation  of  the  $\sigma_{\rm  pix}$  metrics.   The  values  are
typically consistent  and smaller than  a factor 2.  The  selection of
one  of them  is  subjective.  Given  that  the aim  of  DUNES is  the
detection of  very faint excesses,  we have followed  the conservative
approach of taking  the largest noise value for  each individual DUNES
source to assess the presence of an infrared excess.

Finally, when  the sky value  has been determined with  high precision
(using many apertures to improve  the statistics), the signal to noise
ratio can be estimated as:


\begin{equation}
{\rm SNR}(r_{\rm pix})
     = \frac { F_\star {\rm EEF}(r_{\rm pix}) }{{\rm Noise}}
\end{equation}

This  equation  shows  that  there  is an  optimum  extraction  radius
providing the highest SNR possible.  If it is too small, little signal
will be collected,  while if it is too large,  the noise introduced by
the aperture  is considerable.  Optimum  values estimated by  the {\it
  Herschel}    team\footnote{Technical    Note    PICC-ME-TN-037    in
  http://herschel.esac.esa.int}   are   4$\arcsec$,   5$\arcsec$   and
7$\arcsec$-8$\arcsec$ for 70, 100 and 160$\mu$m, respectively. We have
carried out the same exercise using a number of DUNES clean fields and
the  $\sigma_{\rm pix}$ metrics  ($\sigma_{\rm sky}$  is comparatively
more affected  by sky inhomogeneities) and found  essentially the same
results.

\subsection{Correlated noise}
\label{section:correlated_noise}
\begin{table*}[ht!]
\caption{ Image noise  properties  for small  and natural  output
    pixel sizes.  $\sigma_{\rm sky}$ and $\sigma_{\rm  pix}$ have been
    estimated for several fields using  clean square areas of 22'' and
    44'' sizes at 100 and  160 $\mu$m, respectively.  The number after
    the stellar  {\it Hipparcos}  catalogue number is  the wavelength:
    70,  100 and  160 $\mu$m.  Two  different 160  $\mu$m images  were
    available  for  HIP  103389.   The  correlated  noise  is  clearly
    revealed   by  $\sigma_{\rm   sky}$  being   always   larger  than
    $\sigma_{\rm pix} \sqrt{N_{\rm pix}}$ for the small output pixels.
    The correlated factors included in PIC-ME-TN-037 have been used to
    compute the  error ratios  $\Delta F_\star^{\rm natural}  / \Delta
    F_\star^{\rm small} =  \alpha_{\rm corr}^{\rm natural} \sigma_{\rm
      pix}^{\rm   natural}    \sqrt{N_{\rm   pix}^{\rm   natural}}   /
    \alpha_{\rm   corr}^{\rm  small}   \sigma_{\rm   pix}^{\rm  small}
    \sqrt{N_{\rm pix}^{\rm  small}}$.  The ratios are  always close to
    1.0, which means that the factors in PIC-ME-TN-037 account for the
    difference in  correlated noise  between the natural  and standard
    output pixel  sizes. See text  and Table~\ref{tab:cleanFields} for
    more  details on the  correlated noise  factor for  natural output
    pixel sizes.}
\label{tab:correlatedNoise}
\begin{tiny}
\begin{center}
\footnotesize
\begin{tabular}{lccccc}
\hline\noalign{\smallskip}
Unit: Jy & Small pix & & Natural pix & & Error Ratio\\
Image
 & $\sigma_{\rm sky}$ & $\sigma_{\rm pix} \sqrt{ N_{\rm pix} }$
 & $\sigma_{\rm sky}$ & $\sigma_{\rm pix} \sqrt{ N_{\rm pix} }$
 & $\Delta F_\star^{\rm natural} / \Delta F_\star^{\rm small}$ \\
\hline
HIP 103389 70    & 3.32e-03 & 1.02e-03 & 3.84e-03 & 2.31e-03 & 1.01 \\
HIP 103389 100   & 2.02e-03 & 3.62e-04 & 1.50e-03 & 8.64e-04 & 1.06 \\
HIP 107350 100   & 1.34e-03 & 3.42e-04 & 1.46e-03 & 8.73e-04 & 1.14 \\
HIP 114948 100   & 1.23e-03 & 3.48e-04 & 1.01e-03 & 8.18e-04 & 1.05 \\
HIP 103389 160 a & 1.08e-02 & 2.48e-03 & 6.18e-03 & 6.42e-03 & 1.23 \\
HIP 103389 160 b & 3.15e-03 & 1.17e-03 & 3.32e-03 & 2.94e-03 & 1.19 \\
HIP 107350 160   & 5.76e-03 & 1.07e-03 & 4.40e-03 & 2.68e-03 & 1.19 \\
HIP 114948 160   & 4.11e-03 & 1.18e-03 & 3.56e-03 & 3.11e-03 & 1.25 \\
\hline
\end{tabular}
\end{center}
\end{tiny}
\end{table*}

\begin{table*}[ht!]
\caption{Clean  field  correlated  noise estimation.   The  correlated
  noise factors $\alpha_{\rm corr}$ are estimated as the ratio between
  the $\sigma_{\rm  pix}$ value obtained for the  largest output image
  pixel  size  (twice  the  natural   pixel  size)  and  the  size  of
  interest.  No correlation noise  is assumed  for the  largest output
  image pixel  size. The  box size is  approximately constant  for all
  output  pixel sizes:  $\sim$  25'',  50'' for  100  and 160  $\mu$m,
  respectively.
\label{tab:cleanFields}}
\begin{center}
\footnotesize
\begin{tabular}{lllllllll}
\hline\noalign{\smallskip}
$r$  & HIP 544 & 100 $\mu$m & HIP 544 & 160 $\mu$m & HIP 99240 & 100 $\mu$m & HIP 99240 & 160 $\mu$m \\
     & $\sigma_{\rm pix}$ (Jy) & $\alpha_{\rm corr}$ & $\sigma_{\rm pix}$ (Jy) & $\alpha_{\rm corr}$
     & $\sigma_{\rm pix}$ (Jy) & $\alpha_{\rm corr}$ & $\sigma_{\rm pix}$ (Jy) & $\alpha_{\rm corr}$ \\
\hline
3.20 & 1.91e-05 & 3.88 & 5.91e-05 & 3.61 & 4.69e-05 & 3.36 & 1.14e-04 & 3.73 \\
1.47 & 7.79e-05 & 2.07 & 2.50e-04 & 1.85 & 1.89e-04 & 1.81 & 5.05e-04 & 1.83 \\
1.00 & 1.56e-04 & 1.53 & 5.13e-04 & 1.33 & 3.60e-04 & 1.40 & 1.05e-03 & 1.30 \\
0.67 & 3.09e-04 & 1.15 & 8.84e-04 & 1.16 & 6.87e-04 & 1.10 & 1.97e-03 & 1.04 \\
0.50 & 4.75e-04 & 1.00 & 1.36e-03 & 1.00 & 1.01e-03 & 1.00 & 2.73e-03 & 1.00 \\
\hline
\end{tabular}
\end{center}
\end{table*}

As pointed  out before, the  PACS scan map  observations intrinsically
suffer  from correlated  noise. Theoretical  correlated  noise factors
$\alpha_{\rm  corr}$  were  derived  by  \cite{fruchter2002}  for  the
Drizzle  algorithm, which  combines multiple  undersampled  images (in
terms  of the  Nyquist criterion).   They showed  that  the correlated
noise depends on the ratio  $r$ between the linear pixel fraction (the
ratio between the drop and the natural pixel box sizes) and the linear
output  pixel scale  factor  (the  ratio between  the  output and  the
natural  pixel box  sizes). This  procedure,  used by  default in  the
\emph{Herschel} PACS  reduction pipeline, produces  output images with
typical  smaller output  pixel sizes,  better spatial  resolution than
individual frames, but significant correlated noise.

The PACS calibration  team has made extensive tests  on the correlated
noise  measuring the  noise  properties of  fields surrounding  bright
stars  (see  the mentioned  technical  note  PICC-ME-TN-037) and  have
estimated $\alpha_{\rm corr}$ as a  function of the output pixel size.
The value  for output pixel sizes  of 1$\arcsec$ (the size  of our 100
$\mu$m  reduced images)  is  $\alpha_{\rm corr}$  =  2.322, while  for
2$\arcsec$ (160 $\mu$m images)  $\alpha_{\rm corr}$ = 2.656.  However,
these  estimates are  too optimistic  because no  correlated  noise is
assumed for output pixels with a size equal to the natural ones.

We have  analysed the  effect of the  correlated noise on  images with
natural pixel  sizes as it has  a clear effect  on the $\alpha_{\rm
  corr}$ factor we have to  apply for our reduced images. The approach
we have made is the following.

As a  first step,  we have   tried to validate  the PICC-ME-TN-037
  predictions evaluating  the noise properties of the  PACS images of
the  DUNES  stars  HIP  103389,  HIP107350 and  HIP  114948.   Reduced
observations with both small (1''/pix,  70 and 100 $\mu$m and 2''/pix,
160 $\mu$m) and  natural (3.2''/pix, 70 and 100  $\mu$m and 6.4''/pix,
160 $\mu$m) pixel sizes have been considered. Square box sizes of
  22$\arcsec$ and 44$\arcsec$  have been used for 100  and 160 $\mu$m,
  respectively. These values, larger  than the optimal aperture sizes,
  were used to  prevent small number statistics for  the natural pixel
  size   frames.    Table~\ref{tab:correlatedNoise}  summarises   the
results,  from  which  several  conclusions  can  be  drawn.   i)  The
correlated  noise   effect  can  clearly  be   noticed  comparing  the
$\sigma_{\rm pix} \sqrt{ N_{\rm pix} }$ values, which are much smaller
for the  small size  output pixels.  This  means that there  is indeed
significant correlated  noise in the finer sampled  output frames. ii)
Similar  statistical   flux  uncertainties    $\Delta  F_\star  =
  \alpha_{\rm corr} \sigma_{\rm pix} \sqrt{N_{\rm pix}}$ are obtained
for aperture  photometry if the correlation  factors in PICC-ME-TN-037
are  used.  The  agreement is  better  for the  blue detectors.   This
demonstrates  that  the  PICC-ME-TN-037 $\alpha_{\rm  corr}$  formulae
provide  good estimates  of  the differential  increase in  correlated
noise between  natural size and  smaller output pixels.   However, the
amount of correlated noise for  natural size output pixels is unknown.
iii) The sky  value, when averaged over a large  area, is not affected
by correlated noise. It can,  nevertheless, be affected by large scale
sky  inhomogeneities   due  to  residual  $1/f$   noise  or  confusion
(partially resolved background sources).
    
As a second step, the  full correlated noise factors for small and
  natural  pixel sizes  have  been estimated.  Additional tests  were
carried out reducing  the HIP 544 and HIP  99240 images with different
output pixel sizes.  These objects are in fields particularly clean of
additional sources,  which is  critical to really  estimate correlated
noise factors  and not confusion  noise. The output pixel  sizes range
between the standard 1'' and 2'' for 100 and 160 $\mu$m, and twice the
natural pixel size, respectively. The pixel fraction was always set to
the default value of 1.0.  For each image and pixel size, $\sigma_{\rm
  pix}$   was    estimated   on    sky   constant   size    boxes   of
$\approx$25$\arcsec$ and 50$\arcsec$ widths for the 100 and 160 $\mu$m
channels,    respectively.     The    results   are    presented    in
Table~\ref{tab:cleanFields}.  It shows  the median  value $\sigma_{\rm
  pix}$ of each frame estimated  as the median of several measurements
($\sim$6-8) in  boxes placed next  to the central object,  to minimise
sky coverage border effects.
Correlated noise factors in the
table have been  computed assuming no correlated noise  for the images
with  output  pixels  twice   the  natural  size  ($r  =0.50$).   This
assumption  is not  strictly  correct.  However,  larger output  pixel
sizes could not  be studied because the box  sizes required would have
been too  large compared to the  high density coverage  portion in the
DUNES small scan maps. In addition, very large output pixel sizes make
rejection of background sources increasingly difficult. We believe the
small amount  of correlated  noise not considered  for the  very large
pixels compensates  with the  additional background noise  included in
the box averages.

The  correlated  noise  factors  in  Table  \ref{tab:cleanFields}  are
roughly  consistent with  the predictions  by  \cite{fruchter2002}. In
particular, the values  obtained for the fine pixel  maps (1''/pix and
2''/pix for 100 and  160 $\mu$m) bracket the theoretical expectations.
Taking into  account all the  tests carried out, the  correlated noise
factor that has  been used for the analysis of  the whole DUNES sample
and  all wavelengths is: $\alpha_{\rm corr,DUNES}$  = 3.7.  
  It is  the same  for all 70,  100 and  160 $\mu$m because  the ratio
  between natural to standard output pixel sizes is always 3.2.

\section{Results}
\label{section:results}
\subsection{PACS}

\subsubsection{PACS Photometry}
\label{section:photometry}

\begin{figure}[!h]
\centering 
\includegraphics[scale=0.33,angle=270]{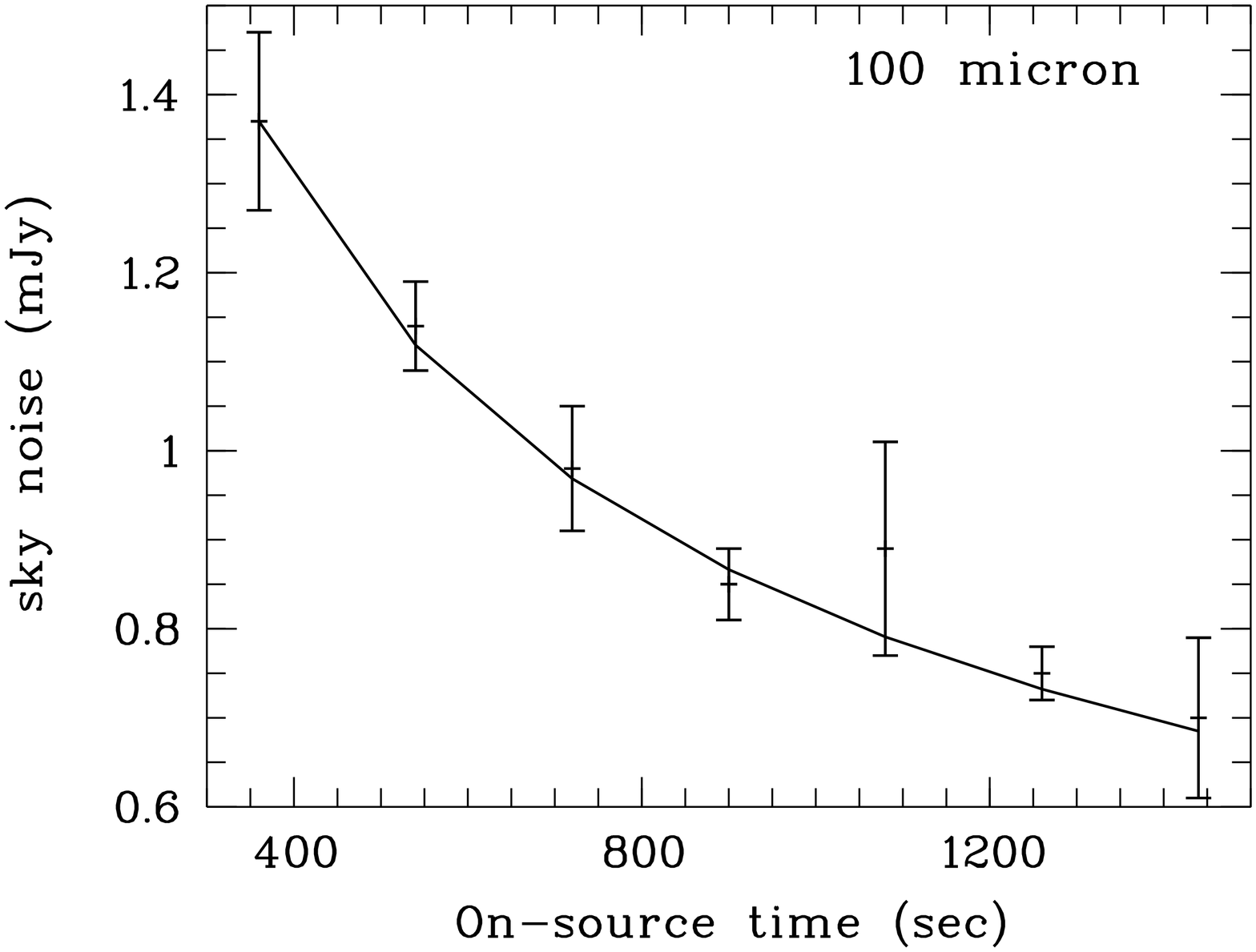}
\includegraphics[scale=0.33,angle=270]{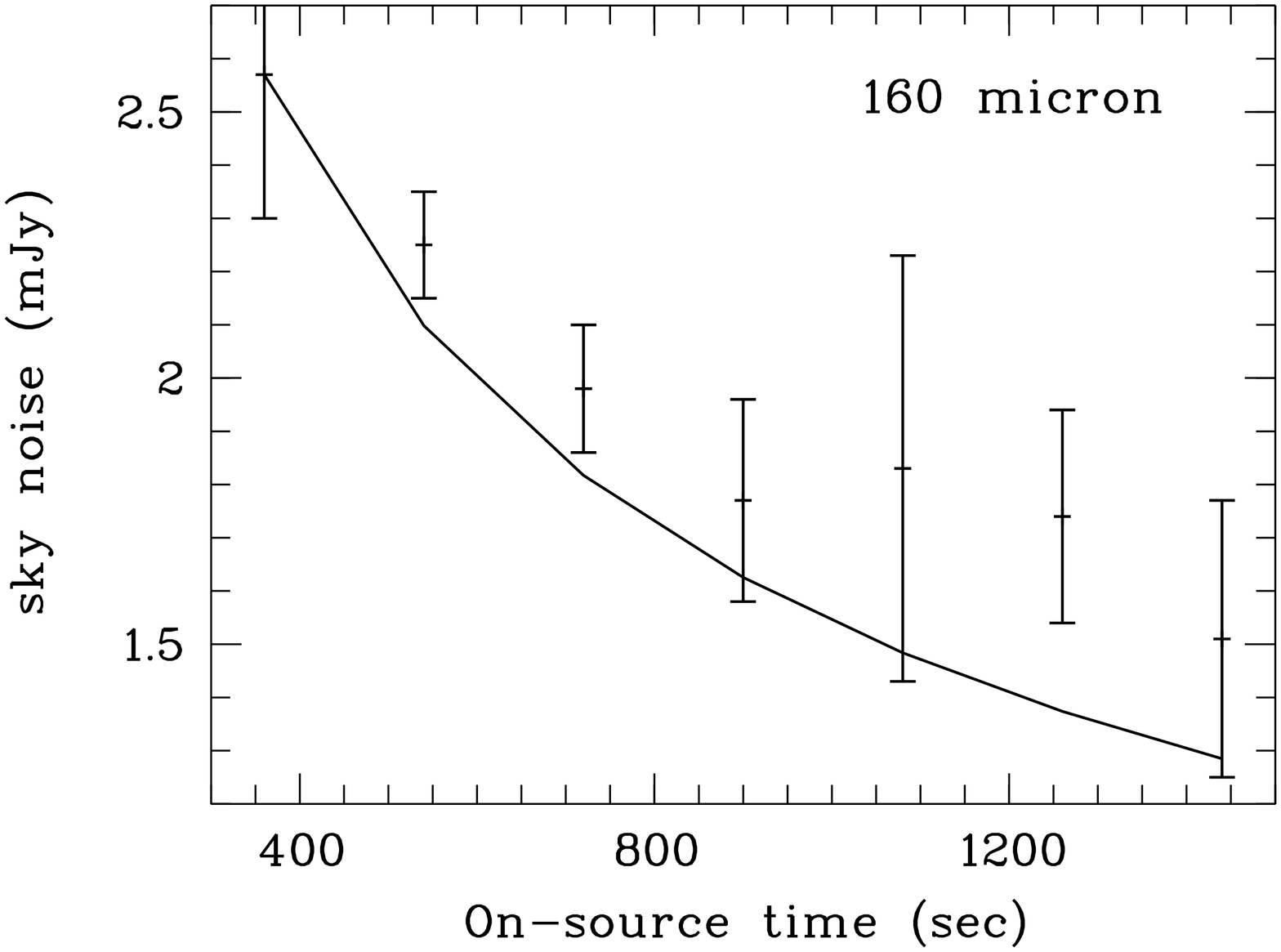}
\caption{Top:  mean value  of the  sky noise  estimates at  100 $\mu$m
  versus on-source integration time.   Error bars are the rms standard
  deviation  of  the sky  noise  in the  images  taken  with the  same
  on-source  observing  time.  The  solid  curve  represents  the  noise
  behaviour assuming that the $S/N$ ratio varies as  the
  square root of the time, normalized  by the mean value of the images
  with an  on-source exposure time of  360 sec. Bottom:  the same for
  the 160 $\mu$m images.}
\label{fig:noise}
\end{figure}

PACS photometry  of the sources identified as  the far-IR counterparts
of the optical stars was  carried out using two different methods. The
first method consisted in estimating PACS fluxes
primarily using  circular aperture  photometry with the  optimal radii
4$\arcsec$, 5$\arcsec$,  and 8$\arcsec$ at  70 $\mu$m, 100  $\mu$m and
160 $\mu$m,  respectively. For extended  sources, the beam  radius was
chosen  large  enough  to  cover  the whole  extended  emission.   The
corresponding beam aperture correction  as given in the technical note
PICC-ME-TN-037 was taken into account. The reference background region
was usually  taken in a ring  of width 10$\arcsec$ at  a separation of
10$\arcsec$  from the  circular  aperture size.   Nonetheless we  took
special  care to  choose the  reference sky  region for  those objects
where the ``default''  sky was or could be  contaminated by background
objects.  In  addition, we also  carried out complete curve  of growth
measurements   with  increasing   apertures   and  the   corresponding
skies. Sky noise for each PACS  band was calculated from the rms pixel
variance of ten sky apertures of  the same size as the source aperture
and  randomly distributed  across the  uniformly covered  part  of the
image  (pixel sky  noise from  the  curves of  growth are  essentially
identical).  Final  error estimates  take into account  the correlated
noise  factor estimated  by  us (see  previous  section) and  aperture
correction factors.  Figure \ref{fig:noise}  (top) shows a plot of the
mean sky  noise value at 100  $\mu$m obtained for all  images with the
same on-source integration time versus the on-source integration time.
Error  bars are the  rms standard  deviation of  the sky  noise values
measured in  the images  taken with the  same on-source time;  we note
that the number  of images is not the same  for each integration time,
so that those  error bars are only indicative  of the noise behaviour.
The plot also shows a curve of the noise assuming that the $S/N$ ratio
varies with the  square root of the time.  The  curve is normalized to
the mean sky  noise value of the images  with the shortest integration
time, 360 sec, showing that the PACS 100 $\mu$m images are essentially
background limited.  Figure \ref{fig:noise}  (bottom) is the same plot
at 160 $\mu$m; the curve  is also normalized to the shortest on-source
integration time. The  160 $\mu$m noise behaviour is  flatter than the
S/N  $\propto t^{1/2}  $ curve,  suggesting that  it is  influenced by
structured background diffuse emission,  and that is confusion limited
for integration times longer than around 900 seconds.  With the second
method we  carried out photometry  using rectangular boxes  with areas
equivalent to the  default circular apertures; in this  case, we chose
box  sizes large  enough  to  cover the  whole  emission for  extended
sources.  Sky level and sky  rms noise from this method were estimated
from measurements in ten fields,  selected as clean as possible by the
eye, of  the same size  as the photometric source  boxes.  Photometric
values  and errors  take into  account beam  correction  factors.  The
estimated fluxes from both  methods, circular and rectangular aperture
photometry, agree within the  errors.  PSF photometry of point sources
using  the DAOPHOT  software package  was also  carried out  for those
cases where a nearby object is  present and prevents us from using any
of the two  methods above.  The fluxes  using aperture photometry
  and  DAOPHOT  are  consistent  within the  uncertainties  for  point
  sources  in  non-crowded fields.  However,  the  errors provided  by
  DAOPHOT  are too  optimistic  by a  typical  factor of  an order  of
  magnitude. This  is a consequence of correlated  noise, which cannot
  easily be handled by  DAOPHOT. Using $alpha_{\rm corr}$ $\sigma_{\rm
    pix}$ as  the flux uncertainty for  each pixel does  not solve the
  problem. The  errors for DAOPHOT  photometry have thus  been estimated
  using   the   formulae   derived   for   standard   DUNES   aperture
  photometry. The  noise introduced  by source crowding  is considered
  negligible  as compared  to  the other  major contributors:  thermal
  noise  background,  stellar  flux  determination and  PACS  absolute
  photometric calibration uncertainties.  The absolute uncertainties 
in this  version of HIPE are  2.64\% (70 $\mu$m),  2.75\% (100 $\mu$m)
and 4.15\% (160 $\mu$m), as indicated in the cited technical note.

\subsubsection{Pointing: excess/non-excess sources}
 \label{section:pointing}

\begin{figure}
\centering
\includegraphics[angle=270,scale=0.33]{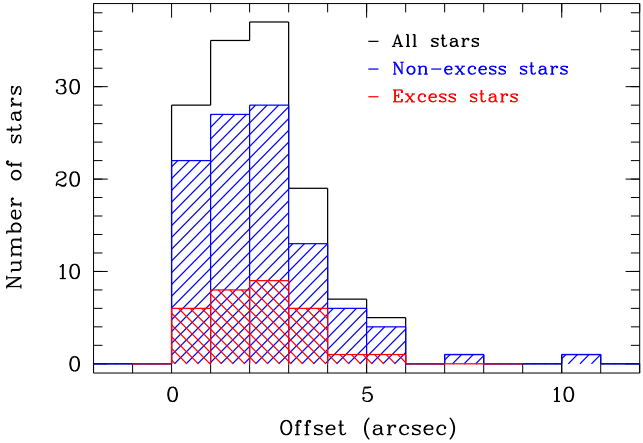}
\caption{Histograms of the offset position between the optical and the PACS 
100 $\mu$m coordinates. Histograms are shown for the whole DUNES sample of 
stars, the non-excess stars and excess star candidates. The spurious sources 
(see section \ref{section:non-excess}) are included as non-excess 
stars in this figure.}
\label{fig:position_offset}
\end{figure}

 PACS at 100 and 160  $\mu$m are very sensitive to background objects,
 usually red galaxies and, therefore, there is a non-negligible chance
 of  contamination  (Section   \ref{section:alignment})  Thus,  it  is
 necessary to check the agreement  between the optical position of the
 stars  and the  one of  the  objects identified  as their  $Herschel$
 counterparts  - as well  as in  the cases  of non-excess  sources the
 agreement  between  the  measured   PACS  fluxes  and  the  predicted
 photospheric   levels   (Section   \ref{section:non-excess}).   Table
 \ref{table:positions}   gives    the   J2000.0   optical   equatorial
 coordinates and the PACS positions  at 100 $\mu$m, corrected from the
 proper motions  of the stars as  given by \cite{2008yCat.1311....0V}.
 Figure \ref{fig:position_offset}  shows histograms of  the positional
 offset between the optical and  PACS 100 $\mu$m positions for all the
 stars, as well  as separately for the non-excess  (including here the
 spurious  sources,  see below)  and  excess  sources.   In all  three
 stellar  samples $\sim$ 65  \% of  the stars  have offsets  less than
 2$\farcs$4,    which   is    the    expected   $Herschel$    pointing
 accuracy\footnote{http://herschel.esac.esa.int/twiki/bin/view/Public/SummaryPointing},
 while  there are 5  non-excess stars  and only  one excess  star with
 positional offsets $> 2\sigma$. In this respect we note that based on
 a grid  of known 24  $\mu$m sources, \cite{berta2010}  found absolute
 astrometric offsets in the GOODS-N field as high as 5$\arcsec$.

 The non-excess sources with offsets $>$ 2$\sigma$ are: HIP 28442, HIP
 34065, HIP 54646, HIP 57939 and  HIP 71681 ($\alpha$ CenB - HIP 71683
 is $\alpha$ CenA and has  an offset of 4$\farcs$2).  These non-excess
 sources, excluding  $\alpha$ Cen, are  faint with no or  dubious (the
 case  of HIP  34065) 160  $\mu$m detection,  but their  estimated 100
 $\mu$m   fluxes  agree  well   with  the   photospheric  predictions,
 $|$F$_{PACS100}$ - F$_{star}| <$ 1.6  mJy. HIP 57939 has an extremely
 high proper  motion; the rest  are multiple stars.  HIP  28442, which
 shows a very  large offset, the largest one,  is an outlier. However,
 it has  a very large  parallax error  (21 mas) and  is a member  of a
 quadruple   star,  CCDM   J06003-3103ABC;  its   optical   and  2MASS
 coordinates differ  around 6$\arcsec$ -  in fact, the  offset between
 the   PACS   100   $\mu$m   and   2MASS  coordinates   is   only   of
 $\sim$4$\arcsec$.   Further, the  accuracy  of its  proper motion  is
 somehow dubious. After  the proper motion values as  given in the LHS
 catalogue \citep{luyten1979} the offset  between the optical and {\it
   Herschel}  positions would  just be  $\approx$ 4$\arcsec$,  but the
 revised version  of that catalogue  \citep{bakos2002} presents proper
 motions similar  to those of  {\it Hipparcos}. Thus, the  real offset
 remains unsolved.  In  the case of $\alpha$ Cen  the offset values in
 Table  \ref{table:positions} do  not  take into  account its  orbital
 motion.  Correcting  from that orbital  motion we find an  offset for
 $\alpha$ Cen A relative to  the pointed position of 1$\farcs$7 at 100
 $\mu$m, i.e., well below  the 1$\sigma$ pointing accuracy (Wiegert et
 al.  in preparation).  We do not have orbital  motion information for
 the rest  of the  multiple sources. Finally,  the offset  between the
 nominal  optical position and  the 100  $\mu$m peak  of the  star HIP
 40843  (Figure  \ref{fig:controversial_1})  is 7$\farcs$1,  but  this
 result  most likely reflects  a case  of coincidental  alignment (see
 section          \ref{section:alignment}         and         Appendix
 \ref{section:controversial}).
 
 The excess-source with  offset $>$ 2$\sigma$ is HIP  171.  Again this
 object  is   a  binary  with  a  separation   between  components  of
 0$\farcs$83, the component  B being a binary itself \citep{bach2009}.
 We do not have information on  the orbital motion so that the PACS 100
$\mu$m  position cannot  be corrected, but  its 100 $\mu$m  flux is very
 well in  agreement with the  photospheric prediction of  the multiple
 system,   $|$F$_{PACS100}$  -  F$_{star}|$   $<$  1.0   mJy  (Section
 \ref{section:excess}).


\begin{table}
\centering
\caption{Optical and PACS 100 $\mu$m equatorial positions (J2000) of the DUNES
  stars  together  with the  positional  offset  between both  nominal
  positions.  Only the first 5 objects of the sample are presented
  here;  the  full  version  of  the  table  is  available  as  online
  material.}
\label{table:positions}
\begin{tiny}
\begin{tabular}{lccc}
\hline\noalign{\smallskip}
HIP    &  ICRS(2000)              &PACS100                   & Offset (arcsec)\\
\hline\noalign{\smallskip}
171    & 00 02 10.16   +27 04 56.1 &00 02 10.57   +27 04 56.0 & 5.5   \\ 
544    & 00 06 36.78   +29 01 17.4 &00 06 36.79   +29 01 15.8 & 1.6   \\ 
910    & 00 11 15.86  --15 28 04.7 &00 11 15.88  --15 28 03.4 & 1.3   \\ 
2941   & 00 37 20.70  --24 46 02.2 &00 37 20.54  --24 46 03.9 & 2.8   \\ 
3093   & 00 39 21.81   +21 15 01.7 &00 39 21.84   +21 14 58.9 & 2.8   \\ 
\hline\noalign{\smallskip}
\end{tabular}
\end{tiny}
\end{table}

\subsection{SPIRE}
\label{section:dunes_spire_phot}
%
%
%
\begin{table*}[!ht]
\centering
\caption{SPIRE fluxes (F$\lambda$) with 1$\sigma$ errors, together
with the photospheric predictions (S$\lambda$). The significance at
each band is given. Figures without errors in the SPIRE columns  give
3$\sigma$ upper limits for the corresponding stars.}
\begin{tabular}{lccccccccc}
\hline\noalign{\smallskip}
HIP    & F250             & S250          & $\chi$250 & F350               & S350           & $\chi$350 & F500             & S500           & $\chi$500 \\
       & (mJy)            & (mJy)         &           & (mJy)              & (mJy)          &           & (mJy)            & (mJy)          &           \\   
\hline
544    & $<$22.5          & 1.20$\pm$0.03 &           & $<$24.3            & 0.61$\pm$0.02  &           & $<$9.2           & 0.30$\pm$0.01  &           \\
7978   &312.30~$\pm$~25.60& 1.37$\pm$0.08 & 12.15     & 179.90~$\pm$~14.60 & 0.70$\pm$0.04  & 12.27     & 78.40~$\pm$~9.80 & 0.34$\pm$0.02  & 7.97      \\
13402  & $<$23.1          & 1.63$\pm$0.03 &           & $<$23.1            & 0.83$\pm$0.02  &           & $<$9.4           & 0.41$\pm$0.01  &           \\
15371  & 59.72~$\pm$~6.70 & 2.03$\pm$0.03 &  8.61     & 24.68~$\pm$~6.89   & 1.04$\pm$0.02  &  3.43     & 20.29~$\pm$~7.66 & 0.51$\pm$0.01  & 2.58      \\
17439  & 53.00~$\pm$~10.40& 0.68$\pm$0.01 &  5.03     & 32.20~$\pm$~8.90   & 0.35$\pm$0.01  &  3.58     & $<$7.2           & 0.17$\pm$0.01  &           \\
22263  & 23.21~$\pm$~6.81 & 1.66$\pm$0.03 &  3.16     & 14.14~$\pm$~6.87   & 0.85$\pm$0.02  &  1.93     & $<$9.6           & 0.42$\pm$0.01  &           \\
32480  & 90.00~$\pm$~15.00& 1.75$\pm$0.02 &  5.88     & 25.00~$\pm$~8.00   & 0.89$\pm$0.01  &  3.01     & $<$24.0          & 0.44$\pm$0.01  &           \\
40843  & $<$24.0          & 1.77$\pm$0.01 &           & $<$23.7            & 0.90$\pm$0.01  &           & $<$26.4          & 0.44$\pm$0.01  &           \\
51502  & 49.65~$\pm$~8.29 & 1.30$\pm$0.02 &  5.83     & 41.01~$\pm$~8.15   & 0.66$\pm$0.01  &  4.95     & 23.10~$\pm$~9.41 & 0.33$\pm$0.01  & 2.42      \\
72603  & $<$24.0          & 1.43$\pm$0.01 &           & $<$24.3            & 0.73$\pm$0.01  &           & $<$30.9          & 0.36$\pm$0.01  &           \\
83389  & $<$20.1          & 0.64$\pm$0.01 &           & $<$21.0            & 0.33$\pm$0.01  &           & $<$24.0          & 0.16$\pm$0.01  &           \\
84862  & $<$20.1          & 1.95$\pm$0.02 &           & $<$21.0            & 0.99$\pm$0.01  &           & $<$24.3          & 0.49$\pm$0.01  &           \\
85235  & $<$22.2          & 1.00$\pm$0.02 &           & $<$23.1            & 0.51$\pm$0.01  &           & $<$29.4          & 0.25$\pm$0.01  &           \\
85295  & $<$19.8          & 1.66$\pm$0.04 &           & $<$21.0            & 0.85$\pm$0.02  &           & $<$24.0          & 0.42$\pm$0.01  &           \\
92043  & 12.12~$\pm$~6.57 & 3.85$\pm$0.05 &  1.26     & $<$21.9            & 1.96$\pm$0.03  &           & $<$24.3          & 0.96$\pm$0.01  &           \\
101997 & $<$19.5          & 0.91$\pm$0.02 &           & $<$21.0            & 0.46$\pm$0.01  &           & $<$24.6          & 0.23$\pm$0.01  &           \\
105312 & $<$19.8          & 0.93$\pm$0.05 &           & $<$20.7            & 0.47$\pm$0.03  &           & $<$24.0          & 0.23$\pm$0.01  &           \\
106696 & $<$19.5          & 0.65$\pm$0.01 &           & $<$20.7            & 0.33$\pm$0.01  &           & $<$24.9          & 0.16$\pm$0.01  &           \\
107649 &113.00~$\pm$~18.00& 1.44$\pm$0.02 &  6.20     & 44.30~$\pm$~9.00   & 0.73$\pm$0.01  &  4.84     & 25.90~$\pm$~8.00 & 0.36$\pm$0.01  & 3.19      \\
108870 & $<$19.8          & 9.86$\pm$0.23 &           & $<$21.0            & 5.03$\pm$0.12  &           & $<$23.4          & 2.47$\pm$0.06  &           \\
\hline
\end{tabular}
\label{table:dunes_spire_phot}
\end{table*}

The method of flux measurement in  the SPIRE maps was dependent on the
expected source brightness and extent (compared to the instrument PSF)
in  each  band,  following  the  recommendations  of  the  SPIRE  data
reduction     guide\footnote{http://herschel.esac.esa.int/hcss-doc-9.0}
(see  SPIRE DRG  Fig  5.57, Section  5.7).   In the  case of  extended
sources  (HIP~7978, HIP~32480  and HIP~107649),  flux  measurement was
made via aperture photometry with aperture radii large enough to cover
the source and a sky annulus of 60$\arcsec$--90$\arcsec$.  In the case
of point  sources brighter than 30~mJy  (HIP~544, HIP~13402, HIP~17439
and  HIP~22263),  the  timeline  fitter  task was  used to estimate  the
photometry  using   aperture  radii  of   22$\arcsec$  at  250~$\mu$m,
30$\arcsec$  at  350~$\mu$m  and  42$\arcsec$  at  500~$\mu$m  with  a
background  annulus of 60$\arcsec$--90$\arcsec$  for all  three bands.
Finally, in the case of sources fainter than 30~mJy or non-detections,
the SUSSEXtractor  tool was  used to estimate  the flux  or 3-$\sigma$
upper  limits from  the sky  background  and rms,  as appropriate.   A
summary    of     the    SPIRE    photometry     is    presented    in
Table \ref{table:dunes_spire_phot} and the flux values are plotted in Figure 
E.1.

 \section{Analysis}
 \label{section:analysis}

\subsection{Non-excess sources}
\label{section:non-excess}

\begin{table*}
\centering
\caption{PACS fluxes with 1$\sigma$ errors of non-excess sources, together with the photospheric 
predictions (S$\lambda$). The significance at each band is given. Figures without errors 
in the PACS160 column 
 give 3$\sigma$ upper limits. Figures in the column L$_{\rm d}$/L$_\star$ give upper limits 
of the fractional luminosity of the dust. The last column gives 
{\it Spitzer} fluxes at 70 $\mu$m. Units for fluxes and photospheric predictions are mJy. Only the first 5 lines of the table are given here. The full version is available as online material.}
\label{table:nonexcess_sources}
\begin{scriptsize}
\begin{tabular}{llrrrrrrrrrrr}
\hline\noalign{\smallskip}
HIP    &  SpT    &PACS70          &  S70   &$\chi_{70}$& PACS100       & S100 &$\chi_{100}$& PACS160      &S160 &$\chi_{160}$& L$_{\rm d}$/L$_\star$& MIPS70\\
 \hline\noalign{\smallskip} 
   910 & F5V     &                &  29.51$\pm$0.19&        & 17.66$\pm$ 1.38&  14.46$\pm$0.09&  2.32 & $<$7.5          &  5.61$\pm$ 0.04&     & 8.9e-07&37.50$\pm$4.48  \\
  2941 & K1V+... &                &  24.09$\pm$0.43&        & 11.20$\pm$ 1.82&  11.80$\pm$0.21& -0.33 & $<$5.7          &  4.61$\pm$ 0.08&     & 2.0e-06&25.80$\pm$11.93 \\
  3093 & K0V     &21.93 $\pm$1.69 &  22.50$\pm$0.34& -0.33  &  8.21$\pm$ 1.23&  11.02$\pm$0.17& -2.26 & 3.76$\pm$ 3.43&  4.31$\pm$ 0.07&-0.16& 1.7e-06&14.90$\pm$5.69  \\
  3497 & G3V     & 7.97 $\pm$0.95 &   8.72$\pm$0.10& -0.78  &  5.23$\pm$ 1.01&   4.27$\pm$0.05&  0.95 & $<$4.8          &  1.67$\pm$ 0.02&     & 2.8e-06& 5.20$\pm$4.41  \\
  3821 & G0V SB  &                & 131.05$\pm$2.06&        & 60.80$\pm$ 2.04&  64.21$\pm$1.01& -1.50 &15.75$\pm$ 2.66& 25.08$\pm$ 0.39&-3.47& 3.3e-07&122.3$\pm$10.74 \\

\hline\noalign{\smallskip}
\end{tabular}
\end{scriptsize}
\end{table*} 

\begin{figure}
\centering
\includegraphics[scale=0.4]{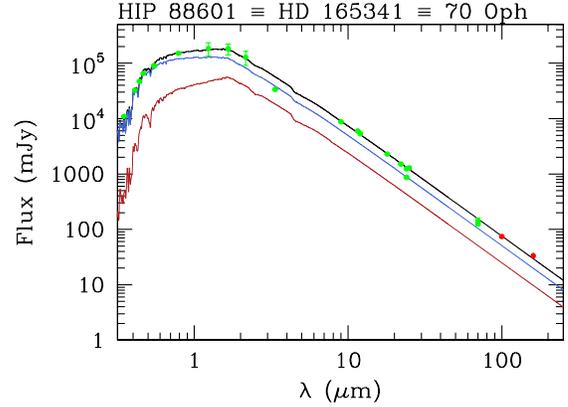}
\caption{Spectral  energy  distribution  of the non-excess star HIP  88601.  Plotted  are
  optical, near-IR, WISE, and {\it Spitzer} MIPS (green symbols),  as well as the
  PACS 100 $\mu$m and 160 $\mu$m (red symbols) fluxes. The photospheric fits of each
  individual component  together with  the added contribution  of both
  stars (black) are shown as continuous lines.}
\label{fig:hip88601}
\end{figure}

\begin{figure}
\centering
\includegraphics[angle=270,scale=0.33]{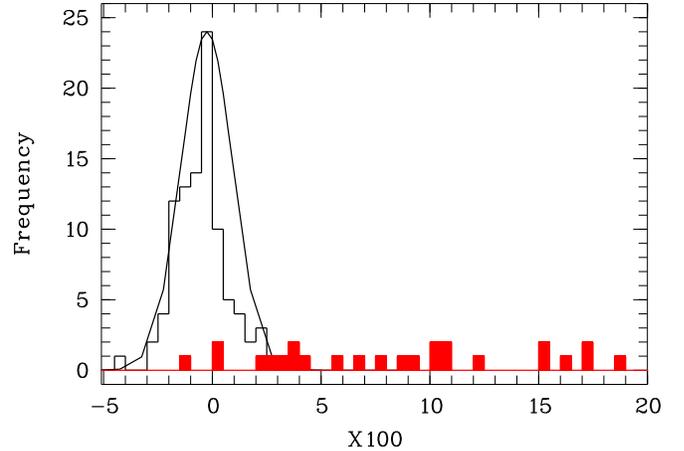}
\caption{Histogram of the 100 $\mu$m significance for the non-excess 
(empty histogram) and excess (red filled histogram) sources. The 
continuous line is a gaussian with $\sigma$ = 1.18, which is the standard 
deviation of the $\chi$100 values of non-excess sources. Excess sources 
with $\chi$100$<$3 are cold disc candidates (see section \ref{section:cold_disc}).
.}
\label{fig:non_excess}
\end{figure}

We consider that a star has an infrared excess at any PACS wavelength when the
significance, $\chi\lambda$ =  (PACS$\lambda$ - S$\lambda$)/$\sigma$, 
is larger than 3, where PACS$\lambda$ is the measured flux, S$\lambda$ 
is the predicted photospheric flux and $\sigma$ is the  total error. The
predicted fluxes  are based on a Rayleigh-Jeans extrapolation from
the  40 $\mu$m fluxes estimated from the PHOENIX/GAIA LTE atmospheric
models (see Appendix  \ref{appendix:predictions}). The sources  for which 
no clear  excesses are detected at  any of the $Herschel$      PACS      
bands      are     listed      in      Table \ref{table:nonexcess_sources},    
where  the    PACS    fluxes, photospheric predictions, and significance  
of the detections  at each PACS  band are given. 
 Figures without errors in the PACS160 column give 3$\sigma$ upper limits 
Errors of the PACS fluxes are the quadratic sum of the photometric 
errors and the absolute calibration uncertainties; for the photometric errors, 
  we  have taken  the  conservative  approach  of choosing  the
largest error values estimated  either from the circular ($\sigma_{\rm
  pix}$   metric)  or   from  the   rectangular   aperture  photometry
($\sigma_{\rm  sky}$ metric). Errors of the predicted fluxes are estimated by means of the least reduced $\chi^2$ procedure described in Appendix  \ref{appendix:predictions}. The significance values in Table \ref{table:nonexcess_sources} are estimated taking as the total error the quadratic sum of the PACS and predicted flux errors.  {\it  Spitzer} 70  $\mu$m MIPS  fluxes
estimated again  for this  work are  given in the  last column  of the
table. The  total number of the  non-excess sources  
are 95  out of 133 ($\sim$71\%).  The spectral type  distribution of this type of objects
(see Table \ref{table:sample_description}) is  16 F-type  stars ($\sim$59\%  of  the total  DUNES F-type  stellar
sample),  37 G-type  stars ($\sim$71\%  of the  G-type) and  42 K-type
stars ($\sim$78\% of  the K-type).  As an example  of the photospheric
fits, Figure  \ref{fig:hip88601} shows the observed SED  of the binary
star HIP  88601 (V 2391  Oph, 70 Oph  AB), where the  fit
takes   into    account   the   contribution    of   both   components
\citep{eggenberger2008}.  A histogram of the significance $\chi$100  of the non-excess sources 
is shown in Figure \ref{fig:non_excess}. 
The median value of $\chi$100 is --0.44, and the mean value is --0.50 with a standard deviation of 1.18. A 
gaussian curve with this $\sigma$ value is also plotted.
If we directly consider the differences between observed and predicted fluxes,
we obtain a mean value of the 100 $\mu$m flux offset of --0.54 mJy with a standard
deviation of 1.40 mJy ($\alpha$ Cen is not included); the median value is --0.60 mJy. 
We note that  the standard
deviation of the 100 $\mu$m flux offsets 
is approximately of the same order as the corresponding sky noise value. The difference in
flux 
suggests that  we might be detecting a small far-IR
deficit between the observed  and predicted fluxes 
This  trend, if real,  might be  reflecting the
fact  that  the  extrapolation  of  the photospheric  fits  (based  on
atmospheric models) to the PACS bands does not take into account that in
solar-type  stars  the   brightness  temperature  decreases  with  the
wavelength as the free-free opacity  of H$^-$ increases. In the Sun
the  origin of the  far-IR radiation  moves to  higher regions  in the
photosphere,  the so-called  temperature minimum  region  
\citep{2003ASPC..286..419A}.   The  apparent  weak far-IR  deficit  we
observe  in the  DUNES sample  might at  least partly  be due  to this
temperature minimum  effect in solar-type stars. In  fact, an in-depth
analysis  of  $\alpha$ Cen  A  using  the  DUNES {\it  Herschel}  data
strongly argues for the  first measurement of this temperature minimum
effect in a star other the Sun \citep{liseau2013}.


Two stars  in Table  \ref{table:nonexcess_sources}, HIP 40693  and HIP
72603,  have  {\it  Spitzer}  fluxes  in excess  of  the  photospheric
emission.  HIP 40693  (HD 69830) has a well  characterized warm debris
disc, as shown by  the MIPS IRS excess between 8 and  35 $\mu$m but no
excess  at  70 $\mu$m  \citep{beichman2005b,beichman2011};  we do  not
detect any 100 or 160 $\mu$m excess with PACS.  The {\it Spitzer} MIPS
70  $\mu$m of  HIP 72603 (Table \ref{table:nonexcess_sources}) suggests the  presence of  
a  far-IR excess; however, this is  clearly not
supported by the {\it Herschel} data since the observed PACS 70 $\mu$m
is in very  good agreement with the predicted  photospheric fluxes, as
well the  PACS 100  and 160 $\mu$m  results.  The 100  $\mu$m aperture
photometry     flux     of     HIP     82860    given     in     Table
\ref{table:nonexcess_sources} presents a marginal excess ($\chi_{100}$
=  2.7)  but  it  is  most  likely contaminated  by  a  bright  nearby
galaxy. PSF photometry  gives 13.2 mJy. Both HIP  82860 and the nearby
bright background galaxy cannot spatially be resolved at 160 $\mu$m. A
similar   situation   is   found   with  HIP   40843   (see   Appendix
\ref{section:controversial}   and   Table  \ref{table:controversial_sources}),
whose apparent  excesses with {\it  Spitzer} and PACS are  most likely
due to contamination by a nearby galaxy. 

There are  7 stars (Table  \ref{table:controversial_sources}) with 160
$\mu$m significance $\chi  160 >$ 3.0; 2 of them  also have 100 $\mu$m
significance  $\chi 100  >$  3.0. However,  the  genuineness of  those
excesses are questionable since  there are extended, background structures or nearby bright
objects which impact  on the reliability of the  160 $\mu$m estimates.
A description of  these objects with contourplots and  images is given
in Appendix D. Summarizing these two last paragraphs,
the stars HIP 40693, HIP 72603, and HIP 82860 are listed in Table 
\ref{table:nonexcess_sources} as non-excess stars with \emph{Herschel},
while the 7 stars in Table \ref{table:controversial_sources} (included the
mentioned HIP 40843) are neither considered excess stars because their    
$\chi$ values larger than 3 are questionable.

\begin{figure}
\centering
\includegraphics[angle=270,scale=0.33]{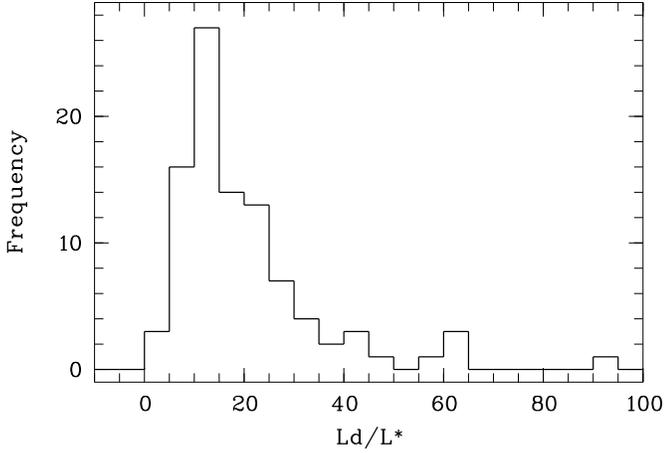}
\caption{Histogram of the upper limit of the fractional luminosity of
  the dust of the non-excess sources. Units: 10$^{-7}$}
\label{fig:Ldust_histo}
\end{figure}

\subsubsection{Dust luminosity upper limits of non-excess sources}
\label{section:limits}

Upper limits of the dust fractional luminosities, $L_{\rm d}/L_\star$,
of     the     non-excess    sources     are     given    in     Table
~\ref{table:nonexcess_sources}.  Those values have been estimated from
the 3$\sigma$ statistical uncertainty of the 100 $\mu$m flux using the
expression   (4)  by  \cite{beichman2006}  and  assuming  a  black  body
temperature of 50  K, which is a representative  value for 100 $\mu$m.
Figure  \ref{fig:Ldust_histo}  presents  a  histogram of  the  $L_{\rm
  dust}/L_\star$ upper  limits.  
The mean  and median values  of these
upper   limits    are   2.0$\times10^{-6}$   and   1.6$\times10^{-6}$,
respectively.  There are  19 stars (8 F-type, 6  G-type, and 5 K-type)
out of the 95 non-excess  stars with $L_{\rm d}/L_\star < 10^{-6}$,
i.e., a few times  the  EKB luminosity. The  two stars  with the
lowest upper limits, $L_{\rm }/L_\star < 5.0 \times 10^{-7}$, are
located  at distances less  than 6.1  pc, i.e.   they are  very nearby
stars (HIP 3821 and HIP 99240).  These  upper  limits  represent  an  increase  in  the
sensitivity  of around  one order  of  magnitude with  respect to  the
detection  limit  with  {\it  Spitzer} at  different  spectral  ranges
\citep[e.g.][]{trilling2008,lawler2009,tanner2009}.

\begin{figure}[!ht]
\centering
\includegraphics[angle=270,scale=0.36]{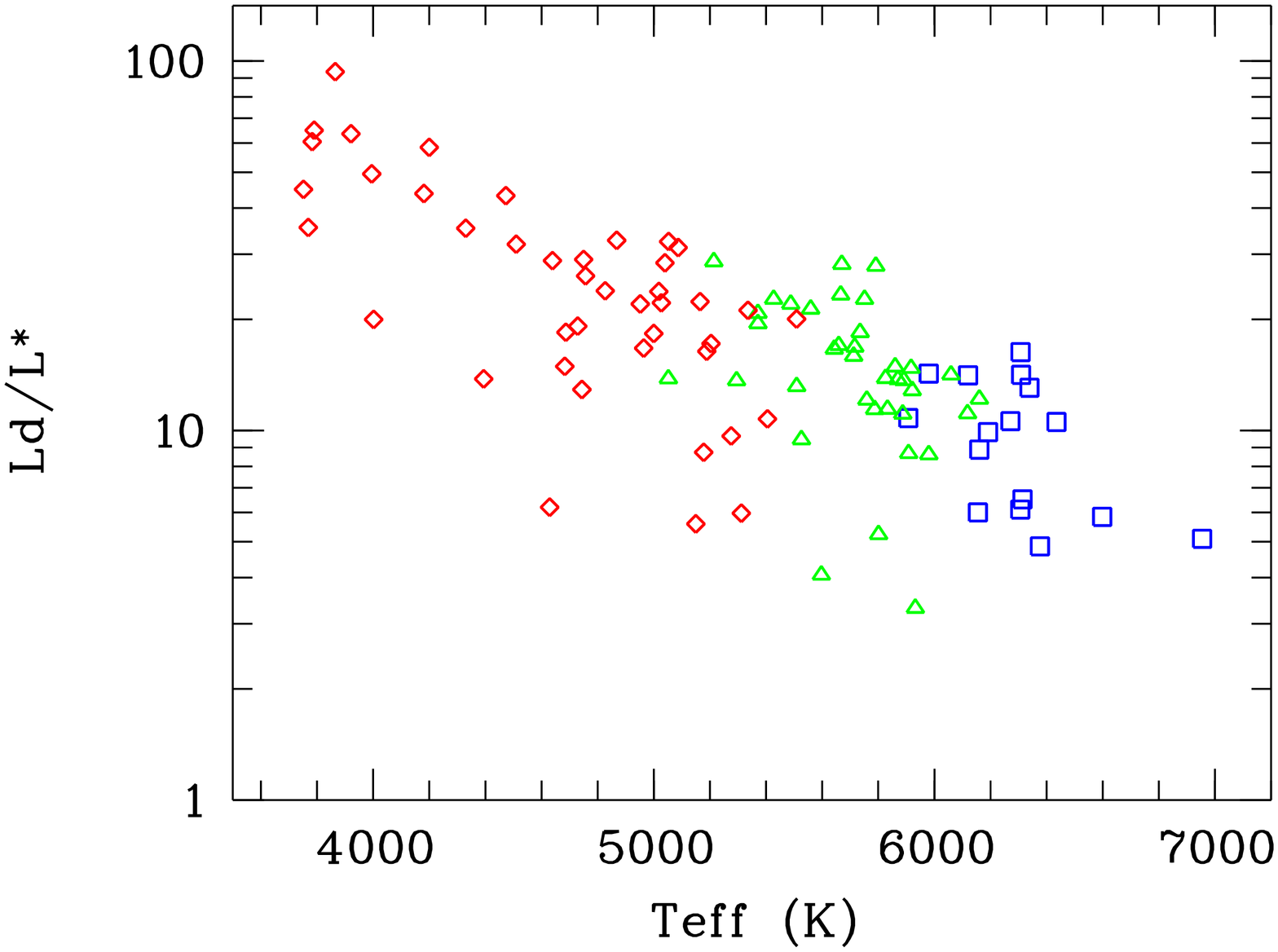}
\includegraphics[angle=270,scale=0.33]{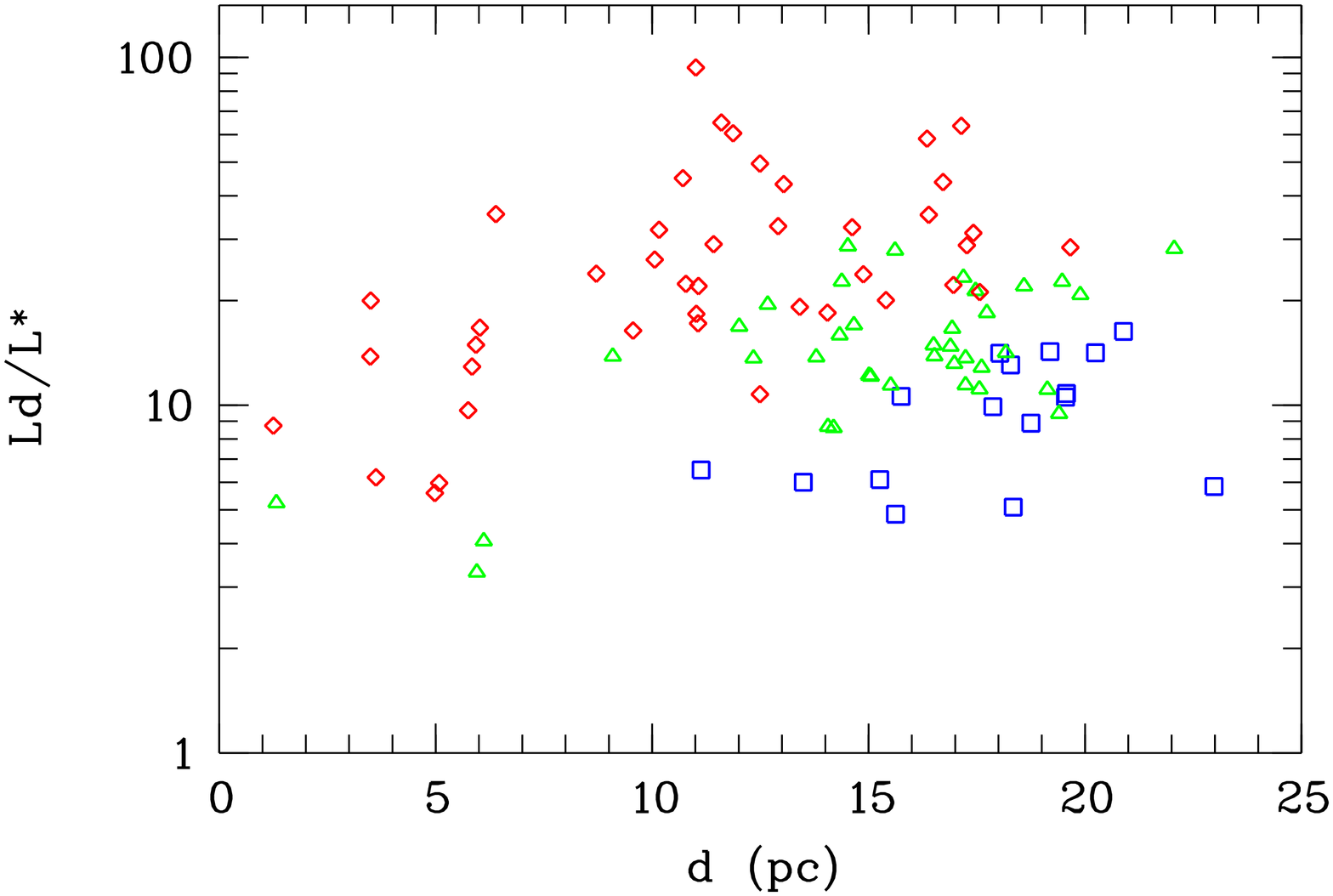}
\includegraphics[angle=270,scale=0.33]{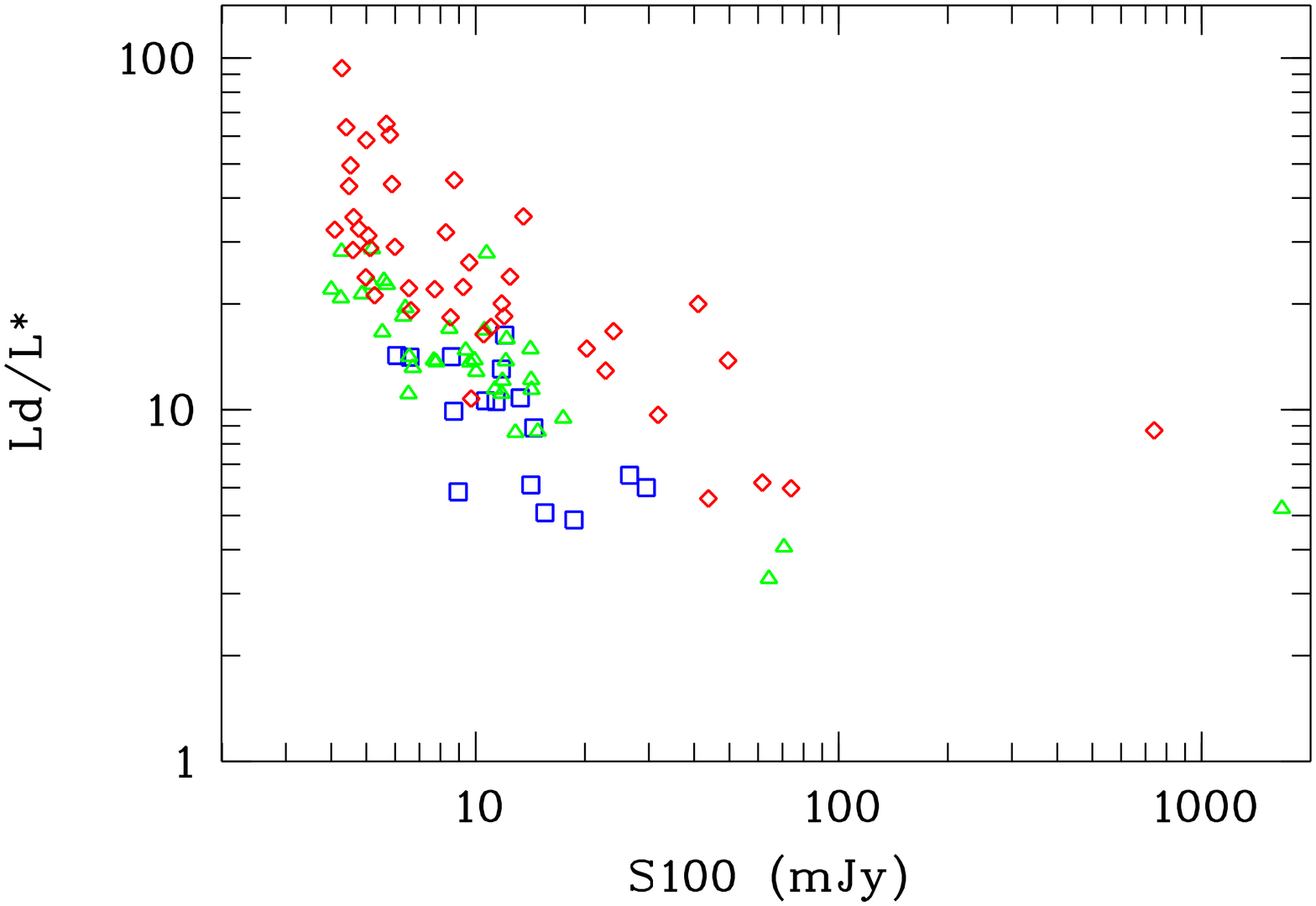}
\caption{Upper limit of  the fractional  luminosity of  the dust (units: 10$^{-7}$) for the
non-excess sources
  versus  effective temperature of the stars (top),  distance  (bottom)  and stellar  flux
  (bottom). Blue squares: F-type stars; green triangles: G-type stars; red diamonds: K-type
  stars.}
\label{fig:non_excess_spt}
\end{figure}
 
Figure \ref{fig:non_excess_spt} presents the $L_{\rm d}/L_\star$ upper
limits as a function of  the effective temperature of the stars, i.e.,
spectral types  (top plot)  and of the  distance to the  stars (middle
plot). Similar  plots have  been presented by  \cite{trilling2008} and
\cite{bryden2009}.   Our   plots   show   that  while   the   $L_{\rm
  d}/L_\star$ upper  limits tend to  increase for the  later K-type
stars,  the   closer  stars  have  low   upper  limit  values,
irrespectively  of  their temperatures.   The  bottom  plot of  Figure
\ref{fig:non_excess_spt}  reflects  that the
flux  contrast  between  the  stellar photosphere  and  a  potentially
existing   debris  disc   is   determined  by   the  bias   introduced
simultaneously by the distances and spectral types.

 \subsection{Excess sources}
\label{section:excess}

\begin{table*}[!ht]
\caption{Overall description of the DUNES sample and summary results. Percentages in parenthesis refer to the amount of stars in the 
corresponding spectral types or the whole samples. Single numbers in parenthesis refer to the number of new debris disc stars 
identified in this survey. }
\label{table:sample_description}
\begin{tabular}{lllll}
\hline\noalign{\smallskip}
Sample                               & F stars &G stars  & K stars & Total   \\
\hline
Solar-type stars observed by DUNES (the DUNES sample)  &  27       &    52      &   54       &  133     \\    
20 pc DUNES subsample                      & 20      & 50       & 54      & 124    \\
Non-excess stars in the whole DUNES sample & 16(59\%)& 37(71\%) & 42(78\%)& 95(71\%)\\
Affected by field contamination      &  2      &  3       &  2      &  7      \\
Excess stars in the whole DUNES sample     &  9(2)    & 12(3)    & 10(5)   & 31(10) \\
Excess stars in the 20 pc subsample  &  4(20\%) & 11(22\%) & 10(18.5\%)& 25(20.2\%)\\
Resolved debris discs                &  5(4)    & 6(4)     &  5(5)     & 16 (13) \\
\hline
\end{tabular}
\end{table*}

A  total of 31  out of  the 133  DUNES targets  show excess  above the
photospheric predictions: 9 F-type, 12  G-type and 10  K-type stars 
(Table \ref{table:sample_description}).
The excess  sources  with the
estimated PACS fluxes, the photospheric predictions and
the significance of the excess at each PACS band are listed in Table  \ref{table:excess_sources}.  
We also include the
MIPS70 $\mu$m flux of each object. In general PACS70 and MIPS70 fluxes
are in  good agreement, although  in the case  of HIP 4148  the larger
MIPS excess is likely due  to contamination by nearby objects. 
Figure \ref{fig:non_excess} shows $\chi$100 and $\chi$160 histograms of the
excess sources (up to the value of 20). Stars with $\chi100<3.0$ correspond to
the cold disc candidates (see below section \ref{section:cold_disc}), while 
stars with $\chi160<3.0$ correspond to the steep SED sources (see below section 
\ref{section:steep_sources}).  
 Figure E.1 
shows the observed SEDs of the stars.  The number
of  excess  sources detected  with  {\it  Herschel}  data reflects  an
increase of 10 sources  with respect to
the number  of previously  known 70 $\mu$m  MIPS {\it  Spitzer} excess
sources (HIP  72603 is  excluded since  it does not  have a  70 $\mu$m
excess with  {\it Herschel}).  We  note again that  HIP 40693 is  a 24
$\mu$m warm excess,  but without 70 $\mu$m excess;  this object is not
listed in Table \ref{table:excess_sources}.  HIP 171 has been reported
as  having  an excess  at  24  $\mu$m but  no  70  $\mu$m MIPS  excess
\citep{koerner2010};  in   this  case,  we   consider  it  as   a  new
detection. We note that most of the new excess sources are K-type stars;  
this  trend   clearly  reflects   the  higher
sensitivity of  {\it Herschel} to detect lower  contrast ratios between
the stellar and dust-disc fluxes, particularly at 100 and 160
$\mu$m.

In order to cleanly assess  the increase of the incidence rate provided
by  {\it Herschel} with  respect to  {\it Spitzer},  we note  that the
figures  of the  previous paragraph  are biased  since  we selectively
included  9  stars between  20  and 25  pc  with  planets and/or  {\it
  Spitzer}  debris  discs  in   the  133  DUNES  sample  (see  section
\ref{section:sample}).  Correcting  the figures from  this bias, i.e.,
considering the 20 pc DUNES sample  of 124 stars, and also taking into
account that the  Spitzer excess of HIP 72603 is  not supported by our
PACS data,  the number  of previously known  stars with  {\it Spitzer}
excesses at 70 $\mu$m is 15,  while the total number of {\it Herschel}
excess  sources,  either at  100  and/or  160  $\mu$m, are  25.   This
represents an  increase of the  incidence rate from the  {\it Spitzer}
12.1\% $\pm$ 5\% to the  {\it Herschel}  20.2\% $\pm$ 2\% rate, i.e.,  around 1.7  times
larger.  The gain  in the debris disc incidence  rate varies very much
with the spectral type.  The 20 pc DUNES sample is formed by 20 F-type
stars,  50 G-type  stars and  54 K-type  stars. According  to spectral
types, the {\it Spitzer} discs  are surrounding 2 F-type stars ($\sim$
10.0 \%), 9  G-type stars ($\sim$ 18.0 \%) and  5 K-type stars ($\sim$
9.3 \%). The  same values for {\it Herschel} are: 4  (20.0 \%) for the
F-type stars, 11 (22\%) for the  G-type stars and 10 (18.5 \%) for the
K-type stars (Table  \ref{table:excess_sources}).  We  note that the fraction of  stars with {\it Spitzer}
excesses  in  our sample  is  a bit lower than what  has  been found  in
different FGK  star programmes  specifically focused to  detect debris
discs   with   the {\it Spitzer}/MIPS    photometer  
\citep[e.g.][]{trilling2008,hillenbrand2008}. This is possibly due to the 
highest spatial resolution of our {\it Herschel} images, which partly avoids the
contamination suffered by the largest {\it Spitzer} beam.

The  results described  in  the  previous paragraph  point   to  an
incidence rate of debris  discs around main-sequence, solar-type stars
of  around 20 \%, irrespectively  of spectral type. This result
can be considered  as a lower limit to the true  number of such discs
and it must be taken very cautiously since it is affected by different
sorts of biases,  as well the previous ones with  {\it Spitzer} were.  We
have shown in section \ref{section:limits} how the $L_{\rm d}/L_\star$
upper  limit depends  on the  combined effect  of the  stars' spectral
types and  distances.  This is  a strong bias clearly  penalizing late
type  stars  at  distances  larger  than  around  10  pc  (see  Figure
\ref{fig:non_excess_spt}).  In  addition, our  20  pc sample  is not  
complete for  K-type stars  for distances
larger than around 15 pc due to {\it {\it Hipparcos}} completeness. If
we restrict the DUNES sample up to 15 pc to avoid this incompleteness,
our incidence  rate is strongly  affected, mainly with respect  to the
F-type stars.  The reason is that  most of the nearby F-type stars are
bright  enough to detect  the stellar  photosphere with  the shallower
DEBRIS   integration  time  and,   according  to   the  DUNES/DEBRIS
agreement, those stars have been observed by that {\it Herschel} OTKP.

 \subsubsection{Background contamination and coincidental alignment. }
 \label{section:alignment}

Some of the  PACS images reveal large scale  field structures denoting
the  presence of  interstellar cirrus.  Good examples  are  some stars
located close  to the galactic  plane like HIP 71683/81  ($\alpha$ Cen
A/B), HIP 124104/07  (61 Cygni A/B) or HIP  71908 ($\alpha$ Cir). These
structures make it difficult to estimate reliable PACS fluxes and even
can mimic an excess over the predicted photospheric flux  
(see Appendix \ref{section:controversial} for some examples).

\begin{figure}[!h]
\centering
\includegraphics[scale=0.33]{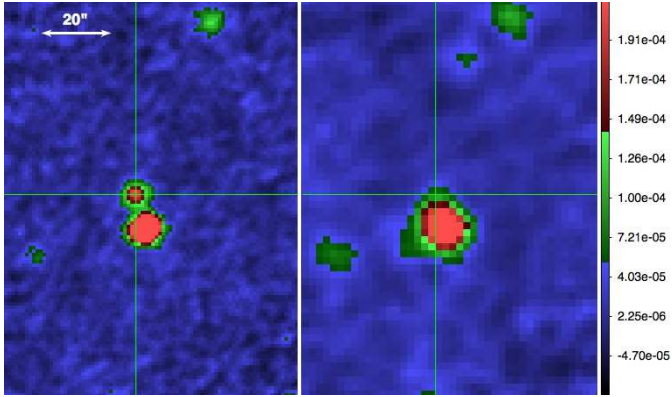}
\caption{PACS 100 (left) and 160 $\mu$m (right) images of HIP 82860. The 
crosshair denotes the position of the 100 $\mu$m flux peak. The angular scale 
is shown in the 100 $\mu$m image by a 20$\arcsec$ segment. Flux scale units are
Jy/pixel. North is up and East 
to the left.}
\label{fig:hip82860}
\end{figure}

In addition, as indicated before, the PACS 100 and 160 $\mu$m images
are very sensitive to background objects.  Therefore, the possibility
of coincidental alignment of such sources with our stars, hindering a
reliable flux measurement or artificially introducing an excess,
cannot be excluded. To assess this potential contamination one needs
to take into account the correlation between the optical and {\it
  Herschel} positions, the photospheric predictions at the different
wavelengths and the {\it Herschel} observed fluxes, as well as the
density of extragalactic sources. HIP 82860 is a concrete example of
such a case of contamination.  The estimated 100 $\mu$m flux agrees
well with the predicted photospheric flux (Table
\ref{table:nonexcess_sources}) but we cannot reliably measure the 160
$\mu$m flux due to the presence of a bright, red background galaxy
(42.2 and 56.0 mJy at 100 and 160 $\mu$m, respectively) located at a
distance of $\sim 10 \arcsec$ from the star (Figure
\ref{fig:hip82860}).  That distance and the 160 $\mu$m ratio between
the star and the galaxy (the 160 $\mu$m predicted flux of HIP 82860 is
5.5 mJy) prevent us from resolving both objects, even using
deconvolution techniques.  Further examples of such potential
contamination by extended structures or background galaxies are
presented in Appendix \ref{section:controversial}, where PACS images
of seven objects with significances $\chi$160 $>$ 3 (some cases also
with $\chi$100 $>$ 3) are described. We remark that none of those
objects are identified as excess sources in this work.

Nonetheless,  we  need to  evaluate  the  impact  of contamination  by
coincidental alignment  in our identified  debris disc stars.   In the
following  we  make  some  probabilistic estimates  to  quantitatively
assess the  chances of  misidentifications of background  objects with
debris discs. We follow  the results obtained by \cite{berta2011}, who
studied the cosmic infrared background in a few large areas of the sky
and carried  out number  counts, i.e., source  densities, in  the PACS
bands and  flux range from $\sim$ 1  mJy to few hundreds  mJy. This is
the range of  interest for our observations. We  base our estimates on
Figure  7  of  \cite{berta2011},  which provides  differential  number
counts per square  degree for the PACS bands in  the GOODS-S field. We
firstly note that  many of our identified {\it  Herschel} debris discs
have very large  excesses, several tens of mJy in  the PACS bands, and
that some of them even show IR excess emission over the photosphere in
the mid-IR wavelength  range of the {\it Spitzer}  IRS instrument (see
Figure  E.1). In  those  cases,  the probability  of  confusion due  to
background   objects  is   practically  negligible.   The  problematic
misidentifications arise for those with low excesses at 160 $\mu$m, of
the order  of few  mJy, with  small $\chi$160 values  and very  low or
non-excess at  100 $\mu$m.  Specifically we identify  6 objects (Table
\ref{table:excess_sources}): HIP 171, HIP 27887, HIP 29271, HIP 49908,
HIP 92043, and HIP 109378.   The 160 $\mu$m differential number counts
for  sources  with  flux  level  $\sim$  5-6  mJy  from  Figure  7  of
\cite{berta2011}  is $\sim$ 2000  sources per  square degree  and mJy,
while for  fluxes $\sim$  12-13 mJy is  $\sim$ 400 sources  per square
degree and mJy.  This flux range recovers the observed excesses of the
above objects.   The size of the  sky area to estimate  the density of
sources is  taken as the  one for which  two different objects  can be
resolved.  In  this respect, we  note that $\alpha$  Cen A and  B with
flux ratio  of about 2  and a separation  of $\sim$ 3$\farcs$1  on the
PACS 100  $\mu$m images can clearly  be distinguished, but  not at 160
$\mu$m. In parallel, we have introduced two fake objects with the same
flux, 7 mJy,  at different angular separations in  the PACS 100 $\mu$m
and  160 $\mu$m  images of  one of  our fields.   Using a  2D gaussian
treatment we could  recover both fake objects at  angular distances of
3$\arcsec$ and 5$\arcsec$ at  100 and 160 $\mu$m, respectively.  These
figures are consistent with  the $\alpha$ Cen observational result and
slightly smaller than the {\it  Herschel} beam sizes.  Thus, taking the
conservative  approach of the  source density  for the  lowest excess,
0.0121  is  the number  of  160  $\mu$m sources  in  a  field of  area
corresponding to the estimated  angular separation, which implies that
the probability of a coincidental  alignment of a background galaxy is
1.2  \%.  Considering the  133 DUNES  stars, the  binomial probability
that all six  objects mentioned above are background  galaxies is just
0.4\%, however the chance that one is a false disc detection is 32\%.

We note that the estimated probabilities depend on a number of
assumptions.  The main one is to assume that all DUNES fields have
equal source densities as the GOODS-S field. This is simply not true
since the number counts change significantly from field to
field. However, \cite{sibthorpe2012} have found no statistically
significant cosmic variance based on 100 $\mu$m data from the DEBRIS
survey; further, assuming that this result is also valid at 160 $\mu$m
and flux levels below the DEBRIS detection limit, they estimate a
confusion probability based on the \cite{berta2011} 160 $\mu$m data
which is in excellent agreement with our own estimate above. Thus, we
conclude that the probability  that all faint debris discs
in Table \ref{table:excess_sources} are background galaxies is rather 
low, although we certainly cannot exclude false
identifications.  At present we cannot distinguish between real cases
and false alarms, although some observational prospectives can be
traced (Krivov et al. 2013, submitted). On the other hand, a detailed
analysis of the number counts and the source density in the DUNES
fields as a function of flux levels, colours and the galactic latitude
will be published in a forthcoming paper (del Burgo et al. in
preparation).

\begin{landscape}
\begin{table}
\caption{Excess sources. PACS fluxes (PACS$\lambda$) together with the photospheric predictions (S$\lambda$) and the significance at each band. The fractional luminosities and the black body 
temperatures and radii of the dust are also given. {\it Spitzer} fluxes
at 70 $\mu$m are also given.}
\label{table:excess_sources}
\addtolength{\tabcolsep}{-2pt}
\begin{tabular}{llrcrrcrrcrrrlrl}
\hline\noalign{\smallskip}
HIP   &  SpT &PACS70          & S70      &$\chi_{70}$& PACS100    & S100 &$\chi_{100}$& PACS160&S160 &$\chi_{160}$& L$_{\rm d}$/L$_\star$ &T$_{\rm d}$ &R$_{\rm d}$  & MIPS70 & Notes\\  
      &      &  (mJy)         & (mJy)    &          & (mJy)      &(mJy) &           &(mJy)   &(mJy)&           &                    & (K)     &  (AU)      & (mJy)         \\
 \hline\noalign{\smallskip} 
   171&G3V   &                &22.90$\pm$0.44&     & 11.70$\pm$ 1.26&11.22$\pm$0.21& 0.38& 12.48$\pm$ 2.36& 4.38$\pm$0.08& 3.43&$\leq$21.6e-06&$\leq$25&$\geq$ 97.1& 44.10$\pm$13.63&H,p,c \\
   544&K0V   &                &15.24$\pm$0.34&     & 54.12$\pm$ 1.00& 7.47$\pm$0.17&47.15& 23.03$\pm$ 2.06& 2.92$\pm$0.07& 9.86&4.8e-05&      90       &  7.5      & 102.6$\pm$7.83 &e     \\
  4148&K2V   & 13.66$\pm$ 1.37& 8.98$\pm$0.13& 3.42& 11.33$\pm$ 1.17& 4.40$\pm$0.06& 5.92& 15.64$\pm$ 1.72& 1.72$\pm$0.02& 8.09&9.4e-06&      32       & 41.2      & 37.10$\pm$6.59 &p     \\
  7978&F8V   &896.20$\pm$26.90&17.48$\pm$0.99&32.67&897.10$\pm$26.90& 8.56$\pm$0.48&33.03&635.90$\pm$31.80& 3.35$\pm$0.19&19.89&3.1e-04&      60       & 26.6      & 863.4$\pm$58.68&e     \\
 13402&K1V   &                &20.72$\pm$0.37&     & 51.77$\pm$ 1.11&10.15$\pm$0.18&37.50& 36.94$\pm$ 2.94& 3.97$\pm$0.07&11.21&1.7e-05&      52       & 17.9      & 67.70$\pm$7.06 &e     \\
 14954&F8V   &                &26.43$\pm$1.17&     & 39.45$\pm$ 1.75&12.95$\pm$0.58&15.14& 31.75$\pm$ 1.16& 5.06$\pm$0.22&23.01&4.2e-06&      40       & 95.0      & 42.50$\pm$4.76 &e, !     \\
 15371&G1V   & 44.50$\pm$ 2.50&25.78$\pm$0.33& 7.49& 40.40$\pm$ 2.50&12.63$\pm$0.16&11.11& 42.60$\pm$ 2.50& 4.93$\pm$0.06&15.07&1.0e-05&      40       & 47.7      & 45.40$\pm$4.95 &e     \\
 17420&K2V   & 15.99$\pm$ 1.81& 9.36$\pm$0.13& 3.66& 14.79$\pm$ 0.84& 4.58$\pm$0.06&12.15& 10.65$\pm$ 1.30& 1.79$\pm$0.03& 6.82&9.2e-06&      45       & 20.7      & 23.60$\pm$5.44 &H,p,s \\
 17439&K1V   & 74.80$\pm$ 4.10& 8.61$\pm$0.14&16.14& 75.00$\pm$ 4.20& 4.22$\pm$0.07&16.85& 74.60$\pm$ 4.70& 1.65$\pm$0.03&15.52&8.1e-05&      48       & 21.3      & 88.50$\pm$7.46 &e     \\
 22263&G3V   &                &21.13$\pm$0.38&     & 77.60$\pm$ 2.00&10.35$\pm$0.18&33.62& 47.00$\pm$ 3.00& 4.04$\pm$0.07&14.32&2.9e-05&      70       & 15.4      & 113.6$\pm$8.53 &e     \\
 27887&K3V   & 14.60$\pm$ 1.43&10.01$\pm$0.19& 3.21&  8.05$\pm$ 0.95& 4.90$\pm$0.10& 3.32&  8.02$\pm$ 1.50& 1.92$\pm$0.04& 4.07&3.8e-06&      29       & 46.2      & 17.20$\pm$4.94 &H,p   \\
 28103&F1V   &                &56.33$\pm$0.31&     & 45.46$\pm$ 1.42&27.60$\pm$0.15&12.58&  9.37$\pm$ 1.84&10.78$\pm$0.06&-0.77&6.3e-05&     100       & 18.3      & 93.90$\pm$7.76 &p,s   \\
 29271&G5V   &                &35.50$\pm$0.42&     & 17.80$\pm$ 1.30&17.40$\pm$0.21& 0.31& 14.35$\pm$ 2.00& 6.80$\pm$0.08& 3.78&$\leq$29.7e-06&$\leq$22&$\geq$147.3& 42.60$\pm$10.50&H,e,c \\
 32480&G0V   &264.00$\pm$ 4.10&22.33$\pm$0.27&58.94&252.30$\pm$ 3.18&10.94$\pm$0.13&75.90&182.09$\pm$ 3.77& 4.27$\pm$0.05&47.17&6.9e-05&      60       & 28.5      & 262.8$\pm$18.29&e     \\
 42438&G1.5Vb&                &18.12$\pm$0.25&     & 17.31$\pm$ 0.80& 8.88$\pm$0.12&10.54&  6.73$\pm$ 2.04& 3.47$\pm$0.05& 1.60&1.1e-05&      99       &  7.8      & 41.20$\pm$4.16 &p,s   \\
 43726&G3V   &                &13.48$\pm$0.26&     & 15.77$\pm$ 0.76& 6.60$\pm$0.13&12.07&  6.09$\pm$ 1.42& 2.58$\pm$0.05& 2.47&1.6e-05&      99       &  7.9      & 32.50$\pm$4.13 &p,s   \\
 49908&K8V   &                &55.55$\pm$2.11&     & 22.50$\pm$ 0.90&24.59$\pm$1.03&-2.33 & 16.00$\pm$ 1.70&9.61$\pm$0.40& 3.16&$\leq$21.6e-06&$\leq$22&$\geq$ 56.6& 38.70$\pm$4.69 &H,e,c \\
 51459&F8V   &                &31.14$\pm$0.53&     & 19.71$\pm$ 1.42&15.26$\pm$0.26& 3.13&  9.17$\pm$ 2.79& 5.96$\pm$0.10& 1.15&9.1e-07&      50       & 38.6      & 33.80$\pm$4.43 &H     \\
 51502&F2V   &                &16.57$\pm$0.22&     & 47.25$\pm$ 2.00& 8.12$\pm$0.11&19.57& 72.45$\pm$ 2.22& 3.17$\pm$0.04&31.21&1.3e-05&      30       &145.0      & 39.60$\pm$3.79 &e, !   \\
 62207&G0V   &                &13.45$\pm$0.11&     & 55.06$\pm$ 2.39& 6.59$\pm$0.05&20.28& 44.49$\pm$ 3.17& 2.57$\pm$0.02&13.22&2.1e-05&      66       & 18.3      & 55.70$\pm$5.20 &e     \\
 65721&G5V   &                &42.07$\pm$0.36&     & 40.73$\pm$ 0.66&20.61$\pm$0.17&30.48& 26.97$\pm$ 1.39& 8.05$\pm$0.07&13.61&3.8e-06&      45       & 66.1      & 79.00$\pm$8.09 &p     \\
 71181&K3V   & 15.06$\pm$ 1.30& 9.98$\pm$0.16& 3.91&  8.79$\pm$ 1.00& 4.89$\pm$0.08& 3.90&  1.63$\pm$ 1.63& 1.91$\pm$0.03&-0.17&8.0e-06&      70       &  7.9      & 29.20$\pm$8.04 &H,p,s \\
 72848&K2V   &                &21.45$\pm$1.06&     & 19.60$\pm$ 2.00&10.51$\pm$0.52& 4.55& 13.30$\pm$ 1.12& 4.11$\pm$0.20& 8.21&2.8e-06&      40       & 34.2      & 33.50$\pm$6.41 &H,e   \\
 73100&F7V   &                &14.42$\pm$0.23&     & 13.84$\pm$ 0.81& 7.06$\pm$0.11& 8.37& 12.46$\pm$ 2.27& 2.76$\pm$0.04& 4.27&2.7e-06&      30       &144.8      & 24.70$\pm$3.17 &p,c, !   \\
 85235&K0V   &                &12.83$\pm$0.19&     & 34.70$\pm$ 1.18& 6.29$\pm$0.09&24.08& 28.00$\pm$ 2.52& 2.46$\pm$0.04&10.13&2.0e-05&      45       & 24.6      & 58.00$\pm$4.92 &e     \\
 92043&F6V   & 59.00$\pm$ 3.50&48.70$\pm$0.68& 2.94& 30.20$\pm$ 2.40&23.90$\pm$0.33& 2.63& 21.90$\pm$ 3.80& 9.34$\pm$0.13& 3.31&7.0e-07&      30       &213.3      & 69.80$\pm$8.93 &H,e,c \\
103389&F7V   & 44.00$\pm$ 2.30&13.45$\pm$0.23&13.28& 26.30$\pm$ 1.70& 6.59$\pm$0.11&11.59&  7.70$\pm$ 2.50& 2.57$\pm$0.04& 2.05&1.7e-05&      90       & 13.6      & 46.60$\pm$4.93 &p,s, !   \\
107350&G0V   &                &12.51$\pm$0.17&     & 15.10$\pm$ 1.30& 6.13$\pm$0.08& 6.90&  4.40$\pm$ 2.30& 2.39$\pm$0.03& 0.87&1.2e-05&     100       &  8.1      & 28.40$\pm$3.15 &p,s   \\
107649&G2V   &284.00$\pm$ 1.50&18.38$\pm$0.20&177.0&311.00$\pm$ 1.00& 9.01$\pm$0.10&301.9&211.00$\pm$ 1.50& 3.52$\pm$0.04&138.3&1.0e-04&      55       & 28.7      & 278.2$\pm$21.46&e     \\
109378&G0    &                & 9.51$\pm$0.13&     &  8.50$\pm$ 1.00& 4.66$\pm$0.07& 3.84& 12.40$\pm$ 1.60& 1.82$\pm$0.03& 6.61&5.4e-06&      22       &160.2      & 7.90 $\pm$2.26 &H,p,c, ! \\
114948&F7V   &                &14.08$\pm$0.19&     & 40.80$\pm$ 1.60& 6.90$\pm$0.09&21.19& 13.30$\pm$ 2.20& 2.69$\pm$0.04& 4.82&2.7e-05&      90       & 13.1      & 68.70$\pm$5.51 &p,s, !   \\
\hline\noalign{\smallskip}                                                                                                       
\multicolumn{16}{l}{Column Notes: H = new {\it Herschel} debris disc, p = point-like object, e= resolved object, c = cold disc, s = steep source, ! = stars of the 20-25 pc subsample}
\end{tabular}
\noindent 
\end{table} 
\end{landscape}

 \subsubsection{Dust properties}
 \label{section:dust}

 The excess infrared emission from the debris discs originates in
 small dust grains produced in collisions of large bodies.  A common
 approach to estimate the temperature and the irradiated luminosity of
 those grains is to assume that they behave as black body grains.
 This approach also allows us to estimate a representative orbital
 distance of the black body-like dust \citep{1993prpl.conf.1253B}.
 This procedure presents several caveats. Observed SEDs of some discs
 clearly reveal the presence of dust at multiple temperatures
 \cite[e.g.][]{hillenbrand2008,lawler2009}. A simple look at the SEDs
 presented in Figure E.1 suggests a range of temperatures of the
 dust. This can either be due to its location at a range of distances
 from the star or to dust at the same distance but with different
 properties. It has already been mentioned the well known fundamental
 degeneracy between distance and size of the dust particles, as small
 grains far from the star can produce the same SED as large grains
 located close in \citep{krivov2010}. At this point we note that the
 distribution of dust over a radial range has directly been proven by
 spatially resolved imaging in scattered and reemitted radiation
 \citep[e.g.][]{ertel2011}.

Nevertheless, the simple black
 body assumption still provides a reasonable approach for most of the
 debris discs and for comparison among solar-type stars, although the
 black body radius underestimates the true one \citep[e.g.][and
   references therein]{wyatt2008,ertelwolf2012}.
 Based on these, black body temperatures, T$_{\rm d}$,
 have been estimated by fitting a black body for those sources with
 excesses at several bands. Where the excess is observed only at one
 band, mainly at 160 $\mu$m, the upper limit of the temperature is
 given.  The fractional luminosity of the dust has been estimated
 integrating the observed excess fluxes; in those cases with excess at
 one wavelength, we have used the expression (4) of
 \cite{beichman2006} assuming a dust temperature of 50 K. 
 The orbital distance at which the dust would be
 located is estimated following equation (3) from
 \cite{1993prpl.conf.1253B}. In general, for those objects previously
 known as debris discs, the dust properties obtained using the {\it
   Herschel} data do not differ significantly from those obtained with
 {\it Spitzer} \cite[e.g.]{beichman2006,trilling2008}. The calculated
 values for the mentioned three dust parameters are given in Table
 \ref{table:excess_sources}.  $L_{\rm d}/L_\star$ spans approximately
 two orders of magnitude, $\sim$ 7$\times$10$^{-7}$ --
 3$\times$10$^{-4}$, and are in a few cases very close to the
 sensitivity limit.  The new {\it Herschel} discs tend to be
 approximately one order of magnitude fainter than the previously
 known ones, mean values are $\sim$ 4$\times 10^{-6}$ against $\sim$
 4$\times 10^{-5}$, respectively (Table \ref{table:excess_sources}).
 T$_{\rm d}$ values range $\sim$ 20 -- 100 K, with the lowest
 temperatures also mainly related to the new {\it Herschel} discs,
 mean value $\sim$ 34 K against $\sim$ 64 K. As expected, the black
 body radius tends to be larger for the {\it Herschel} discs, mean
 distance $\sim$ 82 AU against $\sim$ 38 AU.  A short compendium of
 these results is that the new {\it Herschel} discs trace fainter and
 colder debris discs than the previously known discs.  

\subsubsection{Point-like and extended sources}
\label{section:sizes}

\begin{table*}[ht!]
\centering
\caption{Stars associated with extended emission. Column 2 gives the position angles  on the 
sky (from North to East). Column 3 gives the elliptical diameters (major,minor) of the 
3$\sigma$ contours. Column 4 (FWHM$_c$) gives the x and y FWHM values assuming the 3$\sigma$ contours are Gaussian,
while column 5 (FWHM$_G$) is a 2D Gaussian fit to the whole detected flux. See text for details. 
The last column gives the wavelength at which the sizes have been 
estimated.}
\label{table:sizes}
\begin{tabular}{lccllcrr}
\hline\noalign{\smallskip}
HIP   & Position Angle  & Observed diameters   & FWHM$_c$  &FWHM$_G$             & Band \\
      &                 &  (arcsec)            & (arcsec)  &     (arcsec)        &      \\
 \hline\noalign{\smallskip} 
   544& 102$^{\circ}$    &18.0,16.0 & 8.8,7.8   &8.67$\pm$0.10,8.28$\pm$0.10     & 100 $\mu$m \\
  7978& 51$^{\circ}$     &39.0,28.0 & 15.7,11.3 &15.5$\pm$0.02,9.02$\pm$0.02     & 100 $\mu$m \\
 13402& 50$^{\circ}$     &19.0,14.0 & 9.2,6.7   &8.20$\pm$0.10,7.61$\pm$0.02     & 100 $\mu$m \\
 14954(!)& 30$^{\circ}$  &18.0,14.0 & 9.4,7.3   &9.89$\pm$0.10,7.80$\pm$0.10     & 100 $\mu$m \\
 15371& 98$^{\circ}$     &39.0,18.0 & 24.5,11.6 &20.00$\pm$1.0,12.20$\pm$0.36    & 160 $\mu$m \\
 17439& 105$^{\circ}$    &26.0,14.0 & 13.1,7.1  &13.35$\pm$0.14,8.68$\pm$0.10    & 100 $\mu$m \\ 
 22263& 5$^{\circ}$      &21.0,16.0 & 10,7.6    &10.20$\pm$0.10,8.41$\pm$0.09    & 100 $\mu$m \\ 
 29271&  70$^{\circ}$    &16.0,10.0 & 14.3,8.9  &12.00$\pm$1.22,10.00$\pm$0.73   & 160 $\mu$m \\
 32480& 107$^{\circ}$    &34.0,20.0 & 16.0,9.4  &17.76$\pm$0.07,10.15$\pm$0.04   & 100 $\mu$m \\
 49908&  50$^{\circ}$    &18.0,10.0 & 15.5,8.6  &13.00$\pm$0.94,11.00$\pm$0.60   & 160 $\mu$m \\
 51502(!)& 0$^{\circ}$   &16.0,16.0 & 8.3,8.3   &9.08$\pm$0.12,8.10$\pm$0.11     & 100 $\mu$m \\
 62207&  130$^{\circ}$   &24.0,14.0 & 13.4,7.8  &13.13$\pm$0.19,8.40$\pm$0.11    & 100 $\mu$m \\
 72848& 75$^{\circ}$     &21.0,12.0 & 12.7,7.2  &11.07$\pm$0.33,6.64$\pm$0.20    & 100 $\mu$m \\
 85235& 0$^{\circ}$      &18.0,16.0 & 10.2,9.0  &10.01$\pm$0.23,9.21$\pm$0.21    & 100 $\mu$m \\
 92043&  170$^{\circ}$   &16.0,10.0 & 19.1,11.9 &14.00$\pm$1.10,10.00$\pm$0.79   & 160 $\mu$m \\
107649& 125$^{\circ}$    &17.5,8.0  & 16.1,11.2 &                                & 100 $\mu$m \\
\hline\noalign{\smallskip}
\end{tabular}

(!): Stars of the 20-25 pc subsample
\end{table*} 

Most of the known debris  discs have been characterised by fitting the
observed  SEDs.   Spatially  resolved   images  help  to  break  the inherent SEDs'
degeneracies by showing where most dust is located. More
than    30    debris    discs     are    known    to    be    resolved
(http://www.circumstellardisks.org).   Resolved  imaging at  different
wavelengths serves  not only to confirm the  presence of circumstellar
discs,  but provides important  constraints to  their properties,
 like the  dust location  and reliable temperatures.  Furthermore,
resolved discs display features  as warps, clumps, rings, asymmetries,
etc., which  help in the  study of the dynamics  of the discs  and indirectly
prove    the    presence    of   planets    \citep[][and    references
therein]{wyatt2008}.

 The  {\it Herschel}  PACS observations reveal  a large  number of
  stars  associated with extended  emission at  100 $\mu$m  and/or 160
  $\mu$m.  Their 3$\sigma$ flux contours usually show an elliptical-like
  shape. The size of  the extended  sources is estimated by  fitting   ellipses to  the
  3$\sigma$  contours of each source by eye. Column 2 and 3 of Table  \ref{table:sizes}  gives 
  the position angle (measured from North to East) and the  elliptical
  100 $\mu$m major and minor  diameters. We estimate an uncertainty of 
  $\sim$ 1$\arcsec$ (i.e., 1 pixel at 100 $\mu$m). We have preferably chosen
  that wavelength  due to the
  complexity of  the surrounding fields at  160 $\mu$m in  many of the
  objects.  In  the case of  the HIP 29271,  HIP 49908, and  HIP 92043
  their  sizes are  given at  160 $\mu$m  because they  are  cold disc
  candidates  (see subsection  \ref{section:cold_disc}).  To
  assess if the 3$\sigma$ contours truly denote extended emission,
   we proceed in the following manner.  Firstly, we
  have assumed that  the source brightness profile is well approximated by a Gaussian and 
  measured  the  ratio between the peak flux  and  the 3$\sigma$ flux
  values, assuming  that this ratio  corresponds to a Gaussian  at a
  distance  from the  center given  by  the semiaxes  of the ellipses 
  representing the 3$\sigma$ contours. In this way  we have estimated
  the FWHM of the emission in both axes (column 4 of Table  \ref{table:sizes}). 
  Secondly, we have carried out a two-dimensional  Gaussian analysis of the  
  sources using the IDL procedure MPFIT2DPEAK, fitting the observed brightness 
  profiles with a rotated 2-D Gaussian profile weighted by the uncertainties, without 
  applying any prior assumption as regards the shape of  the emission (column 5 of 
  Table  \ref{table:sizes}).  Both methods
  yield  quite  consistent results  within  the uncertainties.   
  Thirdly, once we have the gaussian sizes
  we need to evaluate whether they reflect truly extended emission.
  FWHM  values of  the {\it Herschel}  PACS PSFs  are $\sim$
  7$\arcsec$ and 12 $\arcsec$ at 100 $\mu$m and 160 $\mu$m, respectively, 
  but  small variations of the PSF due to
  the brightness of  the sources and the observing  strategy are known
  to exist \citep{kennedy2012}. In
  order to assess if the sources are resolved, a Monte-Carlo simulation has
  been  carried out taking  as reference  HIP 544  and HIP  72848, two
  relatively faint sources  with small FWHM values, which in the case of
  HIP 72848 is only  resolved in one direction. The  standard star $\alpha$ Boo is
  taken as  representative of  a pure point-like  PSF. The  100 $\mu$m
  PACS image of $\alpha$ Boo has been  rotated so that the new x and y
  axes correspond  to the axes of  the extended emission,  and its PSF
  has  been scaled  to the  flux of  each star.  The new
  $\alpha$  Boo  PSF  has  been  inserted  in  a  grid  of  2627 and 1568
  locations in  the   cleanest  areas  of  the  HIP   544  and  HIP
  72484 images.  For each position  the x  and y  FWHM values  have been
  estimated  using   a  2D  Gaussian  fit.  This   provides  the  FWHM
  distribution  of point  sources with  the same  flux as  the problem
  objects over a noisy background.  No additional noise has been added to the images
  because the telescope thermal noise and sky confusion, already included in the
 frames, are much larger than the Poisson contribution for such faint
 artificial sources. The  distributions of the FWHM values obtained with both Monte-Carlo simulations
  are  approximately  Gaussian. Thus, by using both Gaussian statistics and the empirical 
  distributions constructed above we can estimate the probability that a point source
  randomly provides the FWHM values listed in Table \ref{table:sizes}. The result is that the probability of 
  false positives seems below 0.1\%. Since HIP 544 and HIP 72484 are among the most unfavourable cases, 
  we conclude that the extended nature of all objects identified as such in 
  Table \ref{table:sizes} is fairly secure. We point out that 
  \cite{kennedy2012} also claim to have resolved a disc with a FWHM of 
  8$\farcs$2$\times$6$\farcs$9.
  
  The number of sources listed in Table \ref{table:sizes} associated with 
  extended emission is 16, i.e., $\sim$52\% out of the 31 excess sources. 
  This represents a huge increase of resolved discs with respect to the
3 sources in our sample previously resolved with {\it  Spitzer} or any
other   facility   (HIP  7978,   HIP   32480,   HIP  107649). 
Disc size values in Table  \ref{table:sizes} are upper limits, since proper
deconvolution  is required  to  estimate more  realistically the  true
extension of the debris discs and their structure. This has already been done for some of
the   sources  for which a   deeper  observational   analysis
\citep{liseau2010,eiroa2011,marshall2011} or highly detailed modelling
\citep{loehne2012} has been carried out. Similar work is in progress for a few more sources
(Marshall et al., Stapelfeldt et al., Faramaz et al., in preparation).
\subsubsection{Cold disc candidates}
\label{section:cold_disc}

Some   of    the   identified    debris   disc   sources    in   Table
\ref{table:excess_sources} show  an excess  at 160 $\mu$m  but little  or no
excess at all at 100 $\mu$m.  These sources are HIP 171, HIP 29271, HIP
49908, HIP  73100, HIP  92043, and HIP  109378. We note that the 100 and 160 $\mu$m source 
positions agree within the astrometric errors. The  infrared excesses
have  been  attributed   to  a  new  class  of   cold,  debris  discs,
characterized by low temperatures, $\lesssim$ 30 K, and low fractional
luminosities \citep{eiroa2011}.  While it
cannot  be excluded  that they  suffer from  background contamination,
i.e., they might not be true circumstellar discs, the probability that
one or more of these  discs are real is large (section \ref{section:alignment}).
If true, the nature of these  faint, cold discs cannot be explained by
simply invoking the ``classical''  collisional models of debris discs.
Alternative  scenarios have
been explored  by Krivov  et al. (2013,  submitted). They  argued that
such  discs might be  composed of  nearly unstirred  primordial grains
with sizes somewhere in the millimeter to kilometer range, which would
imply that  planetesimal formation has  stopped before  ``cometary"
or ``asteroidal" sizes  were reached, at least in  the outer regions of
the  systems.  Discs  of  this kind  would  experience  low-velocity
collisions   without  any  significant   production  of   small,  warm
grains. As a result, a bulk  of the material would have a nearly black
body temperature as suggested by  the observed SEDs. 

\begin{figure*}[ht!]
\centering
\includegraphics[scale=0.28]{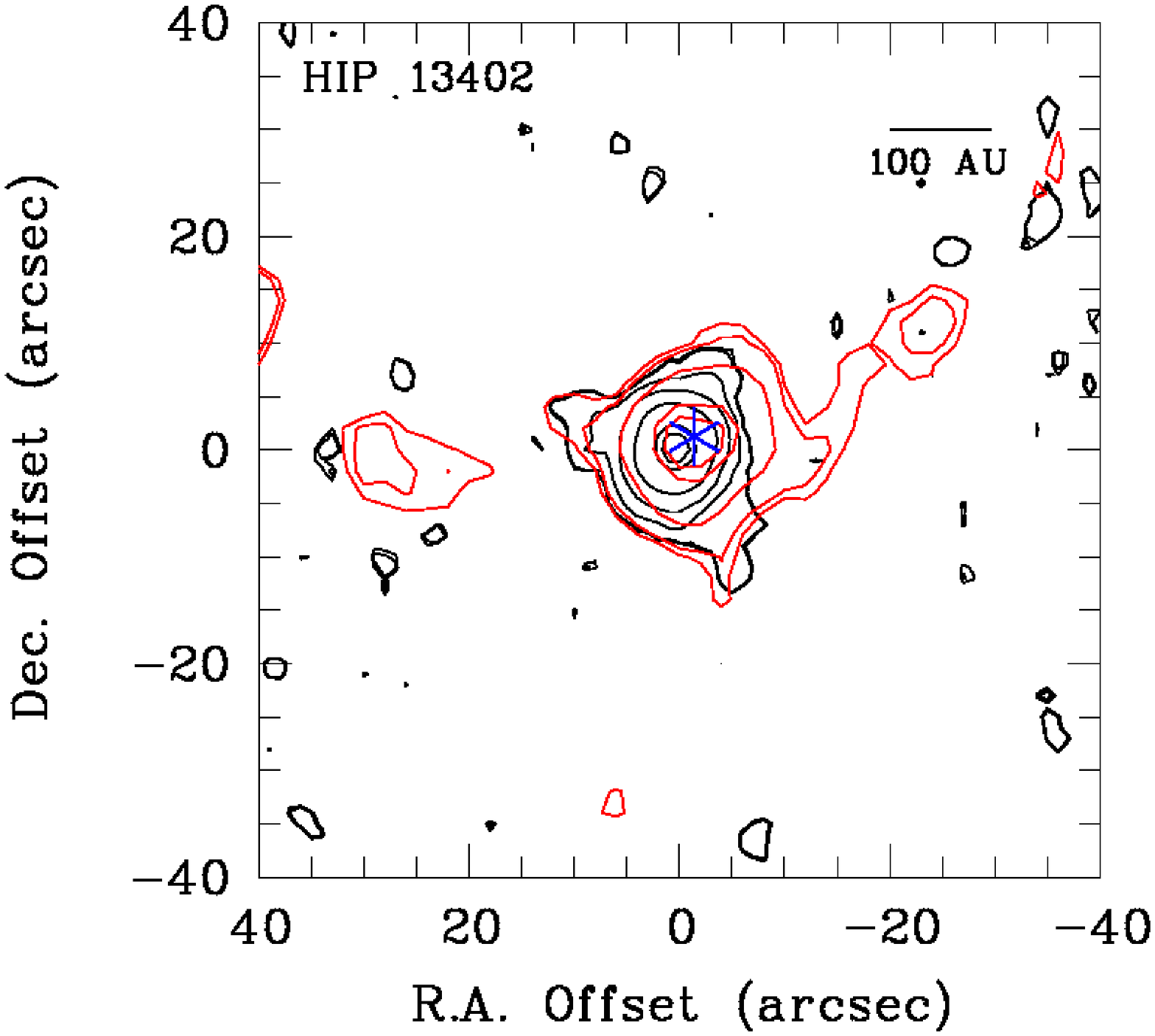}
\includegraphics[scale=0.18]{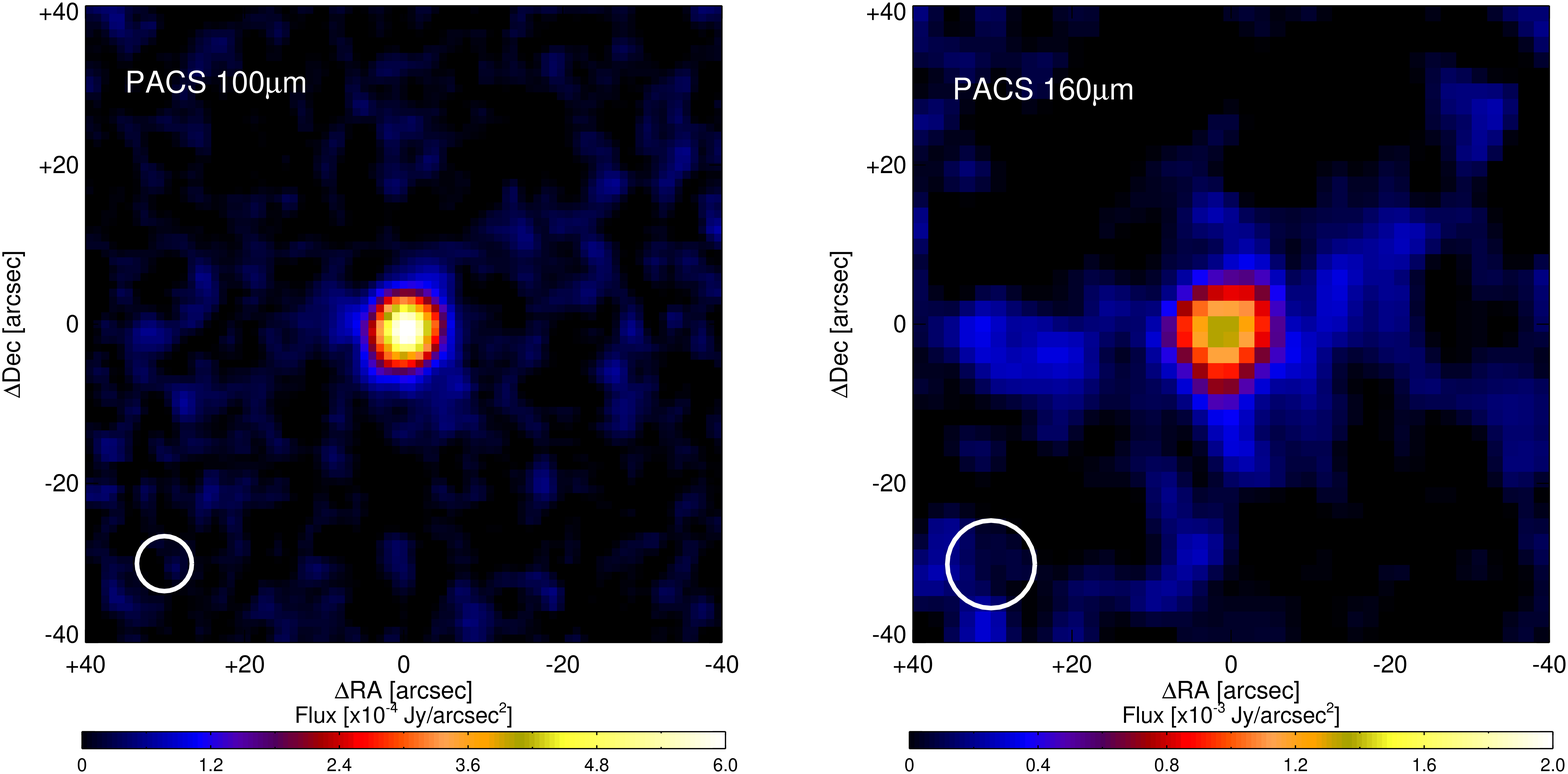}
\includegraphics[scale=0.28]{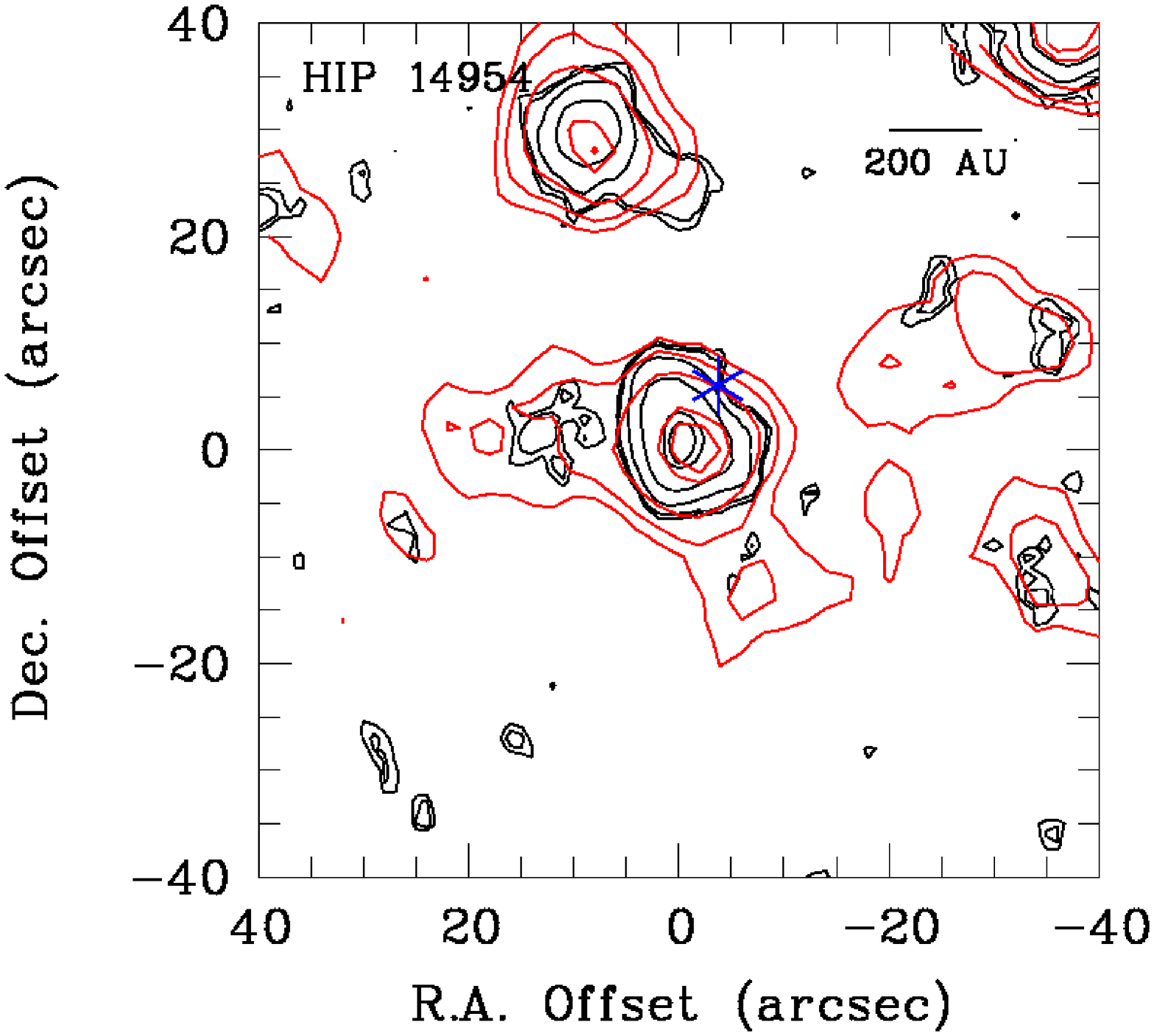}
\includegraphics[scale=0.18]{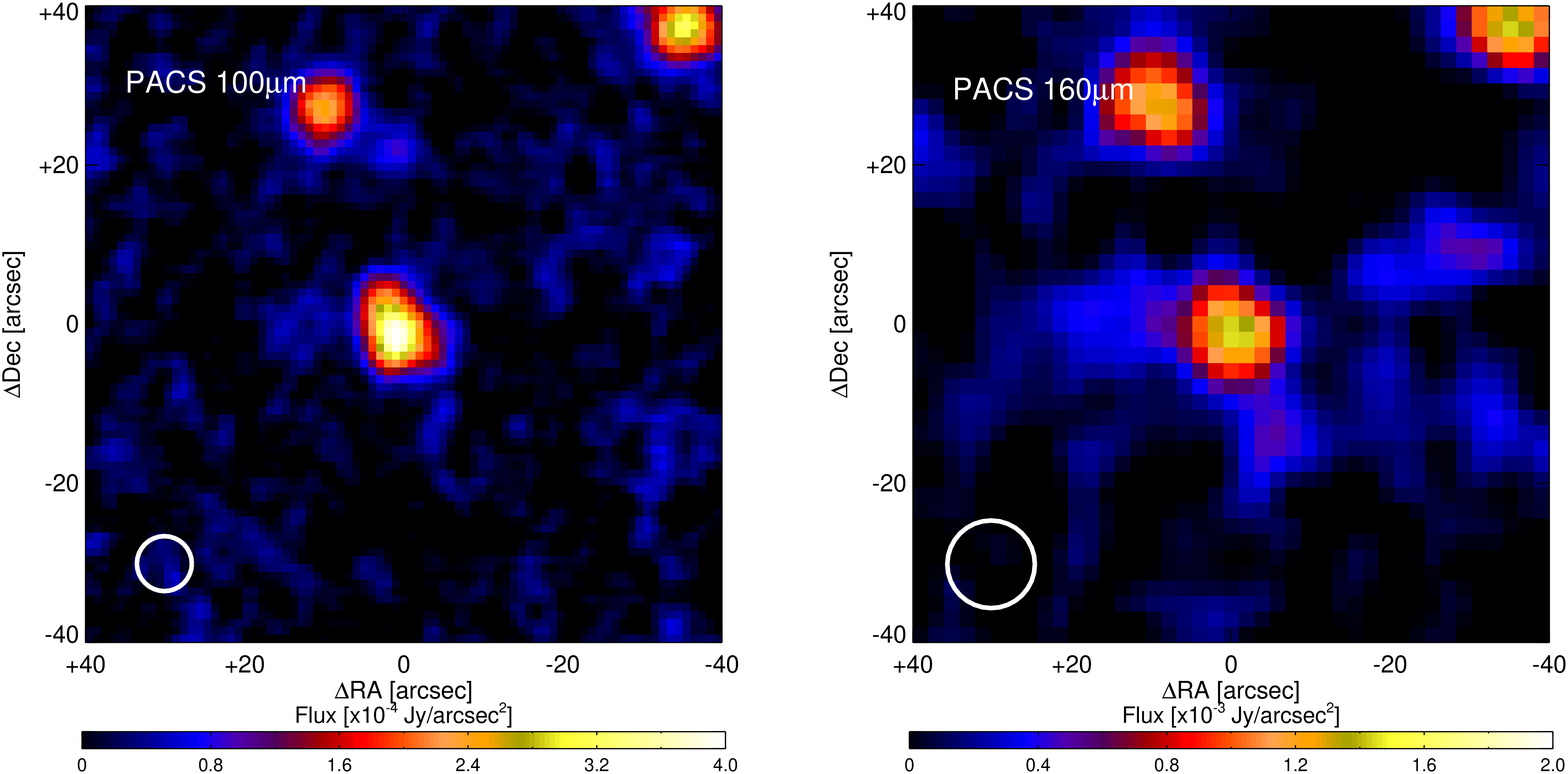}
\caption{
Contours   plots (left) and PACS 100 $\mu$m (middle) and 160 $\mu$m (right) images 
of two resolved debris disc stars. The identification of the stars are given 
in the upper-left corner of the contour plots. Position (0,0) refers to the 100 $\mu$m peak. 
The optical position of the stars with respect to the 100 $\mu$m peak is indicated by a ``star'' symbol. 
North is up and East to the left. Black contours correspond to 100 $\mu$m  
and red contours to 160 $\mu$m. HIP  13402:  contours   are
5\%, 10\%, 20\%, 40\%, 80\%, 90\% of the peak (100 $\mu$m), and 
10\%, 20\%, 40\%, 80\%, 90\% of the peak (160 $\mu$m). 
HIP 14954: contours are  10\%, 20\%, 40\%,  80\%, 90\% of the peak 
at  both bands.  For both objects the lowest contour corresponds to 3$\sigma$.}
\label{fig:excess_images}
\end{figure*} 

\subsubsection{Steep SED sources}
\label{section:steep_sources}

Some stars  in Table \ref{table:excess_sources}  show an SED  with the
largest excess at 70 $\mu$m and  a decrease with the wavelength at 100
and 160 $\mu$m. The stars with  this behaviour are: HIP 28103, HIP
42438, HIP  43726, HIP  71181, HIP 103389,  HIP 107350 and  HIP 114948
(Figure E.1).
In some of these  stars, the observed
flux  at  160  $\mu$m  does  not exceed  the  photosphere,  i.e.,  the
significance  of the  observed fluxes  is $\leq$  3, and  the spectral
index of the  excess is steeper than the one  corresponding to a black
body  in  the Rayleigh-Jeans  regime.   The  IR  excesses are  already
noticeable  in the  wavelength range  of  the IRS  instrument of  {\it
  Spitzer} in  all cases.  All  are point-like sources at  100 $\mu$m,
which implies FWHM  sizes $\lesssim$ 90 -  140 AU, depending on
the  distance to  the star.  The morphology  of the  excess  with well
defined start-  and end-wavelengths  suggests, in principle,  that the
dust is confined in a well defined narrow ring.

\cite{ertel2012}  made  a  detailed  analysis of  this  behaviour  and
demonstrated  that the  particular SED  of the  discs  provides strong
constraints  on the  dust properties.   They showed  that  the naively
expected narrow ring shape of the  disc is not very well constrained by
the  modelling,  and  found that  the  steep  decrease  of the  SED  is
inconsistent with a power-law  exponent of the grain size distribution
of  --3.5,  expected from  a  standard equilibrium  collisional
cascade  \citep{dohnanyi1969}.   In  contrast,  a   steeper  grain  size
distribution or,  alternatively, an upper  grain size in the  range of
few tens of micrometers would  be implied. This suggests  a strong
underabundance of large (millimeter-sized) grains to be present in the
discs.  \cite{donaldson2012} recently  presented another  debris disc
showing  a  similar  behaviour,  namely HD  3003.  This  disc,  however,  is
significantly younger  and more  massive than the  discs found  in our
survey and challenges  the scenarios suggested by Ertel  et al. (2012)
to explain  the phenomenon. An alternative  scenario of enhanced
stirring of  the planetesimal disk  by the companion star  of HD\,3003
has been proposed.

\subsubsection{Short description of individual objects}
\label{section:description_excess}

As it has already been pointed out, a detailed description of individual 
stars or groups of stars is beyond the scope of this paper. Detailed
observational and/or modelling analysis of some sources  
have been the subject of previous DUNES works (papers already mentioned),
and there are some more papers in preparation. However, in order to 
partly illustrate here the achieved results, we present a brief description together 
with the {\it Herschel} images and contour plots of HIP 13402 and HIP 14954 
(Figure \ref{fig:excess_images}), two stars associated with resolved emission. 
The SEDs of these stars are included in Figure E.1.

{\em HIP 13402}. This K1 V star is located at a distance of 10.35 pc and is 
one of the youngest stars in our sample, $\sim$ 130 -- 400 Myr 
(Table \ref{table:stellar_parameters}). Previously identified as a debris disc 
based on the 70 $\mu$m {\it Spitzer} flux \citep{trilling2008}, the excess is 
clearly  present at 100 and 160 $\mu$m  but it has a modest 
fractional luminosity of the dust.  The star appears point-like in the 
 100 $\mu$m {\it Herschel} PACS image but slightly extended at 160 $\mu$m, 
with an observed size  $12\farcs7  \times 11\farcs4$ (Figure \ref{fig:excess_images}). A quadratic  subtraction 
of the stellar PSF gives an intrinsic size of $5\farcs4  \times 2\farcs6$, 
which corresponds to a projected semi-major axis of 28 AU, and therefore 
slightly larger than the black body radius  
(Table \ref{table:excess_sources}).

{\em HIP 14954}. This F8 star is located at a distance of 22.58 pc and is 
one of the debris discs preserved in the DUNES sample because it was already identified 
as such with {\it Spitzer}. The star hosts a gas giant exoplanet and also is the primary component 
of a physical binary. The age is not well constrained and is in the range of $\sim$ 1 -- 5.7 Gyr. 
The fractional luminosity of the dust is among the 
modest values of our debris disc sample (Table \ref{table:excess_sources}). 

\section{Discussion}
\label{section:discussion}


There  are  many  works  in  the literature  searching  for  potential
correlations  between the  debris  disc characteristics  and the  main
properties  of their  associated  stars -like metallicity,  spectral
type, or age-,  the presence of exoplanets around the  stars or if the
stars                are                multiple               systems
\citep[e.g.][]{habing2001,rieke2005,beichman2006,trilling2008,bryden2009,rodriguez2012}.
This search  is motivated  because it might  provide helpful  hints to
deepen  in the  knowledge  of  the conditions  for  the formation  and
evolution  of planetary  systems. In  the following  we  revisit these
analyses in view of the DUNES discs.

\subsection{Debris discs/stellar metallicity}
\label{section:metallicity}

There  exists   a  well   established  relationship  between   a  high
metallicity in solar-type stars  and the incidence of extrasolar giant
planets               orbiting               around               them
\citep[e.g.][]{santos2004,fischervalenti2005}, although  such trend is
not valid  in the  case of  low mass planets,  $M_{\rm p}  \lesssim 30
~M_\oplus$ \citep[e.g.][]{ghezzi2010,2011arXiv1109.2497M}. In the case
of debris disc stars, the results  from various works do not reveal any
correlation between the  presence of discs and the  metallicity of the
stars  \citep[e.g.][]{bryden2006,trilling2008,moor2011}.  The most  recent
and, to our knowledge, comprehensive study on this issue has
been  carried out  by  \cite{maldonado2012}  who, based  on  a set  of
homogeneously determined stellar  parameters, analysed the metallicity
distribution of different samples of stars. These samples included one
of  107 solar-type  stars with  only debris  discs  and  a control
sample of  stars without  known debris discs  and planets.  They found
that both samples have similar metallicity distributions, but there is
a  hint  pointing  out  to  a  deficit of  stars  with  discs  at  low
metallicities or, in  other words, stars with discs  are slightly more
metal rich than stars without discs \citep[Figures
3 and 7  of][]{maldonado2012}. We have repeated this  analysis for the
DUNES  stars,  differentiating  those with  no  detected  disc
emission and  those with associated  discs.  We have removed  from both
groups the  stars with  known exoplanets  in order  to  avoid a
potential contamination. The average  [Fe/H] for the debris disc stars
(26  objects) is -0.10$\pm$0.18  and a  median of  -0.09, while  for the
non-excess  stars  the  corresponding  values are  -0.15$\pm$0.29  and
-0.13,  respectively. Thus, both metallicity  distributions are
practically  undistinguishable. More  conclusive  statements require  further  data,  in
particular  concerning the  incidence  of low-mass  planets in  debris
discs systems,  to confirm or  unconfirm the mentioned hint, which  might shed
light on  the conditions to  form low mass planet  and/or planetesimal
systems \citep[e.g.][Marshall et al.2013, in preparation; Moro-Mart\'in et al. 2013, in preparation]{bryden2006,greaves2007,moromartin2007}.

\begin{figure*}[ht!]
\centering
\includegraphics[angle=270,scale=0.22]{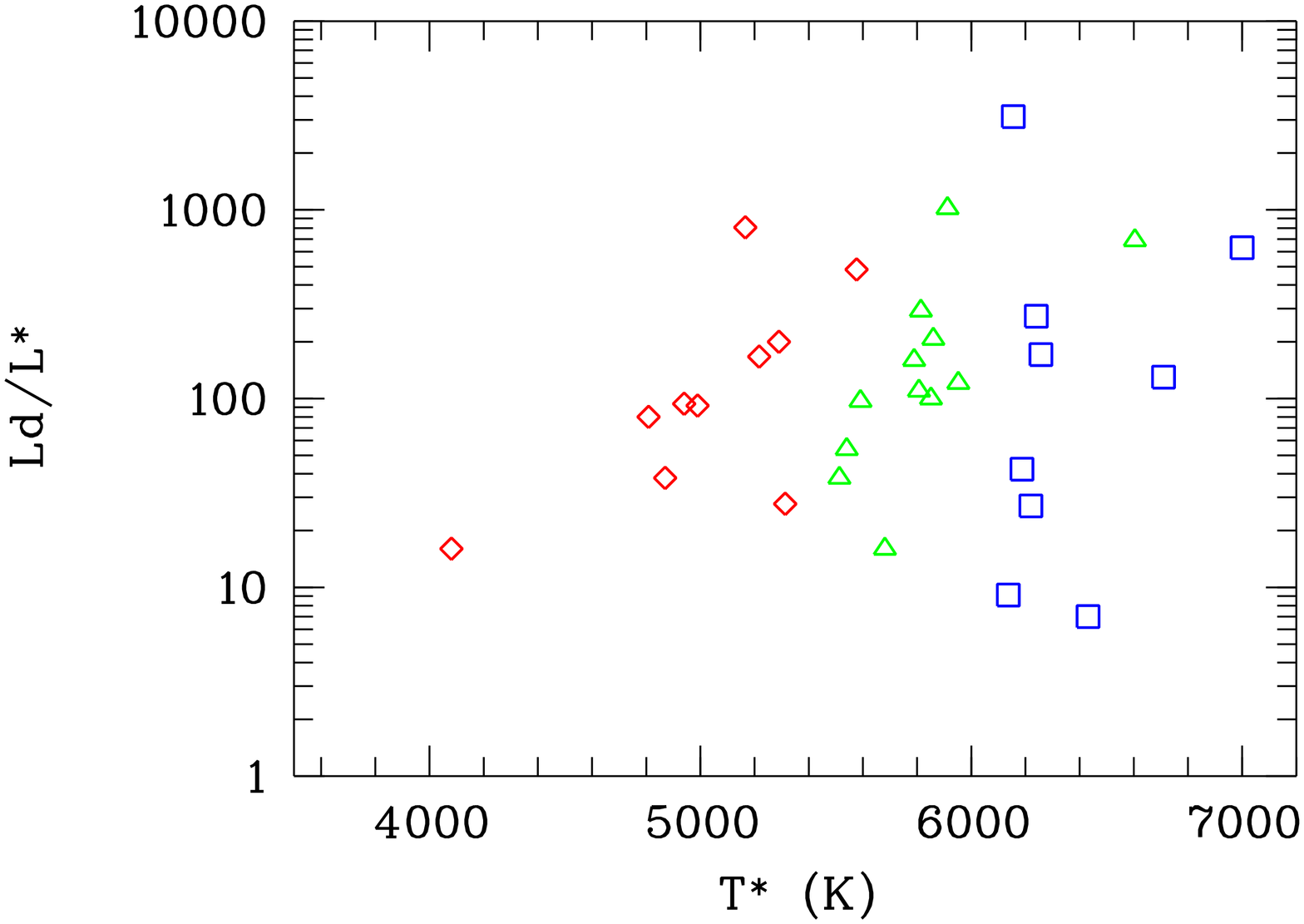}
\includegraphics[angle=270,scale=0.22]{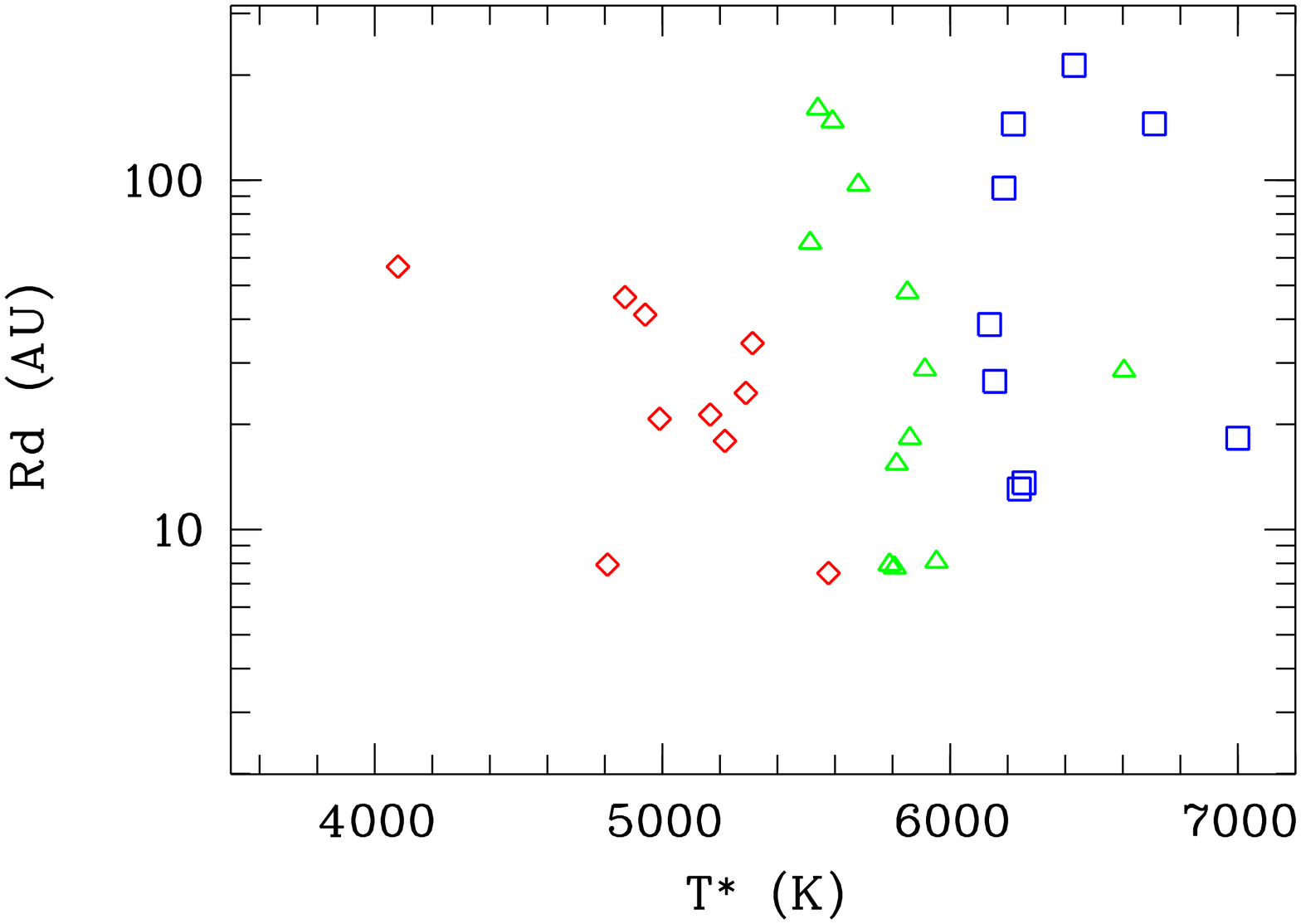}
\includegraphics[angle=270,scale=0.22]{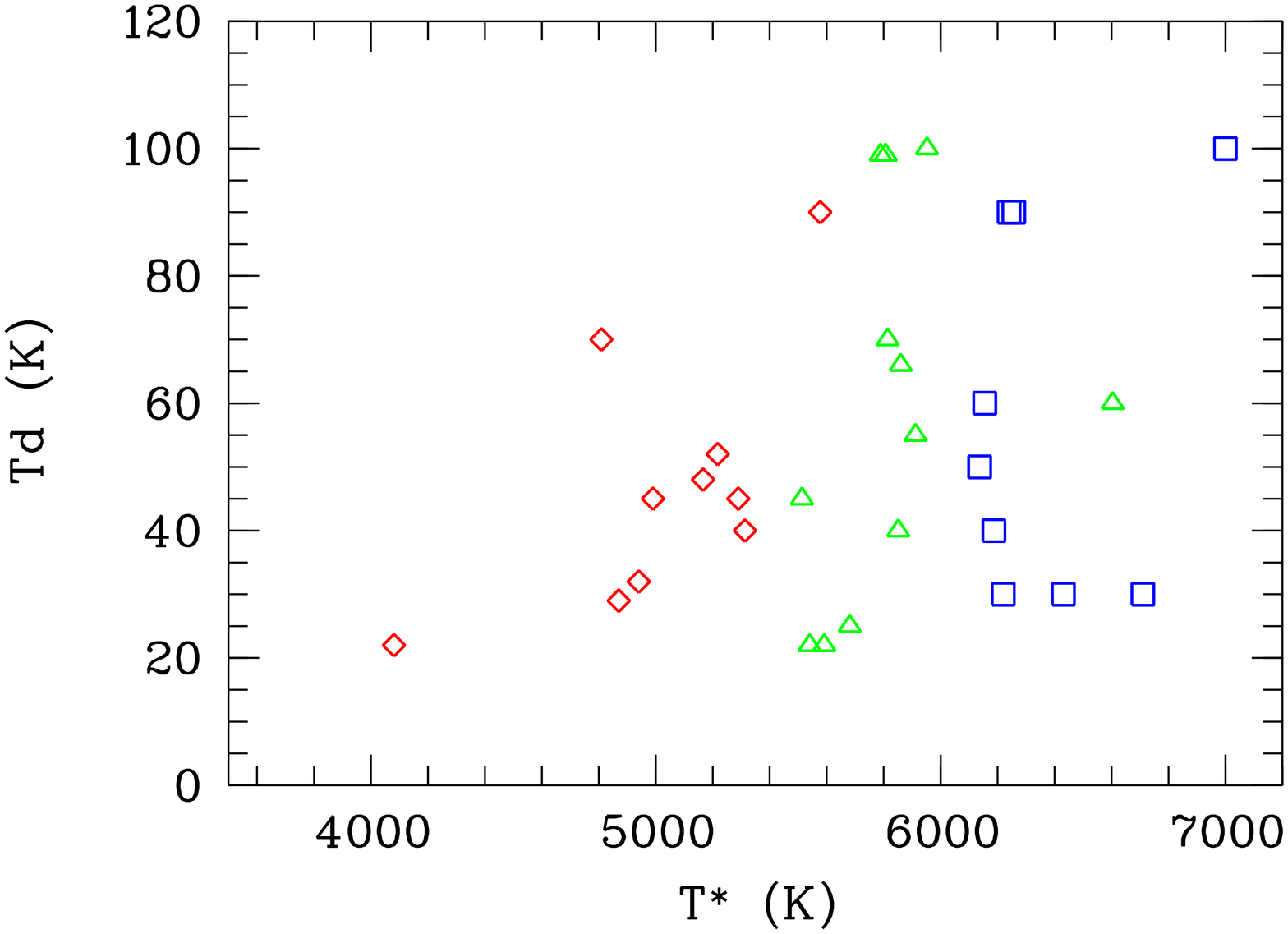}
\includegraphics[angle=270,scale=0.22]{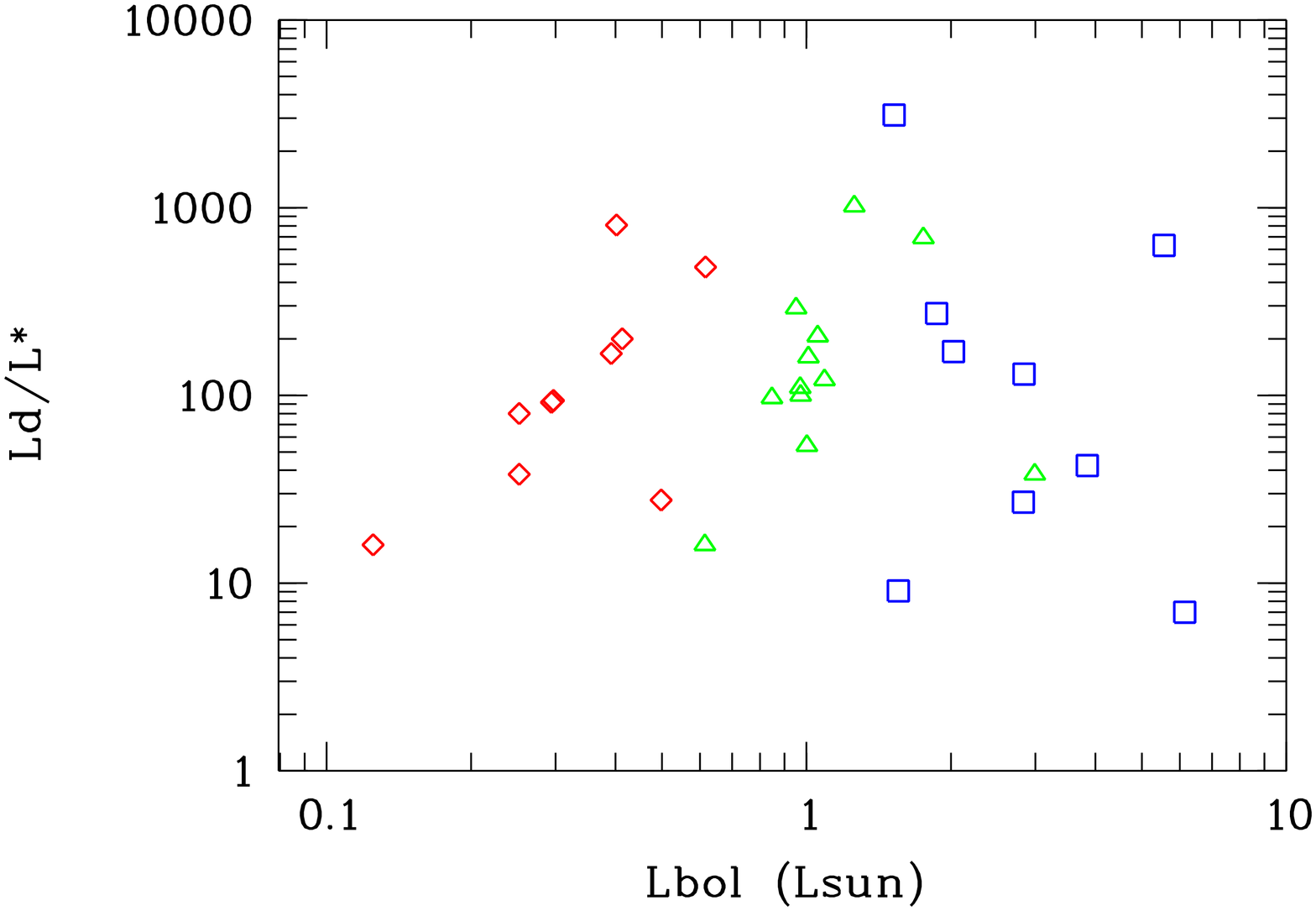}
\includegraphics[angle=270,scale=0.22]{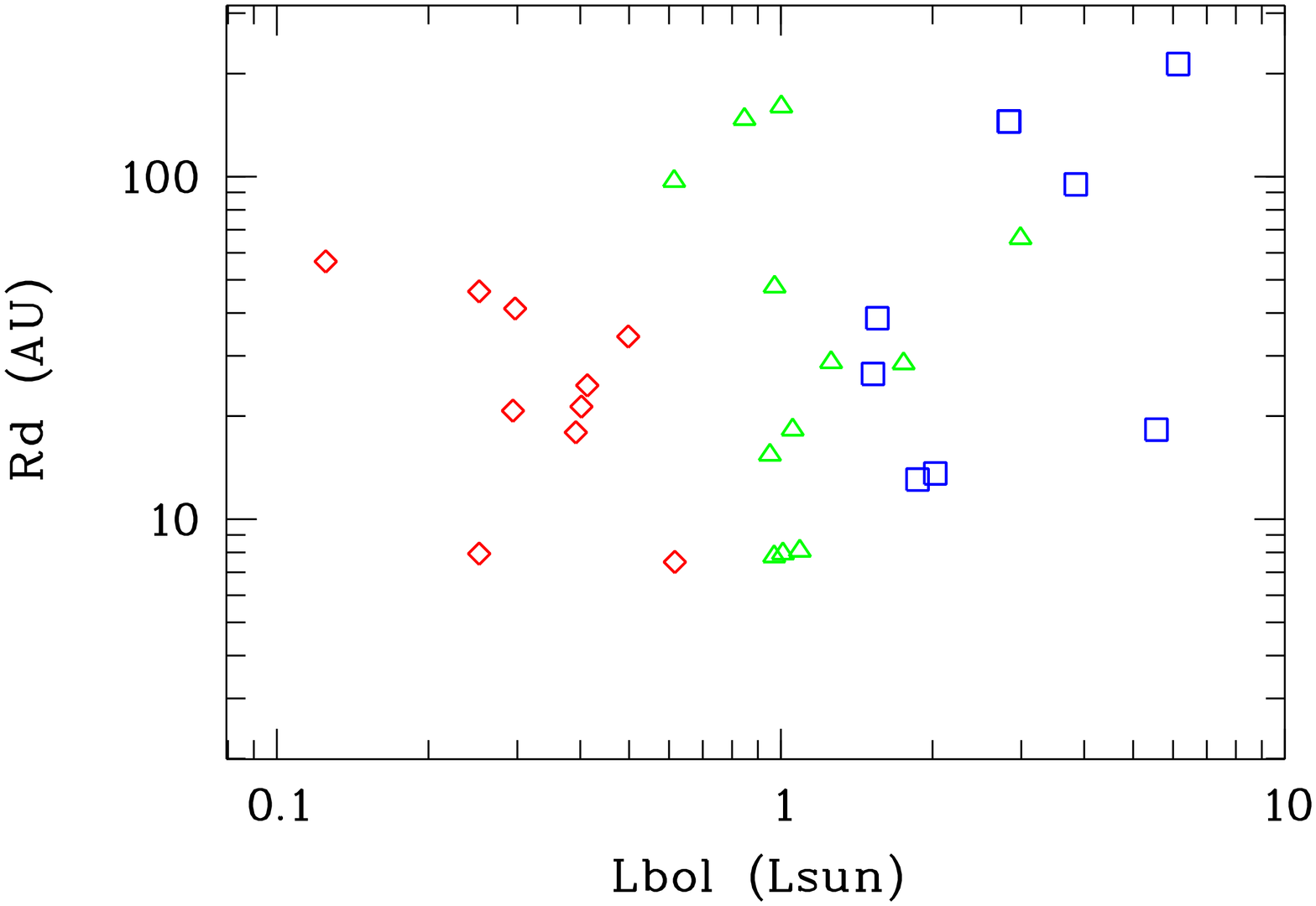}
\includegraphics[angle=270,scale=0.22]{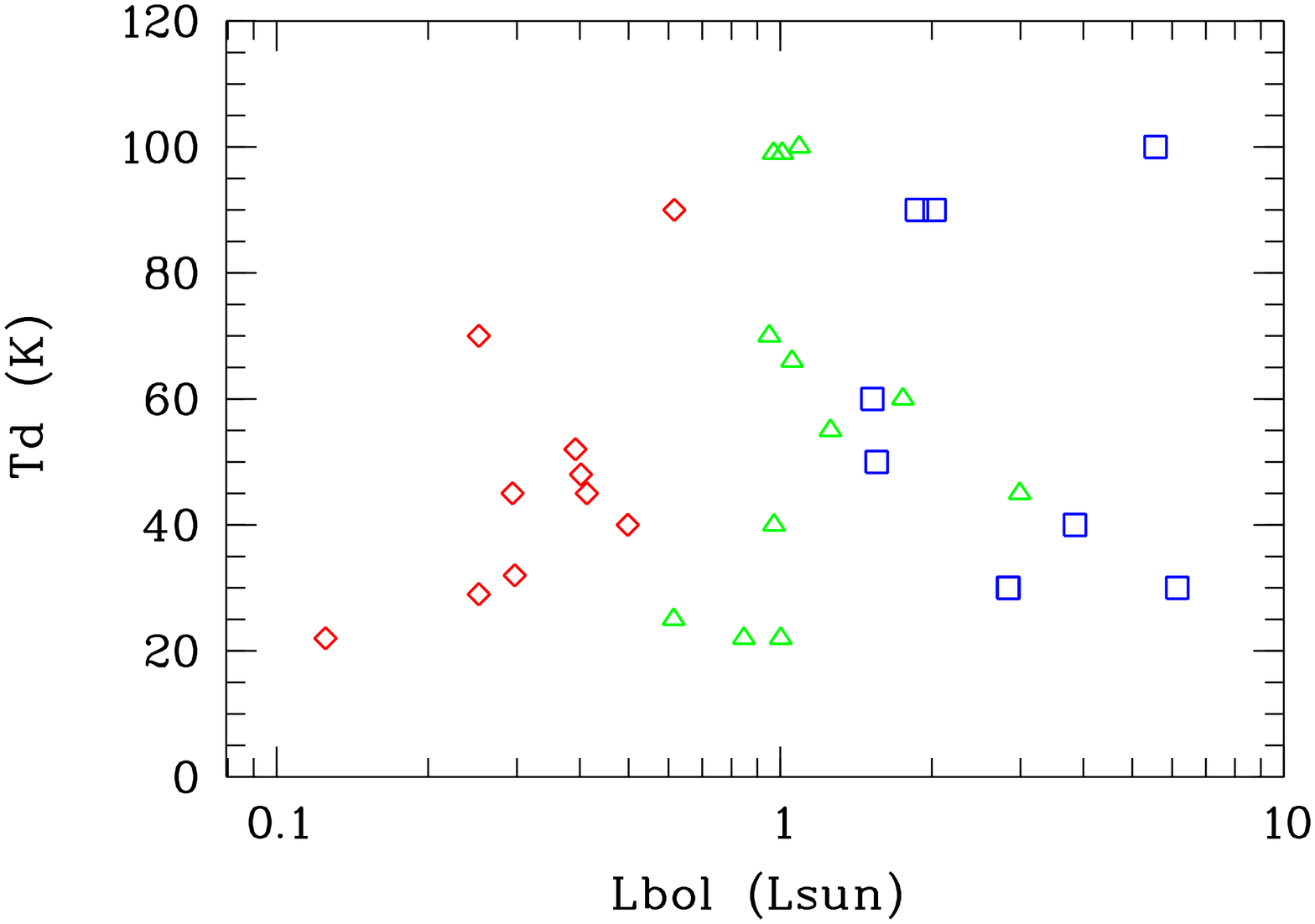}
\includegraphics[angle=270,scale=0.22]{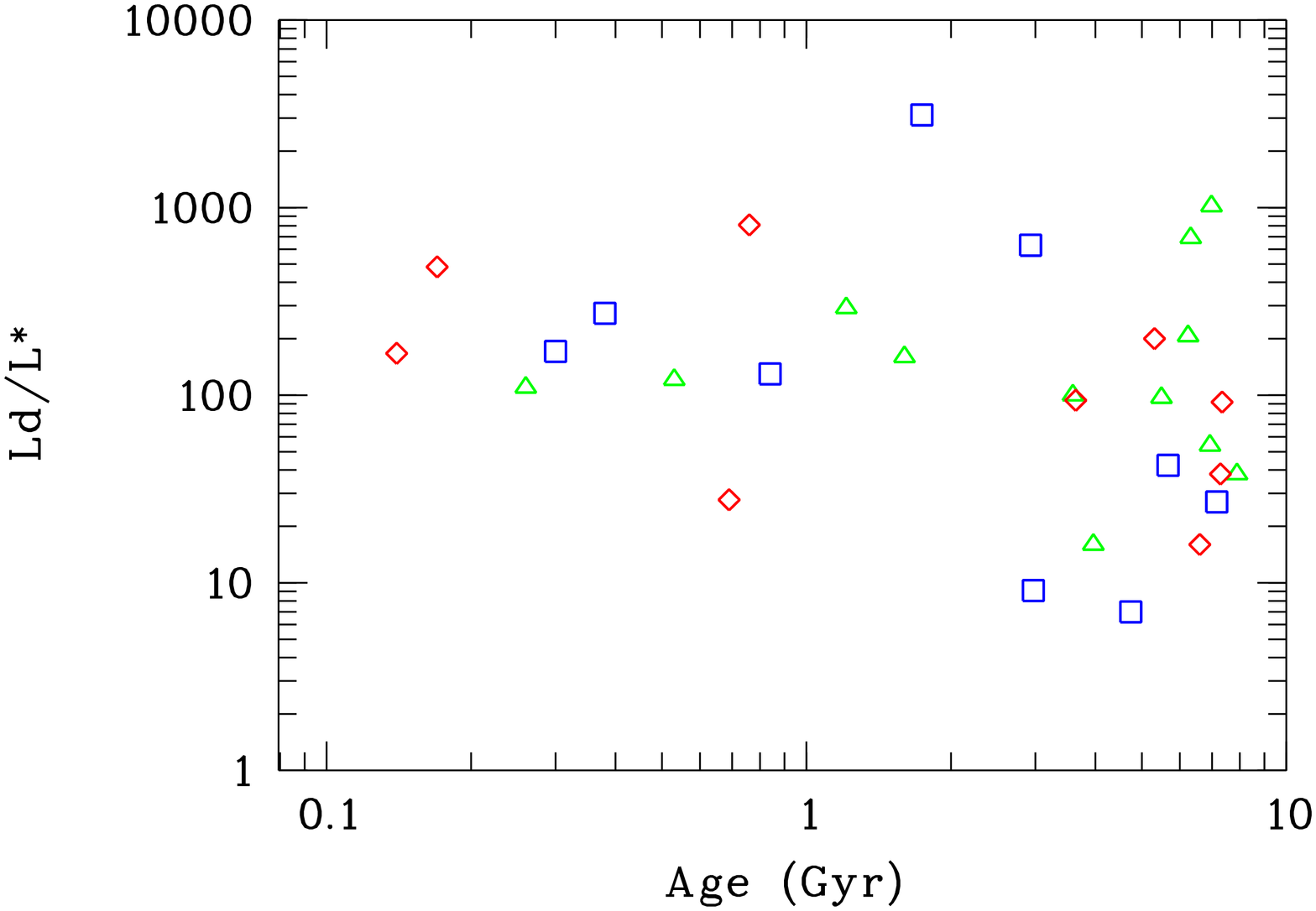}
\includegraphics[angle=270,scale=0.22]{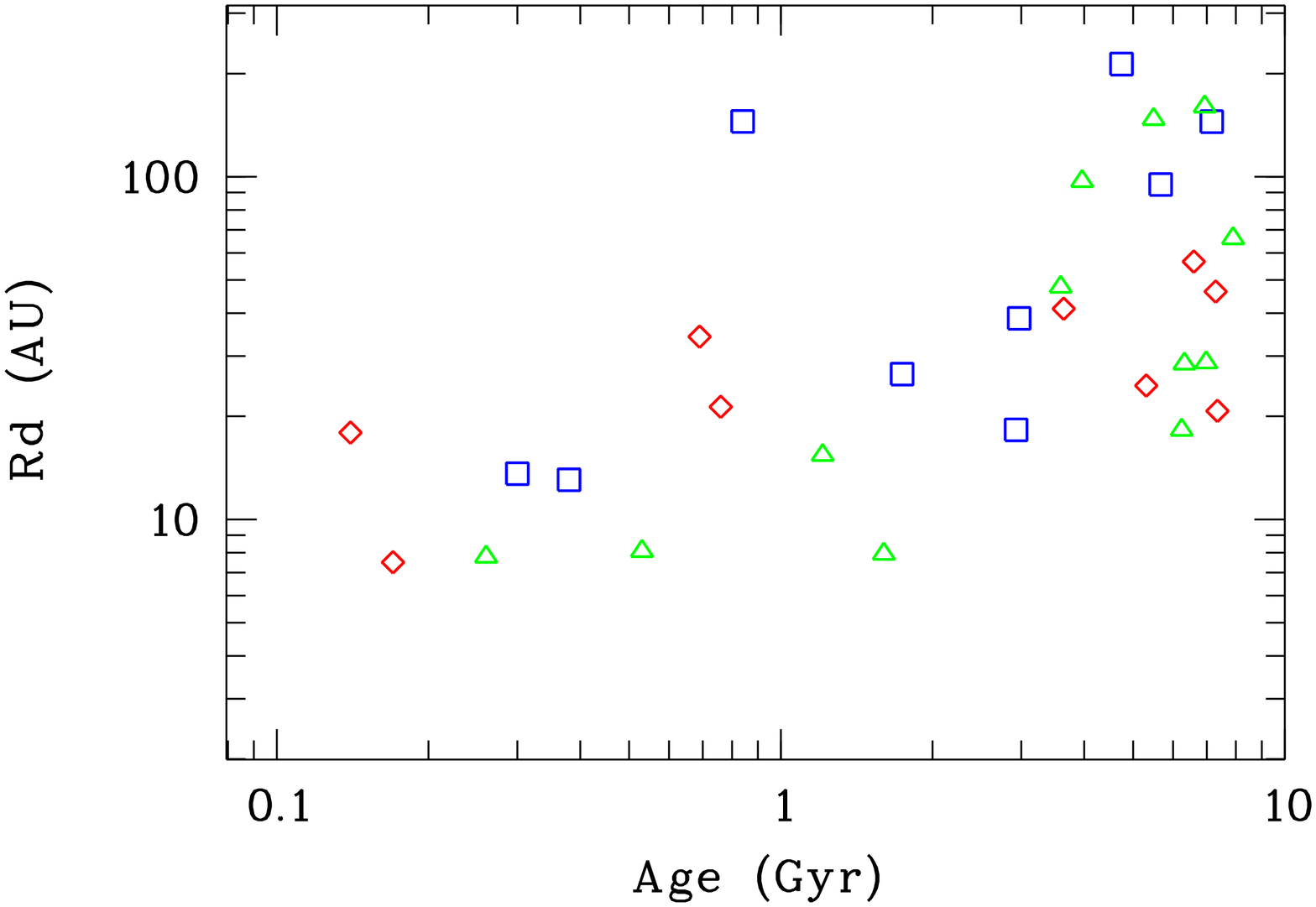}
\includegraphics[angle=270,scale=0.22]{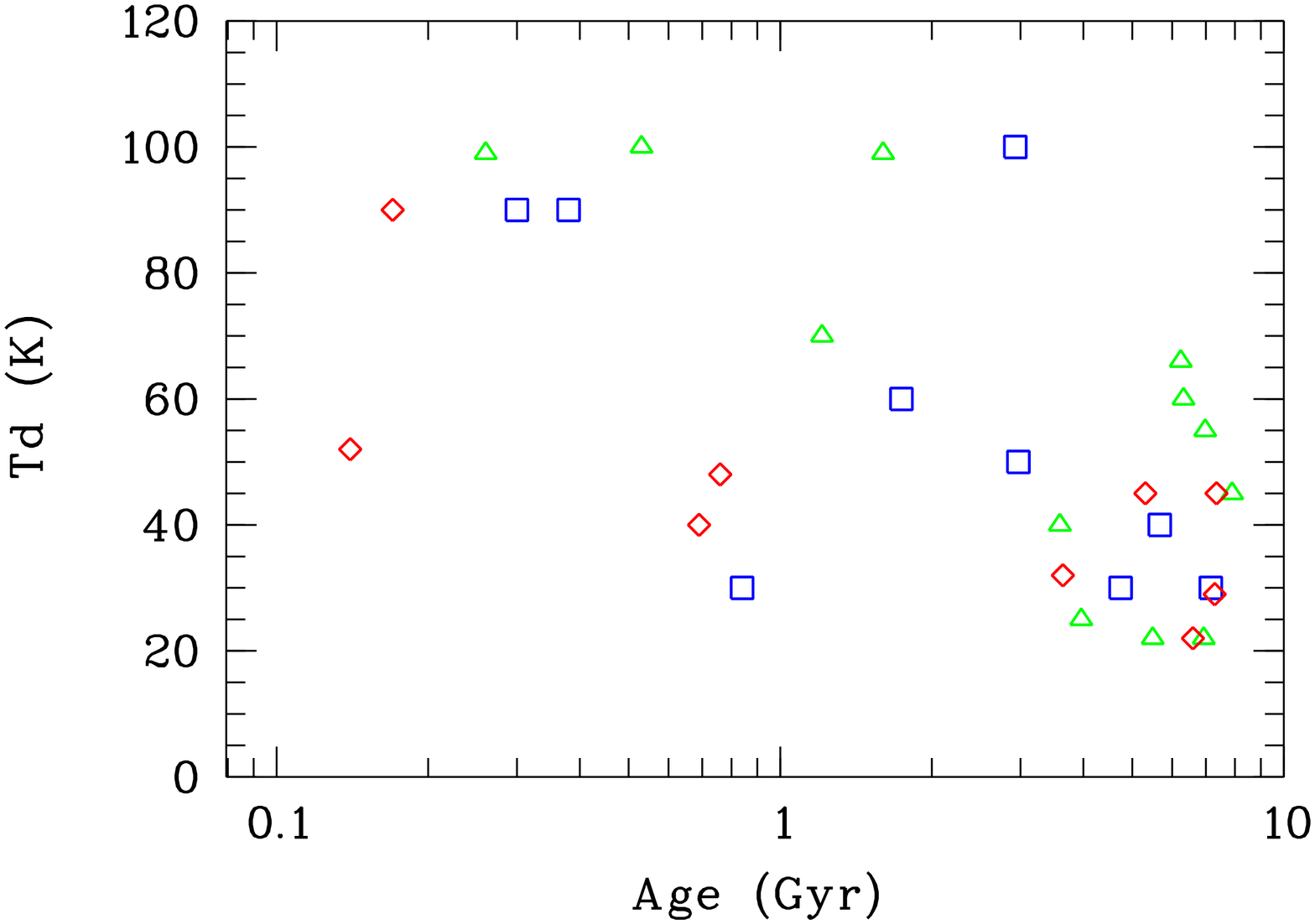}
\caption{Dust properties  versus stellar effective temperature, bolometric luminosity and ages 
(based on the own Ca {\sc ii} activity index). Blue squares: F-type stars; 
green triangles: G-type stars; red diamonds: K-type stars. Units of the fractional luminosity of the dust are
10$^{-7}$. }
\label{fig:dust_starproperties}
\end{figure*}

\subsection{Disc properties/spectral types and bolometric luminosities}
\label{section:spectraltypes}

Figure  \ref{fig:dust_starproperties} (top and middle rows) shows    plots  of   the
fractional  luminosity  of  the   dust,  the  black  body  radius  and
temperature  versus  the  effective  temperature  and  the  bolometric
luminosity of the stars.  The  dust parameters present a large scatter
and  the plots indicate that none of them are  correlated  with the effective 
temperatures and luminosities of the stars, as corroborated by 
Pearson or Spearman correlation tests.   Average
values of  $L_{\rm d}/L_\star$, $T_{\rm d}$  and $R_{\rm d}$ grouping the 
stars by  their F, G and K spectral types agree within  the high
uncertainties  involving these  parameters.  The  only  parameter that
might be weakly correlated with the stars' spectral types is the estimated
average black  body radius,  which might present  a decrease  with the
spectral type:  $<R_{\rm d}>$ = 79$\pm$74 AU for the F-type stars, 53$\pm$54
AU for the G-type, and 28$\pm$16  AU for the K-type stars. The lack of
a  correlation  between the  fractional  luminosity  of  the dust  and
spectral type is also seen in {\it Spitzer} studies \citep[e.g.][]{trilling2008}. We
further note that,  excluding  q$^1$  Eri (HIP  7978),  the range  of
$L_{\rm   d}/L_\star$  does   not  change   much,  although   the  disc
detectability    changes    with    the    spectral    type    (Figure
~\ref{fig:non_excess_spt}).  We also note  the fact that the dust
temperature does not  change with the stars' spectral  type is because
the dust is  located on average at decreasing  distances for the later
stellar spectral types.  Although  the relative values provided by the
simple  black  body  approach  might be  useful  \citep{wyatt2008}  to
provide a basic  view of the discs, it  might be misleading because
realistic grain  properties are  not taken into account.  A firm
interpretation of  our debris disc properties and  their relation with
the  stars'  characteristics  requires  detailed modelling,  which  is
beyond the scope of this work.

\subsection{Debris discs/stellar ages}
\label{section:age}

Stellar age  is one of  the main basic  stellar quantities but  at the
same  time   of  the  most   difficult  to  reliably   determine  for
main-sequence    solar-type    field    stars.    The    theoretical
Hertzsprung-Russell diagram is not very useful since isochrones pile-up
for these spectral types.  There  is a handful of age diagnostics used
as  proxies   \citep{mamajek2008},  but  they   can  give  conflicting
results. However, the use  of consistent age estimates circumvent this
problem, at least partly  \citep[e.g.][]{rieke2005}.  From IRAS, ISO,
and  {\it Spitzer} studies  it is  known  that debris  discs persist  around
solar-type      stars      through      very     long      timescales
\citep{habing2001,decin2003,moor2006,hillenbrand2008}.   The  rate of
debris discs and their dust  luminosity seem to be relatively large up
to ages $\lesssim$ 400-500 Myr. For older stars, there is a small but
not   quite   obvious   decline   up   to   ages   of   several   Gyr
\citep{trilling2008,wyatt2008}. For all ages, the scatter is large.

Figure \ref{fig:dust_starproperties} shows in the bottom panels the
dust fractional luminosity, the black body radius and temperature of
the excess sources as a function of the stellar ages. We have used our
own reduced spectra \citep{martinezarnaiz2010, maldonado2010,
  maldonado2012} to estimate ages in a consistent manner using the
Ca~{\sc ii} $R'_{\rm HK}$ activity index \citep{mamajek2008}. Again
there is a large scatter in $L_{\rm d}/L_\star$ for the stars in the
sample, in particular for ages larger than $\sim$ 3 Gyr where the bulk
of the stars is concentrated.  With respect to $R_{\rm d}$ and $T_{\rm
  d}$, although the values present a scatter, the plots suggest a weak
correlation of $R_{\rm d}$ and an anticorrelation of T$_{\rm d}$ with
the stellar age. These trends hold for the FGK stars as a whole and
individually for each F, G and K spectral types. In fact, a Pearson
correlation test shows that the probability of $R_{\rm d}$ and $T_{\rm
  d}$ to be correlated with the age of all FGK debris disc stars are
$\sim$97\% and larger than 99\%, respectively. The apparent
correlation between $R_{\rm d}$ and stellar age might be a hint for
dynamical inward-out stirring of debris discs.


\begin{table}[ht!]
\centering
\caption{Multiplicity of the excess sources. The separation refers to the
angular distance to the closest identified companion. Column 4 with the black 
body radius of the dust is included with the purpose of a direct comparison.The column "Status" 
informs if the stars are true physical systems (Y), projected but no
true binaries (N) or lack of sufficient information (?).}
\label{table:multiplicity}
\begin{tabular}{lccrl}
\hline\noalign{\smallskip} 
HIP    &Separation&Proj. dist.& R$_{\rm d}$ & Status \\
       &         &  (AU)   & (AU) &  \\ 
\hline\noalign{\smallskip}          
   171 &   0$\farcs$83 & 10.1  & $\geq$97.1 &Y  \\
   544 &               &       &            &?\\ 
 13402 &               &       &            &?\\
 14954 &   3$\farcs$3  &  74.5 & 95.0       &Y  \\
 15371 & 310$\arcsec$  &  3729 & 47.7       &Y  \\ 
 29271 &   3$\farcs$05 &       &            &?    \\
 32480 &  36$\farcs$2  &       &            &N  \\ 
 49908 & 165$\arcsec$  &       &            &N  \\
 51459 & 120$\arcsec$  &  1534 & 38.6       &Y \\
 65721 & 286$\farcs$4  &       &            &N  \\ 
 72848 &               &  0.52 & 34.2       &Y\\ 
 92043 &  12$\farcs$6  &       &            &? \\ 
107350 &  43$\farcs$2  &  795  & 8.1        &Y\\
107649 &  55$\arcsec$  &       &            &N  \\
114948 &               &       &            &?   \\
\hline\noalign{\smallskip}
\end{tabular}
\end{table}

\subsection{Debris discs/stellar multiplicity}
\label{section:multiplicity}

The number  of DUNES stars with  an entry in the  catalogues of binary
and  multiple stars CCDM (Catalog of Components of Double and Multiple Stars),  WDS \citep[Washington Double Star Catalog,][version 2012]{wds},  and SB9
\citep[The 9th Catalogue of Spectroscopic Binary Orbits,][version 2012]{sb9} is 83 which  represents $\sim$ 62 \% of the
sample.  Not all  entries correspond to real multiple  systems. Out of
these 83 entries,  15  correspond  to  stars  identified  as  excess
sources. Table  \ref{table:multiplicity} lists these  stars, where the
angular and linear distances with the closest companions are given for
the cases where  such information is available. In  the last column on
that  table, the stars  with true  identified physical  companions are
given. This association  is based on the proper  motions and distances
for HIP 14954,  HIP 15371, HIP 51459. The stars HIP  171 and HIP 72848
are  spectroscopy  binaries with  well  determined orbital  parameters
\citep{bach2009,halbwachs2003,bonavita2007}. HIP 107350 is accompanied
by   a  substellar   object  with   a   mass  in   the  brown dwarf  regime
\citep{luhman2007}.    HIP   29271    has   a    companion   candidate
\citep{eggenberger2007}  but  no   further  information.  The  optical
companions  of HIP 32480,  HIP 49908,  HIP 65721  and HIP  107649 have
proper motions very different from  the associated stars, so that they
most likely are no real  physical companions.  There is no information
to our knowledge,  beyond the entry in one  of the mentioned catalogues,
for HIP  544, HIP 13402  and HIP 114948.  As a summary, we  can firmly
identify  6 multiple  systems  out  of 31  debris  disc stars.  This
implies  a  rate  of  $\sim$  20   \%,  close  to  the  one  found  by
\cite{rodriguez2012}.  Concerning the stellar  or dust  properties, the
binary stars do not have any  significant trend. We note that our numbers
are not  statistically significant enough to confirm or deny  the claim by
\cite{rodriguez2012}  of  a   lower  fractional  disc  luminosity  for
multiple systems than for single stars.

The comparison of  the binary separation and the  location of the dust
(columns 3 and 4  of Table \ref{table:multiplicity}) is informative to
assess  if the dust  resides in  stable orbits.  The discs  around HIP
15371, HIP  51459 and HIP  107350 are clearly circumstellar  since all
the binary systems  are very wide, while the discs  around HIP 171 and
HIP 7248  are circumbinary. These  discs are located in  stable orbits
\citep[e.g.][for      a     comparison      with      other     debris
discs]{trilling2007,rodriguez2012}.  The only questionable case is HIP
14954, where the  black body radius and binary  separation are similar
and place  the dust in an unstable  location \citep{rodriguez2012}. In order 
to assess stability issues of the debris discs versus the binarity of the stars,
a specific study using realistic grain properties informing on  plausible dust
locations and temperatures is required .

\subsection{Debris discs/planets}
\label{section:planets}

The number  of known stars hosting  exoplanets in the  DUNES sample is
21. This figure  includes HIP 107350 with a companion  of mass M$_{\rm P}$ = 16 M$_{\rm J}$,
i.e., in  the brown  dwarf mass  regime although it  is listed  in the
Extrasolar Planets  Encyclopaedia (http://exoplanet.eu). Six  of these
stars - HIP 7978, HIP 14954, HIP 27887, HIP 65721, HIP 107350, and HIP
109378  - have  debris disc  detected with  {\it Herschel}.  A further
star, HIP 40693 has a warm disc \citep{beichman2005b} but no excess in
our images. HIP 27887 and HIP 109378 are new debris disc detections by
{\it Herschel}. If we consider the 20 pc DUNES sample, the number of stars 
hosting planets is 16, out of which 4 have debris discs (HIP 107350 is not 
included). Thus, the incidence rate of debris discs among exoplanet hosts is 
25\% $\pm$ 5\%, i.e., only marginally larger than the fraction of discs 
 around stars irrespective of whether they host a planet. 
HIP 7978, HIP  107350 and HIP 109378 have been studied
by    \cite{liseau2010},   \cite{ertel2012}    and   \cite{eiroa2011},
respectively. All DUNES and DEBRIS stars hosting exoplanets, including
the  disc non-detections,  currently  are the  subject  of a  detailed
treatment  in  two  papers   in  preparation  (Marshall  et  al.,  and
Moro-Mart\'in et al.).

\section{Summary and conclusions}
\label{section:summary}
The purpose of this paper has been to present the observational results of the {\it Herschel} 
Open Time Key  Project DUNES. The stellar sample consisted of 124 main-sequence, nearby 
(distances less than 20 pc), solar-type FGK stars, plus 9 stars with previously
known debris discs or exoplanets, located at distances between 20 and 25 pc.    
Infrared excesses have been detected around a total of 31 stars out of the 133 
stars and 25 stars out of the 124 stars in the 20 pc subsample. 
This represents a total increase of 10 new debris discs with respect to the ones previously known 
in the whole sample, and an increase in the incidence of debris discs in the 
20 pc subsample from 12.1\%$\pm$5\%  to 20.2\%$\pm$2\%, i.e., around 1.7 times larger. The gain in 
incidence rate varies with the spectral type, being larger for the K-type stars.  
The achieved mean sensitivity is a function of stellar spectral type and distance, 
and the 3$\sigma$ mean upper limit of the fractional luminosity of the dust for non-excess 
sources is 
$L_d/L_\star\sim$2.0$\times$10$^{-6}$, with the lowest values 
$L_d/L_\star \lesssim 4.0 \times 10^{-7}$ corresponding to the closest stars. 
This is a gain of around one order of magnitude against the detection 
limit of {\it Spitzer}. Among the debris disc stars, a few of them show 
$L_d/L_\star$, a few times larger than the EKB value, albeit these disks 
are larger and colder than the predicted EKB dust disk. 

The number of stars with spatially resolved emission is 16, which is a rate of 52\% among 
the identified debris discs, and a huge gain with respect to the previously known resolved
sources (3 objects). In addition, few sources show  excess emission at 
160 $\mu$m and very faint or no excess at 100 $\mu$m which is attributed to a 
new class of cold and faint debris discs. Although it cannot be excluded that 
some of these sources suffer from coincidental alignment with background galaxies, 
the probability that some of these cold disc candidates are true debris discs is
very large. In addition, we have found that some discs show far-IR spectral indexes
steeper than the black body Rayleigh-Jeans index. Both types of cold and steep-SED 
debris discs cannot easily be accommodated to the classical equilibrium collisional
cascade scenario of debris discs.

An analysis of the debris disc parameters with stellar properties shows a weak trend 
of a correlation of the black body dust radii (the location of the dust) and an 
anticorrelation of the dust temperatures with the stellar age. This trend holds for all 
FGK spectral types as a whole, and for each F, G and K spectral types separately. No other correlation is 
found with the (possible) exception of a hint showing a decrease 
of the average black body dust radii from the F to the K spectral type stars.
In-depth observational and modelling analysis of the DUNES debris discs will be 
published elsewhere.

The DUNES survey results provide a legacy value useful to the broad community, 
accomplishing in that way one of the rules for {\it Herschel} Key Programmes. 

\begin{acknowledgements}

C. Eiroa, J. Maldonado, J. P. Marshall, G. Meeus and B. Montesinos were  
supported by  the  Spanish grants AYA2008-01727 and AYA2011-26202
  A. Bayo was partly supported by the Marie Curie Actions of the European Commision (FT7-COFUND). 
J. Sanz was supported by Spanish grants AYA2008-02038 and  AYA2011-30147-C03-03.
A. V. Krivov and T. L\"ohne acknowledge support by the German DFG, grants Kr 2164/10-1 and 
Lo 1715/1-1. NASA support for this work (D. Ardila, Ch. Beichmann, G. Bryden, W. Danchi, A. Roberge, K. Stapelfeldt) was provided through an award issued by
JPL/Caltech.

\end{acknowledgements}


\bibliographystyle{aa}
\bibliography{dunes_results_2013March15}

\begin{appendix}
\section{DUNES Virtual Observatory tool}
\label{DUNES_VO}
The achievement of the  DUNES objectives requires a detailed knowledge
of the properties and environment  of the targets to be studied. There
exists a huge  amount of astrophysical data and  information about the
DUNES   objects,   distributed   in   a   number   of   archives   and
services. Gathering information in a wide variety of types and formats
from a large  number of heterogeneous astronomical data  services is a
tedious, very  time consuming task even  for a modest  data set.  With
this aim we have developed  a Virtual Observatory\footnote{The VO is a project designed to
provided the astronomical community with the data access and the research tools 
necessary to enable the exploration of the digital, multi-wavelength universe
resident in the astronomical data archives. http://www.ivoa.net} (VO) application for
accessing,  visualizing  and  downloading  the  information  on  DUNES
targets available in astronomical  archives and services. Given a list
of  objects, identified  by their  names or  coordinates,  a real-time
exploration       of      Vizier\footnote{http://vizier.u-strasbg.fr/c
  gi-bin/VizieR} using VO protocols is performed to gather photometric
data  as  well  as  physical  parameters.   This  information  can  be
complemented with  searches of images,  spectra and catalogues  in all
the Virtual Observatory services. Moreover, ad hoc access to other non
VO-compliant services of interest (like NStED or Spitzer/FEPS) is also
provided.   The  obtained  information  can be  downloaded  in  ASCII,
VOTable (standard format for tabular data in the Virtual Observatory) 
or HTML format. For heavy queries, the tool implements a batch
mode informing  the user via e-mail  when the search  is complete. The
message includes a link to  the data through which the information can
be downloaded. 

In addition,  one of the goals  of the DUNES consortium  is to provide
the astronomical community with a legacy VO-compliant archive, as also
requested  by rules  of the  {\em Herschel}  OTKPs. The  DUNES Archive
System\footnote{http://sdc.cab.inta-csic.es/dunes}   is   designed  to
ensure that other research groups gain easy access to  
both {\it  Herschel} reduced data and ancillary data
(photometry  and physical  parameters  gathered from  VO services), as well
as to the DUNES VO discovery tool, the DUNES project web page
as well as  to a section including news on the  archive. A HelpDesk to
pose  questions  to archive  staff  is  also  available. 
\end{appendix}

\begin{appendix}
\section{Stellar fluxes and parameters}
\label{appendix:stellar_properties}

Table 3, with several subtables,  presents the magnitudes and fluxes of the DUNES stars which have been 
used to trace their spectral energy distributions. Optical, near-IR,
WISE, AKARI, IRAS and {\it Spitzer} data are included. 

Table 4 gives some relevant parameters of the stars. T$_{\rm eff}$, $\log g$ and 
[Fe/H] are average values of photometric and spectroscopic estimates mainly from 
\cite{gray2003,santos2004,takeda2005,valentifischer2005,gray2006,fuhrmann2008,
sousa2008,holmberg2009}. Rotational velocity values are taken from 
\cite{martinezarnaiz2010}. Bolometric luminosities and stellar radii have been
estimated from the absolute magnitudes and bolometric corrections using the measurements by \cite{flower1996}. 
The activity index $\log R'_{\rm HK}$ has been taken 
from \cite{martinezarnaiz2010} while we have derived the X-ray luminosities based on ROSAT, XMM and {\it Chandra} data. 
The table also provides ages based on the  $\log R'_{\rm HK}$ index
and on the X-Ray luminosities as estimated by \cite{maldonado2010}. 
There is a wide range of age estimates in the literature using 
different tracers for the DUNES stars. Stellar ages of our targets are difficult to estimate 
on the basis of isochrones given that the stars are located on the 
main-sequence (see Figure \ref{HR_diagram}) and that they are sensitive to T$_{\rm eff}$  and
metallicity \citep{holmberg2009}. Thus, we have opted to give in Table 4 the age estimates based 
on our own coherent data set and procedure.  
 
\end{appendix}

\begin{appendix}
\section{Prediction of photospheric fluxes at the PACS and SPIRE wavelengths.}
\label{appendix:predictions}
\subsection{Models}
\label{Section:Models}

The behaviour of three families of model atmospheres was
studied in order to choose the best option for the photospheric work
of the project: PHOENIX/GAIA (Brott \& Hauschildt, 2005), ATLAS9
\citep{castelli2003} and MARCS \citep{gustafsson2008}.  
It was found that for $T_{\rm eff}\geq 5000$
K the three sets of models are virtually identical. In the interval
4000--5000 K the models start to show some differences which are more
pronounced towards lower temperatures and shorter wavelengths, the
models being identical for $\lambda>4$ $\mu$m. For lower temperatures
--only seven DUNES stars have $T_{\rm eff}$ below 4000 K-- the three
sets of models present larger discrepancies, with ATLAS9 being more
different when compared with the other two families. 

The PHOENIX/GAIA set of models was finally chosen because of its finer grid in
effective temperatures, sampling of the individual model spectra and
overall behaviour. The models were computed in LTE, the
opacity treated with the opacity sampling formalism, and more than 300
million lines were included. The synthetic spectra have a variable
amount of wavelength points, between 50,000 and 55,000, cover the
interval 0.001--50 $\mu$m, $v_{\rm turb}$ was set to 2 km/s, the
mixing length parameter is 1.5 and the geometry is plane parallel, or
spherical in those cases where that one is not correct. 

Due to their extremely large resolution, the synthetic spectra were
smoothed with a gaussian filter with FWHM=0.005  after taking the
decimal logarithm of the wavelengths in angstroms. Following that, the
wavelength scale was put back in physical units.

A grid of 1980 spectra (55 temperatures $\times$ 6 gravities $\times$
6 metallicities) was available. The ranges covered are 3000 K $< T_{\rm
eff}<$9800 K, $3.0 < \log g < 5.5$ (step 0.5 dex) and $-2.0 <{\rm
[Fe/H]}< +0.5$ (step 0.5 dex). 

In general, the synthetic spectrum for a given star is not contained
in the grid, therefore, an interpolation in three
dimensions had to be done. Since the PHOENIX/GAIA models only run up
to $\lambda\!=\!40$ $\mu$m, an extension up to 4 mm using the
Rayleigh-Jeans approximation was attached to the original model.

\subsection{Normalization of the models to the observed SED}
\label{Section:Normalization}

The normalization of the model photosphere to the observed SED was
done using the procedure outlined by \cite{bertone2004}.  The
monochromatic fluxes of the SED, $s(\lambda)$ (in units of Jy), were
compared with those of the synthetic model, $m(\lambda)$ (in the same units),
at the corresponding wavelengths, deriving a residual function:

\begin{equation}
X(\lambda_i)= \ln s(\lambda_i) - \ln m(\lambda_i) + k 
\end{equation}

The offset constant is such that 

\begin{equation}
\sum_i X(\lambda_i)=0
\end{equation}

\noindent so $k=<\ln m(\lambda)-\ln s(\lambda)>$.

Five subsets of the full SED were chosen to carry out five
normalizations, namely, VI+nIR, BVI+nIR, VI+nIR+WISE, nIR+WISE and
VI+nIR+WISE. The near infrared photometry (nIR) consists of 2MASS
JHK$_{\rm s}$ (only magnitudes with quality flags ``A'' or ``B'' were
considered), plus additional JHKL points, when available. WISE band W1
(3.35 $\mu$m) was used in most of the cases despite of being nominally
saturated\footnote{http://wise2.ipac.caltech.edu/docs/release/allsky/expsup/sec6\_3d.html}
because PSF photometry was carried out on the images and therefore the
values provided in the all-sky release turned out to be usable
(Stapelfeldt, private communication), only the brightest targets
showed unacceptable values of W1 magnitudes; WISE W2 (4.60 $\mu$m)
photometry was never used; its flux level, when looking at the SED as a
whole, always deviates from the overall behaviour; WISE W3 (11.56
$\mu$m) was always used unless it was brighter than the saturation
magnitude; WISE W4 (22.09 $\mu$m) was also always used unless the
shape of the SED indicated that an excess could start around that
wavelength.

Since each normalization was done with a different number of points
(e.g. degrees of freedom), a reduced $\chi^2$ was computed for each
one in order to make a comparison of all of them.  The selected
normalization was that with the least reduced $\chi^2$ and was used to
predict the fluxes at the PACS and SPIRE wavelengths.  The
  uncertainties in the individual photospheric fluxes were estimated
  by computing the total $\sigma$ of the normalization, in logarithmic
  units; in this calculation the observed flux at each wavelength
  involved in the normalization process was compared with its
  corresponding predicted flux. The result is that the normalized
  model $\log S(\lambda)$ can be allowed to move up and down a
  quantity $\pm\sigma$. That value of $\sigma$ was then translated
  into individual --linear-- uncertainties of the fluxes at the
  relevant {\it Herschel} wavelengths.

\end{appendix}


\begin{appendix}
\section{Spurious sources} 
\label{section:controversial}

\begin{figure*}[h!]
\centering
\includegraphics[scale=0.28]{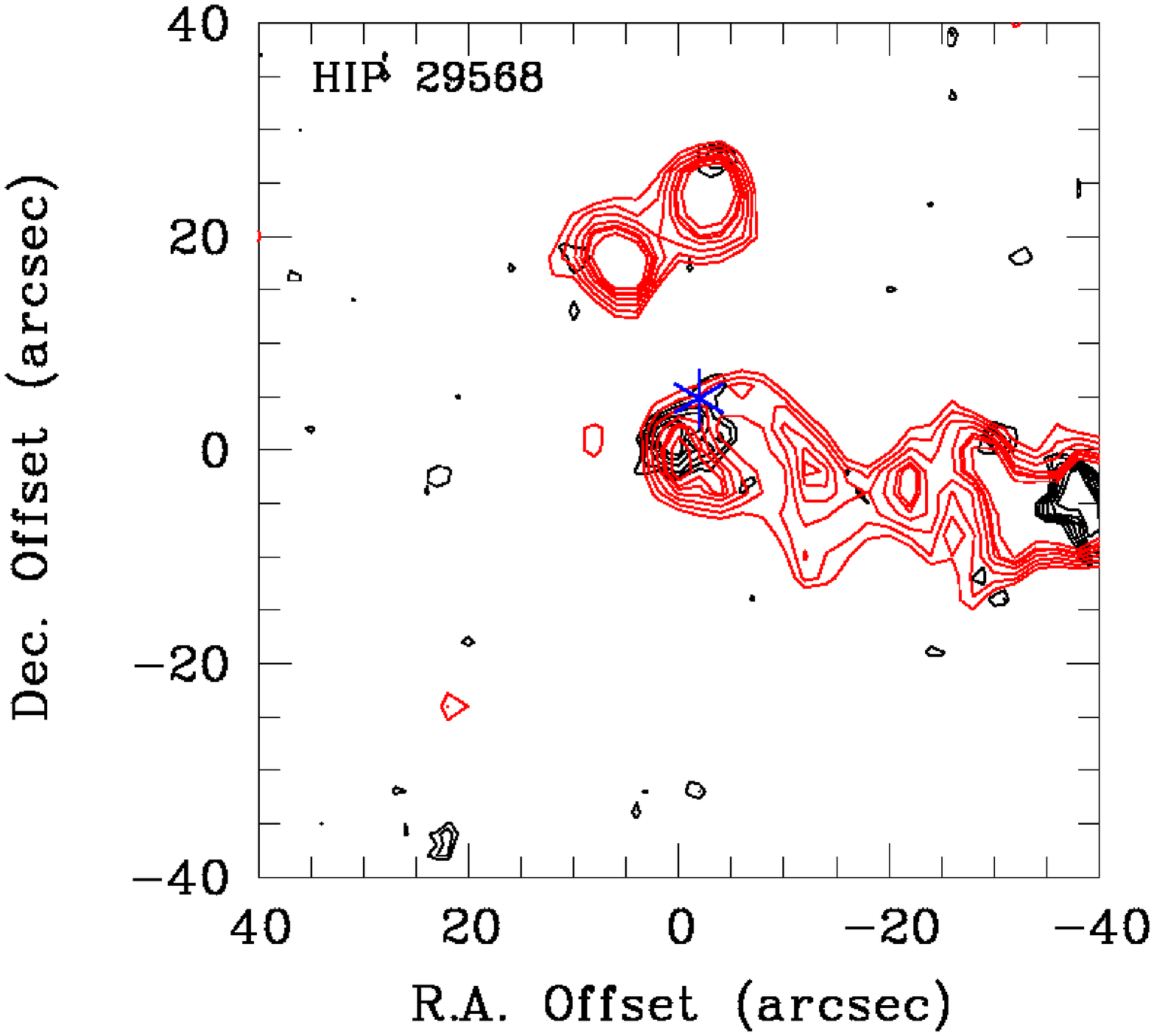}
\includegraphics[scale=0.18]{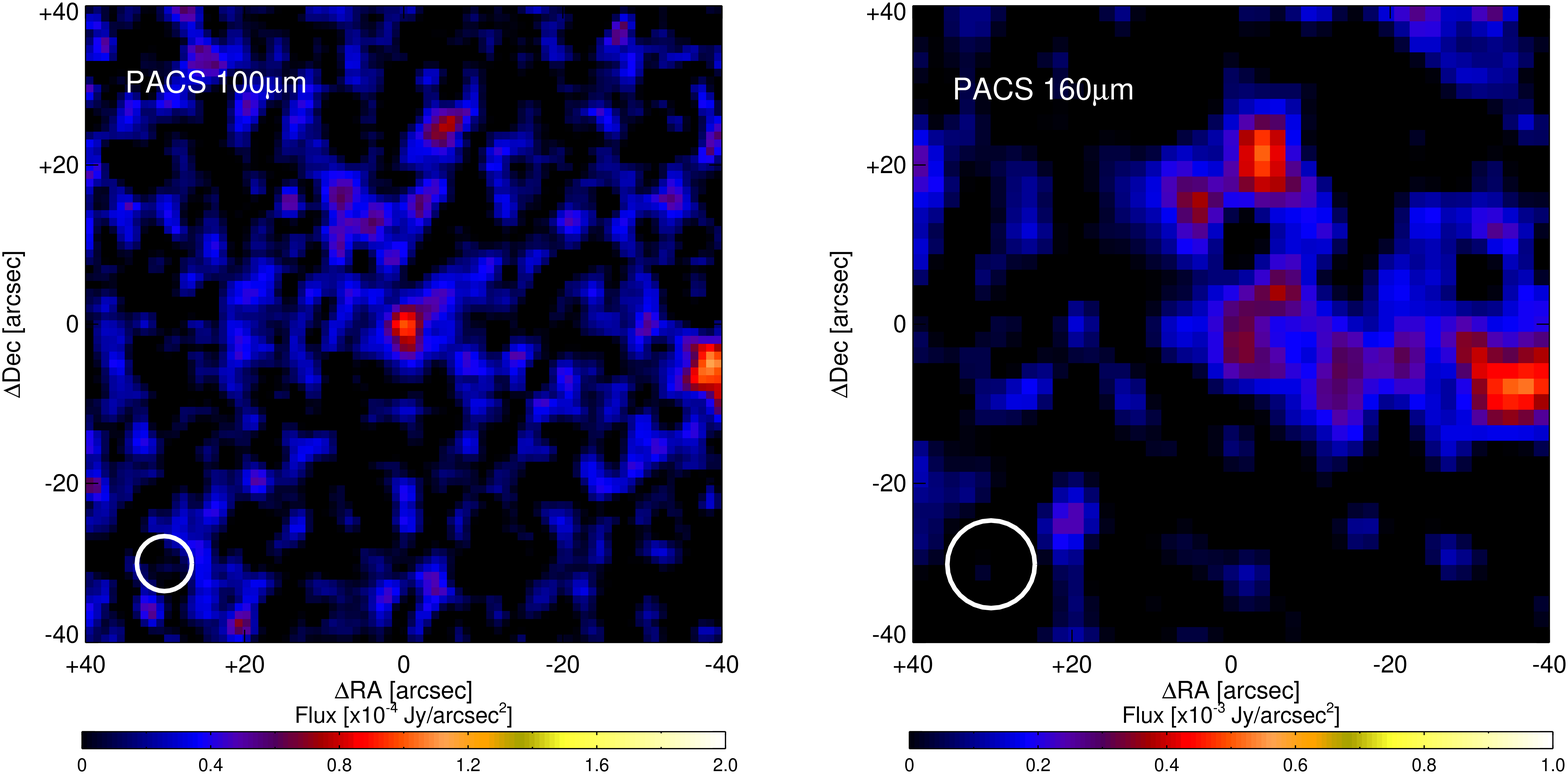}
\includegraphics[scale=0.28]{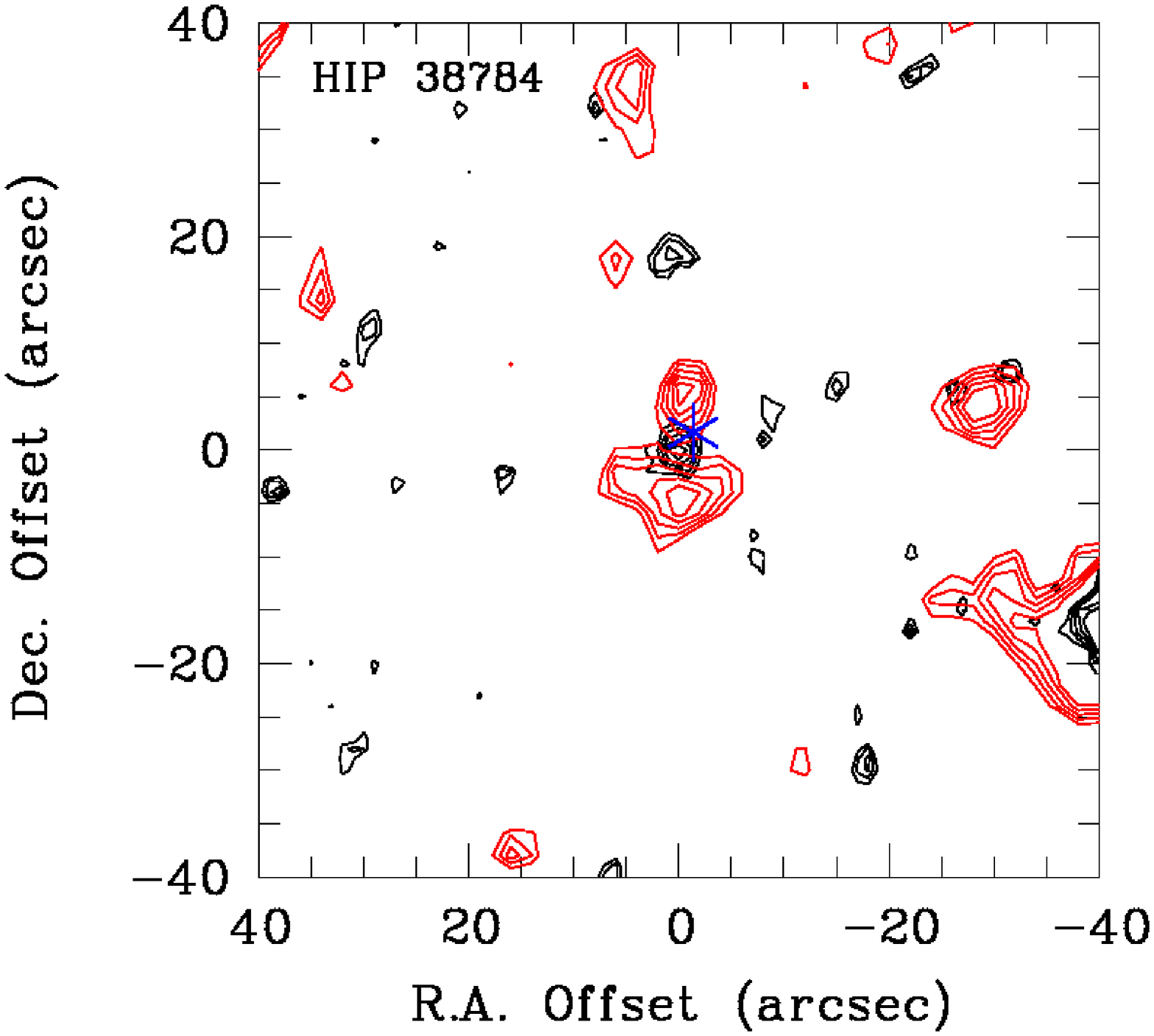}
\includegraphics[scale=0.18]{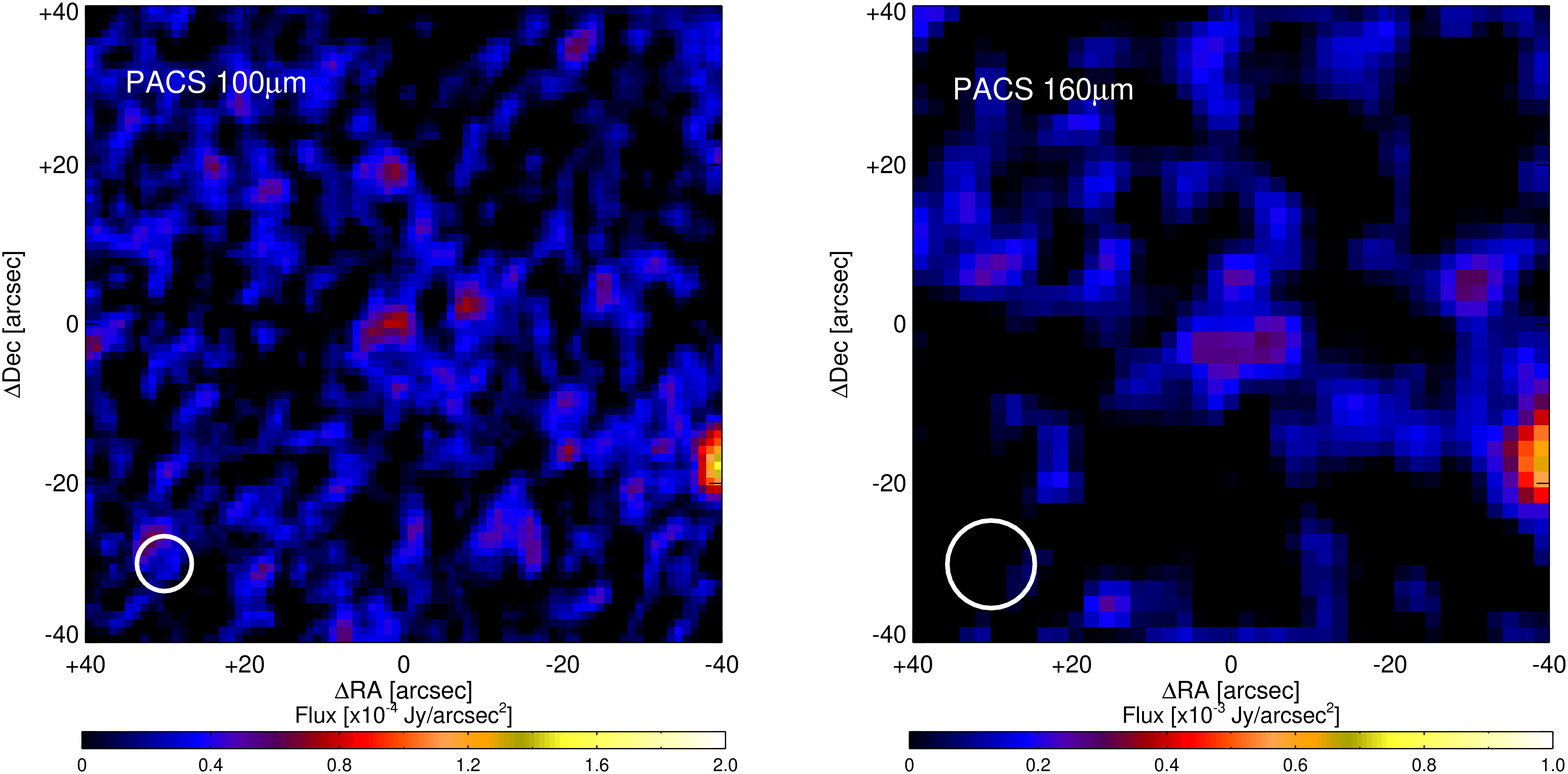}
\includegraphics[scale=0.28]{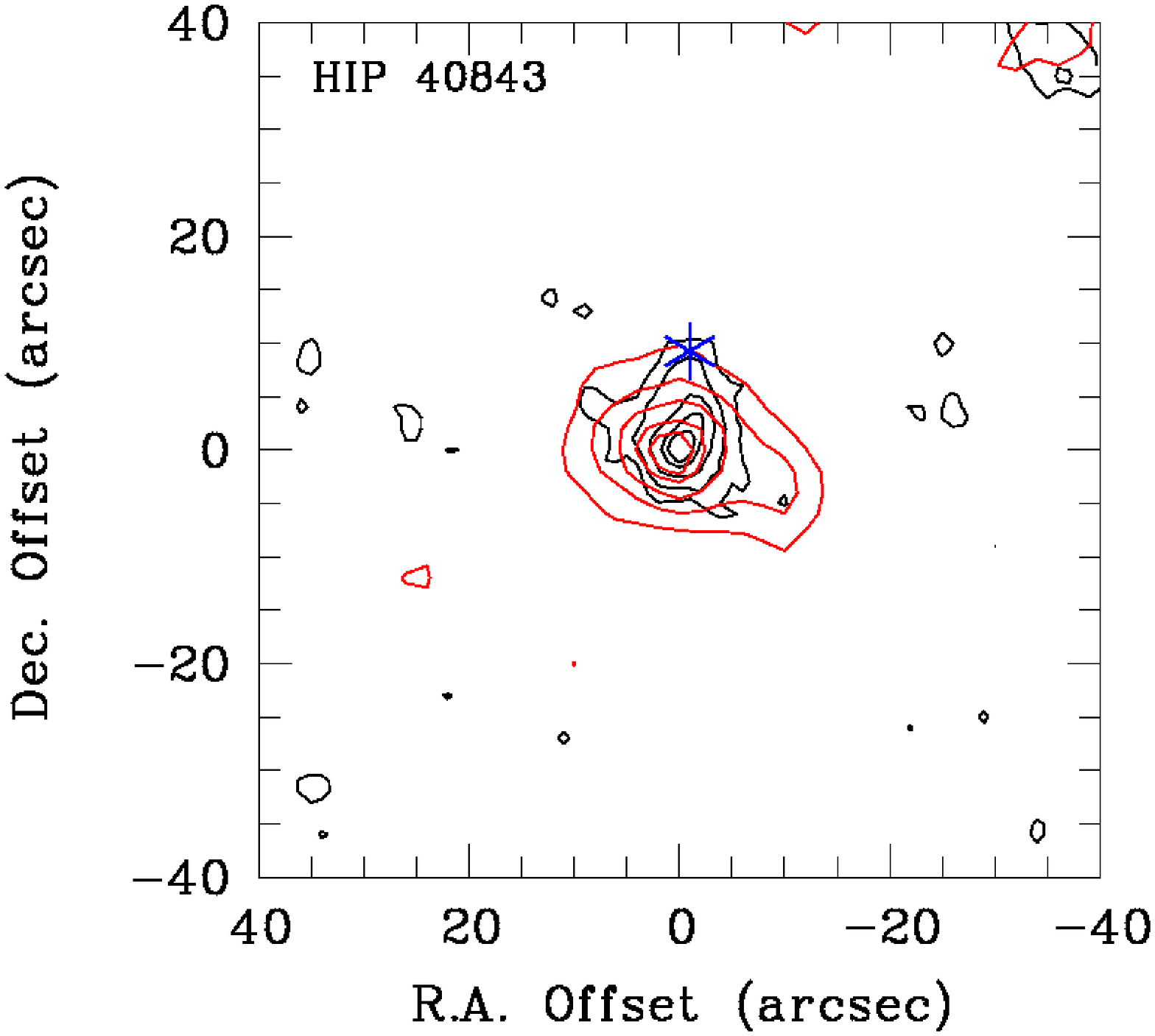}
\includegraphics[scale=0.18]{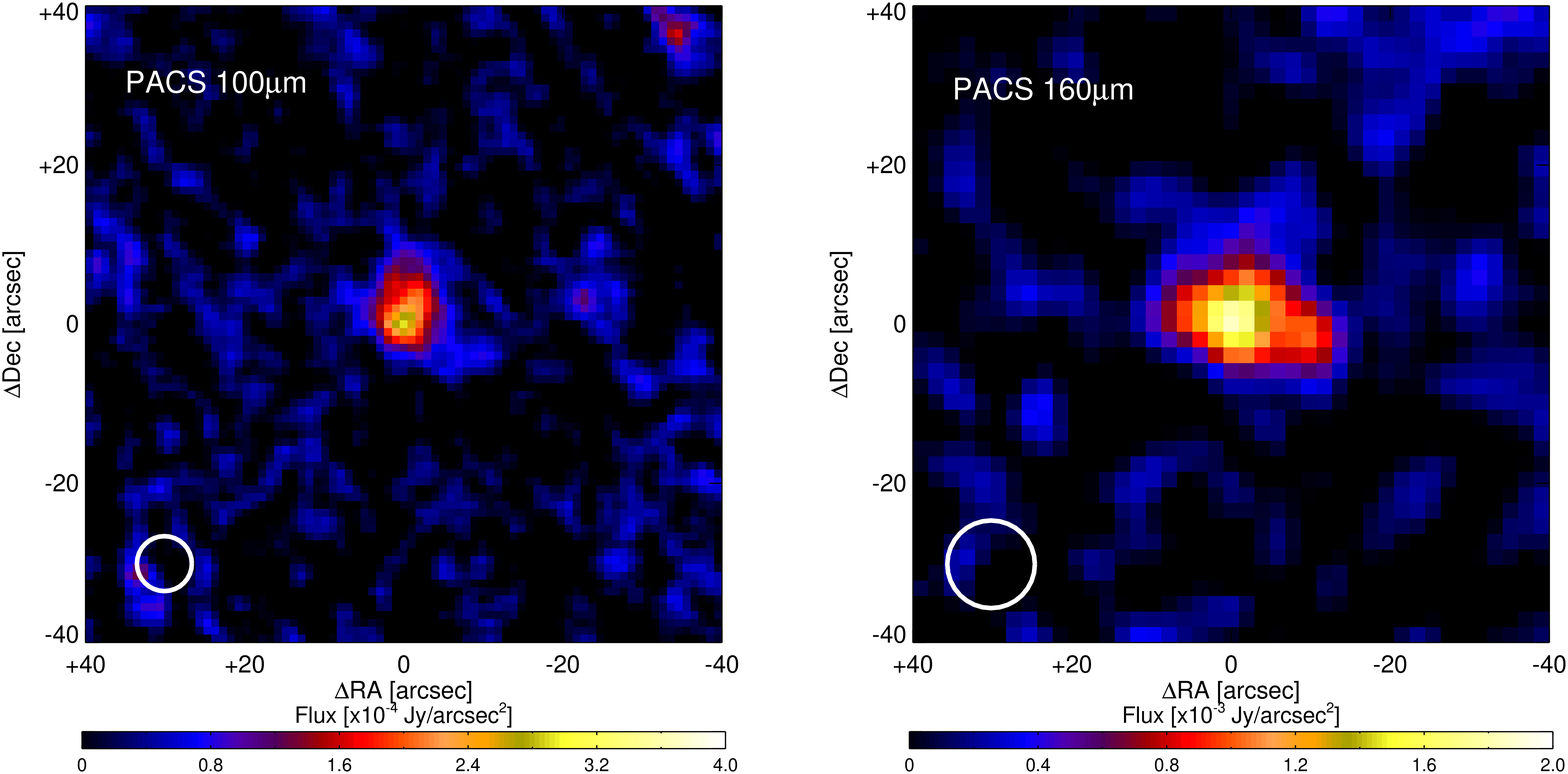}
\caption{
Contour plots (left) and PACS 100 $\mu$m (middle) and 160 $\mu$m
(right) images of stars for which contamination impacts the apparent
excesses of the stars. The identification of the stars are given in
the upper-left corner of the contour plots. Position (0,0) refers to
the 100 $\mu$m peak. The optical position of the stars with respect to
the 100 $\mu$m peak is indicated by a ``star'' symbol.  North is up
and East to the left. Black contours correspond to 100 $\mu$m and red
contours to 160 $\mu$m. HIP 29568: contours are 40\%, 50\%, 60\%,
70\%, 80\%, 90\% of the peak at both bands. HIP 38784: contours are
50\%, 60\%, 70\%, 80\%, 90\% of the peak at both bands.  HIP 40843:
contours 20\%, 40\%, 60\%, 80\%, 90\% of the peak at both bands.}
\label{fig:controversial_1}
\end{figure*}

\begin{figure*}[h!]
\centering
\includegraphics[scale=0.28]{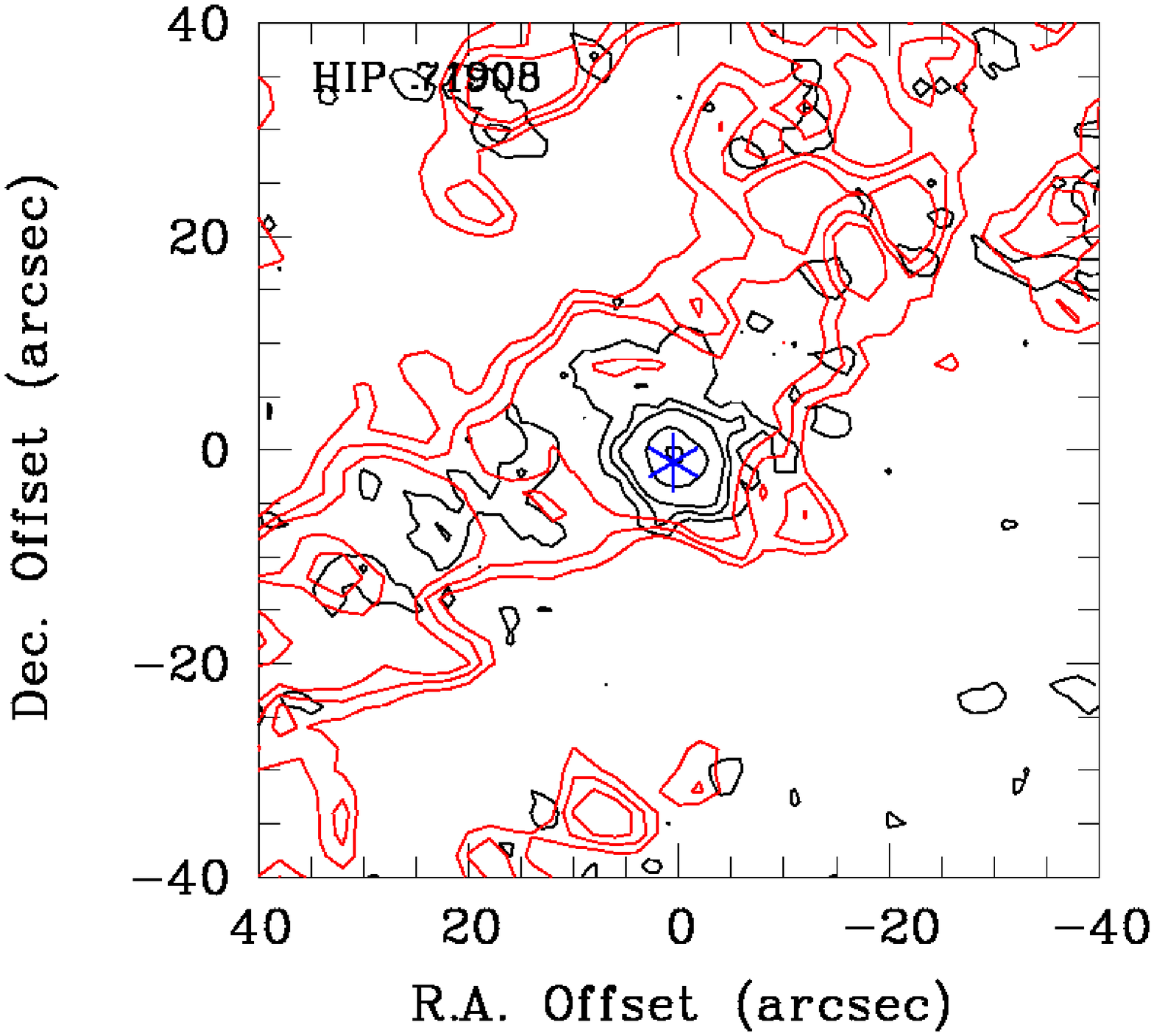}
\includegraphics[scale=0.18]{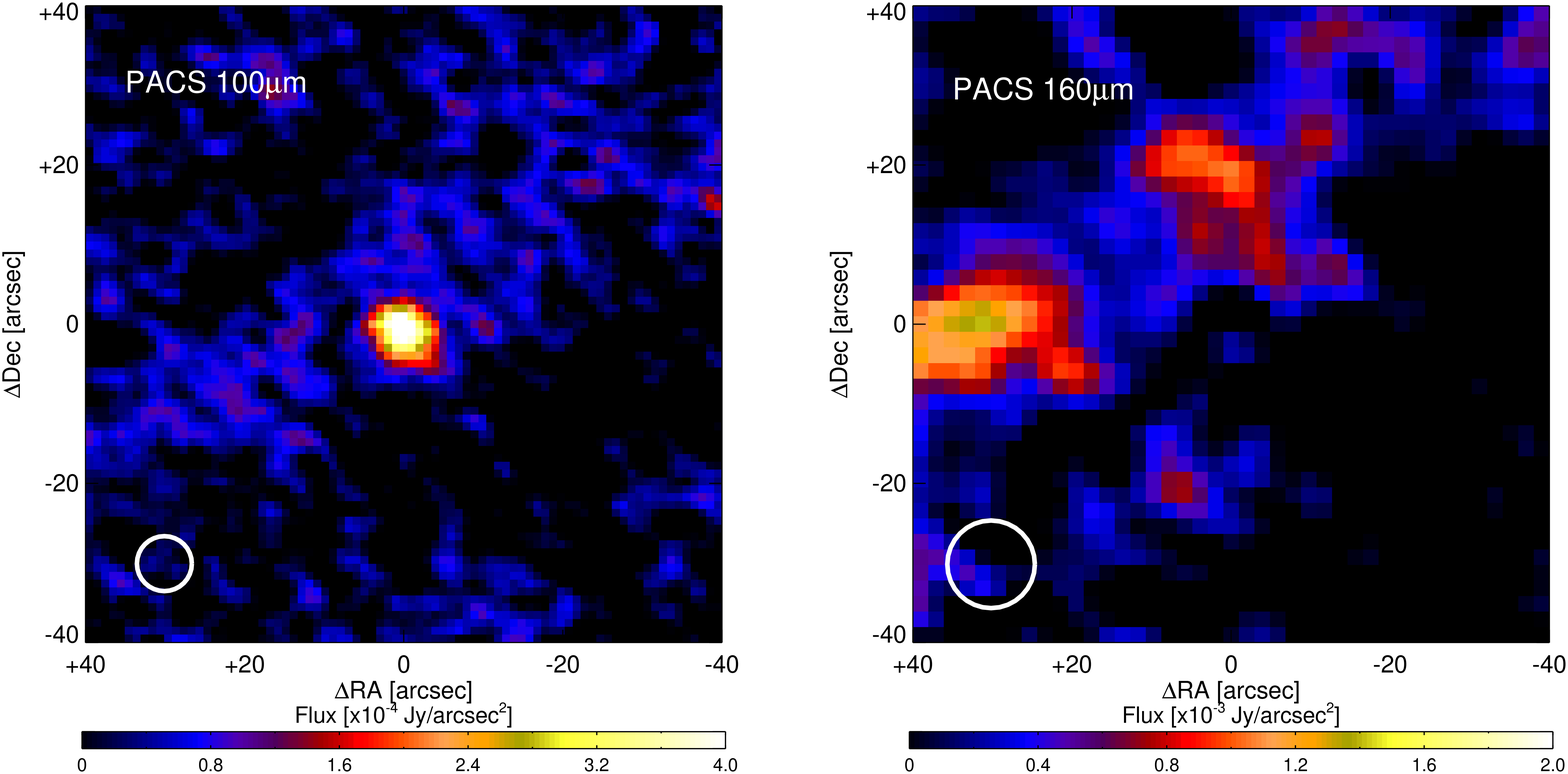}
\includegraphics[scale=0.28]{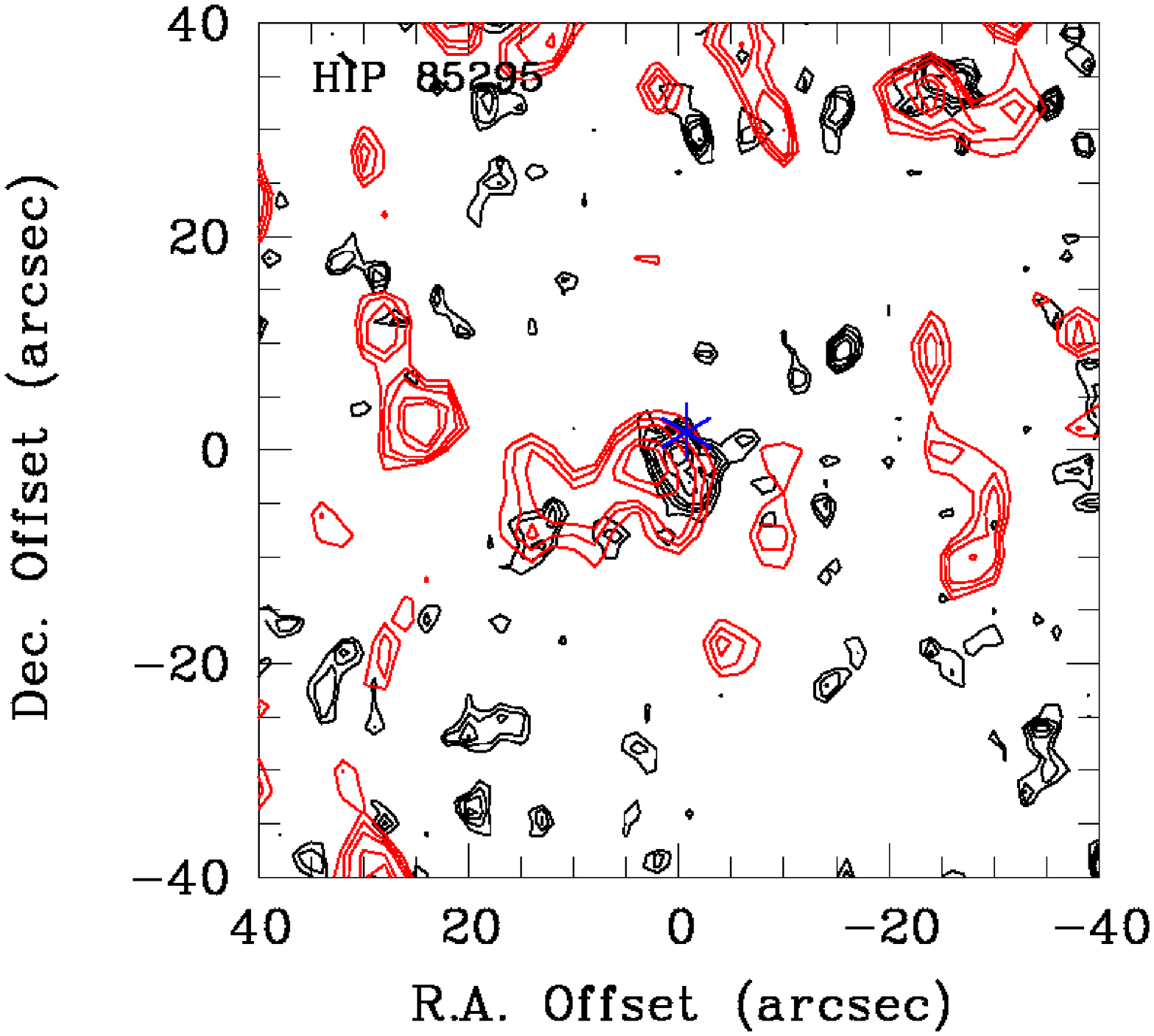}
\includegraphics[scale=0.18]{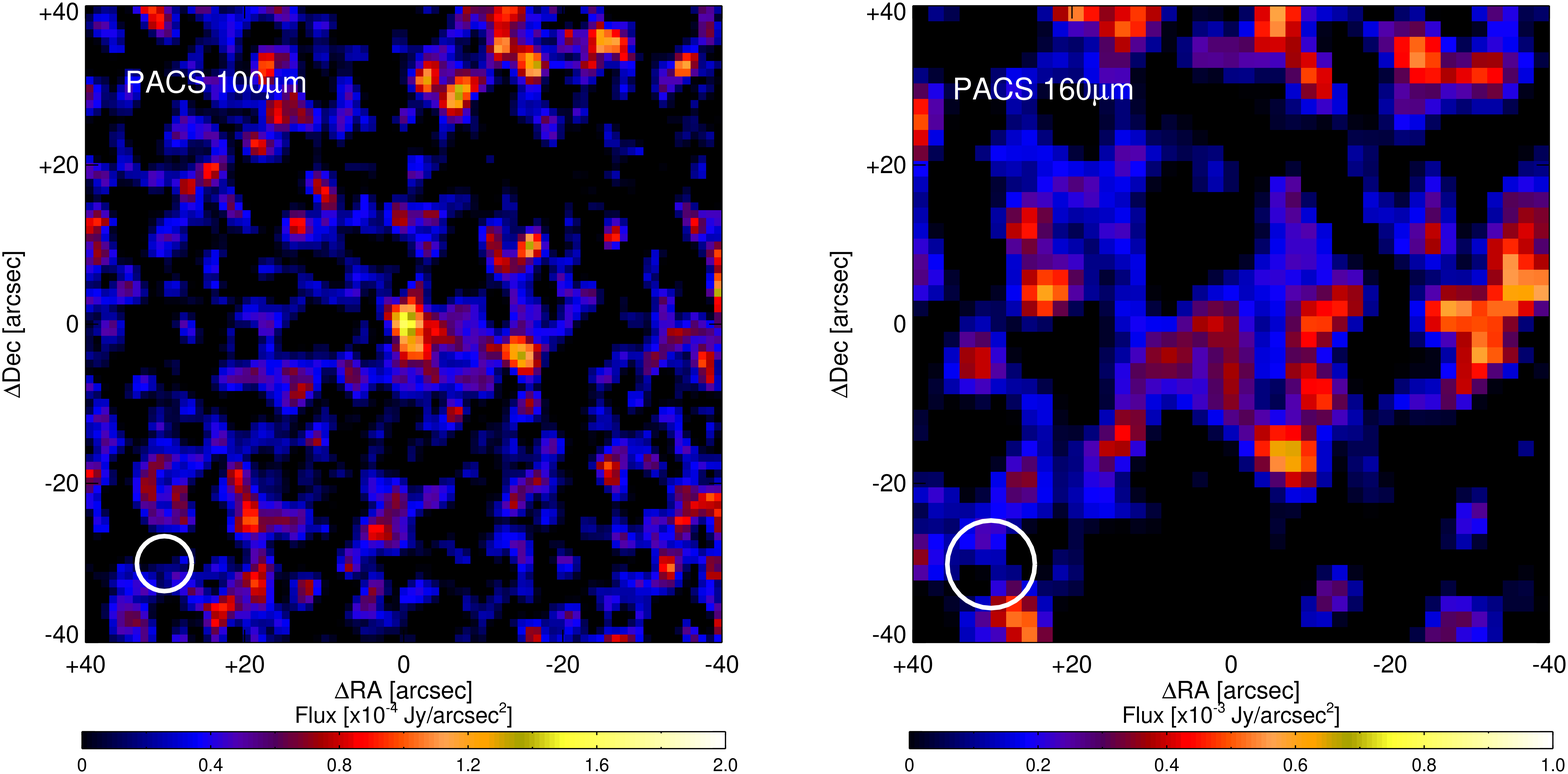}
\includegraphics[scale=0.28]{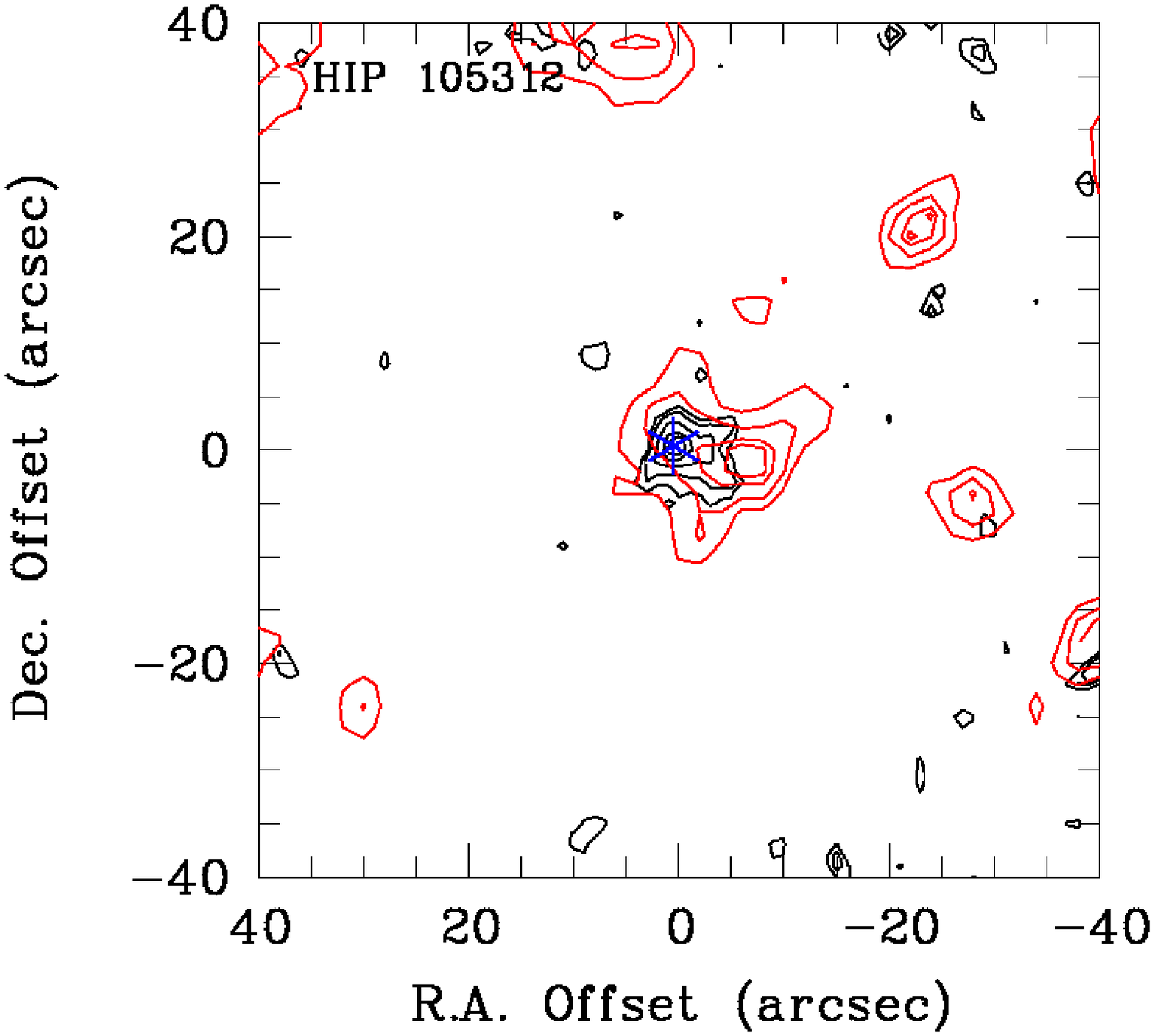}
\includegraphics[scale=0.18]{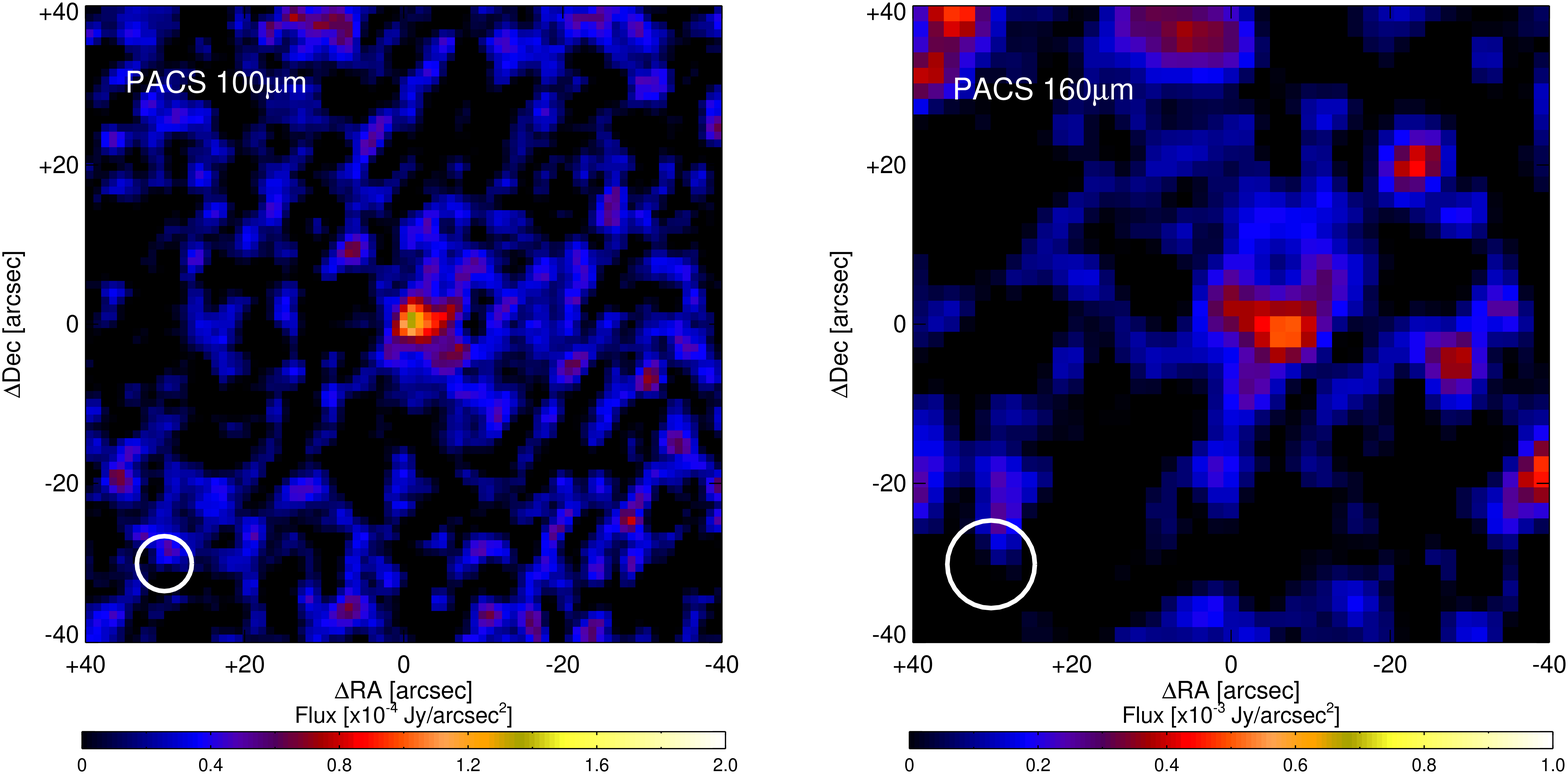}
\includegraphics[scale=0.28]{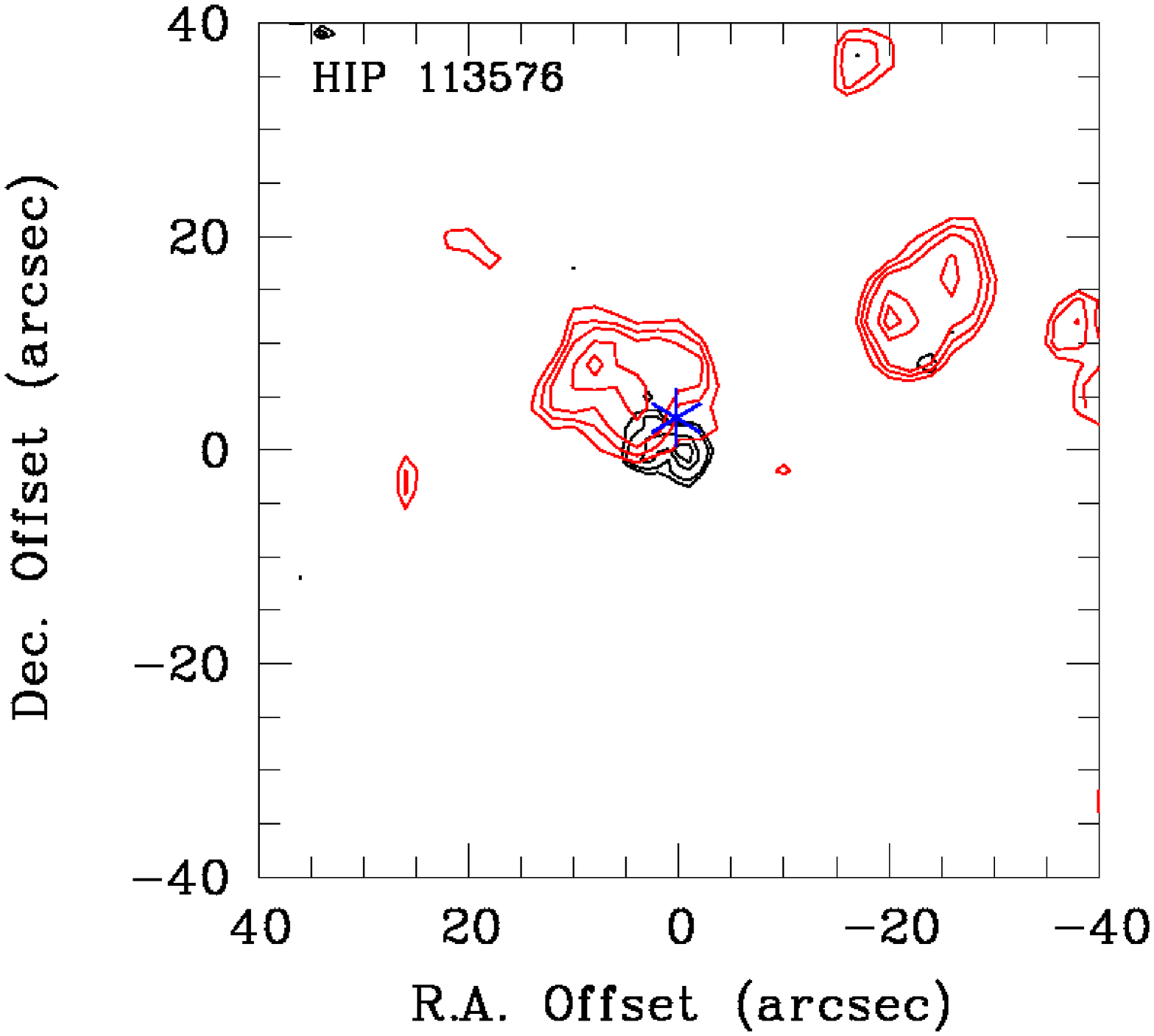}
\includegraphics[scale=0.18]{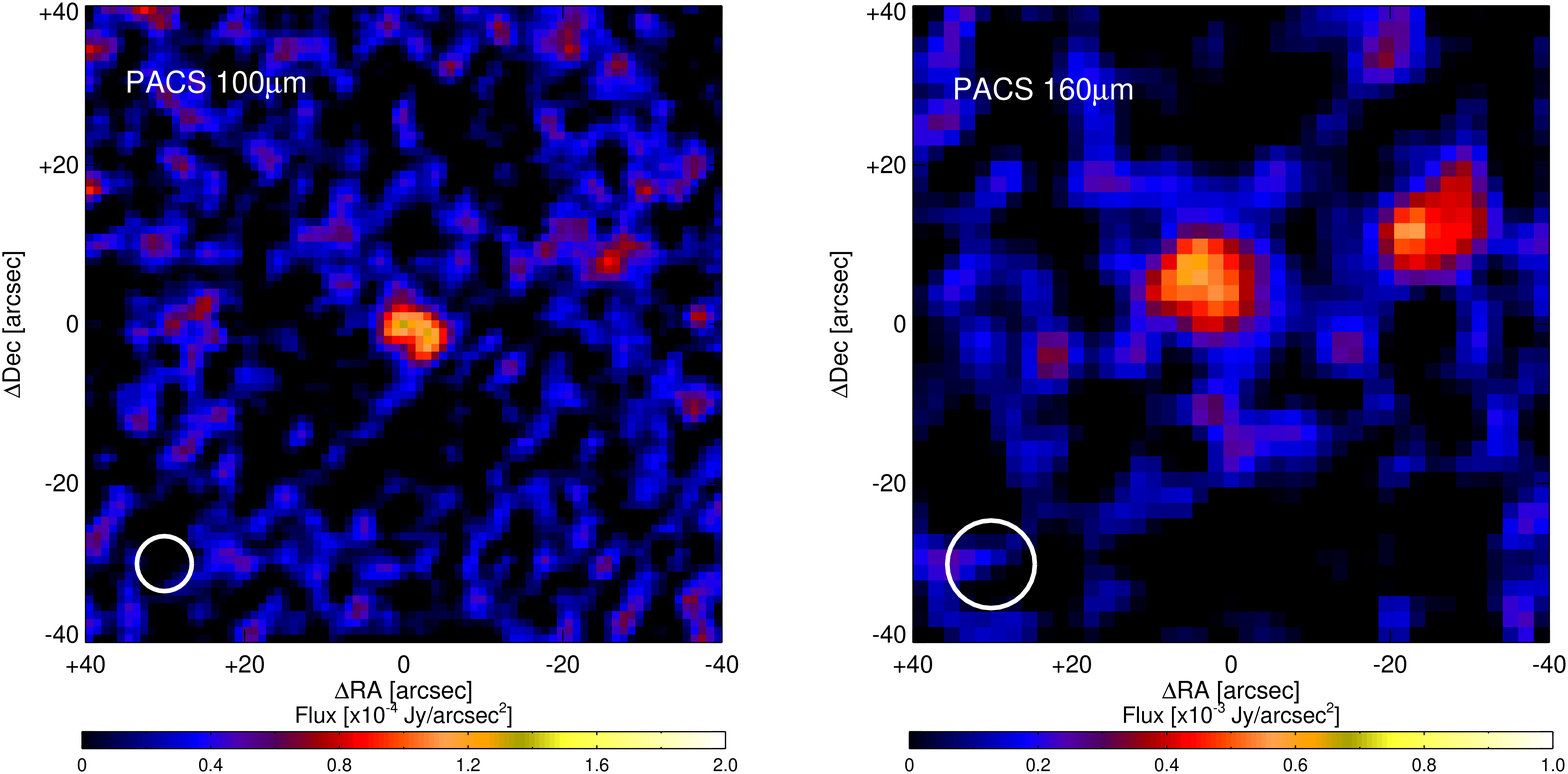}
\caption{The same as \ref{fig:controversial_1}. In this case contours
  are 20\%, 40\%, 60\%, 80\%, 90\% of the peak at both 100 and 160
  $\mu$m bands for all the images.}
\label{fig:controversial_2}
\end{figure*}

There is a number of  objects whose fluxes seemingly denote an excess,
which can in  a first instance be attributed  to a circumstellar disc,
but whose morphologies and surrounding fields suggest they are not true
debris discs associated  with the stars. Very likely,  these stars are
affected by coincidental alignment  or contamination from a structured
background.  In the following,  we briefly  describe these  cases and
show   their    PACS   images    and   contour   plots    in   Figures
\ref{fig:controversial_1}  and  \ref{fig:controversial_2}.   
Table  \ref{table:controversial_sources}   lists  these
sources   together  with   their  PACS   and   predicted  photospheric
fluxes. The  significance of the ``apparent'' excesses  is also given,
as well as the {\it Spitzer} MIPS 70 $\mu$m fluxes.

\begin{table*}
\setlength{\tabcolsep}{4.5pt}
\centering
\caption{DUNES stars whose apparent excesses are very likely due to contamination by 
background galactic extended structures or extragalactic objects. PACS and predicted photospheric fluxes, and the significance of the measured 
apparent excesses are given. The last column gives {\it Spitzer} MIPS fluxes at 70 $\mu$m.} 

\label{table:controversial_sources}
\begin{tabular}{lllcllcllcll}
\hline\noalign{\smallskip}
HIP    & SpT &  PACS70        &S70        &$\chi_{70}$& PACS100        &S100        &$\chi_{100}$&PACS160       &S160        &$\chi_{160}$ & MIPS70     \\  
       &     &  (mJy)         & (mJy)     &          & (mJy)          &(mJy)       &           &(mJy)         &(mJy)       &            &(mJy)      \\
\hline\noalign{\smallskip}
29568  & G5V   &              & 10.48$\pm$0.19 &        &7.83$\pm$1.16  & 5.13$\pm$0.10 & 1.92    & 8.28$\pm$2.11& 2.01$\pm$0.04 & 3.00     &15.3$\pm$2.3           \\ 
38784  & G8V   & 8.66$\pm$2.65& 9.47 $\pm$0.10 & -0.31  &4.19$\pm$0.74  & 4.64$\pm$0.05 & -0.61   & 5.98$\pm$1.28& 1.81$\pm$0.02 & 3.26     &10.6$\pm$3.0           \\
40843  & F6V   &              & 22.59$\pm$0.13 &        &33.17$\pm$2.76 &11.07$\pm$0.06 & 8.03    &53.35$\pm$3.17& 4.32$\pm$0.02 & 15.47    &32.5$\pm$5.3           \\
71908  & F1Vp  &              & 68.58$\pm$4.16 &        &38.54$\pm$1.58 &33.60$\pm$2.04 & 3.13    &              &13.13$\pm$0.80 &          &                       \\ 
85295  & K7V   &              & 21.14$\pm$0.46 &        &12.94$\pm$1.12 &10.36$\pm$0.22 & 2.30    &14.15$\pm$3.29& 4.05$\pm$0.09 & 3.07     &13.8$\pm$4.6           \\
105312 & G5V   &12.44$\pm$1.72& 11.86$\pm$0.66 &  0.34  &7.84$\pm$0.76  & 5.81$\pm$0.32 & 2.67    &9.20$\pm$1.74 & 2.27$\pm$0.13 &3.98      &13.6$\pm$5.7           \\
113576 & K5/M0V&              & 19.01$\pm$0.97 &        &9.43$\pm$0.79  & 9.32$\pm$0.48 & 0.14    &12.66$\pm$1.50& 3.64$\pm$0.19 &6.01      &19.1$\pm$2.9           \\
\hline\noalign{\smallskip}
\end{tabular}
\end{table*} 

{\em HIP 29568}.  The 160 $\mu$m flux has a significance of
$\chi_{160}$ = 3.00 (Table \ref{table:controversial_sources}), being a
cold disc candidate.  However, the image shows a lot of background
structure (Figure \ref{fig:controversial_1}) which makes the flux
estimate doubtful.

{\em HIP 38784}.  This is a faint star which apparently shows a small
excess at 160 $\mu$m. However, reduced images from HIPE versions 7.2
(Figure \ref{fig:controversial_1}) and 4.2 are not quite consistent,
as neither are the individual scans likely due to the faintness of the
source.

{\em HIP 40843}.  This star was identified by {\it Spitzer} as an
excess source. However, the positional offset between the star and the
peak of the PACS image is large.  Both the 100 and 160 $\mu$m images
are extended but practically in perpendicular directions (Figure
\ref{fig:controversial_1}).  The fluxes of the table correspond to the
whole extended emission.  However, given the position of the star and
the different orientations at 100 and 160$\mu$m, we consider it a case
of coincidental alignment of a background galaxy.  In fact, there is a
very faint, secondary 100 $\mu$m peak embedded in the extended
emission. This peak is at the position $\alpha$ (2000.0) = 8:20:03.78,
$\delta$ (2000.0) = 27:13:4.6, i.e., and offset  of 1$\farcs$4 wrt the optical
positions on the star, and of 8$\farcs$1 wrt the main 100 $\mu$m peak. A 2D 
gaussian fit supports the presence of two peaks.

{\em HIP 71908}. The star lies on top of an emission strip at both 100
and 160  $\mu$m  (Fig \ref{fig:controversial_2}), which  prevents us
from estimating an  enough accurate flux  in the red band.  This star is
located at  the galactic plane.  The marginal 100  $\mu$m significance,
$\chi$100 =3.13, is likely not real.

{\em HIP 85295}. There is an offset between the 100 $\mu$m peak, which
coincides with the optical position, and the 160 $\mu$m peak emission
(Fig \ref{fig:controversial_2}). At this wavelength, the object seems
to be formed by two different ones separated by $\sim$ 6$\arcsec$.
The western one is close to the 100 $\mu$m peak.  Thus, the apparent
excess emission at 160 $\mu$m is likely due to contamination by a
background galaxy.

{\em HIP  105312}. The 100 $\mu$m  peak agrees well with
the  optical position,  but  the 160  $\mu$m  emission, which  appears
resolved,   is   displaced   a   bit  towards   the   west.    (Figure
\ref{fig:controversial_2}).

{\em HIP  113576}. This is a case  of a very clear  offset between the
100  and  the 160 $\mu$m emission, which  likely falsifies the presence
of an excess (Figure \ref{fig:controversial_2}).

\end{appendix}

\begin{appendix}
\section{SEDs of the excess sources} 
\label{section:SEDs}
%
%
\begin{figure*}
\caption{SEDs of DUNES stars with excesses.}
\includegraphics[scale=0.4]{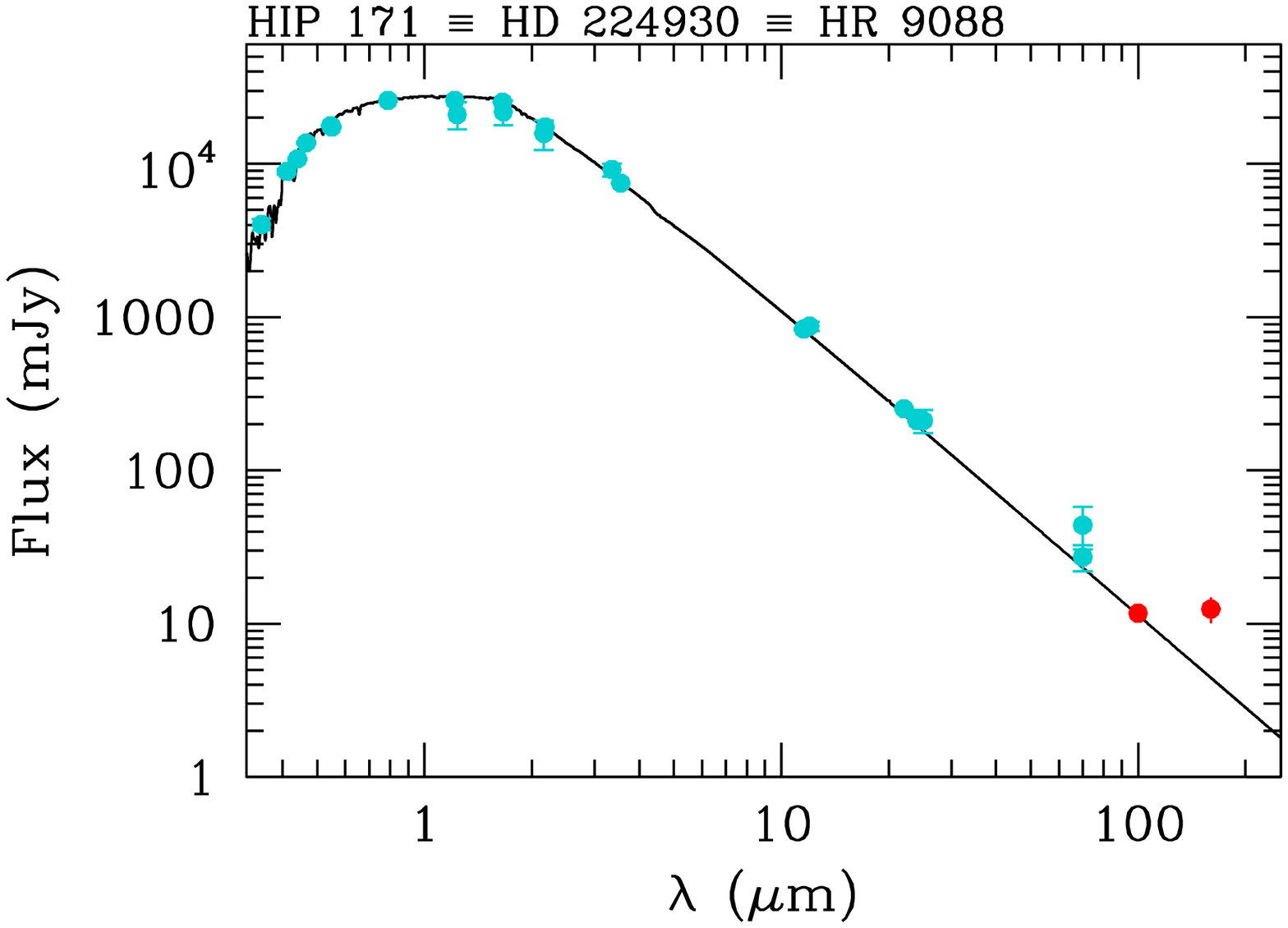}\includegraphics[scale=0.4]{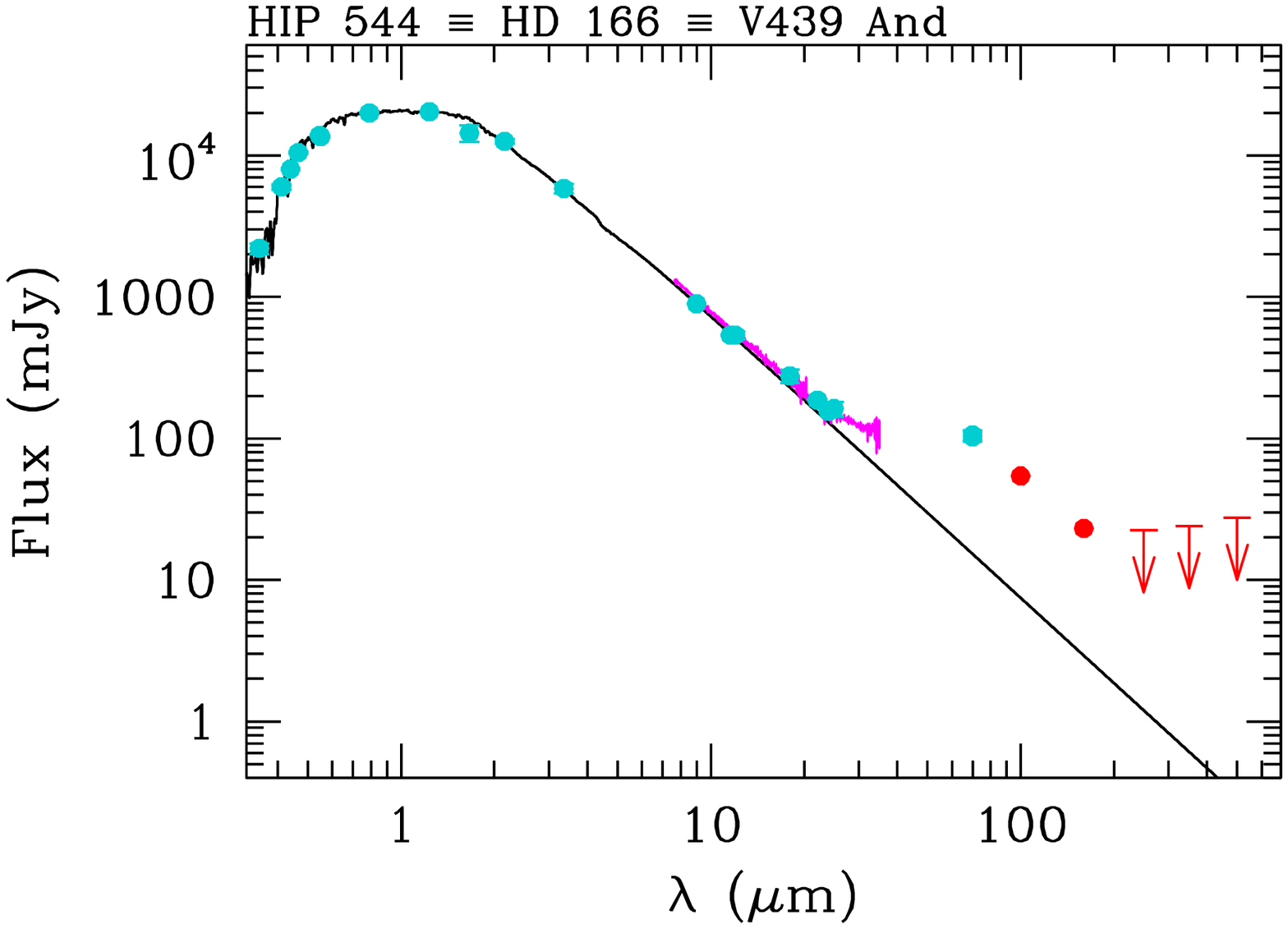}
\includegraphics[scale=0.4]{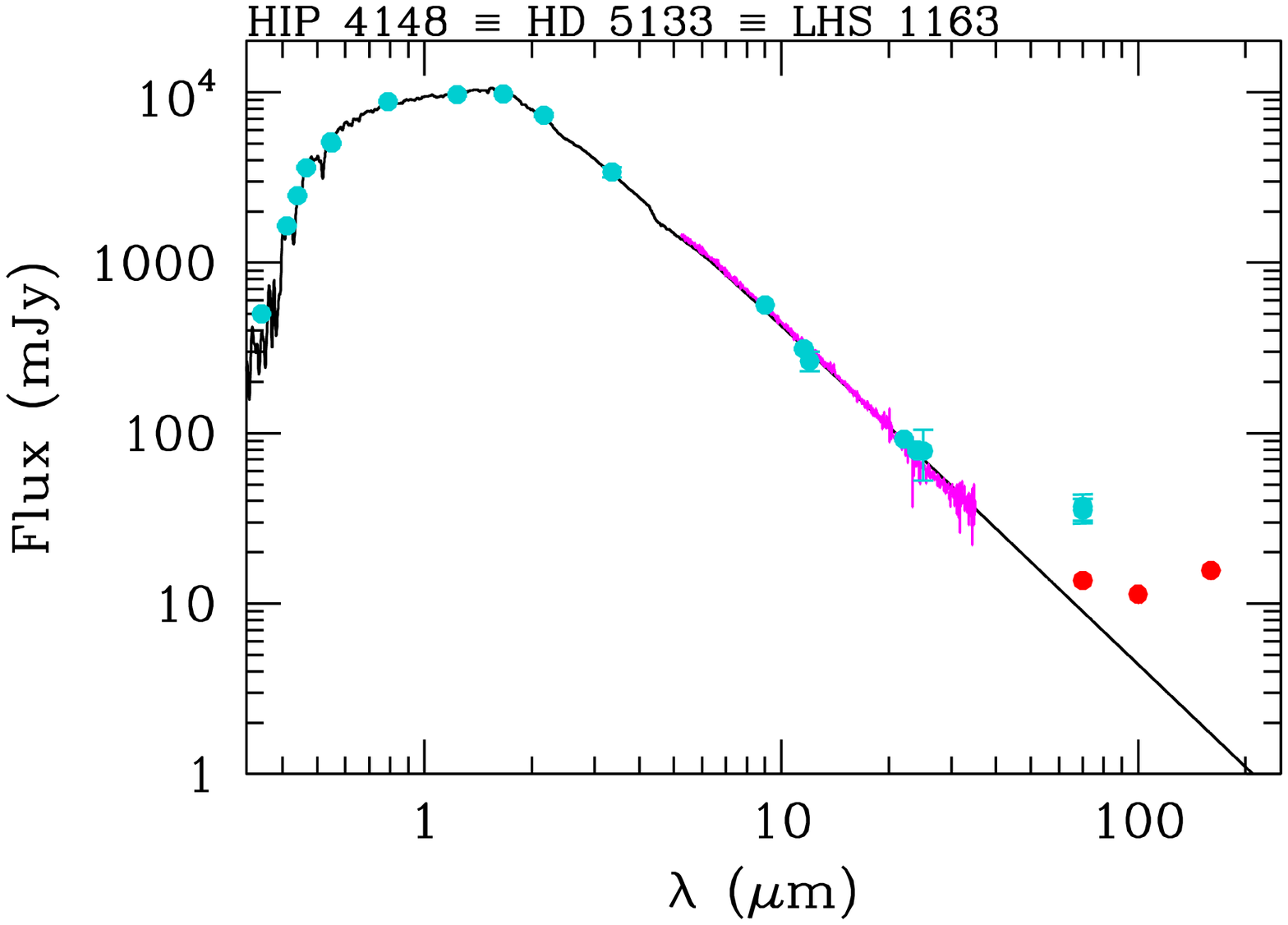}\includegraphics[scale=0.4]{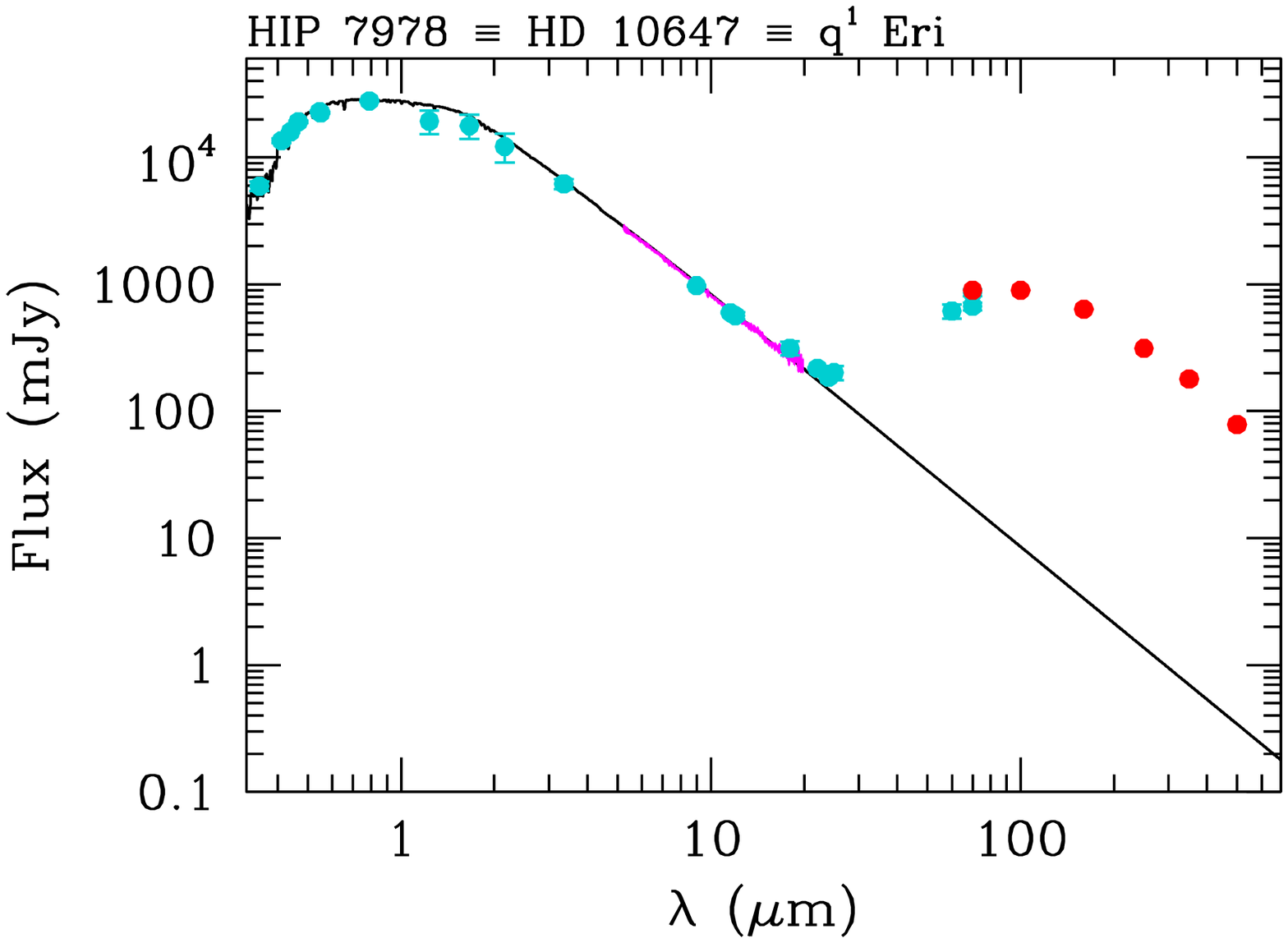}
\includegraphics[scale=0.4]{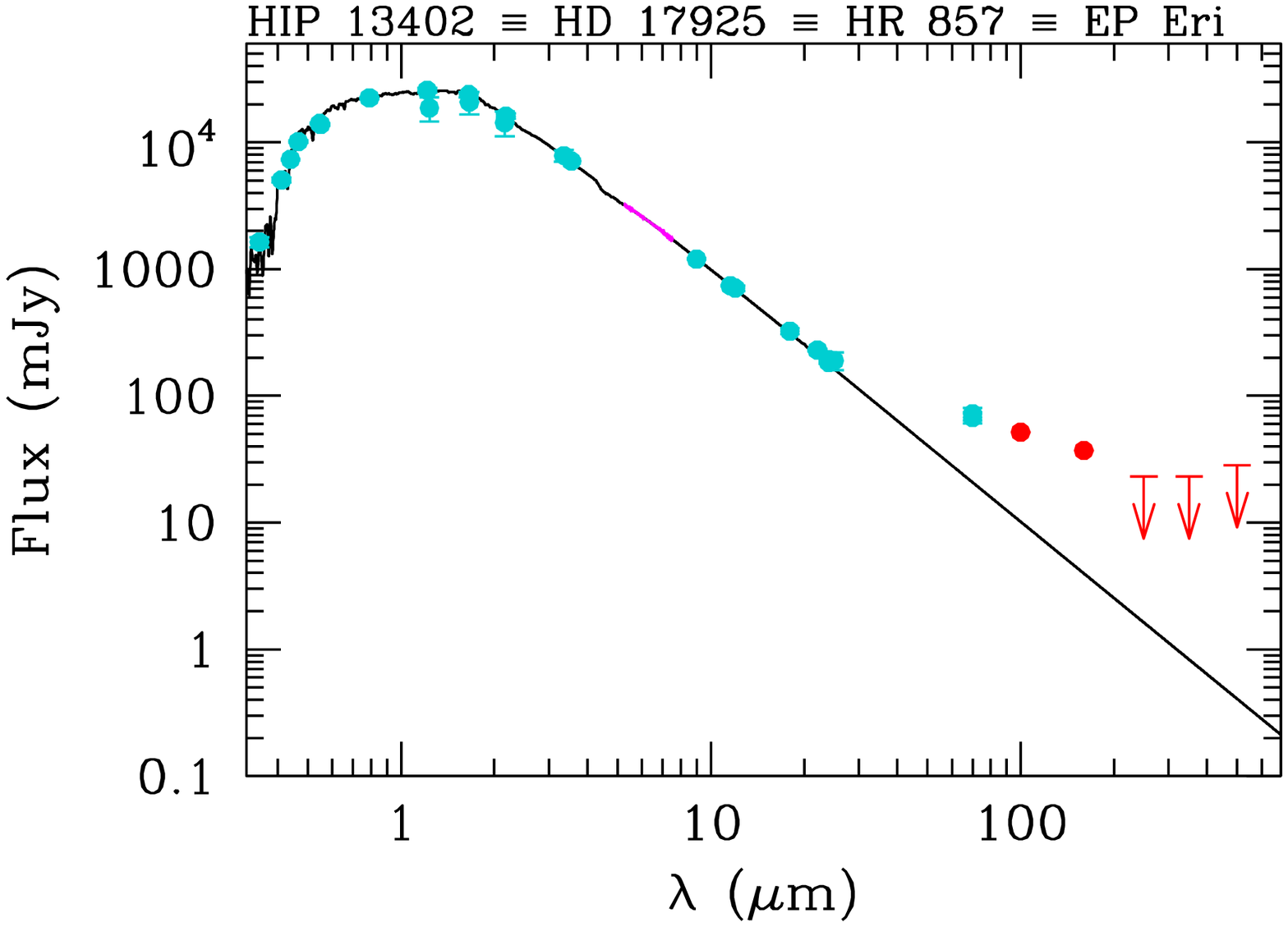}\includegraphics[scale=0.4]{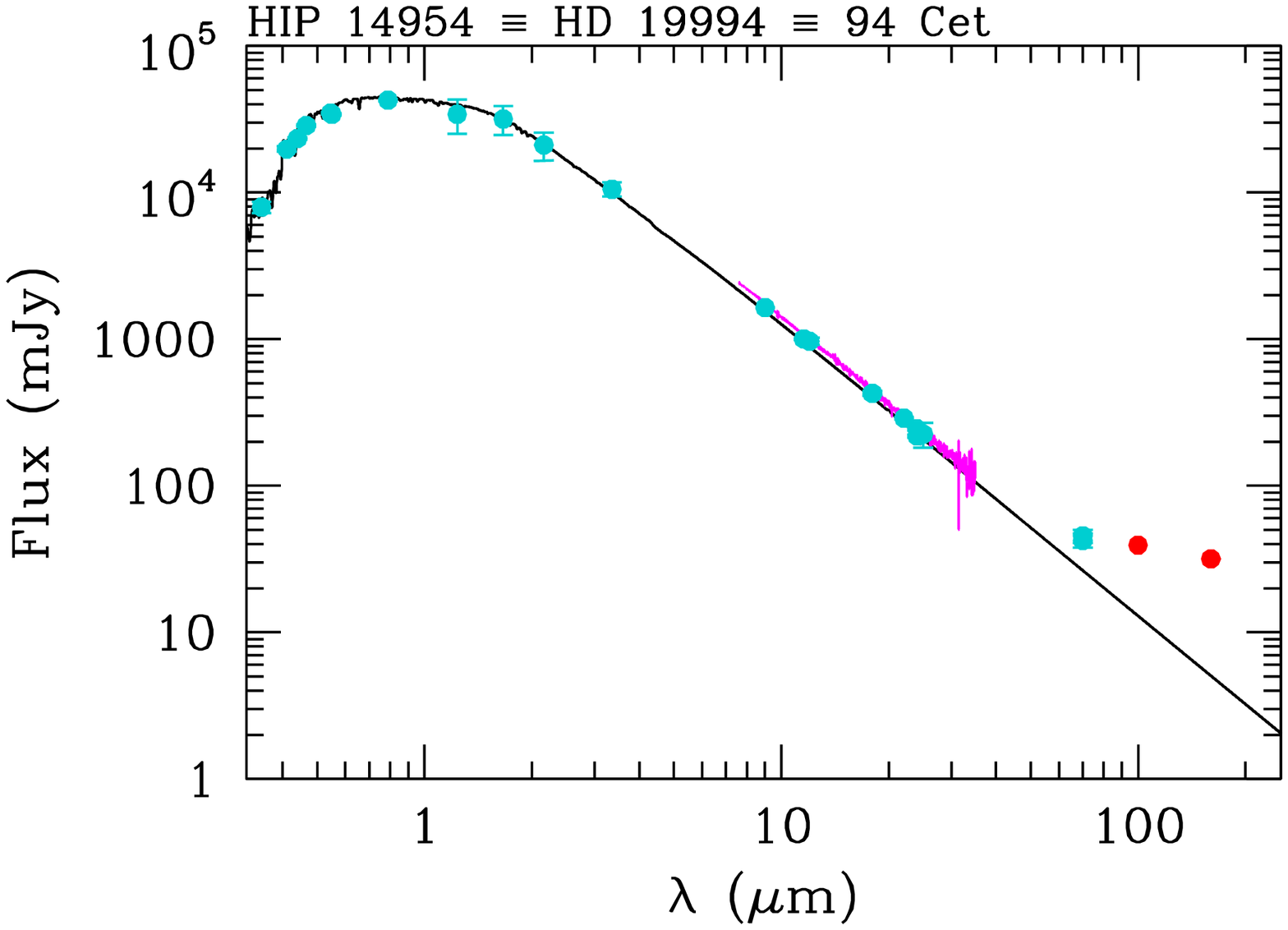}
\includegraphics[scale=0.4]{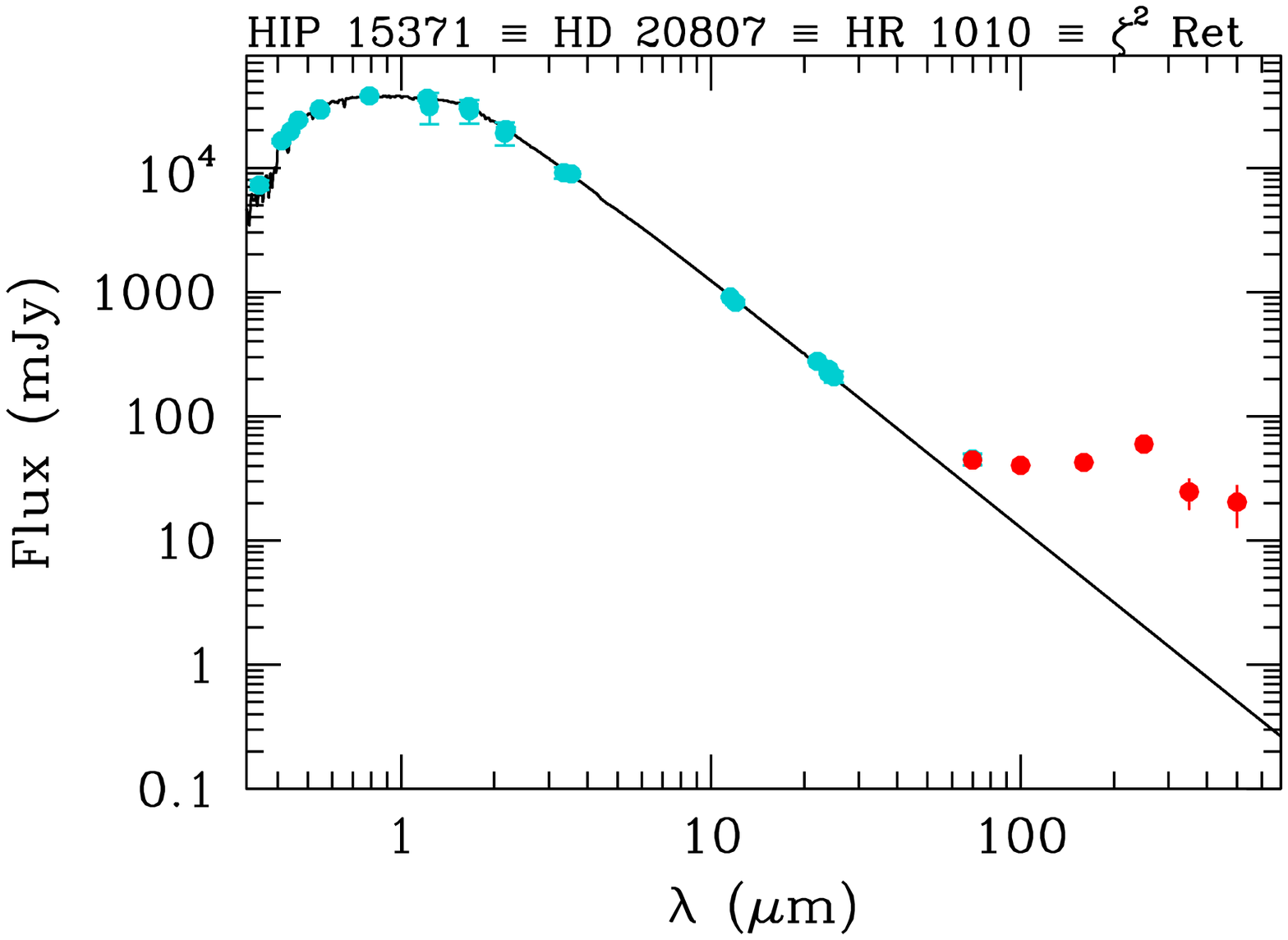}\includegraphics[scale=0.4]{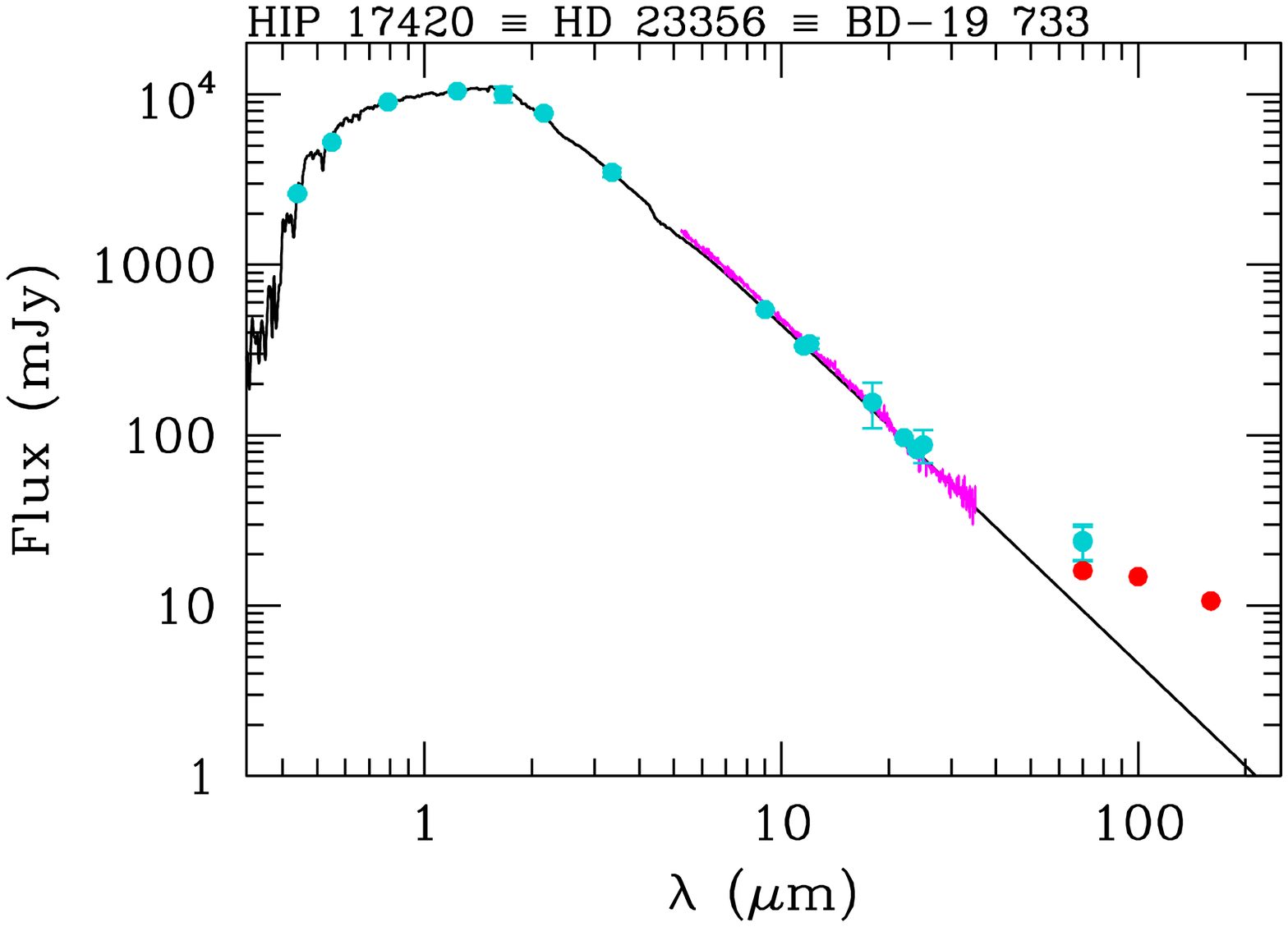}
\end{figure*}
\clearpage

\addtocounter{figure}{-1}
\begin{figure*}
\caption{SEDs of DUNES stars with excesses (continued).}
\includegraphics[scale=0.4]{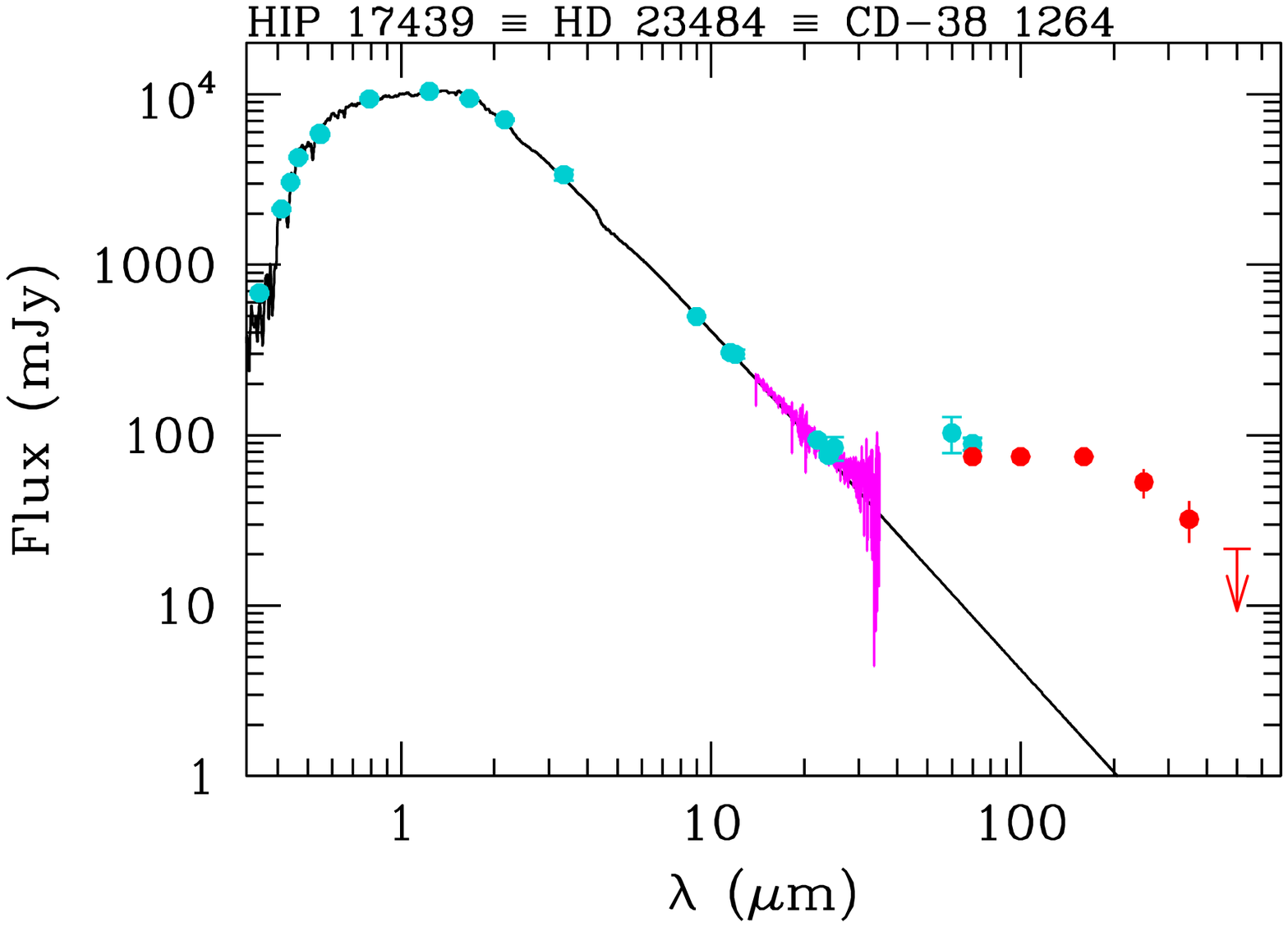}\includegraphics[scale=0.4]{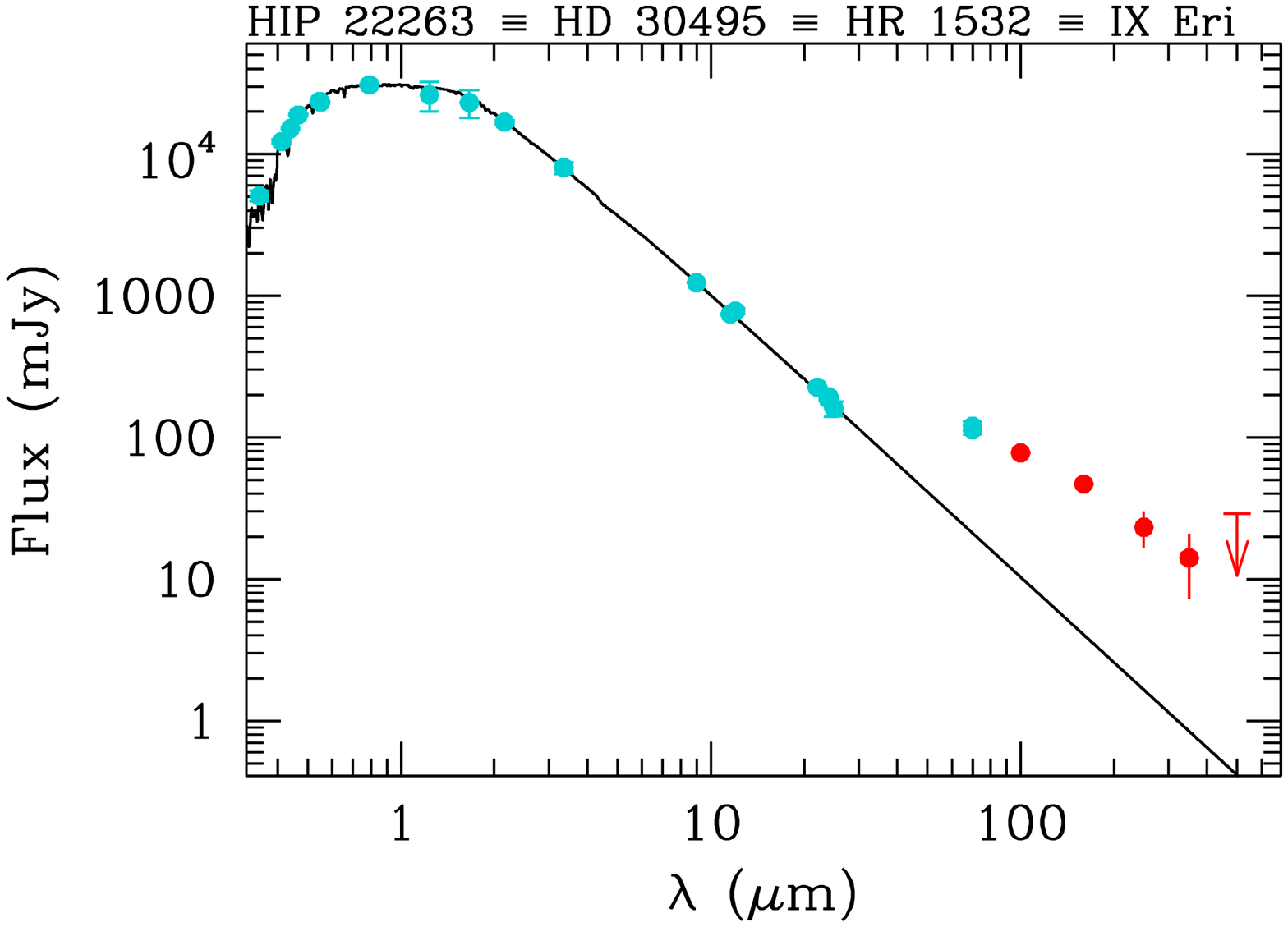}
\includegraphics[scale=0.4]{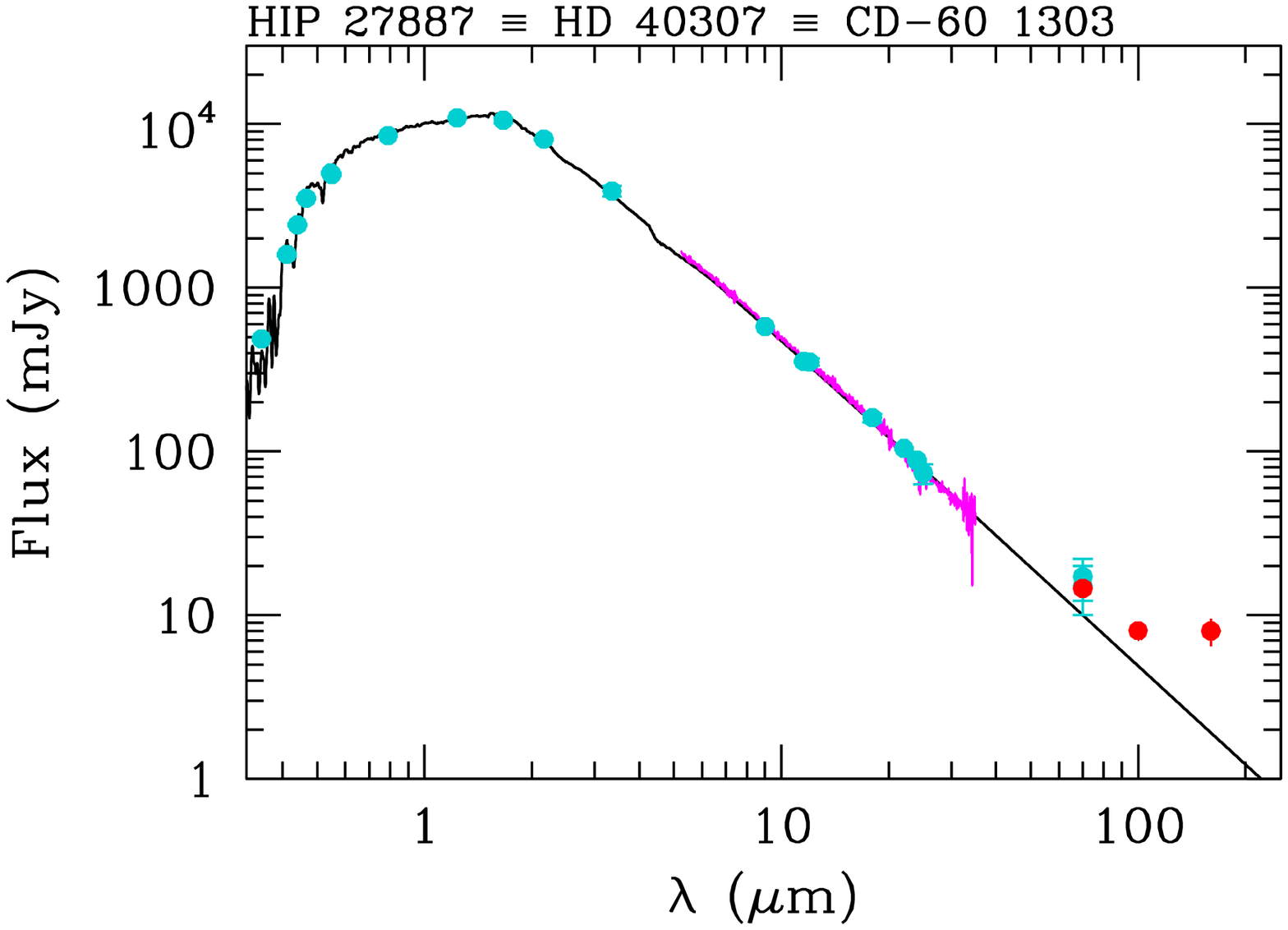}\includegraphics[scale=0.4]{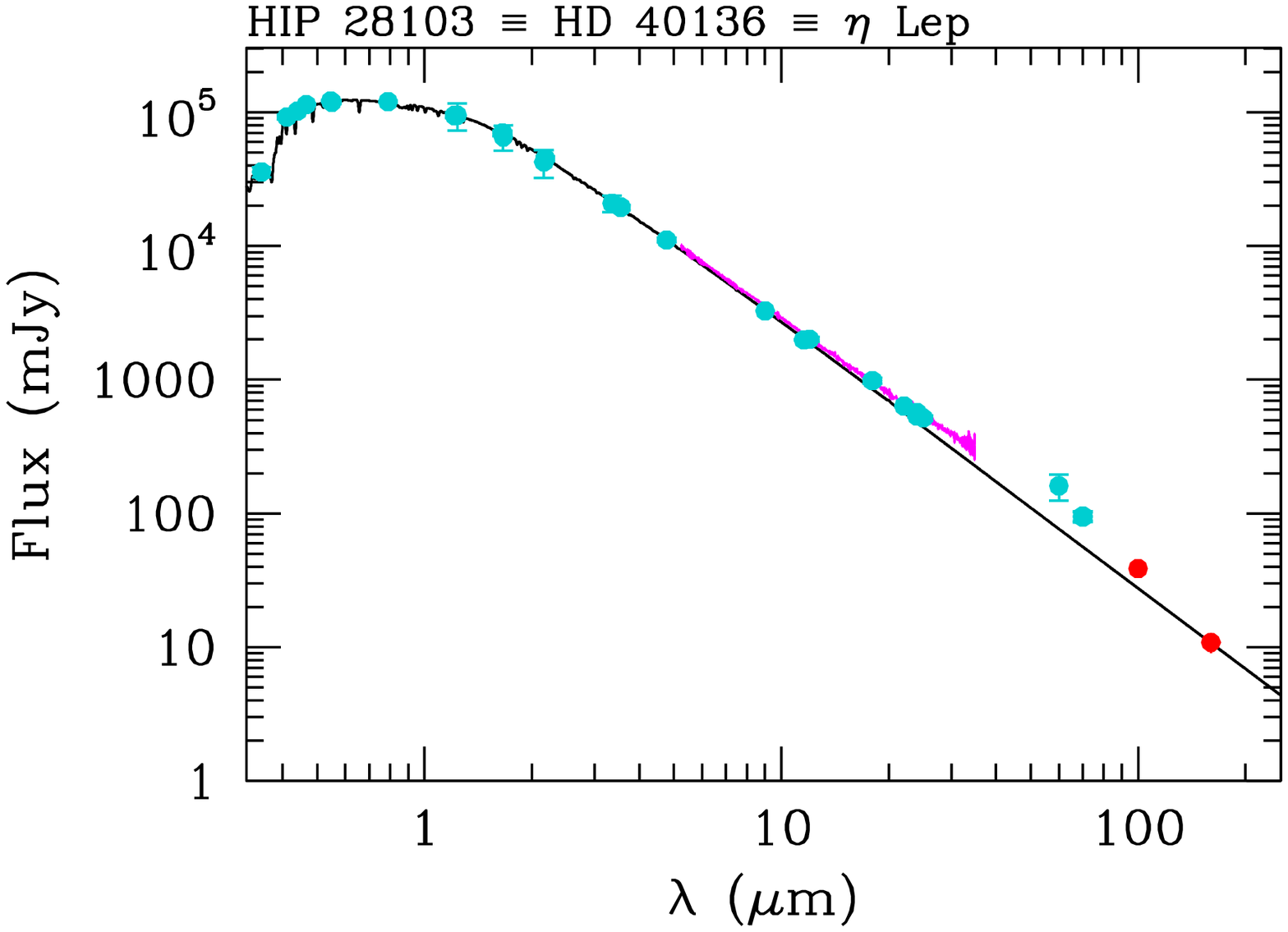}
\includegraphics[scale=0.4]{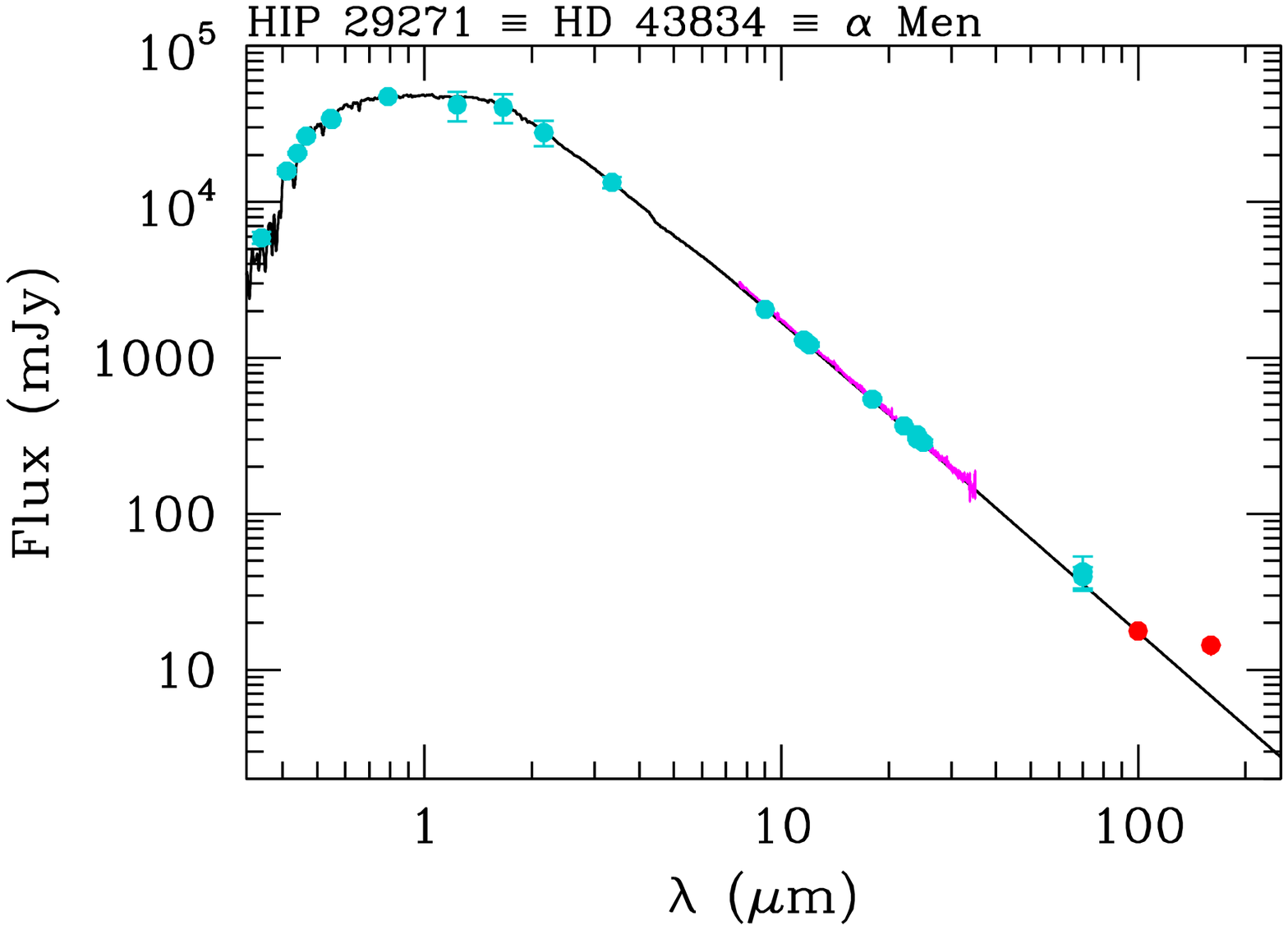}\includegraphics[scale=0.4]{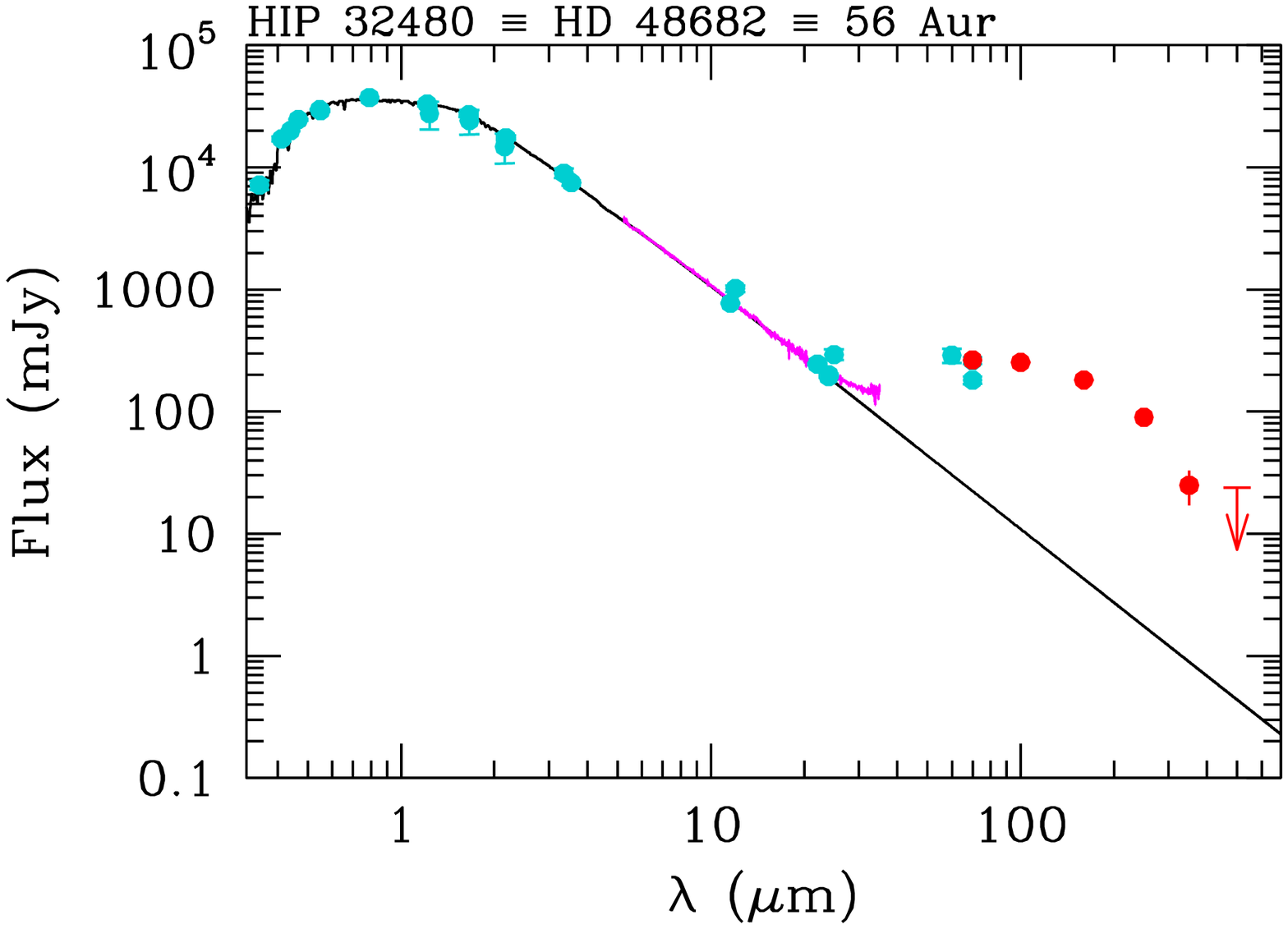}
\includegraphics[scale=0.4]{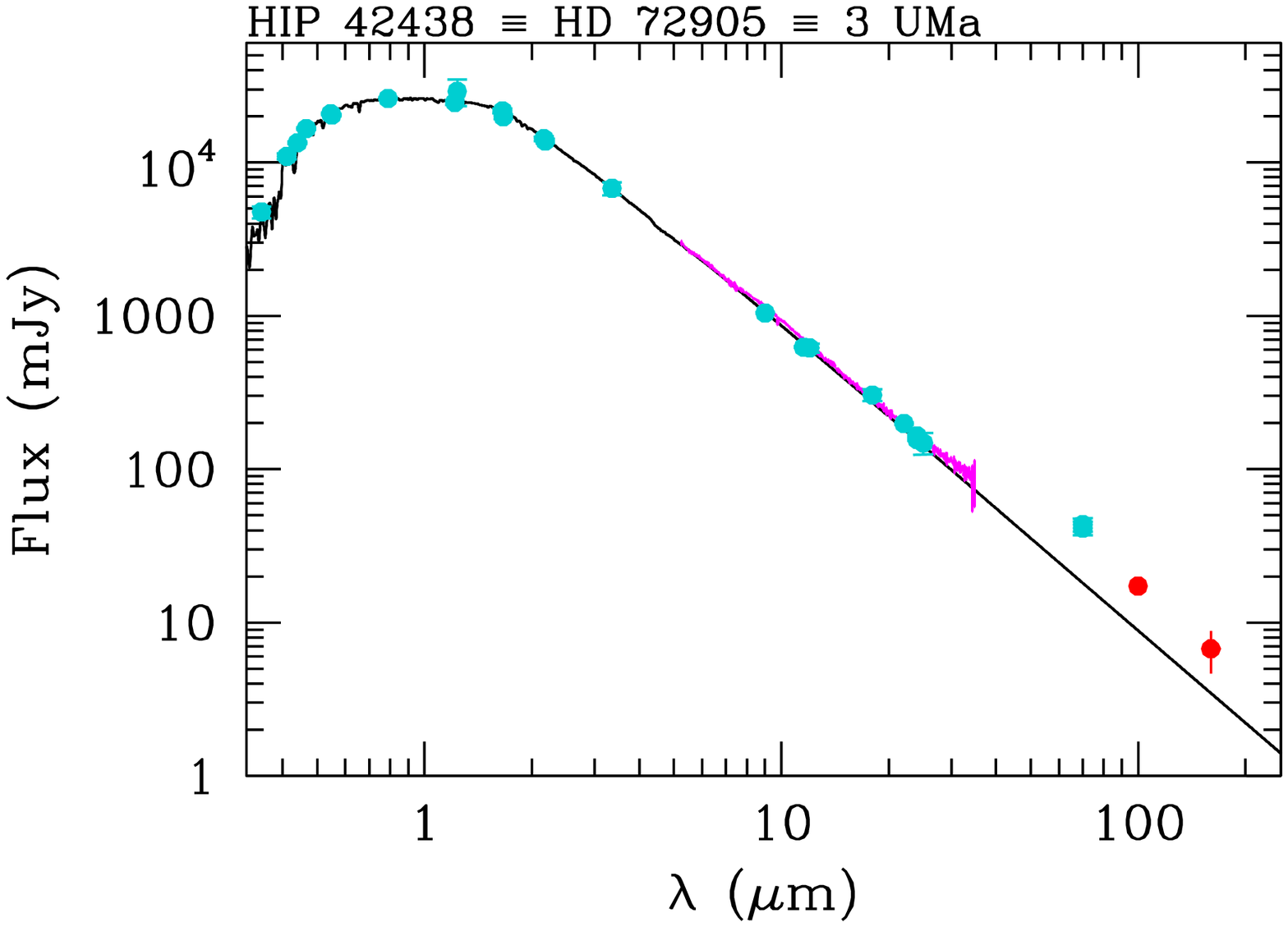}\includegraphics[scale=0.4]{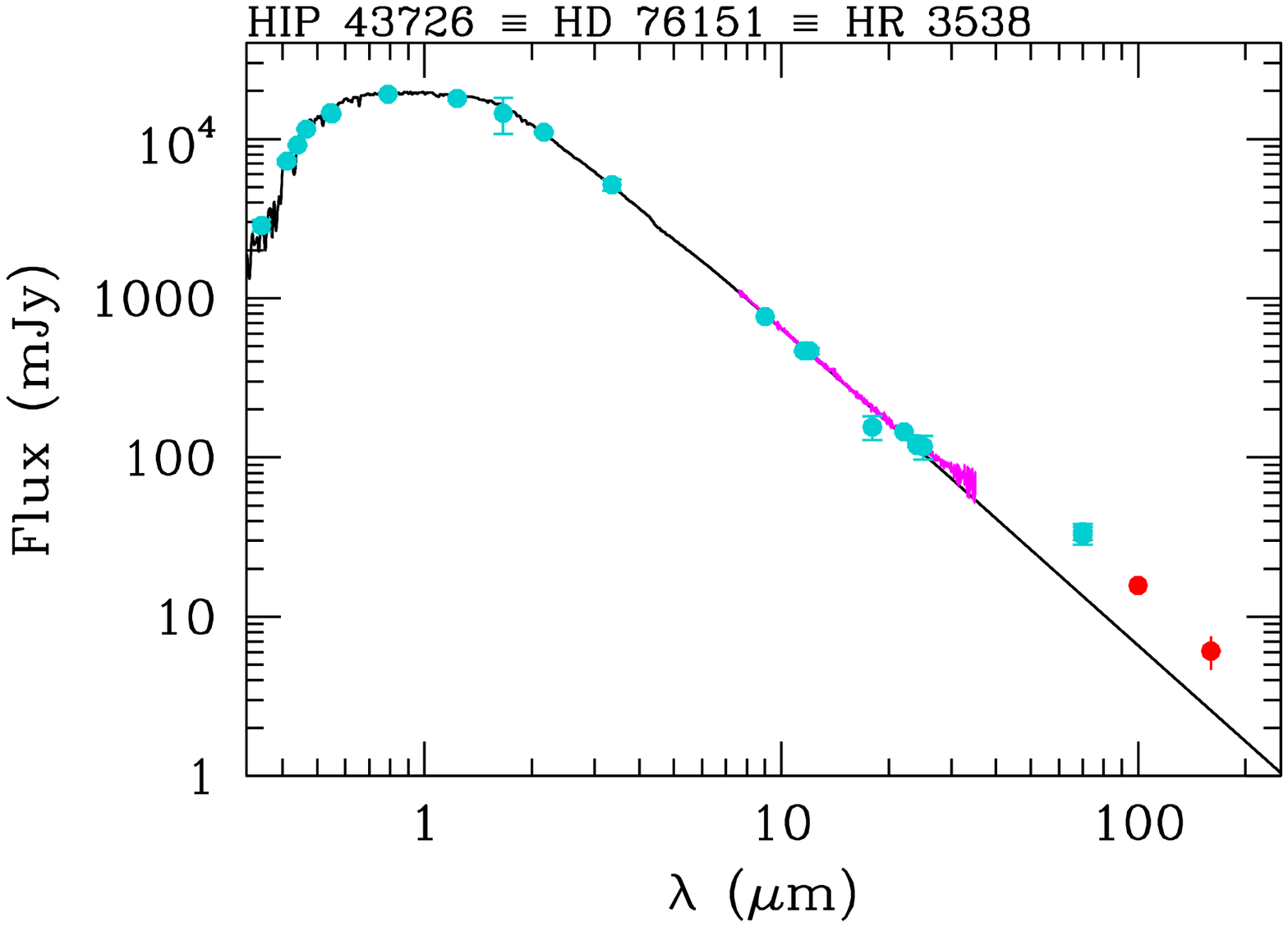}
\end{figure*}
\clearpage

\addtocounter{figure}{-1}
\begin{figure*}
\caption{SEDs of DUNES stars with excesses (continued).}
\includegraphics[scale=0.4]{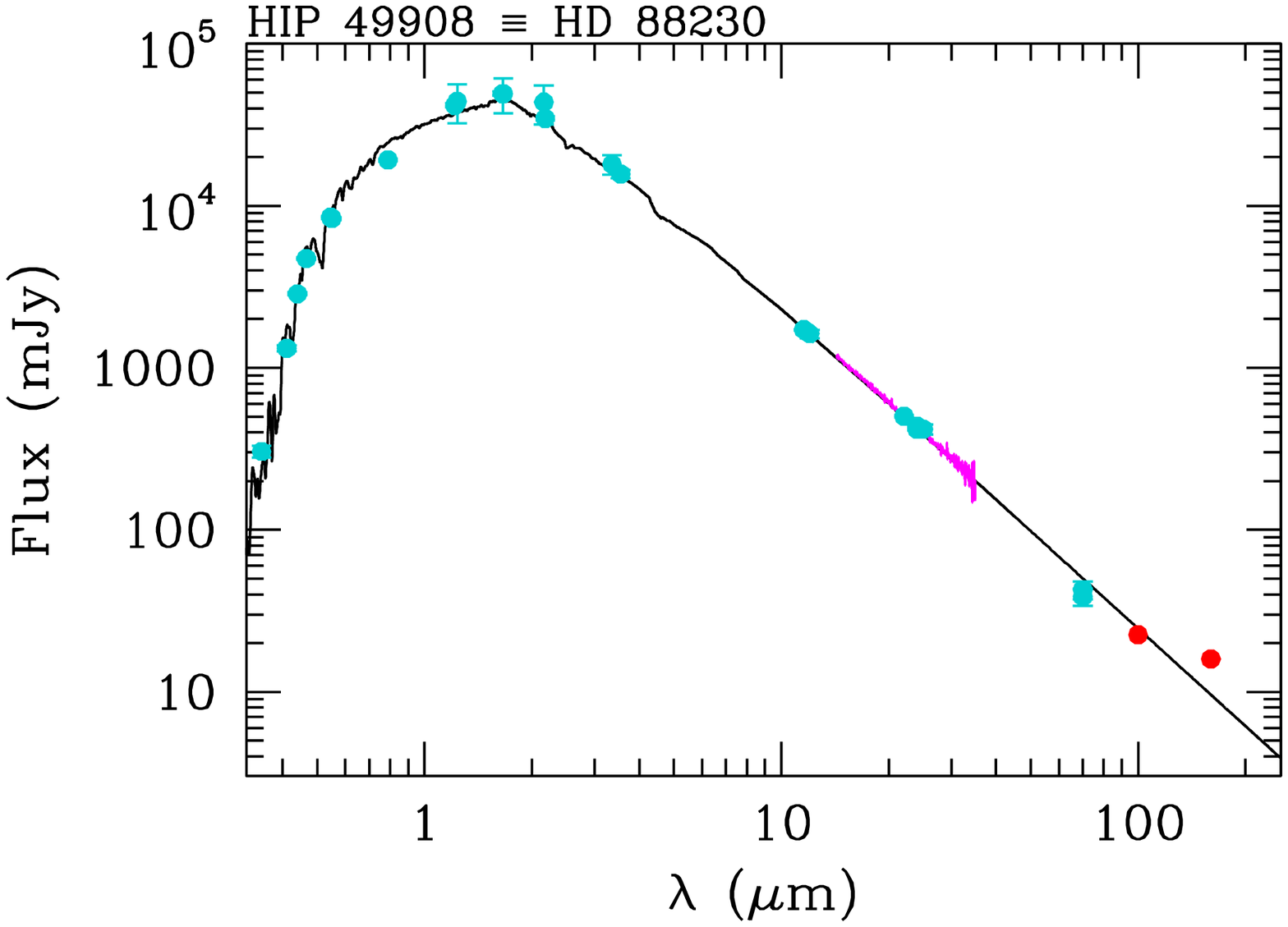}\includegraphics[scale=0.4]{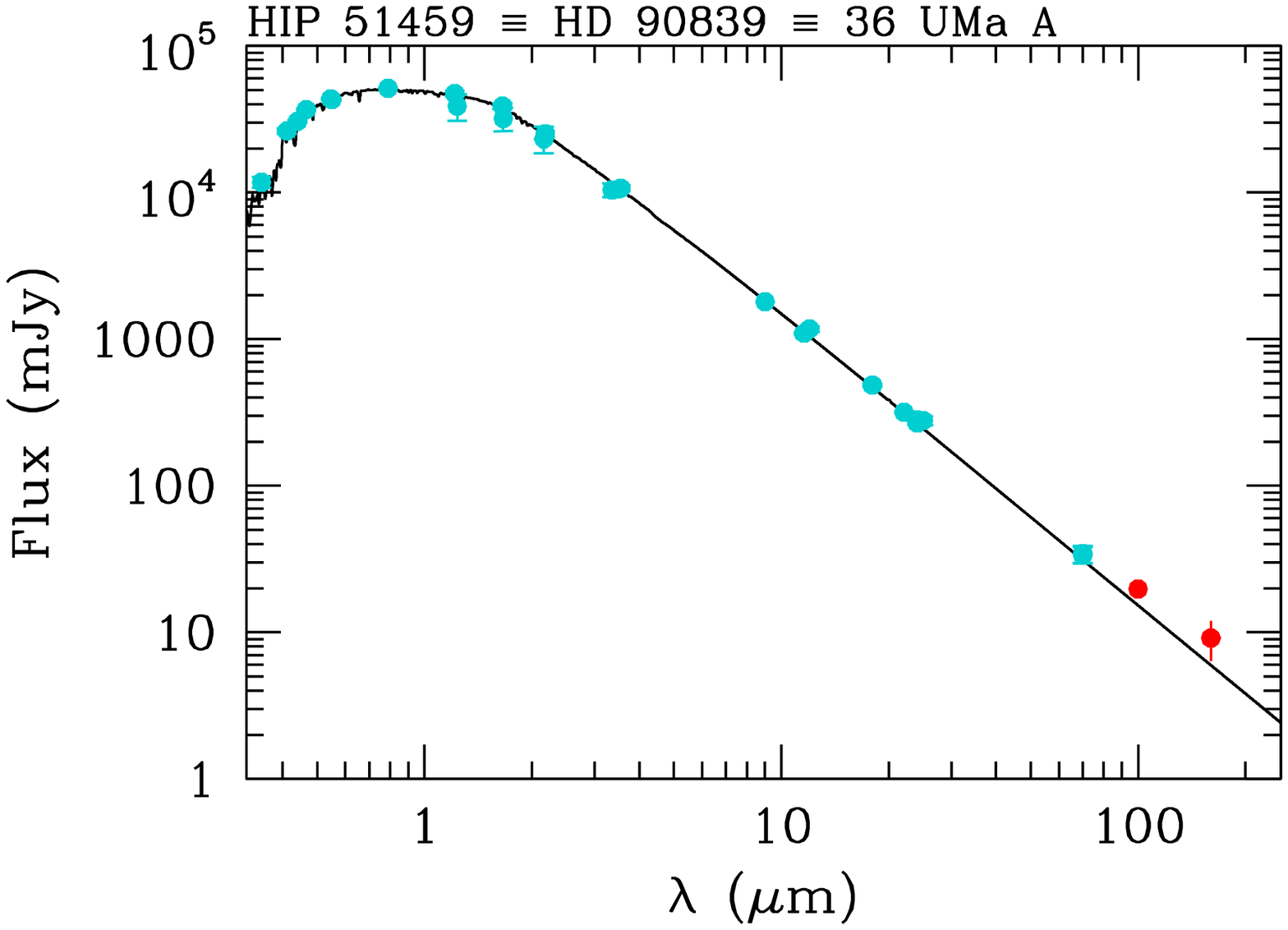}
\includegraphics[scale=0.4]{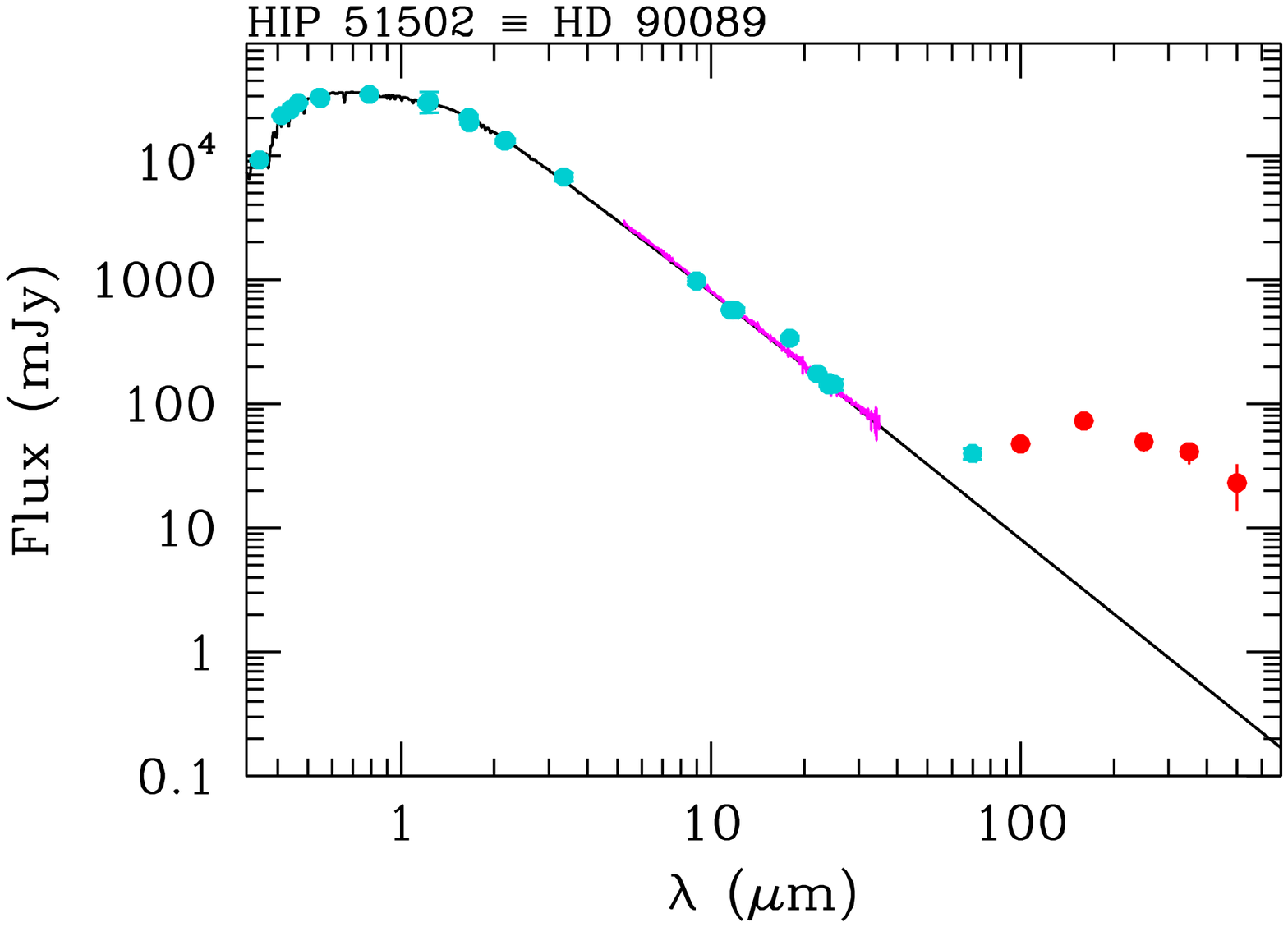}\includegraphics[scale=0.4]{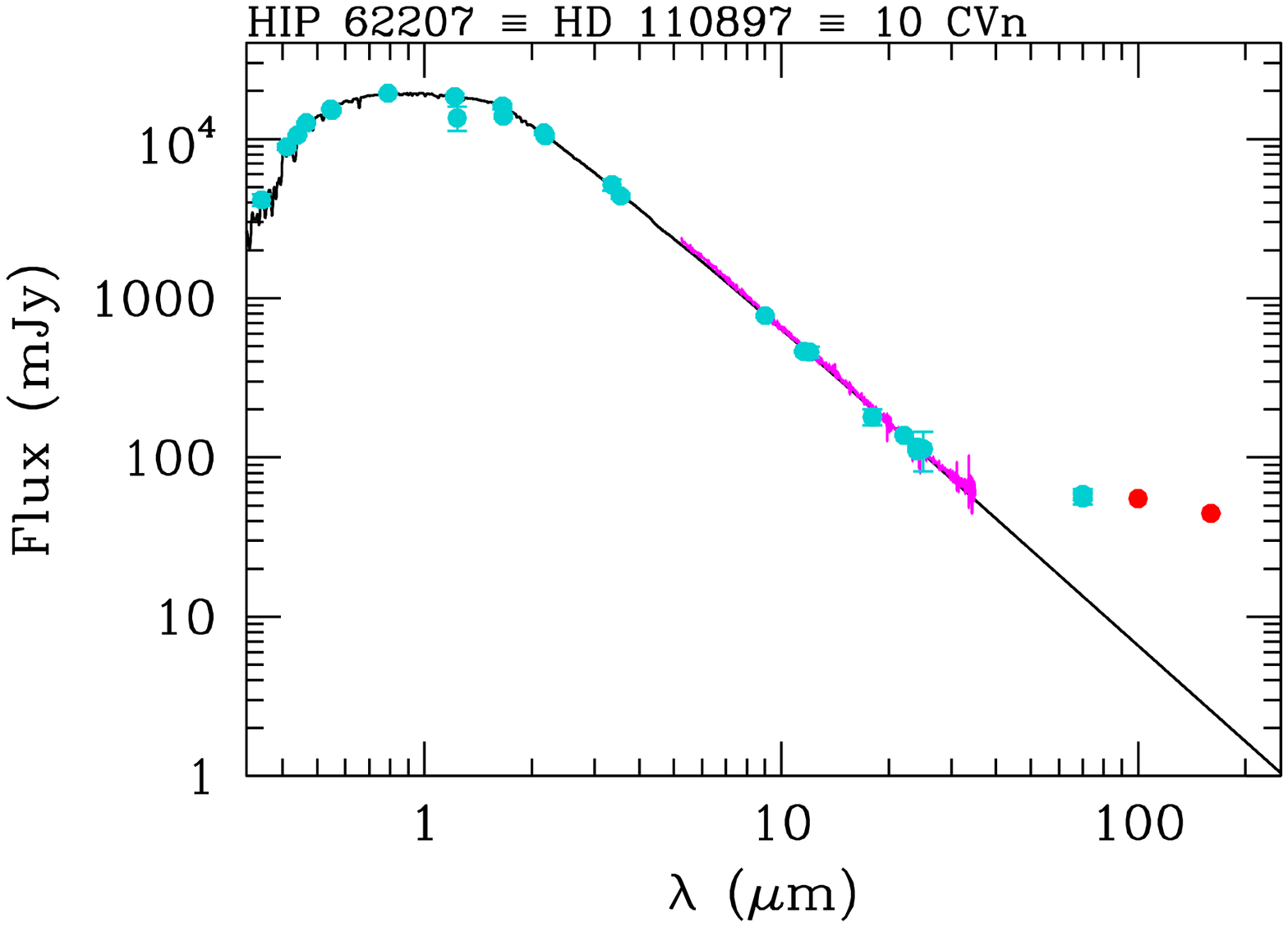}
\includegraphics[scale=0.4]{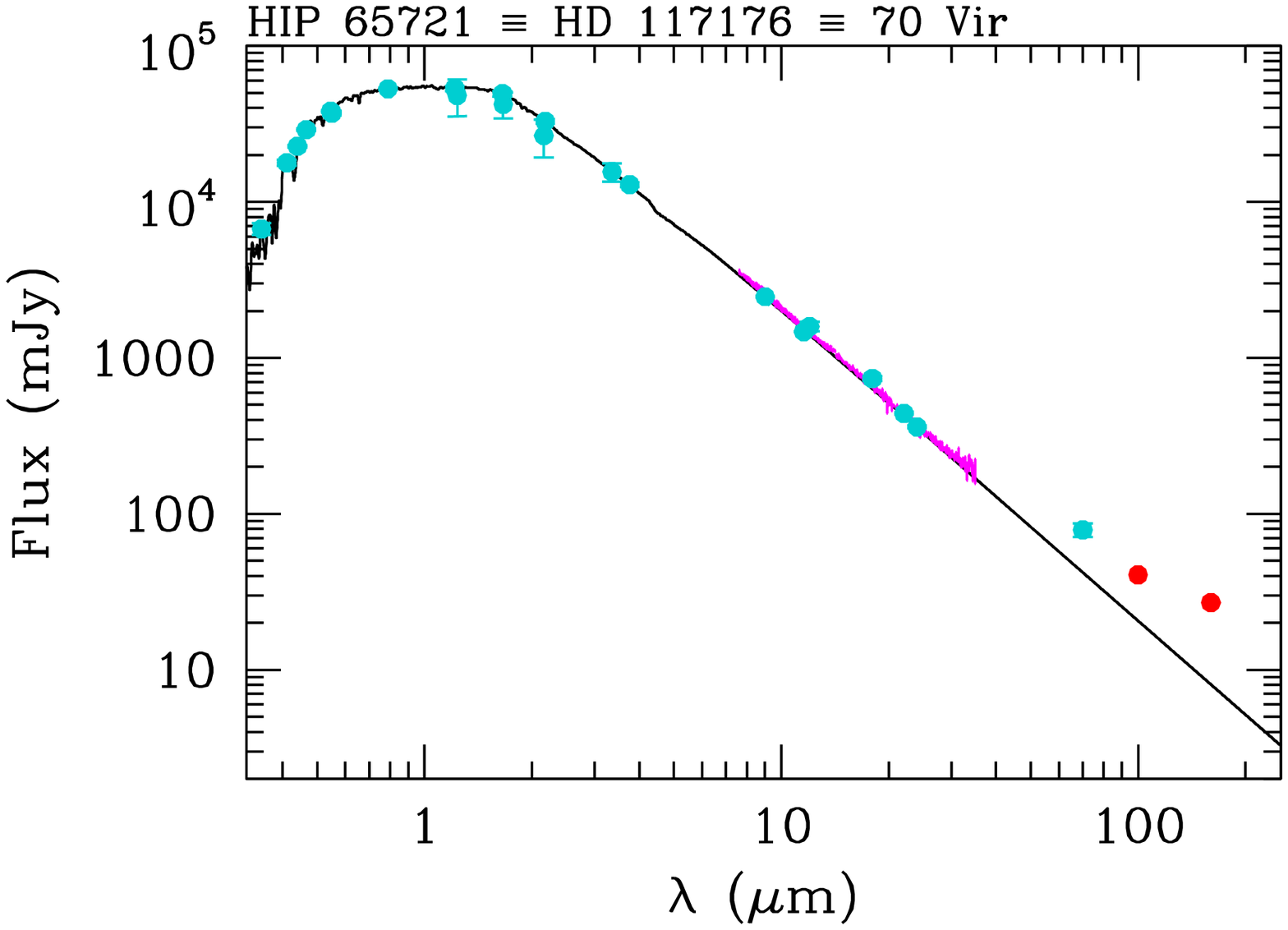}\includegraphics[scale=0.4]{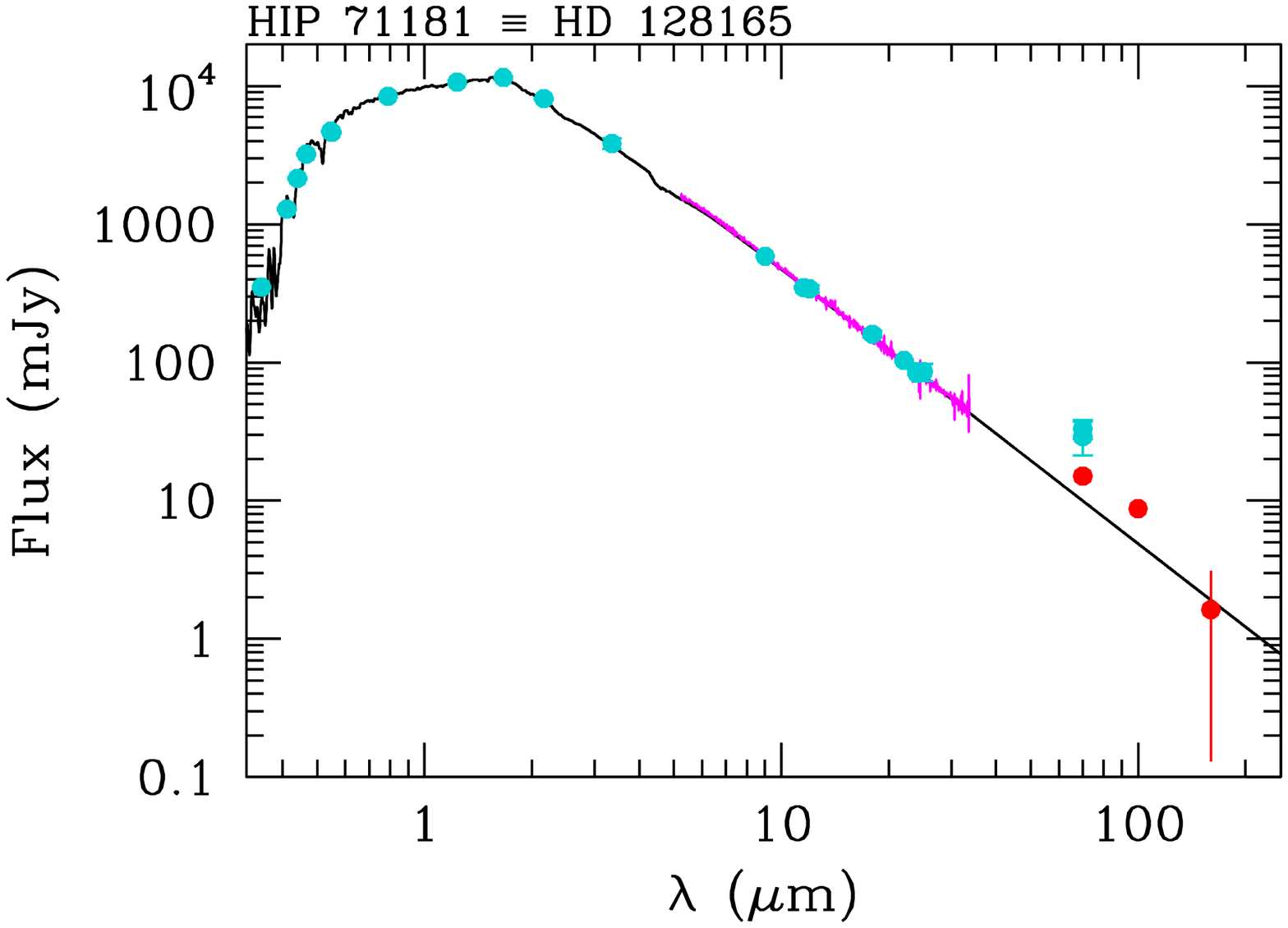}
\includegraphics[scale=0.4]{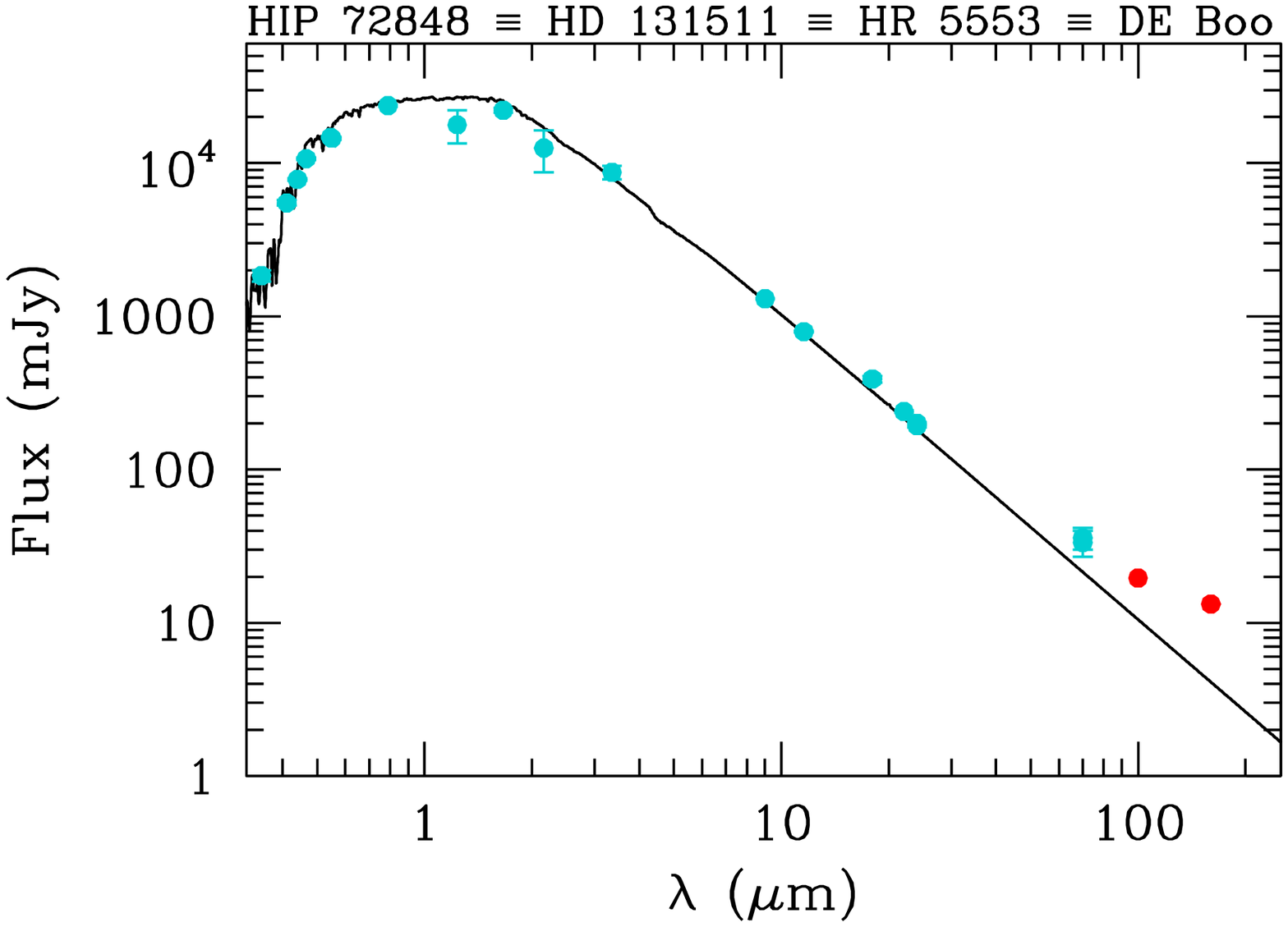}\includegraphics[scale=0.4]{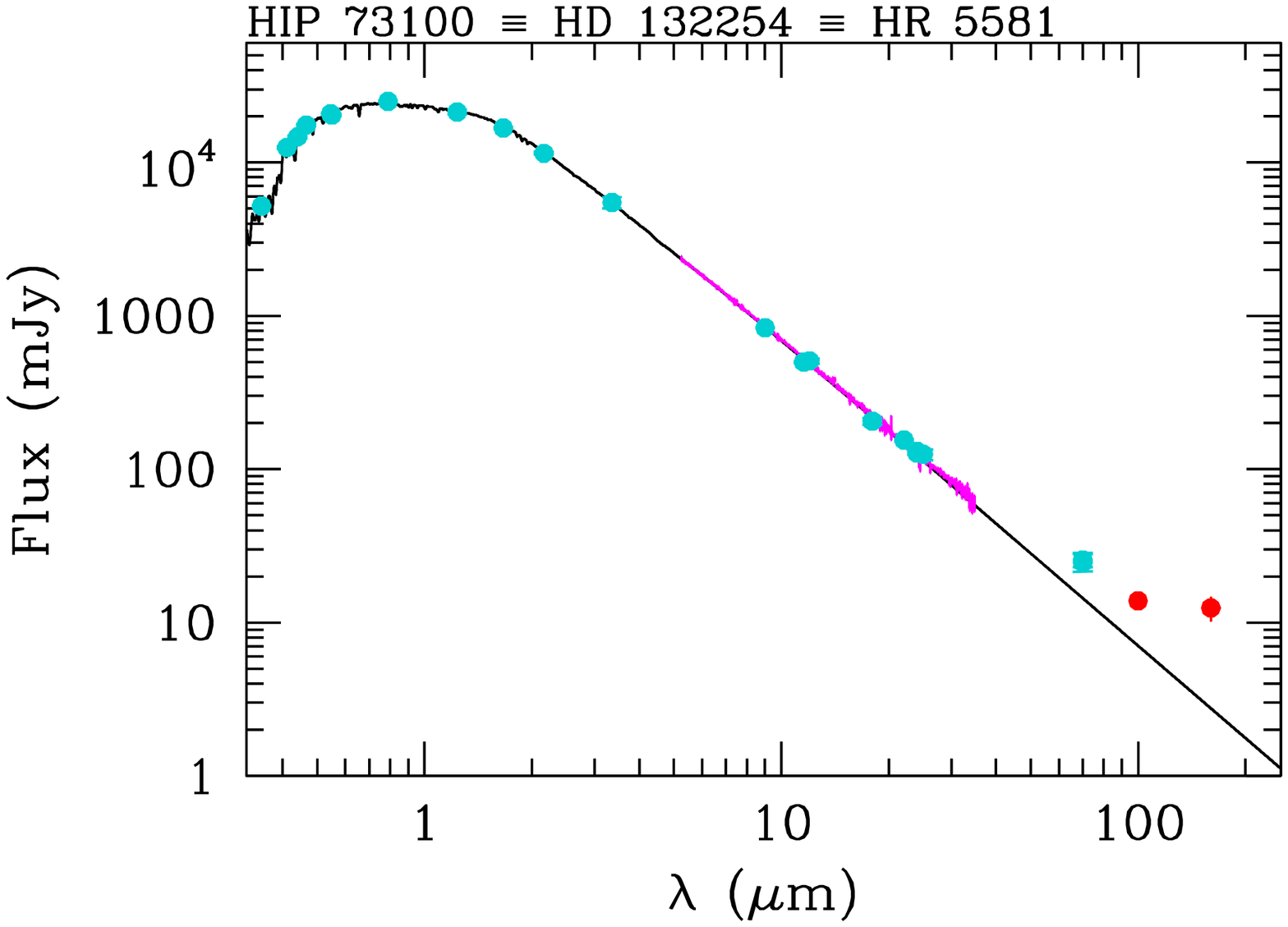}
\end{figure*}
\clearpage

\addtocounter{figure}{-1}
\begin{figure*}
\caption{SEDs of DUNES stars with excesses (continued).}
\includegraphics[scale=0.4]{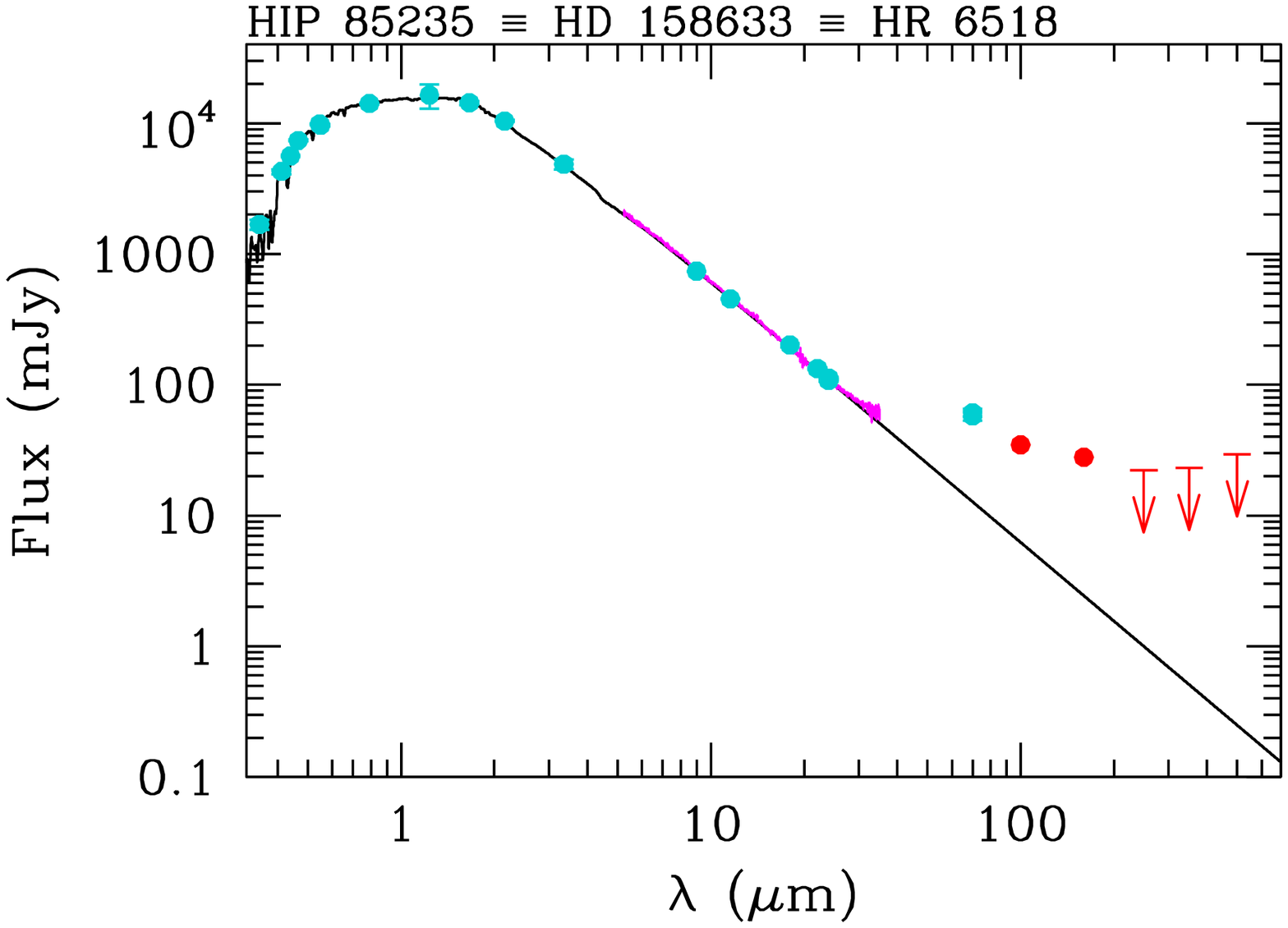}\includegraphics[scale=0.4]{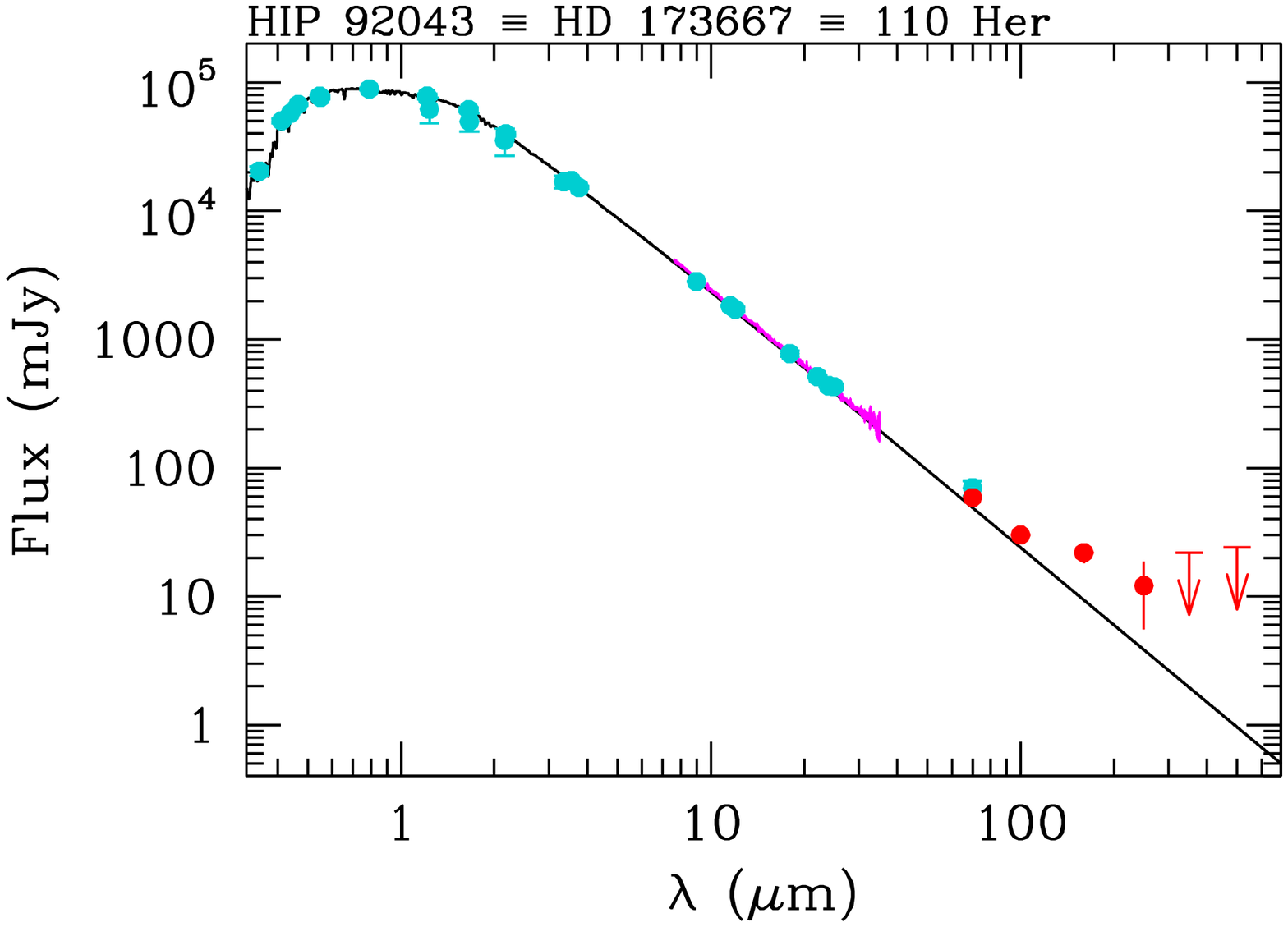}
\includegraphics[scale=0.4]{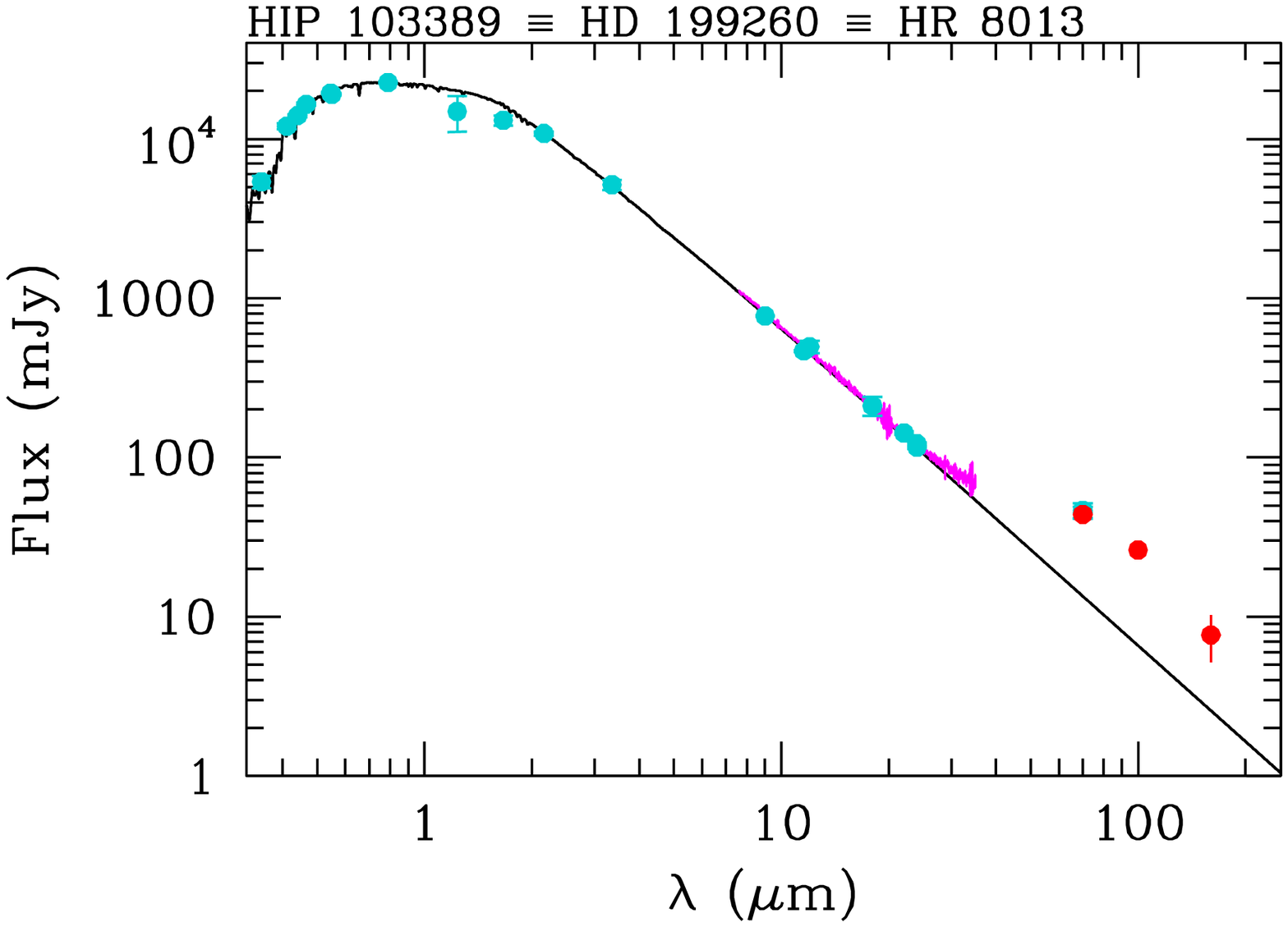}\includegraphics[scale=0.4]{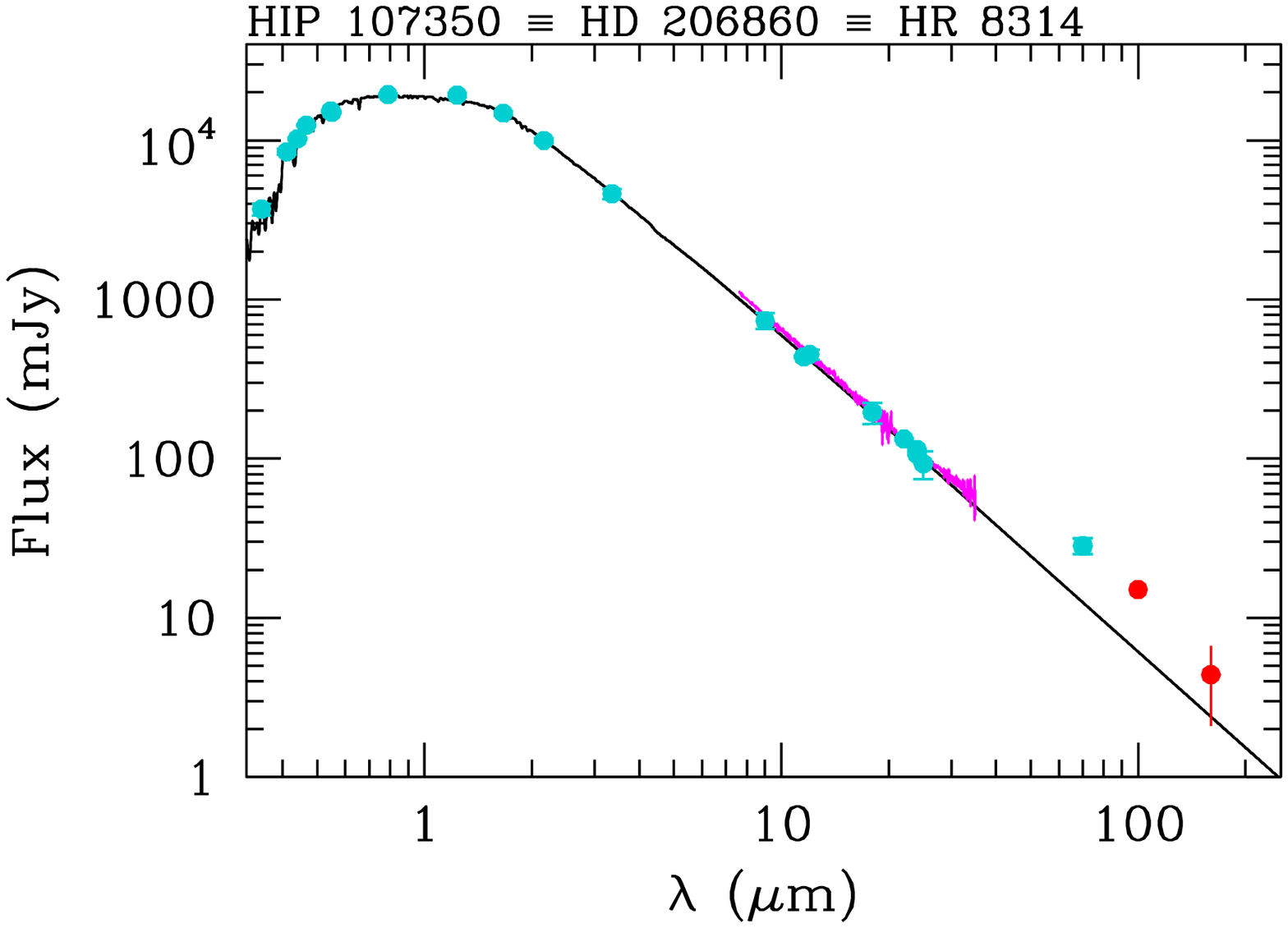}
\includegraphics[scale=0.4]{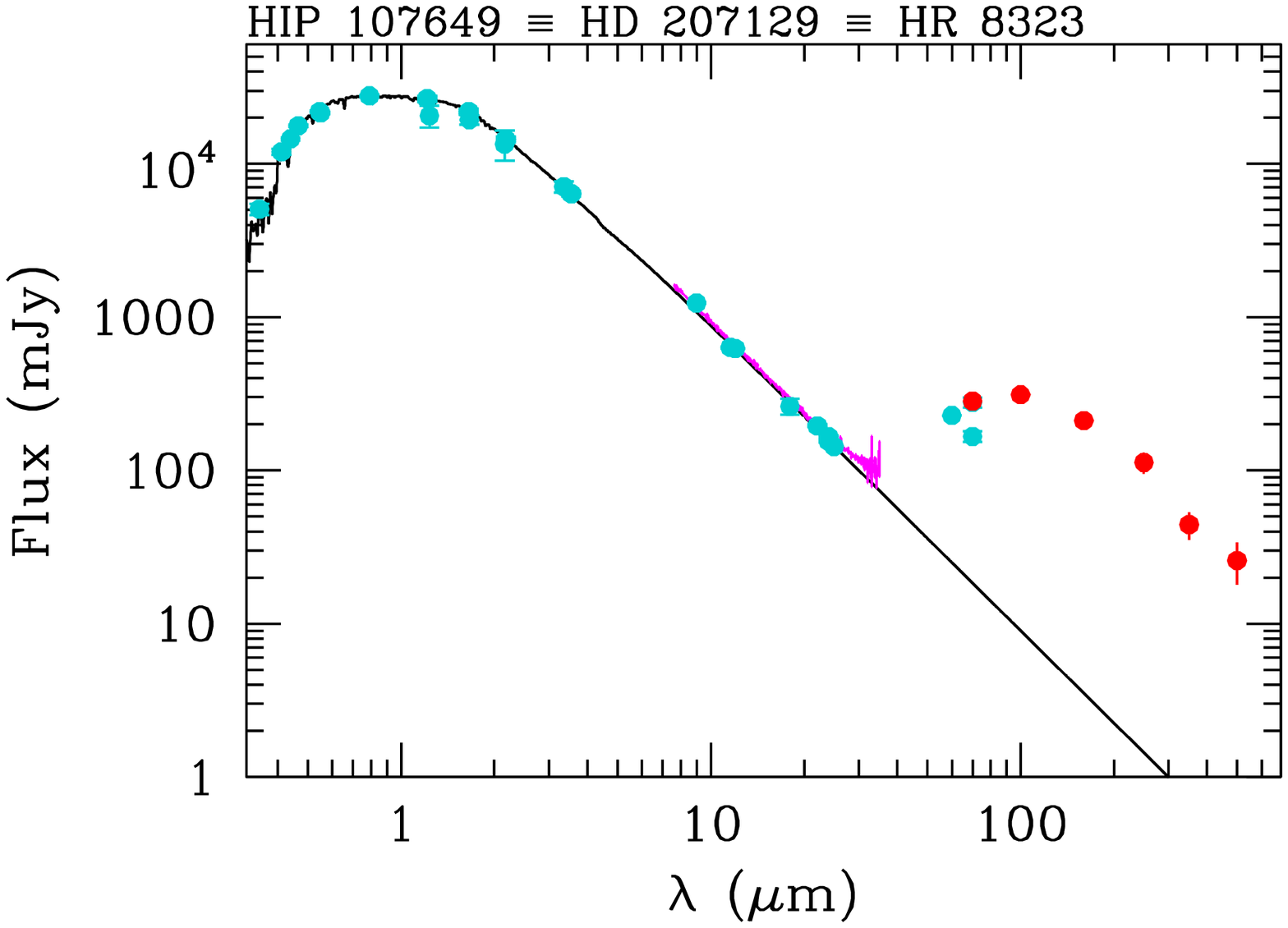}\includegraphics[scale=0.4]{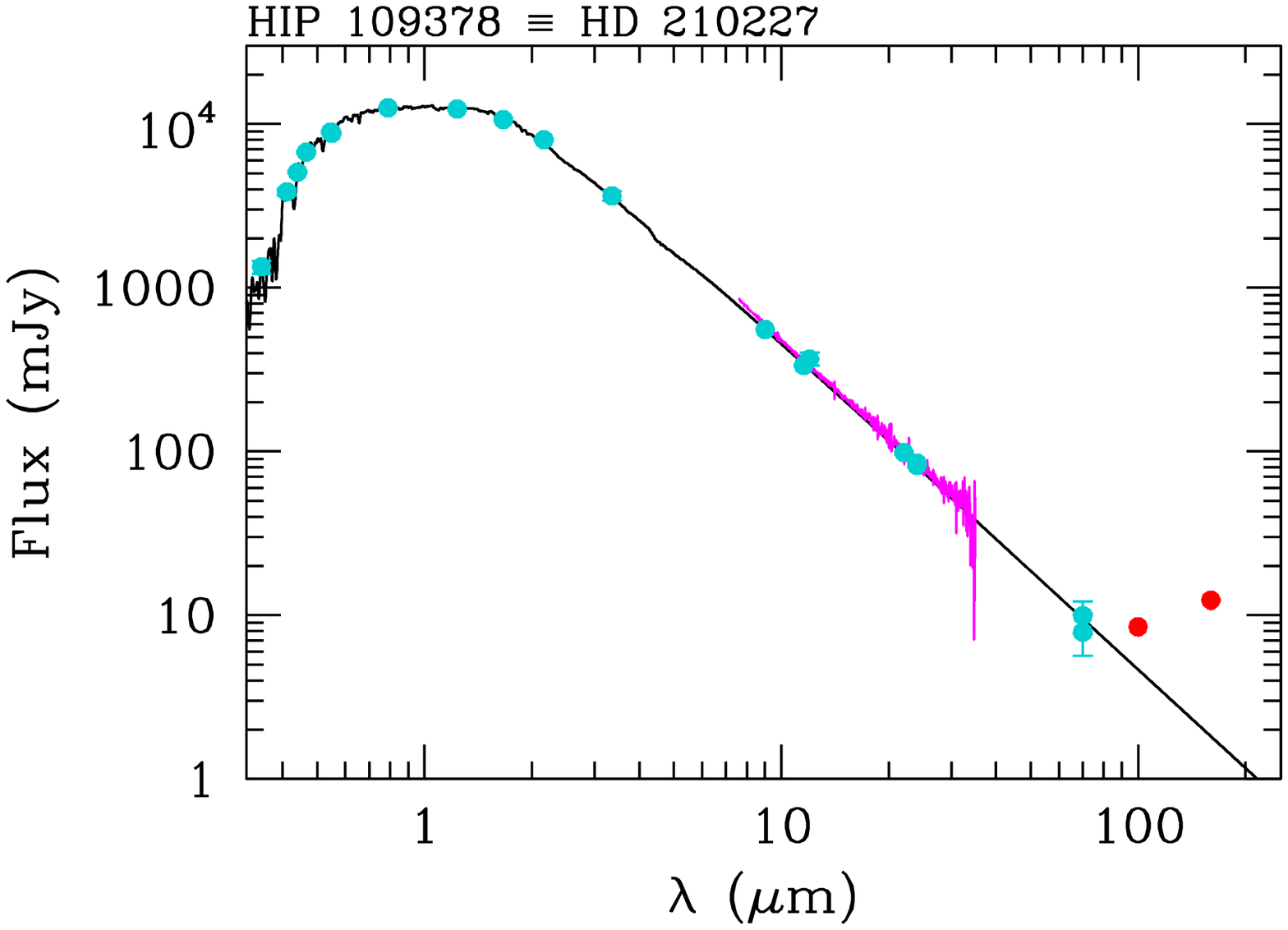}
\includegraphics[scale=0.4]{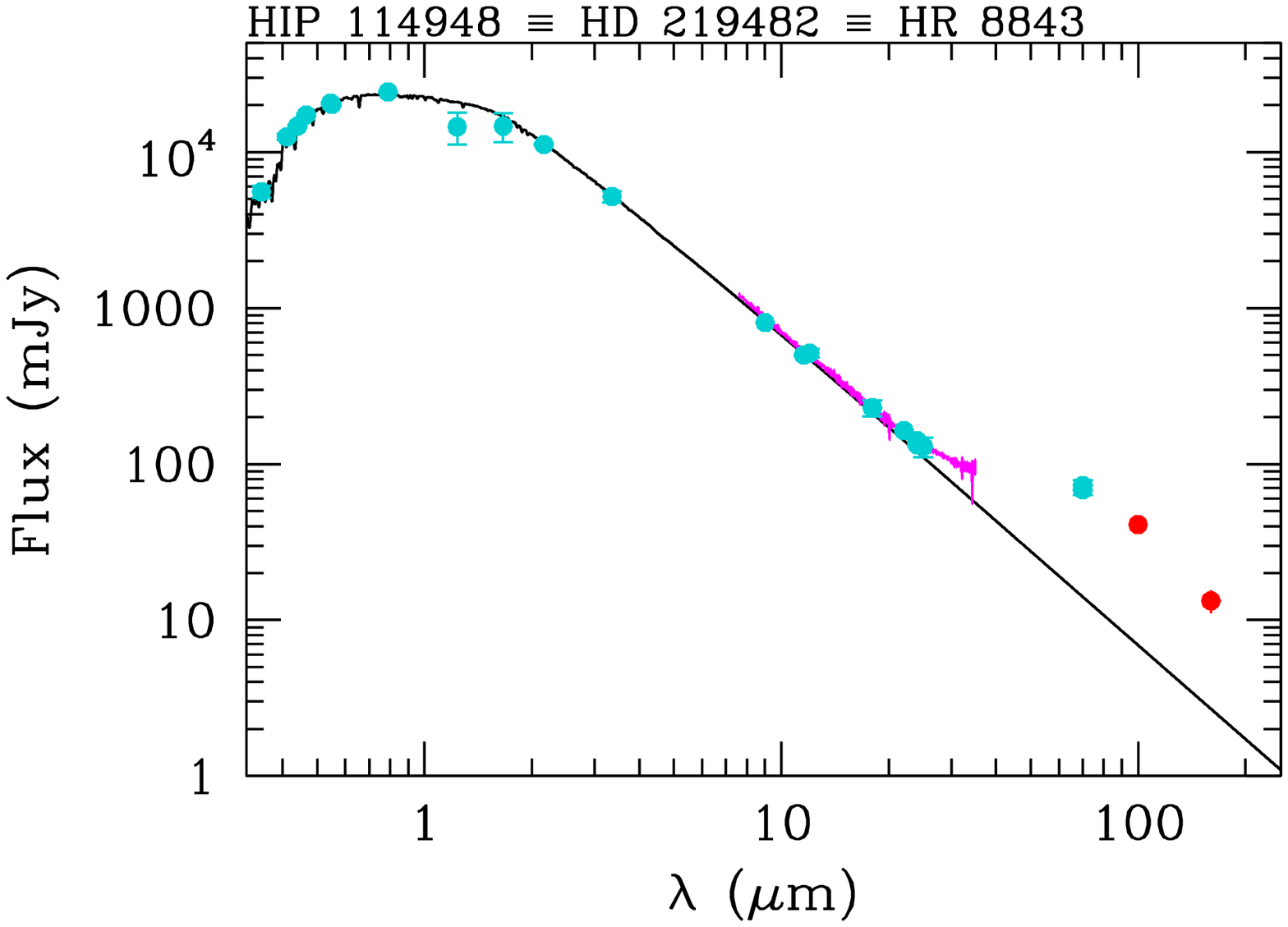}
\end{figure*}
\clearpage
\end{appendix}

\Online

\onllongtabL{2}{
\begin{center}
\label{longtable:sample_stars}
\begin{scriptsize}
\begin{longtable}{lllllllllll}
\multicolumn{10}{c}{\tablename\ \thetable{}. The DUNES stellar sample.}\\
\hline
HIP &  HD    &Name        &      SpT     & SpT range &               ICRS (2000)       &      Galactic    &$\pi$(mas) &d(pc)      \\
\hline
\endfirsthead
\multicolumn{10}{c}{\tablename\ \thetable{}. Continued.}\\
\hline
HIP &  HD    &Name        &      SpT     & SpT range &               ICRS (2000)       &      Galactic    &$\pi$(mas) &d(pc)      \\
\hline
\endhead
\hline
\multicolumn{11}{l}{Continued\dots}\\
\endfoot
\endlastfoot                                                                                                                           
171    &224930 &HR 9088             &G3V       &G2V - G5V        &00 02 10.156 +27 04 56.13&109.6056 -34.5113 &   82.17$\pm$ 2.23& 12.17  \\ 
544    &166    &V439 And            &K0V       &G8V - K0V        &00 06 36.785 +29 01 17.40&111.2636 -32.8326 &   73.15$\pm$ 0.56& 13.67  \\
910    &693    &6 Cet               &F5V       &F5V - F8V        &00 11 15.858 -15 28 04.73&082.2269 -75.0650 &   53.34$\pm$ 0.64& 18.75  \\
2941   &3443   &HR 159              &K1V+...   &G5V - G9V        &00 37 20.720 -24 46 02.18&068.8453 -86.0493 &   64.93$\pm$ 1.85& 15.40  \\
3093   &3651   &54 Psc              &K0V       &K0V - K2V        &00 39 21.806 +21 15 01.71&119.1726 -41.5331 &   90.42$\pm$ 0.32& 11.06  \\
3497   &4308   &LHS 1139            &G3V       &G3V - G6V        &00 44 39.267 -65 38 58.28&304.0542 -51.4639 &   45.34$\pm$ 0.32& 22.06  \\
3821   &4614   &$\eta$ Cas          &G0V SB    &F9V - G0V        &00 49 06.291 +57 48 54.68&122.6200 -05.0551 &  167.98$\pm$ 0.48&  5.95  \\
3909   &4813   &19 Cet              &F7IV-V    &F7IV/V - F9V     &00 50 07.590 -10 38 39.58&121.7965 -73.5132 &   63.48$\pm$ 0.35& 15.75  \\
4148   &5133   &LHS 1163            &K2V       &K2V - K3V        &00 53 01.135 -30 21 24.90&296.8997 -86.7528 &   70.56$\pm$ 0.61& 14.17  \\
7513   &9826   &$\upsilon$ And      &F8V       &F8V - F9V        &01 36 47.842 +41 24 19.64&132.0009 -20.6662 &   74.12$\pm$ 0.19& 13.49  \\
7978   &10647  &HR 506              &F8V       &F8V - F9V        &01 42 29.315 -53 44 27.00&286.8974 -61.7694 &   57.36$\pm$ 0.25& 17.43  \\
8768   &11507  &LHS 1307            &K5/M0V    &K5V - M1.5V      &01 52 49.173 -22 26 05.49&197.6581 -75.3115 &   90.86$\pm$ 1.16& 11.01  \\
10138  &13445  &LHS 13              &K0V       &K1 - K1.5V       &02 10 25.934 -50 49 25.42&275.9263 -61.9610 &   92.74$\pm$ 0.32& 10.78  \\
10798  &14412  &LHS 1387            &G8V       &G8V              &02 18 58.505 -25 56 44.48&214.4545 -70.4102 &   78.93$\pm$ 0.35& 12.67  \\
11452  &15285  &BD+03 339           &K7V:      &K5V - M1.5V      &02 27 45.861 +04 25 55.75&162.9900 -50.7924 &   58.33$\pm$ 1.08& 17.14  \\
11964  &16157  &CC Eri              &K7V SB    &K6Ve -M0VP       &02 34 22.567 -43 47 46.88&258.4808 -63.4144 &   86.18$\pm$ 0.78& 11.60  \\
12777  &16895  &$\theta$ Per         &F7V       &F7V - F8V        &02 44 11.987 +49 13 42.41&141.1655 -09.6097 &   89.87$\pm$ 0.22& 11.13  \\
13402  &17925  &EP Eri              &K1V       &K1V - K3V        &02 52 32.128 -12 46 10.97&192.0737 -58.2540 &   96.60$\pm$ 0.40& 10.35  \\
14954  &19994  &94 Cet              &F8V       &F8V - G0IV       &03 12 46.437 -01 11 45.96&181.5041 -47.3417 &   44.29$\pm$ 0.28& 22.58  \\
15330  &20766  &$\zeta^1$ Ret       &G2V       &G2.5 - G3/5V     &03 17 46.163 -62 34 31.16&279.0962 -47.2141 &   83.28$\pm$ 0.20& 12.01  \\
15371  &20807  &$\zeta^2$ Ret       &G1V       &G0V - G2V        &03 18 12.819 -62 30 22.90&278.9698 -47.2170 &   83.11$\pm$ 0.19& 12.03  \\
15799  &21175  &CCDMJ03236-4005AB   &K0V       &K0V - K1V        &03 23 35.261 -40 04 34.99&245.2416 -56.0919 &   57.40$\pm$ 0.67& 17.42  \\
16134  &21531  &LHS 1548            &K5V       &K5V - K7V        &03 27 52.406 -19 48 16.14&209.9077 -53.5362 &   80.04$\pm$ 0.99& 12.49  \\
17420  &23356  &LTT 1759            &K2V       &K2IV/V - K2.5V   &03 43 55.343 -19 06 39.23&210.8832 -49.7380 &   71.69$\pm$ 0.67& 13.95  \\
17439  &23484  &CD-38 1264          &K1V       &K1V -K2V         &03 44 09.173 -38 16 54.38&241.3261 -52.3674 &   62.39$\pm$ 0.52& 16.03  \\
19849  &26965  &DY Eri              &K1V       &G9V - K1V        &04 15 16.320 -07 39 10.34&200.7528 -38.0478 &  200.62$\pm$ 0.23&  4.98  \\
19884  &27274  &LHS 1650            &K5V       &K4.5V - K5V      &04 15 56.902 -53 18 35.30&262.3295 -44.3763 &   76.66$\pm$ 0.48& 13.04  \\
22263  &30495  &IX Eri              &G3V       &G1.5V - G2.5IV/V &04 47 36.291 -16 56 04.04&215.3638 -34.8052 &   75.32$\pm$ 0.36& 13.28  \\
23311  &32147  &LHS 200             &K3V       &K3V - K4V        &05 00 49.000 -05 45 13.23&205.0915 -27.1757 &  114.84$\pm$ 0.50&  8.71  \\
25110  &33564  &HR 1686             &F6V       &F6V - F7V        &05 22 33.532 +79 13 52.14&133.7345 +22.6480 &   47.88$\pm$ 0.21& 20.89  \\
27887  &40307  &CD-60 1303          &K3V       &K2.5V - K3V      &05 54 04.240 -60 01 24.50&268.8143 -30.3404 &   76.95$\pm$ 0.37& 13.00  \\
28103  &40136  &$\eta$ Lep          &F1V       &F0IV - F2V       &05 56 24.293 -14 10 03.72&219.7553 -18.4714 &   67.21$\pm$ 0.25& 14.88  \\
28442  &40887  &CCDM J06003-3103ABC &K3/K4V    &K5V - K6.5V      &06 00 19.525 -31 01 43.61&236.9472 -23.8110 &   61.00$\pm$21.39& 16.39  \\
29271  &43834  &$\alpha$ Men        &G5V       &G6V - G7V        &06 10 14.474 -74 45 10.96&285.7582 -28.8016 &   98.06$\pm$ 0.14& 10.20  \\
29568  &43162  &V352 CMa            &G5V       &G5V - G6.5V      &06 13 45.296 -23 51 42.98&230.8535 -18.5230 &   59.80$\pm$ 0.49& 16.72  \\
32439  &46588  &HR 2401             &F8V       &F8V              &06 46 14.150 +79 33 53.32&134.5962 +26.4095 &   55.95$\pm$ 0.27& 17.87  \\
32480  &48682  &56 Aur              &G0V       &F9V - G0V        &06 46 44.338 +43 34 38.73&172.3557 +17.5241 &   59.82$\pm$ 0.30& 16.72  \\
33277  &50692  &37 Gem              &G0V       &G0V              &06 55 18.668 +25 22 32.52&190.4185 +12.0639 &   58.00$\pm$ 0.41& 17.24  \\
34017  &52711  &LHS 1893            &G4V       &G0V - G4V, G8IV  &07 03 30.458 +29 20 13.49&187.4376 +15.3223 &   52.27$\pm$ 0.41& 19.13  \\
34065  &53705  &HR 2667             &G3V...    &G0V - G3V        &07 03 57.317 -43 36 28.94&254.0840 -16.2198 &   60.55$\pm$ 1.04& 16.52  \\
35136  &55575  &HR 2721             &G0V       &F9V - G0V        &07 15 50.138 +47 14 23.87&170.3427 +23.5149 &   59.20$\pm$ 0.33& 16.89  \\
36439  &58855  &22 Lyn              &F6V       &F6V              &07 29 55.957 +49 40 20.86&168.3340 +26.3536 &   49.41$\pm$ 0.36& 20.24  \\
38382  &64096  &9 Pup               &G2V       &G0V - G2V        &07 51 46.303 -13 53 52.92&232.2713 +06.6205 &   60.59$\pm$ 0.59& 16.50  \\
38784  &62613  &LHS 5130            &G8V       &G8V - K0V        &07 56 17.228 +80 15 55.95&133.6760 +29.4348 &   58.17$\pm$ 0.36& 17.19  \\
40693  &69830  &LHS 245             &K0V       &G8V - K0V        &08 18 23.947 -12 37 55.81&234.5634 +12.8237 &   80.04$\pm$ 0.35& 12.49  \\
40843  &69897  &$\chi$ Cnc          &F6V       &F6V - F7V        &08 20 03.862 +27 13 03.74&195.7841 +30.4668 &   54.73$\pm$ 0.32& 18.27  \\
42430  &73752  &CCDM J08391-2240AB  &G3/G5V    &G5IV/V - G8III/V &08 39 07.901 -22 39 42.81&245.8887 +11.3140 &   51.55$\pm$ 0.63& 19.40  \\
42438  &72905  &$\pi^{01}$ UMa       &G1.5Vb    &G0.5V - G1.5V    &08 39 11.704 +65 01 15.27&150.5517 +35.7042 &   69.66$\pm$ 0.37& 14.36  \\
43587  &75732  &55 Cnc              &G8V       &G8V - K0V        &08 52 35.811 +28 19 50.95&196.7946 +37.6989 &   81.03$\pm$ 0.75& 12.34  \\
43726  &76151  &LTT 3283            &G3V       &G2V - G5V        &08 54 17.948 -05 26 04.06&233.2058 +24.1642 &   57.52$\pm$ 0.39& 17.39  \\
44897  &78366  &LTT 12401           &F9V       &F9V - G0IV/V     &09 08 51.071 +33 52 55.98&190.5283 +42.1549 &   52.11$\pm$ 0.33& 19.19  \\
45333  &79028  &16 UMa              &F9V       &F9V - G0 IV/V    &09 14 20.543 +61 25 23.94&153.6241 +40.4672 &   51.10$\pm$ 0.32& 19.57  \\
45617  &79969  &IP Cnc              &K3V       &K3V              &09 17 53.456 +28 33 37.86&198.1045 +43.1732 &   57.92$\pm$ 0.76& 17.27  \\
46580  &82106  &LHS 2147            &K3V       &K3V              &09 29 54.825 +05 39 18.49&227.6707 +37.5676 &   77.48$\pm$ 0.64& 12.91  \\
47592  &84117  &LTT 3558            &G0V       &F8V - G0V        &09 42 14.417 -23 54 56.05&256.6978 +21.5239 &   66.61$\pm$ 0.21& 15.01  \\
49081  &86728  &LHS 2216            &G1V       &G1V - G5V        &10 01 00.658 +31 55 25.22&195.0155 +52.8533 &   66.46$\pm$ 0.32& 15.05  \\
49908  &88230  &NSV 4765            &K8V       &K6V - M0V        &10 11 22.141 +49 27 15.26&165.8661 +52.1456 &  205.21$\pm$ 0.54&  4.87  \\
51459  &90839  &36 UMa              &F8V       &F6V - F8V        &10 30 37.580 +55 58 49.93&154.2918 +51.7025 &   78.25$\pm$ 0.28& 12.78  \\
51502  &90089  &HR 4084             &F2V       &F2V - F4V        &10 31 04.644 +82 33 30.91&128.0322 +33.1244 &   46.51$\pm$ 1.39& 21.50  \\
53721  &95128  &47 UMa              &G0V       &G0V - G1V        &10 59 27.973 +40 25 48.92&175.7824 +63.3691 &   71.11$\pm$ 0.25& 14.06  \\
54646  &97101  &LHS 2367            &K8V       &K7V - M1V        &11 11 05.172 +30 26 45.66&198.3695 +67.8103 &   84.23$\pm$ 0.86& 11.87  \\
56452  &100623 &HR 4458             &K0V       &K0V              &11 34 29.487 -32 49 52.82&284.7773 +27.3133 &  104.61$\pm$ 0.37&  9.56  \\
57507  &102438 &HR 4525             &G5V       &G5V - G8V        &11 47 15.808 -30 17 11.44&286.8400 +30.5720 &   57.23$\pm$ 0.41& 17.47  \\
57939  &103095 &LHS 44              &G8Vp      &G8V - K1V        &11 52 58.769 +37 43 07.23&168.5310 +73.7783 &  109.99$\pm$ 0.41&  9.09  \\
58345  &103932 &LHS 319             &K4V       &K4V              &11 57 56.207 -27 42 25.37&288.6842 +33.6802 &   98.45$\pm$ 0.57& 10.16  \\
62145  &110833 &G 199-36            &K3V       &K2V - K3V        &12 44 14.545 +51 45 33.49&125.6002 +65.3317 &   67.20$\pm$ 0.66& 14.88  \\
62207  &110897 &10 CVn              &G0V       &F7V - G1V        &12 44 59.405 +39 16 44.11&128.8344 +77.7753 &   57.55$\pm$ 0.32& 17.38  \\
62523  &111395 &LW Com              &G7V       &G5V - G7V        &12 48 47.048 +24 50 24.82&288.1762 +87.6355 &   59.06$\pm$ 0.45& 16.93  \\
64792  &115383 &59 Vir              &G0Vs      &F8V - G0V/IV     &13 16 46.516 +09 25 26.96&322.7902 +71.3124 &   56.95$\pm$ 0.26& 17.56  \\
64797  &115404 &IDS 13118+1733      &K2V       &K1V - K2.5V      &13 16 51.053 +17 01 01.84&334.4012 +78.3056 &   90.32$\pm$ 0.74& 11.07  \\
65026  &115953 &CCDMJ13198+4747AB   &K0        &K0, M0V - M3V    &13 19 45.653 +47 46 40.95&109.7971 +68.6214 &   93.40$\pm$ 2.21& 10.71  \\
65721  &117176 &70 Vir              &G5V       &G4V - G5V        &13 28 25.809 +13 46 43.64&337.6654 +74.1007 &   55.60$\pm$ 0.24& 17.99  \\
67275  &120136 &$\tau$ Boo           &F7V       &F6IV - F7V/IV    &13 47 15.743 +17 27 24.86&358.9368 +73.8890 &   64.03$\pm$ 0.19& 15.62  \\
67422  &120476 &CCDM J13491+2659AB  &K2        &K2V - K4V        &13 49 04.001 +26 58 47.68&035.5632 +77.1755 &   74.58$\pm$ 0.78& 13.41  \\
67620  &120690 &LHS 2814            &G5V       &G5Va             &13 51 20.328 -24 23 25.33&319.9539 +36.4935 &   51.35$\pm$ 0.45& 19.47  \\
68184  &122064 &G 239-8             &K3V       &K3V              &13 57 32.058 +61 29 34.30&109.6142 +53.8932 &   99.36$\pm$ 0.32& 10.06  \\
68682  &122742 &HR 5273             &G8V       &G6V - G8V        &14 03 32.351 +10 47 12.40&352.4578 +66.4473 &   58.88$\pm$ 0.62& 16.98  \\
69965  &125276 &LHS 2891            &F7Vw      &F5V - F9V        &14 19 00.895 -25 48 55.53&326.5293 +33.0120 &   55.45$\pm$ 0.82& 18.03  \\
70319  &126053 &LHS 2907            &G1V       &G1V - G2V        &14 23 15.285 +01 14 29.65&347.1642 +56.0170 &   58.17$\pm$ 0.53& 17.19  \\
70857  &128642 &BD+81 482           &G5        &G5               &14 29 22.296 +80 48 35.48&118.2735 +35.4147 &   50.27$\pm$ 0.48& 19.89  \\
71181  &128165 &BD+53 1719          &K3V       &K3V              &14 33 28.868 +52 54 31.65&093.5731 +58.0096 &   75.65$\pm$ 0.42& 13.22  \\
71681  &128621 &$\alpha$ Cen B      &K1V       &K1V - K2IV       &14 39 35.063 -60 50 15.10&315.7303 -00.6814 &  796.92$\pm$25.90&  1.25  \\
71683  &128620 &$\alpha$ Cen A      &G2V       &G2V              &14 39 36.494 -60 50 02.37&315.7342 -00.6796 &  754.81$\pm$ 4.11&  1.32  \\
71908  &128898 &$\alpha$ Cir        &F1Vp      &A7p - F1Vp       &14 42 30.420 -64 58 30.49&314.3349 -04.5917 &   60.35$\pm$ 0.14& 16.57  \\
72567  &130948 &HP Boo              &G2V       &F9IV/V - G2V     &14 50 15.811 +23 54 42.64&032.7645 +63.0621 &   55.03$\pm$ 0.34& 18.17  \\
72603  &130819 &8 Lib               &F3V       &F3V - F4IV       &14 50 41.181 -15 59 50.05&340.3091 +38.0711 &   43.52$\pm$ 0.43& 22.98  \\
72848  &131511 &DE Boo              &K2V       &K0.5V - K2V      &14 53 23.767 +19 09 10.08&023.5506 +60.9355 &   86.88$\pm$ 0.46& 11.51  \\
73100  &132254 &LTT 14437           &F7V       &F5V - F8V        &14 56 23.041 +49 37 42.42&084.6437 +57.1726 &   39.83$\pm$ 0.26& 25.11  \\
73182  &131976 &BD -20 4123         &K5V       &M0.5V - M2V      &14 57 26.533 -21 24 41.58&338.2393 +32.6833 &  168.77$\pm$21.54&  5.93  \\
73184  &131977 &KX Lib              &K4V       &K4V - K5V        &14 57 28.001 -21 24 55.71&338.2423 +32.6769 &  171.22$\pm$ 0.94&  5.84  \\
73996  &134083 &45 Boo              &F5V       &F5V - F6V        &15 07 18.066 +24 52 09.10&036.4509 +59.4794 &   51.14$\pm$ 0.31& 19.55  \\
77052  &140538 &$\psi$ Ser          &G5V       &G2V - G8V        &15 44 01.821 +02 30 54.63&009.6985 +41.9676 &   68.22$\pm$ 0.66& 14.66  \\
78459  &143761 &$\rho$ CrB          &G2V       &G0V - G0IV       &16 01 02.662 +33 18 12.63&053.4888 +48.9235 &   58.02$\pm$ 0.28& 17.24  \\
78775  &144579 &LHS 3152            &G8V       &G8V - K0V        &16 04 56.793 +39 09 23.43&062.3693 +48.2726 &   68.87$\pm$ 0.33& 14.52  \\
79248  &145675 &14 Her              &K0V       &K0IV - K0V       &16 10 24.314 +43 49 03.53&069.1704 +46.9449 &   56.91$\pm$ 0.34& 17.57  \\
80725  &148653 &CCDM J16289+1825AB  &K2V       &K1V - K2V        &16 28 52.661 +18 24 50.62&035.0885 +39.4996 &   50.87$\pm$ 0.80& 19.66  \\
82860  &153597 &19 Dra              &F6Vvar    &F6V              &16 56 01.689 +65 08 05.26&095.6783 +36.4627 &   65.54$\pm$ 0.33& 15.26  \\
83389  &154345 &LHS 3260            &G8V       &G8V - K0V        &17 02 36.404 +47 04 54.76&073.0223 +37.6635 &   53.80$\pm$ 0.32& 18.59  \\
84862  &157214 &W Her               &G0V       &G0V - G2V        &17 20 39.295 +32 28 21.15&055.8658 +32.3091 &   69.80$\pm$ 0.25& 14.33  \\
85235  &158633 &LHS 3287            &K0V       &G8V - K0V        &17 25 00.099 +67 18 24.15&097.5588 +33.1704 &   78.11$\pm$ 0.30& 12.80  \\
85295  &157881 &LHS 447             &K7V       &K5V - M1         &17 25 45.233 +02 06 41.12&024.7576 +19.9755 &  129.86$\pm$ 0.73&  7.70  \\
86036  &160269 &26 Dra              &G0V       &G0V - G2V        &17 34 59.594 +61 52 28.40&090.9931 +32.6496 &   70.47$\pm$ 0.37& 14.19  \\
86796  &160691 &$\mu$ Ara           &G5V       &G3V/VI - G5V     &17 44 08.701 -51 50 02.59&340.0607 -11.4981 &   64.47$\pm$ 0.31& 15.51  \\
88601  &165341 &70 Oph              &K0V SB    &K0V              &18 05 27.285 +02 30 00.36&029.8934 +11.3670 &  196.72$\pm$ 0.83&  5.08  \\
88972  &166620 &HR 6806             &K2V       &K2V              &18 09 37.416 +38 27 28.00&065.3335 +24.2075 &   90.71$\pm$ 0.30& 11.02  \\
89042  &165499 &HR 6761             &G0V       &G0V - G3IV/V     &18 10 26.163 -62 00 07.91&332.2098 -19.1639 &   56.78$\pm$ 0.52& 17.61  \\
91009  &234677 &BY Dra              &K7Vvar    &K4V - K6V        &18 33 55.773 +51 43 08.91&080.5567 +23.5794 &   61.15$\pm$ 0.68& 16.35  \\
92043  &173667 &110 Her             &F6V       &F5V - F7IV       &18 45 39.726 +20 32 46.72&050.7901 +10.4300 &   52.06$\pm$ 0.25& 19.21  \\
95995  &184467 &LHS 3466            &K1V       &K1V - 2V         &19 31 07.974 +58 35 09.64&090.2567 +18.0113 &   58.96$\pm$ 0.65& 16.96  \\
96100  &185144 &$\sigma$ Dra        &K0V       &G9V - K0V        &19 32 21.590 +69 39 40.24&101.3033 +21.8770 &  173.77$\pm$ 0.18&  5.75  \\
96441  &185395 &13 Cyg              &F4V       &F3V - F4V        &19 36 26.534 +50 13 15.96&082.6677 +13.8470 &   54.54$\pm$ 0.15& 18.34  \\
97944  &188088 &V4200 Sgr           &K3/K4V    &K2V/VI - K4V     &19 54 17.745 -23 56 27.85&017.1729 -23.9128 &   71.18$\pm$ 0.42& 14.05  \\
98959  &189567 &LHS 484             &G2V       &G2V - 5V         &20 05 32.765 -67 19 15.22&328.4657 -31.9895 &   56.41$\pm$ 0.44& 17.73  \\
99240  &190248 &$\delta$ Pav        &G5IV-Vvar &G8IV             &20 08 43.610 -66 10 55.44&329.7673 -32.4165 &  163.71$\pm$ 0.17&  6.11  \\
99461  &191408 &LHS 486             &K2V       &K2V - K2.5V      &20 11 11.938 -36 06 04.36&005.2332 -30.9247 &  166.25$\pm$ 0.27&  6.02  \\
101955 &196795 &OQ Del              &K5V       &K5V - M0/1V      &20 39 37.710 +04 58 19.34&050.6634 -21.3253 &   59.80$\pm$ 3.42& 16.72  \\
101997 &196761 &LHS 3570            &G8/K0V    &G7.5V - K0V      &20 40 11.756 -23 46 25.92&021.2637 -33.7781 &   69.53$\pm$ 0.40& 14.38  \\
103389 &199260 &HR 8013             &F7V       &F6V - F8V        &20 56 47.331 -26 17 46.96&019.5852 -38.0955 &   45.52$\pm$ 0.38& 21.97  \\
104214 &201091 &61 Cyg A            &K5V       &K5V              &21 06 53.952 +38 44 57.99&082.3197 -05.8181 &  286.82$\pm$ 6.78&  3.49  \\
104217 &201092 &61 Cyg B            &K7V       &K7V - M0V        &21 06 55.264 +38 44 31.40&082.3171 -05.8262 &  285.88$\pm$ 0.54&  3.50  \\
105312 &202940 &HR 8148             &G5V       &G5V - G7V        &21 19 45.625 -26 21 10.38&021.1811 -43.0983 &   55.65$\pm$ 0.62& 17.97  \\
106696 &205390 &NLTT 51629          &K2V       &K1.5V - K2V      &21 36 41.244 -50 50 43.39&346.4153 -46.4331 &   68.40$\pm$ 0.58& 14.62  \\
107350 &206860 &HN Peg              &G0V       &G0IV -  G0V      &21 44 31.329 +14 46 18.98&069.8594 -28.2693 &   55.91$\pm$ 0.45& 17.89  \\
107649 &207129 &HR 8323             &G2V       &G0V              &21 48 15.751 -47 18 13.02&350.8781 -49.1061 &   62.52$\pm$ 0.35& 15.99  \\
108870 &209100 &$\epsilon$ Ind A    &K5V       &K3V - K4V        &22 03 21.658 -56 47 09.52&336.1926 -48.0449 &  276.06$\pm$ 0.28&  3.62  \\
109378 &210277 &LTT 8887            &G0        &G0V, G8IV/V - G7V&22 09 29.866 -07 32 55.15&052.3504 -46.9651 &   46.38$\pm$ 0.48& 21.56  \\
109422 &210302 &$\tau$ PsA          &F6V       &F6V              &22 10 08.780 -32 32 54.27&014.1234 -54.8142 &   54.71$\pm$ 0.28& 18.28  \\
110109 &211415 &HR 8501             &G1V       &G0V - G3V        &22 18 15.614 -53 37 37.46&339.0111 -51.3872 &   72.54$\pm$ 0.36& 13.79  \\
113357 &217014 &51 Peg              &G5V       &G2IV - G5V       &22 57 27.980 +20 46 07.79&090.0626 -34.7273 &   64.07$\pm$ 0.38& 15.61  \\
113576 &217357 &LHS 3885            &K5/M0V    &K5V - M0.5V      &23 00 16.122 -22 31 27.65&037.1386 -64.4142 &  121.69$\pm$ 0.69&  8.22  \\
114948 &219482 &HR 8843             &F7V       &F6V - F8V        &23 16 57.686 -62 00 04.32&320.6226 -51.7597 &   48.69$\pm$ 0.33& 20.54  \\
116745 &222237 &LHS 3994            &K3V       &K3V              &23 39 37.389 -72 43 19.76&310.1686 -43.3827 &   87.56$\pm$ 0.51& 11.42  \\
120005 &79211  &CCDM J09144+5241B   &K2        &K7V, M0V - M0.5V &09 14 24.696 +52 41 11.09&164.9346 +42.6723 &  156.45$\pm$ 8.58&  6.39  \\
\noalign{\smallskip}\hline\noalign{\smallskip}
\end{longtable}
\end{scriptsize}
\end{center}
}

%
%

%

\onllongtabL{3}{
\begin{center}
\renewcommand{\thetable}{\arabic{table}\alph{subtable}}
\setcounter{subtable}{1}



\setcounter{subtable}{1}

\begin{longtable}{lrlllllllll}
\multicolumn{11}{c}{\tablename\ \thetable{}. Johnson (BV) and Cousins (I), Str\"omgren and 2MASS photometry.}

\\

\hline
HIP    &  \multicolumn{1}{c}{V}     &  \multicolumn{1}{c}{B--V} &  \multicolumn{1}{c}{V--I}  & $b-y$ &  \multicolumn{1}{c}{$m_1$} 
       &  \multicolumn{1}{c}{$c_1$} &  \multicolumn{1}{c}{J}    &  \multicolumn{1}{c}{H}     &  \multicolumn{1}{c}{K$_{\rm s}$} &  \multicolumn{1}{c}{Q} \\   
\hline
\endfirsthead

\multicolumn{11}{c}{\tablename\ \thetable{}. Continued.}\\\hline
HIP    & \multicolumn{1}{c}{V}      &  \multicolumn{1}{c}{B--V} &  \multicolumn{1}{c}{V--I}  & $b-y$ &  \multicolumn{1}{c}{$m_1$} 
       &  \multicolumn{1}{c}{$c_1$} &  \multicolumn{1}{c}{J}    &  \multicolumn{1}{c}{H}     &  \multicolumn{1}{c}{K$_{\rm s}$} &  \multicolumn{1}{c}{Q} \\   
\hline
\endhead

\hline
\multicolumn{11}{l}{Continued\dots}\\
\endfoot

\\
\endlastfoot
171      & 5.80    &0.69  & 0.82 & 0.432& 0.184& 0.218 & 4.702$\pm$0.214 & 4.179$\pm$ 0.198 & 4.068$\pm$ 0.236& CCD\\    
544      & 6.07    &0.75  & 0.80 & 0.460& 0.290& 0.311 & 4.733$\pm$0.019 & 4.629$\pm$ 0.144 & 4.314$\pm$ 0.042& EBE\\
910      & 4.89    &0.49  & 0.59 & 0.328& 0.130& 0.405 & 4.153$\pm$0.268 & 3.800$\pm$ 0.208 & 3.821$\pm$ 0.218& DCD\\
2941     & 5.57    &0.72  & 0.78 & 0.435& 0.254& 0.287 & 4.437$\pm$0.266 & 3.976$\pm$ 0.224 & 4.027$\pm$ 0.210& DDC\\
3093     & 5.88    &0.85  & 0.83 & 0.507& 0.384& 0.335 & 4.549$\pm$0.206 & 4.064$\pm$ 0.240 & 3.999$\pm$ 0.036& CDE\\
3497     & 6.55    &0.66  & 0.72 & 0.406& 0.192& 0.303 & 5.366$\pm$0.024 & 5.101$\pm$ 0.016 & 4.945$\pm$ 0.020& AEA\\
3821     & 3.46    &0.59  & 0.66 & 0.372& 0.185& 0.275 & 2.109$\pm$0.570 & 2.086$\pm$ 0.504 & 1.988           & DDF\\
3909     & 5.17    &0.51  & 0.59 & 0.332& 0.161& 0.360 & 4.412$\pm$0.308 & 4.012$\pm$ 0.280 & 4.015$\pm$ 0.264& DDD\\
4148     & 7.15    &0.94  & 1.00 & 0.531& 0.466& 0.269 & 5.537$\pm$0.029 & 5.049$\pm$ 0.029 & 4.894$\pm$ 0.031& AAA\\
7513     & 4.10    &0.54  & 0.58 & 0.346& 0.176& 0.415 & 3.175$\pm$0.210 & 2.957$\pm$ 0.178 & 2.859$\pm$ 0.274& CCD\\
7978     & 5.52    &0.55  & 0.62 & 0.350& 0.173& 0.340 & 4.791$\pm$0.226 & 4.399$\pm$ 0.234 & 4.340$\pm$ 0.276& DDD\\
8768     & 8.89    &1.42  & 1.80 &      &      &       & 6.064$\pm$0.021 & 5.405$\pm$ 0.026 & 5.178$\pm$ 0.020& AAA\\
10138    & 6.12    &0.81  & 0.88 & 0.484& 0.337& 0.287 & 4.785$\pm$0.248 & 4.245$\pm$ 0.280 & 4.125$\pm$ 0.036& DDE\\
10798    & 6.33    &0.72  & 0.83 & 0.442& 0.258& 0.236 & 5.056$\pm$0.024 & 4.694$\pm$ 0.044 & 4.551$\pm$ 0.016& EEA\\
11452    & 8.69    &1.40  & 1.66 &      &      &       & 5.990$\pm$0.020 & 5.332$\pm$ 0.029 & 5.113$\pm$ 0.018& AAA\\
11964    & 8.89    &1.39  & 1.87 &      &      &       & 5.795$\pm$0.018 & 5.126$\pm$ 0.027 & 4.885$\pm$ 0.016& AAA\\
12777    & 4.10    &0.51  & 0.59 & 0.325& 0.160& 0.392 & 3.031$\pm$0.244 & 2.863$\pm$ 0.206 & 2.697$\pm$ 0.288& DCD\\
13402    & 6.05    &0.86  & 0.91 & 0.509& 0.393& 0.300 & 4.830$\pm$0.230 & 4.230$\pm$ 0.220 & 4.167$\pm$ 0.242& DDD\\
14954    & 5.07    &0.58  & 0.63 & 0.361& 0.185& 0.422 & 4.174$\pm$0.280 & 3.768$\pm$ 0.238 & 3.748$\pm$ 0.236& DDD\\
15330    & 5.53    &0.64  & 0.72 & 0.404& 0.199& 0.288 & 4.462$\pm$0.302 & 4.041$\pm$ 0.268 & 3.994$\pm$ 0.252& DDD\\
15371    & 5.24    &0.60  & 0.67 & 0.382& 0.179& 0.298 & 4.271$\pm$0.300 & 3.874$\pm$ 0.232 & 3.860$\pm$ 0.228& DDD\\
15799    & 6.90    &0.84  & 0.88 & 0.507& 0.355& 0.332 & 5.333$\pm$0.021 & 4.948$\pm$ 0.038 & 4.767$\pm$ 0.016& AAA\\
16134    & 8.39    &1.34  & 1.55 & 0.771& 0.746& 0.053 & 5.888$\pm$0.017 & 5.251$\pm$ 0.024 & 5.068$\pm$ 0.021& AAA\\
17420    & 7.10    &0.93  & 0.97 &      &      &       & 5.457$\pm$0.018 & 5.025$\pm$ 0.116 & 4.837$\pm$ 0.020& AEA\\
17439    & 6.99    &0.87  & 0.91 & 0.508& 0.400& 0.295 & 5.462$\pm$0.019 & 5.088$\pm$ 0.016 & 4.934$\pm$ 0.016& AAA\\
19849    & 4.43    &0.82  & 0.89 & 0.484& 0.332& 0.301 & 3.013$\pm$0.238 & 2.594$\pm$ 0.198 & 2.498$\pm$ 0.236& DCD\\
19884    & 7.64    &1.12  & 1.23 & 0.629& 0.675& 0.172 & 5.624$\pm$0.019 & 5.098$\pm$ 0.029 & 4.923$\pm$ 0.020& AAA\\
22263    & 5.49    &0.63  & 0.69 & 0.398& 0.213& 0.321 & 4.466$\pm$0.254 & 4.116$\pm$ 0.236 & 3.999$\pm$ 0.036& DDE\\
23311    & 6.22    &1.05  & 1.10 & 0.602& 0.635& 0.238 & 4.389$\pm$0.244 & 3.797$\pm$ 0.214 & 3.706$\pm$ 0.228& DCD\\
25110    & 5.08    &0.51  & 0.58 & 0.321& 0.161& 0.431 & 4.416$\pm$0.278 & 4.201$\pm$ 0.276 & 3.911$\pm$ 0.020& DDE\\
27887    & 7.17    &0.94  & 0.98 & 0.541& 0.462& 0.261 & 5.412$\pm$0.019 & 4.968$\pm$ 0.040 & 4.793$\pm$ 0.016& AEA\\
28103    & 3.71    &0.34  & 0.39 & 0.218& 0.162& 0.622 & 3.063$\pm$0.246 & 2.985$\pm$ 0.232 & 2.993$\pm$ 0.254& DDD\\
28442    & 7.85    &1.14  & 1.23 &      &      &       & 5.659$\pm$0.020 & 5.070$\pm$ 0.027 & 4.902$\pm$ 0.018& AAA\\
29271    & 5.08    &0.71  & 0.75 & 0.440& 0.265& 0.337 & 3.951$\pm$0.232 & 3.508$\pm$ 0.228 & 3.446$\pm$ 0.200& DDC\\
29568    & 6.37    &0.71  & 0.76 & 0.430& 0.246& 0.299 & 5.129$\pm$0.018 & 4.863$\pm$ 0.036 & 4.726$\pm$ 0.016& AAA\\
32439    & 5.44    &0.53  & 0.60 & 0.338& 0.164& 0.346 & 4.512$\pm$0.212 & 4.262$\pm$ 0.146 & 4.141$\pm$ 0.034& CBA\\
32480    & 5.24    &0.58  & 0.65 & 0.358& 0.184& 0.373 & 4.406$\pm$0.274 & 4.071$\pm$ 0.246 & 4.132$\pm$ 0.294& DDD\\
33277    & 5.74    &0.57  & 0.70 & 0.377& 0.179& 0.313 & 4.913$\pm$0.276 & 4.509$\pm$ 0.204 & 4.293$\pm$ 0.036& DCE\\
34017    & 5.93    &0.60  & 0.74 & 0.384& 0.182& 0.328 & 4.885$\pm$0.268 & 4.612$\pm$ 0.242 & 4.539$\pm$ 0.075& DDE\\
34065    & 5.56    &0.62  & 0.73 & 0.392& 0.180& 0.325 & 4.405$\pm$0.224 & 3.990$\pm$ 0.212 & 4.037$\pm$ 0.036& DCE\\
35136    & 5.54    &0.58  & 0.65 & 0.374& 0.167& 0.297 & 4.655$\pm$0.250 & 4.289$\pm$ 0.220 & 4.115$\pm$ 0.018& DDE\\
36439    & 5.35    &0.47  & 0.54 & 0.308& 0.142& 0.390 & 4.104$\pm$0.328 & 3.937$\pm$ 0.230 & 4.178$\pm$ 0.015& DDE\\
38382    & 5.16    &0.60  & 0.67 & 0.380& 0.196& 0.316 & 4.109$\pm$0.248 & 3.695$\pm$ 0.256 & 3.730$\pm$ 0.226& DDD\\
38784    & 6.55    &0.72  & 0.77 & 0.450& 0.261& 0.293 & 5.235$\pm$8.888 & 4.933$\pm$ 0.038 & 4.862$\pm$ 0.021& FAA\\
40693    & 5.95    &0.75  & 0.79 & 0.458& 0.296& 0.312 & 4.953$\pm$0.268 & 4.364$\pm$ 0.224 & 4.165$\pm$ 0.036& DDE\\
40843    & 5.13    &0.49  & 0.56 & 0.315& 0.149& 0.384 & 4.127$\pm$0.222 & 3.942$\pm$ 0.192 & 3.868$\pm$ 0.246& DCD\\
42430    & 5.05    &0.72  & 0.73 & 0.447& 0.277& 0.394 & 4.048$\pm$0.302 & 3.594$\pm$ 0.248 & 3.589$\pm$ 0.246& DDD\\
42438    & 5.63    &0.62  & 0.66 & 0.396& 0.197& 0.293 & 4.348$\pm$0.214 & 4.282$\pm$ 0.036 & 4.173$\pm$ 0.036& CEE\\
43587    & 5.96    &0.87  & 0.82 & 0.536& 0.357& 0.415 & 4.768$\pm$0.244 & 4.265$\pm$ 0.234 & 4.015$\pm$ 0.036& DDE\\
43726    & 6.01    &0.66  & 0.70 & 0.413& 0.236& 0.332 & 4.871$\pm$0.037 & 4.625$\pm$ 0.276 & 4.456$\pm$ 0.023& EDA\\
44897    & 5.95    &0.59  & 0.66 & 0.377& 0.196& 0.323 & 4.844$\pm$0.037 & 4.601$\pm$ 0.027 & 4.545$\pm$ 0.020& EAA\\
45333    & 5.18    &0.61  & 0.67 & 0.386& 0.182& 0.375 & 4.231$\pm$0.246 & 3.914$\pm$ 0.242 & 3.782$\pm$ 0.346& DDD\\
45617    & 7.20    &0.99  & 0.90 & 0.585& 0.515& 0.240 & 5.382$\pm$0.029 & 4.893$\pm$ 0.023 & 4.767$\pm$ 0.017& AAA\\
46580    & 7.20    &1.00  & 1.07 & 0.570& 0.552& 0.231 & 5.429$\pm$0.026 & 5.002$\pm$ 0.055 & 4.788$\pm$ 0.023& AAA\\
47592    & 4.93    &0.53  & 0.58 & 0.338& 0.164& 0.374 & 4.076$\pm$0.338 & 3.728$\pm$ 0.242 & 3.722$\pm$ 0.250& DDD\\
49081    & 5.37    &0.68  & 0.74 & 0.418& 0.232& 0.390 & 4.267$\pm$0.328 & 4.040$\pm$ 0.258 & 3.821$\pm$ 0.036& DDE\\
49908    & 6.60    &1.33  & 1.29 & 0.790& 0.741& 0.032 & 3.894$\pm$0.290 & 3.298$\pm$ 0.260 & 2.962$\pm$ 0.288& DDD\\
51459    & 4.82    &0.54  & 0.58 & 0.341& 0.171& 0.333 & 4.033$\pm$0.218 & 3.758$\pm$ 0.198 & 3.644$\pm$ 0.222& DCD\\
51502    & 5.25    &0.40  & 0.46 & 0.265& 0.143& 0.456 & 4.416$\pm$0.208 & 4.364$\pm$ 0.036 & 4.272$\pm$ 0.036& CEE\\
53721    & 5.03    &0.62  & 0.69 & 0.391& 0.202& 0.343 & 3.960$\pm$0.296 & 3.736$\pm$ 0.224 & 3.750$\pm$ 0.340& DDD\\
54646    & 8.31    &1.26  & 1.42 & 0.769& 0.755& 0.055 & 5.764$\pm$0.020 & 5.130$\pm$ 0.020 & 4.979$\pm$ 0.018& AAA\\
56452    & 5.96    &0.81  & 0.88 & 0.483& 0.325& 0.238 & 4.784$\pm$0.228 & 4.138$\pm$ 0.214 & 4.022$\pm$ 0.036& DCE\\
57507    & 6.48    &0.68  & 0.74 & 0.426& 0.221& 0.280 & 5.217$\pm$0.018 & 4.935$\pm$ 0.023 & 4.798$\pm$ 0.018& EAA\\
57939    & 6.42    &0.75  & 0.88 & 0.484& 0.222& 0.155 & 4.937$\pm$0.196 & 4.500$\pm$ 0.024 & 4.373$\pm$ 0.027& CEA\\
58345    & 6.99    &1.13  & 1.22 & 0.641& 0.719& 0.159 & 4.992$\pm$0.037 & 4.617$\pm$ 0.242 & 4.525$\pm$ 0.300& EDD\\
62145    & 7.01    &0.94  & 0.98 & 0.543& 0.489& 0.283 & 5.318$\pm$8.888 & 4.900$\pm$ 0.020 & 4.780$\pm$ 0.016& FEA\\
62207    & 5.95    &0.56  & 0.65 & 0.374& 0.147& 0.284 & 5.173$\pm$0.186 & 4.667$\pm$ 0.036 & 4.465$\pm$ 0.029& CEA\\
62523    & 6.29    &0.70  & 0.74 & 0.438& 0.241& 0.334 & 5.123$\pm$0.262 & 4.705$\pm$ 0.036 & 4.645$\pm$ 0.020& DEA\\
64792    & 5.19    &0.59  & 0.64 & 0.373& 0.189& 0.384 & 4.392$\pm$0.284 & 4.107$\pm$ 0.208 & 4.033$\pm$ 0.238& DCD\\
64797    & 6.49    &0.93  & 0.97 & 0.532& 0.422& 0.296 & 4.901$\pm$0.037 & 4.584$\pm$ 0.017 & 4.381$\pm$ 0.036& EEE\\
65026    & 8.48    &1.47  & 1.89 &      &      &       & 5.338$\pm$0.026 & 4.721$\pm$ 0.018 & 4.494$\pm$ 0.018& EAA\\
65721    & 4.97    &0.71  & 0.77 & 0.447& 0.231& 0.352 & 3.798$\pm$0.286 & 3.457$\pm$ 0.206 & 3.500$\pm$ 0.294& DCD\\
67275    & 4.50    &0.51  & 0.51 & 0.318& 0.177& 0.439 & 3.617$\pm$0.284 & 3.546$\pm$ 0.242 & 3.507$\pm$ 0.348& DDD\\
67422    & 7.05    &1.11  & 1.16 &      &      &       & 5.077$\pm$0.048 & 4.561$\pm$ 0.026 & 6.333           & EAU\\
67620    & 6.43    &0.70  & 0.78 & 0.433& 0.241& 0.310 & 5.530$\pm$0.308 & 4.829$\pm$ 0.042 & 4.665$\pm$ 0.017& DAA\\
68184    & 6.49    &1.04  & 1.01 &      &      &       & 5.018$\pm$0.260 & 4.402$\pm$ 0.206 & 4.094           & DCF\\
68682    & 6.27    &0.73  & 0.82 & 0.451& 0.267& 0.316 & 5.388$\pm$0.294 & 4.765$\pm$ 0.015 & 4.459$\pm$ 0.024& DEA\\
69965    & 5.87    &0.52  & 0.61 & 0.338& 0.125& 0.300 & 4.956$\pm$0.334 & 4.717$\pm$ 0.076 & 4.512$\pm$ 0.027& DEA\\
70319    & 6.25    &0.64  & 0.71 & 0.405& 0.188& 0.281 & 5.053$\pm$0.037 & 4.814$\pm$ 0.067 & 4.644$\pm$ 0.017& EEA\\
70857    & 6.88    &0.77  & 0.81 & 0.472& 0.269& 0.275 & 5.425$\pm$0.018 & 5.011$\pm$ 0.031 & 4.917$\pm$ 0.018& EAA\\
71181    & 7.24    &1.00  & 1.04 & 0.571& 0.566& 0.246 & 5.436$\pm$0.037 & 4.872$\pm$ 0.029 & 4.789$\pm$ 0.017& EAA\\
71681    & 1.35    &0.90  & 0.88 &      &      &       & 1.454$\pm$0.133 & 1.886$\pm$ 0.220 & 2.008$\pm$ 0.260& BDD\\
71683    & --0.01  &0.71  & 0.69 & 0.438& 0.248& 0.373 & 1.454$\pm$0.133 & 1.886$\pm$ 0.220 & 2.008$\pm$ 0.260& BDD\\
71908    & 3.18    &0.26  & 0.26 & 0.137& 0.209& 0.782 & 2.544$\pm$0.284 & 2.471$\pm$ 0.196 & 2.425$\pm$ 0.216& DCC\\
72567    & 5.86    &0.58  & 0.67 & 0.374& 0.191& 0.321 & 4.998$\pm$0.218 & 4.688$\pm$ 0.226 & 4.458$\pm$ 0.020& DDA\\
72603    & 5.15    &0.40  & 0.48 & 0.263& 0.157& 0.494 & 4.475$\pm$0.266 & 4.266$\pm$ 0.204 & 4.297$\pm$ 0.280& DCD\\
72848    & 6.00    &0.84  & 0.92 & 0.502& 0.365& 0.296 & 4.882$\pm$0.260 & 4.165$\pm$ 0.036 & 4.316$\pm$ 0.326& DED\\
73100    & 5.63    &0.53  & 0.61 & 0.338& 0.174& 0.410 & 4.685$\pm$0.037 & 4.464$\pm$ 0.018 & 4.408$\pm$ 0.026& EAA\\
73182    & 8.01    &1.52  & 2.22 & 0.944& 0.486& 0.189 & 4.550$\pm$0.262 & 3.910$\pm$ 0.200 & 3.802$\pm$ 0.230& DCD\\
73184    & 5.72    &1.02  & 1.23 & 0.617& 0.669& 0.190 & 3.663$\pm$0.258 & 3.085$\pm$ 0.196 & 3.048$\pm$ 0.224& DCD\\
73996    & 4.93    &0.43  & 0.51 & 0.287& 0.159& 0.446 & 4.246$\pm$0.268 & 4.013$\pm$ 0.218 & 3.863$\pm$ 0.040& DDE\\
77052    & 5.86    &0.68  & 0.74 & 0.424& 0.231& 0.334 & 4.593$\pm$0.352 & 4.045$\pm$ 0.348 & 4.297$\pm$ 0.036& DDE\\
78459    & 5.39    &0.61  & 0.68 & 0.394& 0.178& 0.337 & 4.088$\pm$0.240 & 3.989$\pm$ 0.036 & 3.857$\pm$ 0.036& DEE\\
78775    & 6.66    &0.73  & 0.80 & 0.456& 0.241& 0.229 & 5.182$\pm$0.020 & 4.824$\pm$ 0.017 & 4.755$\pm$ 0.016& AAA\\
79248    & 6.61    &0.88  & 0.92 & 0.537& 0.366& 0.438 & 5.158$\pm$0.029 & 4.803$\pm$ 0.016 & 4.714$\pm$ 0.016& AAA\\
80725    & 6.98    &0.85  & 0.94 &      &      &       & 5.387$\pm$0.019 & 4.971$\pm$ 0.017 & 4.901$\pm$ 0.017& AAA\\
82860    & 4.88    &0.48  & 0.56 & 0.327& 0.154& 0.364 & 4.020$\pm$0.312 & 3.830$\pm$ 0.228 & 3.736$\pm$ 0.292& DDD\\
83389    & 6.76    &0.73  & 0.78 & 0.449& 0.272& 0.285 & 5.411$\pm$0.041 & 5.098$\pm$ 0.020 & 5.003$\pm$ 0.017& AAA\\
84862    & 5.38    &0.62  & 0.70 & 0.404& 0.179& 0.312 & 4.160$\pm$0.266 & 3.905$\pm$ 0.246 & 3.911$\pm$ 0.234& DDD\\
85235    & 6.44    &0.76  & 0.80 & 0.463& 0.279& 0.242 & 4.969$\pm$0.228 & 4.637$\pm$ 0.036 & 4.515$\pm$ 0.018& DEA\\
85295    & 7.54    &1.36  & 1.62 & 0.783& 0.744& 0.045 & 4.934$\pm$0.024 & 4.341$\pm$ 0.044 & 4.370$\pm$ 0.274& AED\\
86036    & 5.23    &0.60  & 0.67 & 0.395& 0.189& 0.330 & 4.239$\pm$0.292 & 3.885$\pm$ 0.226 & 3.744$\pm$ 0.290& DDD\\
86796    & 5.12    &0.69  & 0.71 & 0.432& 0.244& 0.393 & 4.158$\pm$0.284 & 3.724$\pm$ 0.206 & 3.683$\pm$ 0.246& DCD\\
88601    & 4.03    &0.86  & 0.96 & 0.508& 0.390& 0.262 & 2.343$\pm$0.296 & 1.876$\pm$ 0.244 & 1.791$\pm$ 0.304& DDD\\
88972    & 6.38    &0.88  & 1.00 & 0.518& 0.406& 0.295 & 4.952$\pm$0.278 & 4.458$\pm$ 0.192 & 4.232$\pm$ 0.021& DCA\\
89042    & 5.47    &0.59  & 0.65 & 0.377& 0.182& 0.350 & 4.430$\pm$0.306 & 4.022$\pm$ 0.244 & 4.130$\pm$ 0.266& DDD\\
91009    & 8.20    &1.27  & 1.44 &      &      &       & 5.597$\pm$0.023 & 5.020$\pm$ 0.016 & 4.843$\pm$ 0.021& AAA\\
92043    & 4.19    &0.48  & 0.55 & 0.314& 0.150& 0.484 & 3.527$\pm$0.242 & 3.286$\pm$ 0.176 & 3.190$\pm$ 0.256& DCD\\
95995    & 6.60    &0.86  & 0.90 & 0.513& 0.373& 0.275 & 5.020$\pm$0.043 & 4.613$\pm$ 0.036 & 4.463$\pm$ 0.015& EAE\\
96100    & 4.67    &0.79  & 0.85 & 0.472& 0.321& 0.267 & 3.423$\pm$0.270 & 3.039$\pm$ 0.214 & 2.900$\pm$ 0.324& DCD\\
96441    & 4.49    &0.40  & 0.44 & 0.261& 0.157& 0.502 & 3.878$\pm$0.284 & 3.716$\pm$ 0.234 & 3.537$\pm$ 0.296& DDD\\
97944    & 6.22    &1.02  & 1.02 & 0.596& 0.591& 0.244 & 4.747$\pm$0.254 & 4.158$\pm$ 0.240 & 4.043$\pm$ 0.252& DDD\\
98959    & 6.07    &0.65  & 0.72 & 0.405& 0.194& 0.298 & 5.108$\pm$0.266 & 4.724$\pm$ 0.076 & 4.511$\pm$ 0.026& DEA\\
99240    & 3.55    &0.75  & 0.76 & 0.466& 0.292& 0.384 & 2.299$\pm$0.272 & 2.041$\pm$ 0.240 & 1.933$\pm$ 0.272& DDD\\
99461    & 5.32    &0.87  & 0.98 & 0.515& 0.362& 0.271 & 3.518$\pm$0.300 & 2.999$\pm$ 0.422 & 3.008$\pm$ 0.602& DDD\\
101955   & 7.88    &1.24  & 1.44 &      &      &       & 5.509$\pm$0.021 & 4.899$\pm$ 0.016 & 4.739$\pm$ 0.018& AAA\\
101997   & 6.36    &0.72  & 0.78 & 0.440& 0.257& 0.260 & 5.296$\pm$0.266 & 4.810$\pm$ 0.044 & 4.597$\pm$ 0.016& DAA\\
103389   & 5.70    &0.51  & 0.57 & 0.328& 0.164& 0.355 & 5.077$\pm$0.272 & 4.734$\pm$ 0.076 & 4.475$\pm$ 0.036& DEE\\
104214   & 5.20    &1.07  & 1.13 & 0.658& 0.679& 0.132 & 3.114$\pm$0.268 & 2.540$\pm$ 0.198 & 2.248$\pm$ 0.318& DCD\\
104217   & 6.05    &1.31  & 1.27 & 0.791& 0.675& 0.067 & 3.546$\pm$0.278 & 2.895$\pm$ 0.218 & 2.544$\pm$ 0.328& DDD\\
105312   & 6.56    &0.74  & 0.85 & 0.450& 0.247& 0.273 & 5.114$\pm$0.017 & 4.720$\pm$ 0.076 & 4.574$\pm$ 0.021& EEA\\
106696   & 7.14    &0.88  & 0.94 & 0.514& 0.401& 0.263 & 5.518$\pm$0.035 & 5.137$\pm$ 0.053 & 4.970$\pm$ 0.021& AAA\\
107350   & 5.96    &0.59  & 0.66 & 0.377& 0.186& 0.314 & 4.793$\pm$0.035 & 4.598$\pm$ 0.036 & 4.559$\pm$ 0.038& EEA\\
107649   & 5.57    &0.60  & 0.66 & 0.383& 0.190& 0.337 & 4.720$\pm$0.176 & 4.306$\pm$ 0.076 & 4.236$\pm$ 0.240& CED\\
108870   & 4.69    &1.06  & 1.15 & 0.603& 0.618& 0.212 & 2.894$\pm$0.292 & 2.349$\pm$ 0.214 & 2.237$\pm$ 0.240& DCD\\
109378   & 6.54    &0.77  & 0.77 & 0.466& 0.285& 0.369 & 5.275$\pm$0.024 & 4.957$\pm$ 0.031 & 4.799$\pm$ 0.020& AAA\\
109422   & 4.94    &0.49  & 0.54 & 0.313& 0.167& 0.430 & 3.930$\pm$0.302 & 3.639$\pm$ 0.262 & 3.704$\pm$ 0.292& DDD\\
110109   & 5.36    &0.61  & 0.70 & 0.389& 0.178& 0.303 & 4.426$\pm$0.260 & 3.925$\pm$ 0.248 & 3.913$\pm$ 0.272& DDD\\
113357   & 5.45    &0.67  & 0.69 & 0.416& 0.233& 0.371 & 4.655$\pm$0.296 & 4.234$\pm$ 0.270 & 3.911$\pm$ 0.021& DDA\\
113576   & 7.88    &1.38  & 1.63 & 0.798& 0.700& 0.057 & 5.346$\pm$0.021 & 4.696$\pm$ 0.076 & 4.478$\pm$ 0.016& AEA\\
114948   & 5.64    &0.52  & 0.59 & 0.337& 0.161& 0.355 & 5.100$\pm$0.250 & 4.606$\pm$ 0.228 & 4.437$\pm$ 0.015& DDE\\
116745   & 7.09    &0.99  & 1.11 & 0.568& 0.526& 0.243 & 5.249$\pm$0.018 & 4.745$\pm$ 0.061 & 4.581$\pm$ 0.029& AEA\\
120005   & 7.70    &1.42  & 1.72 & 0.858& 0.626& 0.076 & 4.779$\pm$0.174 & 4.043$\pm$ 0.206 & 4.136$\pm$ 0.020& CCE\\
\hline
\end{longtable}


\onecolumn

\addtocounter{table}{-1}
\addtocounter{subtable}{1}

\begin{longtable}{lllllllll}
\multicolumn{9}{c}{\tablename\ \thetable{}. Additional near-infrared photometry.}

\\
\hline
HIP    & HR & \multicolumn{1}{c}{J} & \multicolumn{1}{c}{H}  & \multicolumn{1}{c}{K}   
            & \multicolumn{1}{c}{L} & \multicolumn{1}{c}{L'} & \multicolumn{1}{c}{M}  & Ref. \\     
\hline
\endfirsthead

\multicolumn{9}{c}{\tablename\ \thetable{}. Continued} \\\hline
HIP    & HR & \multicolumn{1}{c}{J} & \multicolumn{1}{c}{H}  & \multicolumn{1}{c}{K}   
            & \multicolumn{1}{c}{L} & \multicolumn{1}{c}{L'} & \multicolumn{1}{c}{M}  & Ref. \\
\hline
\endhead

\hline
\multicolumn{2}{l}{Continued\dots} \\
\endfoot

\\
\endlastfoot
910    &  33   &   3.95  &  3.68  &  3.63  &  3.59  &  3.58  &        & 1    \\ 
       &       &   3.964 &  3.689 &  3.636 &  3.60  &  3.60  &        & 2    \\
       &       &   3.966 &  3.672 &  3.647 &  3.611 &        &        & 3, 4 \\
       &       &   3.943 &  3.679 &  3.639 &  3.617 &        &  3.657 & 5    \\
3821   &  219  &   2.35  &  2.02  &  1.96  &        &        &        & 1    \\
       &       &   2.466 &        &  2.085 &  2.065 &        &        & 6    \\
7513   &  458  &   3.11  &        &  2.84  &        &  2.80  &        & 1    \\
       &       &   3.166 &        &  2.874 &  2.863 &        &        & 6    \\
8768   &       &   6.120 &  5.387 &  5.240 &        &        &        & 3    \\
       &       &   6.015 &  5.397 &  5.206 &  5.126 &        &        & 7    \\
       &       &   6.11  &  5.39  &  5.21  &        &  5.07  &        & 8    \\
10138  &  637  &   4.630 &  4.210 &  4.140 &  4.100 &        &        & 9    \\
12777  &  799  &   3.34  &  3.07  &  2.98  &  2.96  &        &        & 1    \\
13402  &  857  &   4.500 &  4.110 &  4.020 &  3.970 &        &        & 9    \\
15330  &  1006 &   4.382 &  4.036 &  3.980 &  3.947 &        &        & 3, 4 \\
       &       &   4.423 &  4.088 &  4.016 &        &        &        & 5    \\
       &       &   4.340 &  4.020 &  3.970 &  3.940 &        &        & 9    \\
15371  &  1010 &   4.130 &  3.820 &  3.770 &  3.730 &        &        & 9    \\
19849  &  1325 &   2.962 &        &  2.452 &  2.428 &        &        & 6    \\
23311  &  1614 &   4.350 &  3.900 &  3.800 &  3.750 &        &        & 9    \\
25110  &  1686 &   4.190 &  3.970 &  3.930 &  3.910 &        &        & 9, 10\\
28103  &  2085 &   3.10  &  2.95  &  2.90  &  2.87  &        &        & 1    \\
       &       &   3.101 &  2.953 &  2.907 &  2.878 &        &  2.895 & 5    \\
       &       &   3.080 &  2.940 &  2.910 &  2.880 &        &        & 9, 10\\
32439  &  2401 &   4.450 &  4.200 &  4.160 &  4.140 &        &        & 9    \\
32480  &  2483 &   4.230 &  3.970 &  3.930 &        &        &        & 2, 10\\
34017  &  2643 &   4.89  &  4.55  &  4.52  &        &        &        & 1    \\
       &       &   4.937 &        &  4.565 &  4.624 &        &        & 6    \\
34065  &  2667 &   4.455 &  4.148 &  4.081 &        &        &        & 5    \\
35136  &  2721 &   4.475 &  4.180 &  4.126 &        &  4.08  &        & 2    \\
       &       &   4.48  &  4.13  &  4.10  &        &        &        & 10   \\
38382  &  3064 &   3.979 &  3.685 &  3.630 &        &        &        & 11   \\
40693  &  3259 &   4.590 &  4.220 &  4.170 &  4.110 &        &        & 9    \\
40843  &  3262 &   4.180 &  3.940 &  3.920 &  3.900 &        &        & 9    \\
42438  &  3391 &   4.55  &  4.21  &  4.19  &        &        &        & 10   \\
43587  &  3522 &   4.59  &  4.14  &  4.07  &  3.98  &        &        & 1    \\
45333  &  3648 &   4.060 &  3.76  &  3.71  &        &        &        & 10   \\
       &       &   4.060 &  3.760 &  3.710 &  3.690 &        &        & 9    \\
49908  &       &   3.98  &  3.32  &  3.19  &  3.110 &        &        & 12   \\
51459  &  4112 &   3.84  &  3.58  &  3.54  &        &        &        & 9, 10\\
51502  &  4084 &   4.47  &  4.28  &  4.24  &        &        &        & 10   \\
       &       &   4.470 &  4.280 &  4.240 &  4.230 &        &        & 9    \\
57939  &  4550 &   4.92  &  4.44  &  4.38  &  4.34  &  4.34  &  4.41  & 1    \\ 
       &       &   4.957 &  4.466 &  4.400 &  4.35  &  4.38  &  4.41  & 2    \\
       &       &   4.870 &  4.455 &  4.385 &  4.340 &        &        & 13   \\
       &       &   4.89  &  4.43  &  4.37  &  4.34  &        &        & 12   \\
62207  &  4845 &   4.91  &  4.55  &  4.50  &        &        &        & 1    \\
       &       &   4.815 &  4.530 &  4.485 &  4.500 &        &        & 11   \\
65026  &       &   5.25  &  4.70  &  4.505 &        &        &        & 10   \\
65721  &  5072 &   3.71  &  3.32  &  3.25  &        &  3.21  &        & 1    \\
       &       &   3.759 &        &  3.289 &  3.282 &        &        & 6    \\
67275  &  5185 &   3.684 &        &  3.393 &  3.390 &        &        & 6    \\
70319  &  5384 &   5.13  &  4.74  &  4.68  &  4.61  &  4.61  &  4.62  & 1    \\
       &       &   5.027 &  4.713 &  4.662 &  4.637 &        &        & 11   \\
       &       &   5.149 &  4.788 &  4.718 &  4.703 &        &  4.708 & 5    \\
72567  &  5534 &   4.79  &  4.53  &  4.48  &        &        &        & 10   \\
72603  &  5530 &   4.41  &  4.19  &  4.16  &  4.10  &        &  4.13  & 1    \\
       &       &   4.416 &  4.211 &  4.155 &  4.104 &        &  4.153 & 5    \\ 
       &       &   4.40  &  4.18  &  4.130 &        &        &        & 10   \\
       &       &   4.350 &  4.140 &  4.110 &  4.100 &        &        & 9    \\
73184  &  5568 &   3.83  &  3.24  &  3.17  &  3.08  &        &        & 1    \\
73996  &  5634 &   4.12  &  3.90  &  3.88  &  3.89  &        &        & 1    \\         
       &       &   4.094 &  3.907 &  3.866 &        &        &        &      \\
       &       &   4.090 &  3.890 &  3.860 &  3.850 &        &        & 9    \\
       &       &   4.193 &        &  3.912 &  3.918 &        &        & 6    \\
78459  &  5968 &   4.241 &  3.940 &  3.877 &        &        &        & 11   \\
82860  &  6315 &   3.900 &  3.640 &  3.590 &  3.570 &        &        & 9    \\
88972  &  6806 &   4.77  &  4.29  &  4.24  &  4.16  &        &        & 1    \\         
89042  &  6761 &   4.410 &  4.120 &  4.080 &  4.010 &        &        & 9    \\
92043  &  7061 &   3.32  &        &  3.06  &        &  3.03  &        & 1    \\
       &       &   3.300 &  3.080 &  3.040 &  3.010 &        &        & 9    \\
       &       &   3.374 &        &  3.092 &  3.090 &        &        & 6    \\
96441  &  7469 &   3.75  &        &  3.52  &        &  3.47  &        & 1    \\
       &       &   3.803 &        &  3.551 &  3.527 &        &        & 6    \\
98959  &  7644 &   4.958 &  4.617 &  4.553 &        &        &        & 5    \\
99240  &  7665 &   2.317 &  1.990 &  1.924 &  1.877 &        &  1.950 & 5    \\
       &       &   2.326 &  1.953 &  1.897 &  1.870 &        &        & 4    \\
104214 &  8085 &   3.10  &  2.44  &  2.35  &        &        &        & 1    \\
104217 &  8086 &   3.53  &  2.83  &  2.70  &        &        &        & 1    \\
107649 &  8323 &   4.470 &  4.200 &  4.140 &  4.090 &        &        & 9    \\
109422 &  8447 &   4.030 &  3.820 &  3.780 &  3.770 &        &        & 9    \\
113357 &  8729 &   4.36  &  4.03  &  3.97  &        &        &        & 1    \\
\hline
\label{ }

\end{longtable}
\end{center}

\noindent\underline{References}:                                                                            \\
1: UKIRT very bright standards: http://www.jach.hawaii.edu/UKIRT/astronomy/calib/phot\_cal/bright\_stds.html\\
2: UKIRT general standards: http://www.jach.hawaii.edu/UKIRT/astronomy/calib/phot\_cal/brtJHK\_stds.html    \\
3: SAAO infrared standards: http://www.saao.ac.za/fileadmin/files/links/IRstd.txt                           \\
4: \cite{Carter1990}                                                                                        \\
5: \cite{Koornneef1983}                                                                                     \\
6: \cite{Selby1988}                                                                                         \\
7: \cite{Elias1983}                                                                                         \\
8: AAO photometric standards: \cite{Allen1983}                                                              \\
9: \cite{Aumann1991}                                                                                        \\
10: Catalogue of Infrared Observations, Gezari et al. (2000): http://ircatalog.gsfc.nasa.gov/               \\
11: CVF standards: http://www.iac.es/telescopes/pages/es/inicio/utilidades.php\#CVF                         \\
12: \cite{Glass1975}                                                                                        \\
13: \cite{Elias1982}                                                                                        \\


\onecolumn

\addtocounter{table}{-1}
\addtocounter{subtable}{1}

\begin{center}
\renewcommand{\thetable}{\arabic{table}\alph{subtable}}
\begin{longtable}{lcc@{}cccc}
\multicolumn{7}{c}{\tablename\ \thetable{}. AKARI and WISE fluxes.}

\\
\hline
HIP   & \multicolumn{2}{c}{AKARI}   & &     \multicolumn{3}{c}{WISE}         \\\cline{2-3}\cline{5-7} 
      & 9  $\mu$m & 18  $\mu$m  & & 3.35 $\mu$m (W1)  & 11.56 $\mu$m (W3)  & 22.09 $\mu$m (W4) \\
      & (Jy)      & (Jy)        & & (Jy)              &  (Jy)              &  (Jy)             \\
\hline
\endfirsthead

\multicolumn{7}{c}{\tablename\ \thetable{}. Continued.} \\\hline\
HIP   & \multicolumn{2}{c}{AKARI}   & &     \multicolumn{3}{c}{WISE}        \\\cline{2-3}\cline{5-7} 
      & 9  $\mu$m & 18  $\mu$m  & & 3.35 $\mu$m (W1)  & 11.56 $\mu$m (W3)  & 22.09 $\mu$m (W4) \\
      & (Jy)      & (Jy)        & & (Jy)              &  (Jy)              &  (Jy)             \\
\hline
\endhead

\hline
\multicolumn{7}{l}{Continued\dots}\\
\endfoot

\\
\endlastfoot
171    &                       &                       & &  $9.15(+0)\pm 8.86(-1)$  &  $8.36(-1)\pm 1.16(-2)$  &  $2.53(-1)\pm 5.37(-3)$  \\
544    & $1.05(+0)\pm5.95(-3)$ & $2.73(-1)\pm2.94(-2)$ & &  $5.84(+0)\pm 4.42(-1)$  &  $5.36(-1)\pm 6.92(-3)$  &  $1.86(-1)\pm 3.95(-3)$  \\
910    & $2.00(+0)\pm3.24(-2)$ & $4.54(-1)\pm4.68(-2)$ & &  $1.10(+1)\pm 1.17(+0)$  &  $1.04(+0)\pm 1.34(-2)$  &  $3.05(-1)\pm 5.89(-3)$  \\
2941   &                       &                       & &  $9.27(+0)\pm 1.14(+0)$  &  $8.20(-1)\pm 1.06(-2)$  &  $2.50(-1)\pm 4.37(-3)$  \\
3093   & $1.51(+0)\pm1.62(-2)$ & $3.49(-1)\pm1.19(-2)$ & &  $8.84(+0)\pm 8.15(-1)$  &  $7.77(-1)\pm 1.00(-2)$  &  $2.27(-1)\pm 4.60(-3)$  \\
3497   & $6.12(-1)\pm1.77(-2)$ & $1.68(-1)\pm1.16(-2)$ & &  $3.29(+0)\pm 2.28(-1)$  &  $3.00(-1)\pm 4.15(-3)$  &  $9.13(-2)\pm 2.61(-3)$  \\
3821   & $1.03(+1)\pm6.06(-2)$ & $2.24(+0)\pm3.53(-2)$ & &  $2.68(+1)\pm 9.38(-1)$  &  $7.91(+0)\pm 2.91(-2)$  &  $1.72(+0)\pm 1.58(-2)$  \\
3909   & $1.53(+0)\pm2.51(-2)$ & $3.23(-1)\pm2.32(-2)$ & &  $7.99(+0)\pm 6.26(-1)$  &  $7.72(-1)\pm 1.07(-2)$  &  $2.34(-1)\pm 5.60(-3)$  \\
4148   & $6.64(-1)\pm6.69(-3)$ &                       & &  $3.41(+0)\pm 2.26(-1)$  &  $3.13(-1)\pm 4.90(-3)$  &  $9.22(-2)\pm 2.38(-3)$  \\
7513   & $4.15(+0)\pm1.32(-2)$ & $9.26(-1)\pm3.18(-2)$ & &  $2.11(+1)\pm 3.90(+0)$  &  $2.05(+0)\pm 2.83(-2)$  &  $6.05(-1)\pm 1.00(-2)$  \\
7978   & $1.15(+0)\pm1.88(-2)$ & $3.12(-1)\pm3.93(-2)$ & &  $6.20(+0)\pm 5.60(-1)$  &  $6.00(-1)\pm 7.74(-3)$  &  $2.18(-1)\pm 4.02(-3)$  \\
8768   & $5.43(-1)\pm2.03(-2)$ & $1.25(-1)\pm2.23(-2)$ & &  $2.86(+0)\pm 1.90(-1)$  &  $2.74(-1)\pm 3.79(-3)$  &  $8.63(-2)\pm 2.31(-3)$  \\
10138  &                       &                       & &  $7.23(+0)\pm 5.87(-1)$  &  $6.56(-1)\pm 7.85(-3)$  &  $1.92(-1)\pm 3.88(-3)$  \\
10798  & $8.47(-1)\pm1.33(-2)$ & $1.79(-1)\pm2.03(-2)$ & &  $5.02(+0)\pm 3.94(-1)$  &  $4.46(-1)\pm 6.16(-3)$  &  $1.33(-1)\pm 2.94(-3)$  \\
11452  & $5.73(-1)\pm6.70(-3)$ &                       & &  $2.96(+0)\pm 1.97(-1)$  &  $2.74(-1)\pm 3.79(-3)$  &  $8.32(-2)\pm 2.07(-3)$  \\
11964  & $7.51(-1)\pm6.59(-3)$ & $2.23(-1)\pm3.36(-2)$ & &  $3.68(+0)\pm 2.61(-1)$  &  $3.87(-1)\pm 5.35(-3)$  &  $1.23(-1)\pm 2.72(-3)$  \\
12777  & $4.20(+0)\pm3.99(-2)$ & $9.91(-1)\pm3.76(-2)$ & &  $2.12(+1)\pm 1.96(+0)$  &  $2.02(+0)\pm 2.05(-2)$  &  $6.04(-1)\pm 8.90(-3)$  \\
13402  & $1.42(+0)\pm2.04(-2)$ & $3.21(-1)\pm1.68(-2)$ & &  $7.88(+0)\pm 8.07(-1)$  &  $7.40(-1)\pm 9.55(-3)$  &  $2.30(-1)\pm 4.46(-3)$  \\
14954  & $1.94(+0)\pm2.22(-2)$ & $4.23(-1)\pm1.40(-2)$ & &  $1.06(+1)\pm 1.16(+0)$  &  $1.00(+0)\pm 1.20(-2)$  &  $2.89(-1)\pm 5.33(-3)$  \\
15330  &                       &                       & &  $8.56(+0)\pm 8.37(-1)$  &  $7.62(-1)\pm 1.05(-2)$  &  $2.26(-1)\pm 4.57(-3)$  \\
15371  &                       &                       & &  $9.13(+0)\pm 9.94(-1)$  &  $9.09(-1)\pm 1.26(-2)$  &  $2.77(-1)\pm 5.10(-3)$  \\
15799  & $7.46(-1)\pm9.75(-3)$ & $2.04(-1)\pm2.74(-2)$ & &  $4.11(+0)\pm 3.15(-1)$  &  $3.96(-1)\pm 5.11(-3)$  &  $1.18(-1)\pm 2.49(-3)$  \\
16134  & $5.64(-1)\pm5.81(-3)$ &                       & &  $3.01(+0)\pm 1.97(-1)$  &  $2.93(-1)\pm 3.77(-3)$  &  $8.80(-2)\pm 2.11(-3)$  \\
17420  & $6.41(-1)\pm1.45(-2)$ & $1.55(-1)\pm4.64(-2)$ & &  $3.48(+0)\pm 2.12(-1)$  &  $3.33(-1)\pm 4.91(-3)$  &  $9.65(-2)\pm 2.40(-3)$  \\
17439  & $5.87(-1)\pm7.85(-3)$ &                       & &  $3.37(+0)\pm 2.42(-1)$  &  $3.06(-1)\pm 4.23(-3)$  &  $9.40(-2)\pm 1.99(-3)$  \\
19849  &                       &                       & &  $3.17(+1)\pm 7.13(+0)$  &  $2.96(+0)\pm 3.54(-2)$  &  $9.15(-1)\pm 1.69(-2)$  \\
19884  & $6.26(-1)\pm8.36(-3)$ & $1.50(-1)\pm1.37(-2)$ & &  $3.35(+0)\pm 2.47(-1)$  &  $3.21(-1)\pm 4.14(-3)$  &  $9.48(-2)\pm 2.10(-3)$  \\
22263  & $1.46(+0)\pm8.34(-3)$ &                       & &  $8.02(+0)\pm 7.62(-1)$  &  $7.42(-1)\pm 1.03(-2)$  &  $2.26(-1)\pm 4.36(-3)$  \\
23311  & $1.75(+0)\pm2.28(-2)$ &                       & &  $9.81(+0)\pm 9.05(-1)$  &  $8.98(-1)\pm 1.24(-2)$  &  $2.61(-1)\pm 6.02(-3)$  \\
25110  & $1.50(+0)\pm1.56(-2)$ & $3.70(-1)\pm3.60(-2)$ & &  $8.38(+0)\pm 8.11(-1)$  &  $7.58(-1)\pm 1.05(-2)$  &  $2.30(-1)\pm 5.30(-3)$  \\
27887  & $6.83(-1)\pm7.21(-3)$ & $1.59(-1)\pm9.25(-3)$ & &  $3.91(+0)\pm 2.95(-1)$  &  $3.55(-1)\pm 4.57(-3)$  &  $1.04(-1)\pm 2.40(-3)$  \\
28103  & $3.89(+0)\pm1.09(-2)$ & $9.75(-1)\pm4.59(-2)$ & &  $2.09(+1)\pm 3.03(+0)$  &  $1.99(+0)\pm 2.19(-2)$  &  $6.38(-1)\pm 1.12(-2)$  \\
28442  & $6.48(-1)\pm1.71(-2)$ & $1.64(-1)\pm4.68(-3)$ & &  $3.33(+0)\pm 3.35(-1)$  &  $3.10(-1)\pm 7.43(-3)$  &  $9.51(-2)\pm 4.91(-3)$  \\
29271  & $2.42(+0)\pm6.54(-3)$ & $5.39(-1)\pm1.25(-2)$ & &  $1.33(+1)\pm 1.09(+0)$  &  $1.30(+0)\pm 1.20(-2)$  &  $3.67(-1)\pm 6.42(-3)$  \\
29568  & $7.76(-1)\pm1.59(-2)$ &                       & &  $3.74(+0)\pm 2.55(-1)$  &  $3.83(-1)\pm 5.29(-3)$  &  $1.15(-1)\pm 2.32(-3)$  \\
32439  & $1.20(+0)\pm1.14(-2)$ & $2.59(-1)\pm2.40(-2)$ & &  $7.03(+0)\pm 6.55(-1)$  &  $6.15(-1)\pm 9.06(-3)$  &  $1.82(-1)\pm 4.18(-3)$  \\
32480  & $1.60(+0)\pm5.32(-2)$ & $4.58(-1)\pm1.73(-2)$ & &  $8.96(+0)\pm 8.02(-1)$  &  $7.74(-1)\pm 1.07(-2)$  &  $2.46(-1)\pm 4.75(-3)$  \\
33277  & $1.09(+0)\pm4.87(-3)$ & $2.51(-1)\pm7.37(-2)$ & &  $6.11(+0)\pm 5.35(-1)$  &  $5.40(-1)\pm 6.97(-3)$  &  $1.62(-1)\pm 3.72(-3)$  \\
34017  & $9.23(-1)\pm2.63(-2)$ & $2.51(-1)\pm1.14(-2)$ & &  $5.17(+0)\pm 4.48(-1)$  &  $4.64(-1)\pm 5.55(-3)$  &  $1.37(-1)\pm 2.90(-3)$  \\
34065  & $1.39(+0)\pm2.35(-2)$ & $4.25(-1)\pm4.81(-2)$ & &  $8.59(+0)\pm 7.44(-1)$  &  $7.13(-1)\pm 9.19(-3)$  &  $2.18(-1)\pm 4.41(-3)$  \\
35136  & $1.33(+0)\pm1.55(-2)$ & $2.55(-1)\pm3.45(-2)$ & &  $7.63(+0)\pm 7.88(-1)$  &  $6.64(-1)\pm 9.17(-3)$  &  $2.00(-1)\pm 4.06(-3)$  \\
36439  & $1.25(+0)\pm7.20(-3)$ & $2.99(-1)\pm3.19(-2)$ & &  $6.33(+0)\pm 5.37(-1)$  &  $6.11(-1)\pm 8.45(-3)$  &  $1.79(-1)\pm 3.80(-3)$  \\
38382  & $1.89(+0)\pm3.79(-2)$ & $4.62(-1)\pm2.60(-2)$ & &  $1.13(+1)\pm 1.41(+0)$  &  $9.71(-1)\pm 1.43(-2)$  &  $2.87(-1)\pm 5.83(-3)$  \\
38784  & $6.31(-1)\pm1.04(-2)$ & $1.53(-1)\pm8.13(-3)$ & &  $3.43(+0)\pm 2.34(-1)$  &  $3.36(-1)\pm 4.64(-3)$  &  $9.85(-2)\pm 2.18(-3)$  \\
40693  & $1.30(+0)\pm1.34(-2)$ & $4.84(-1)\pm1.02(-1)$ & &  $6.91(+0)\pm 5.99(-1)$  &  $6.71(-1)\pm 9.27(-3)$  &  $2.78(-1)\pm 4.86(-3)$  \\
40843  &                       & $5.30(-1)\pm3.58(-2)$ & &  $8.22(+0)\pm 7.81(-1)$  &  $8.04(-1)\pm 1.04(-2)$  &  $2.30(-1)\pm 4.66(-3)$  \\
42430  & $2.38(+0)\pm1.93(-2)$ & $5.14(-1)\pm2.66(-2)$ & &  $1.19(+1)\pm 7.69(-1)$  &  $1.17(+0)\pm 1.29(-2)$  &  $3.61(-1)\pm 6.98(-3)$  \\
42438  & $1.23(+0)\pm1.34(-2)$ & $3.02(-1)\pm2.65(-2)$ & &  $6.77(+0)\pm 6.62(-1)$  &  $6.26(-1)\pm 8.07(-3)$  &  $1.99(-1)\pm 3.66(-3)$  \\
43587  & $1.41(+0)\pm2.16(-2)$ & $2.90(-1)\pm1.62(-2)$ & &  $7.18(+0)\pm 6.42(-1)$  &  $6.89(-1)\pm 8.89(-3)$  &  $2.06(-1)\pm 4.75(-3)$  \\
43726  & $9.03(-1)\pm1.23(-2)$ & $1.54(-1)\pm2.63(-2)$ & &  $5.17(+0)\pm 4.25(-1)$  &  $4.66(-1)\pm 5.15(-3)$  &  $1.45(-1)\pm 3.47(-3)$  \\
44897  & $8.67(-1)\pm1.12(-2)$ & $2.04(-1)\pm3.18(-2)$ & &  $4.58(+0)\pm 3.63(-1)$  &  $4.31(-1)\pm 5.56(-3)$  &  $1.31(-1)\pm 3.01(-3)$  \\
45333  & $1.85(+0)\pm8.04(-3)$ & $4.12(-1)\pm3.69(-2)$ & &  $1.03(+1)\pm 1.05(+0)$  &  $9.54(-1)\pm 1.23(-2)$  &  $2.82(-1)\pm 5.98(-3)$  \\
45617  & $7.50(-1)\pm1.46(-2)$ & $1.30(-1)\pm3.45(-2)$ & &  $3.51(+0)\pm 2.62(-1)$  &  $3.52(-1)\pm 4.87(-3)$  &  $1.07(-1)\pm 2.26(-3)$  \\
46580  & $7.33(-1)\pm1.38(-2)$ & $1.06(-1)\pm7.80(-4)$ & &  $3.51(+0)\pm 2.23(-1)$  &  $3.57(-1)\pm 4.27(-3)$  &  $1.07(-1)\pm 2.37(-3)$  \\
47592  & $1.96(+0)\pm1.66(-2)$ & $4.70(-1)\pm2.05(-2)$ & &  $1.15(+1)\pm 1.24(+0)$  &  $1.01(+0)\pm 1.39(-2)$  &  $2.97(-1)\pm 6.56(-3)$  \\
49081  & $1.63(+0)\pm1.92(-2)$ & $3.98(-1)\pm1.83(-2)$ & &  $9.35(+0)\pm 9.14(-1)$  &  $8.34(-1)\pm 1.08(-2)$  &  $2.45(-1)\pm 4.51(-3)$  \\
49908  &                       &                       & &  $1.80(+1)\pm 2.48(+0)$  &  $1.72(+0)\pm 1.43(-2)$  &  $5.03(-1)\pm 8.34(-3)$  \\
51459  & $2.12(+0)\pm1.31(-2)$ & $4.80(-1)\pm1.20(-2)$ & &  $1.04(+1)\pm 1.12(+0)$  &  $1.10(+0)\pm 1.51(-2)$  &  $3.18(-1)\pm 6.45(-3)$  \\
51502  & $1.16(+0)\pm7.47(-3)$ & $3.34(-1)\pm1.37(-2)$ & &  $6.73(+0)\pm 5.46(-1)$  &  $5.71(-1)\pm 7.36(-3)$  &  $1.75(-1)\pm 3.38(-3)$  \\
53721  & $2.07(+0)\pm1.86(-2)$ & $4.46(-1)\pm3.82(-2)$ & &  $1.11(+1)\pm 7.28(-1)$  &  $1.10(+0)\pm 1.21(-2)$  &  $3.14(-1)\pm 6.08(-3)$  \\
54646  & $6.27(-1)\pm1.80(-2)$ & $1.73(-1)\pm9.39(-2)$ & &  $3.40(+0)\pm 2.48(-1)$  &  $3.12(-1)\pm 4.31(-3)$  &  $9.22(-2)\pm 2.21(-3)$  \\
56452  & $1.44(+0)\pm7.13(-3)$ & $4.24(-1)\pm4.50(-2)$ & &  $8.25(+0)\pm 8.06(-1)$  &  $7.42(-1)\pm 9.57(-3)$  &  $2.19(-1)\pm 4.44(-3)$  \\
57507  & $6.91(-1)\pm1.68(-2)$ & $1.29(-1)\pm1.16(-2)$ & &  $3.86(+0)\pm 2.71(-1)$  &  $3.44(-1)\pm 4.75(-3)$  &  $1.04(-1)\pm 2.58(-3)$  \\
57939  &                       &                       & &  $5.48(+0)\pm 3.99(-1)$  &  $5.19(-1)\pm 6.69(-3)$  &  $1.54(-1)\pm 3.41(-3)$  \\
58345  & $1.10(+0)\pm1.48(-2)$ & $2.97(-1)\pm3.00(-2)$ & &  $6.20(+0)\pm 4.97(-1)$  &  $5.64(-1)\pm 6.75(-3)$  &  $1.70(-1)\pm 3.92(-3)$  \\
62145  & $7.05(-1)\pm1.07(-2)$ & $1.50(-1)\pm2.44(-2)$ & &  $3.98(+0)\pm 2.86(-1)$  &  $3.61(-1)\pm 4.66(-3)$  &  $1.06(-1)\pm 2.43(-3)$  \\
62207  & $9.17(-1)\pm2.33(-2)$ & $1.78(-1)\pm2.06(-2)$ & &  $5.16(+0)\pm 4.18(-1)$  &  $4.64(-1)\pm 6.41(-3)$  &  $1.38(-1)\pm 3.05(-3)$  \\
62523  & $7.90(-1)\pm1.82(-2)$ &                       & &  $4.20(+0)\pm 2.91(-1)$  &  $4.00(-1)\pm 5.53(-3)$  &  $1.17(-1)\pm 2.47(-3)$  \\
64792  & $1.67(+0)\pm2.26(-2)$ & $3.96(-1)\pm1.64(-2)$ & &  $9.40(+0)\pm 9.80(-1)$  &  $8.56(-1)\pm 1.18(-2)$  &  $2.61(-1)\pm 4.56(-3)$  \\
64797  & $1.38(+0)\pm2.32(-2)$ & $3.65(-1)\pm1.21(-2)$ & &  $6.18(+0)\pm 9.02(-1)$  &  $4.98(-1)\pm 5.97(-3)$  &  $1.56(-1)\pm 3.59(-3)$  \\
65026  & $1.12(+0)\pm1.67(-2)$ & $3.25(-1)\pm3.51(-2)$ & &  $5.45(+0)\pm 4.12(-1)$  &  $5.85(-1)\pm 7.54(-3)$  &  $1.81(-1)\pm 3.67(-3)$  \\
65721  & $2.91(+0)\pm3.67(-2)$ & $7.30(-1)\pm2.42(-2)$ & &  $1.56(+1)\pm 2.15(+0)$  &  $1.47(+0)\pm 9.51(-3)$  &  $4.41(-1)\pm 9.34(-3)$  \\
67275  & $2.63(+0)\pm1.23(-2)$ & $5.59(-1)\pm1.43(-2)$ & &  $1.53(+1)\pm 2.42(+0)$  &  $1.36(+0)\pm 1.63(-2)$  &  $4.00(-1)\pm 8.47(-3)$  \\
67422  & $1.12(+0)\pm1.43(-2)$ & $2.65(-1)\pm3.86(-2)$ & &  $5.53(+0)\pm 6.84(-1)$  &  $4.67(-1)\pm 1.16(-2)$  &  $1.42(-1)\pm 7.06(-3)$  \\
67620  & $8.42(-1)\pm2.25(-2)$ & $2.16(-1)\pm8.64(-2)$ & &  $4.29(+0)\pm 3.28(-1)$  &  $4.03(-1)\pm 5.56(-3)$  &  $1.20(-1)\pm 3.20(-3)$  \\
68184  & $1.33(+0)\pm1.61(-2)$ & $3.19(-1)\pm3.38(-2)$ & &  $7.26(+0)\pm 6.36(-1)$  &  $6.80(-1)\pm 9.39(-3)$  &  $2.01(-1)\pm 3.53(-3)$  \\
68682  & $9.68(-1)\pm2.26(-2)$ & $2.62(-1)\pm2.14(-2)$ & &  $5.38(+0)\pm 4.51(-1)$  &  $4.96(-1)\pm 6.39(-3)$  &  $1.49(-1)\pm 3.29(-3)$  \\
69965  &                       & $2.69(-1)\pm4.20(-2)$ & &  $5.39(+0)\pm 4.58(-1)$  &  $4.82(-1)\pm 6.65(-3)$  &  $1.42(-1)\pm 3.15(-3)$  \\
70319  & $7.44(-1)\pm1.17(-2)$ & $1.29(-1)\pm4.62(-2)$ & &  $4.25(+0)\pm 3.09(-1)$  &  $3.90(-1)\pm 5.02(-3)$  &  $1.16(-1)\pm 2.36(-3)$  \\
70857  & $5.91(-1)\pm9.43(-3)$ & $1.65(-1)\pm3.97(-2)$ & &  $3.39(+0)\pm 2.59(-1)$  &  $3.13(-1)\pm 4.62(-3)$  &  $9.49(-2)\pm 2.27(-3)$  \\
71181  & $6.92(-1)\pm9.04(-3)$ & $1.59(-1)\pm8.79(-3)$ & &  $3.85(+0)\pm 3.34(-1)$  &  $3.49(-1)\pm 4.82(-3)$  &  $1.03(-1)\pm 2.28(-3)$  \\
71681  &                       &                       & &                          &                          &                          \\
71683  &                       &                       & &                          &                          &                          \\
71908  & $5.08(+0)\pm1.93(-2)$ & $1.18(+0)\pm3.31(-2)$ & &  $2.23(+1)\pm 3.86(+0)$  &  $2.58(+0)\pm 2.85(-2)$  &  $7.77(-1)\pm 1.50(-2)$  \\
72567  & $9.19(-1)\pm1.08(-2)$ & $2.80(-1)\pm3.16(-2)$ & &  $5.17(+0)\pm 4.00(-1)$  &  $4.73(-1)\pm 6.53(-3)$  &  $1.42(-1)\pm 2.89(-3)$  \\
72603  & $1.29(+0)\pm1.59(-2)$ & $3.32(-1)\pm2.84(-2)$ & &  $7.21(+0)\pm 6.52(-1)$  &  $6.48(-1)\pm 8.36(-3)$  &  $1.92(-1)\pm 4.25(-3)$  \\
72848  & $1.53(+0)\pm1.11(-2)$ & $3.84(-1)\pm1.88(-2)$ & &  $8.70(+0)\pm 8.99(-1)$  &  $7.96(-1)\pm 1.03(-2)$  &  $2.39(-1)\pm 4.40(-3)$  \\
73100  & $9.87(-1)\pm6.41(-3)$ & $2.03(-1)\pm9.95(-3)$ & &  $5.50(+0)\pm 4.67(-1)$  &  $5.00(-1)\pm 6.90(-3)$  &  $1.55(-1)\pm 2.99(-3)$  \\
73182  &                       &                       & &                          &                          &                          \\
73184  &                       &                       & &  $1.67(+1)\pm 1.85(-1)$  &  $1.65(+0)\pm 1.06(-2)$  &  $5.22(-1)\pm 9.14(-3)$  \\
73996  & $1.56(+0)\pm1.02(-2)$ & $3.44(-1)\pm1.92(-2)$ & &  $8.35(+0)\pm 7.63(-1)$  &  $8.08(-1)\pm 1.12(-2)$  &  $2.40(-1)\pm 4.64(-3)$  \\
77052  & $1.21(+0)\pm1.07(-2)$ & $3.06(-1)\pm2.36(-2)$ & &  $6.84(+0)\pm 6.12(-1)$  &  $6.26(-1)\pm 8.65(-3)$  &  $1.82(-1)\pm 4.03(-3)$  \\
78459  & $1.57(+0)\pm5.03(-3)$ & $3.85(-1)\pm2.86(-2)$ & &  $8.49(+0)\pm 8.14(-1)$  &  $7.96(-1)\pm 1.10(-2)$  &  $2.39(-1)\pm 4.62(-3)$  \\
78775  & $7.11(-1)\pm9.99(-3)$ & $1.35(-1)\pm2.86(-2)$ & &  $3.90(+0)\pm 3.09(-1)$  &  $3.59(-1)\pm 4.62(-3)$  &  $1.06(-1)\pm 2.34(-3)$  \\
79248  & $7.27(-1)\pm2.94(-2)$ & $1.75(-1)\pm1.30(-2)$ & &  $4.02(+0)\pm 3.04(-1)$  &  $3.76(-1)\pm 4.85(-3)$  &  $1.11(-1)\pm 2.36(-3)$  \\
80725  & $6.40(-1)\pm1.44(-2)$ &                       & &  $3.50(+0)\pm 2.45(-1)$  &  $3.27(-1)\pm 4.82(-3)$  &  $9.39(-2)\pm 2.34(-3)$  \\
82860  & $2.06(+0)\pm8.52(-3)$ & $4.84(-1)\pm7.03(-3)$ & &  $1.13(+1)\pm 1.15(+0)$  &  $1.09(+0)\pm 1.41(-2)$  &  $3.14(-1)\pm 4.91(-3)$  \\
83389  & $5.73(-1)\pm5.17(-3)$ & $9.64(-2)\pm7.25(-3)$ & &  $3.05(+0)\pm 2.14(-1)$  &  $2.86(-1)\pm 3.95(-3)$  &  $8.47(-2)\pm 2.03(-3)$  \\
84862  & $1.68(+0)\pm7.42(-3)$ & $3.93(-1)\pm2.39(-2)$ & &  $9.38(+0)\pm 9.61(-1)$  &  $8.64(-1)\pm 1.19(-2)$  &  $2.55(-1)\pm 4.69(-3)$  \\
85235  & $8.69(-1)\pm4.33(-3)$ & $1.99(-1)\pm7.31(-3)$ & &  $4.86(+0)\pm 4.13(-1)$  &  $4.54(-1)\pm 5.85(-3)$  &  $1.33(-1)\pm 2.45(-3)$  \\
85295  &                       & $3.22(-1)\pm2.19(-2)$ & &  $6.82(+0)\pm 6.29(-1)$  &  $6.85(-1)\pm 8.83(-3)$  &  $2.05(-1)\pm 4.54(-3)$  \\
86036  & $1.93(+0)\pm7.07(-3)$ & $4.62(-1)\pm1.19(-2)$ & &  $1.05(+1)\pm 1.16(+0)$  &  $1.01(+0)\pm 1.30(-2)$  &  $2.97(-1)\pm 5.20(-3)$  \\
86796  & $2.08(+0)\pm2.16(-2)$ & $5.24(-1)\pm4.59(-3)$ & &  $1.02(+1)\pm 5.25(-1)$  &  $1.19(+0)\pm 8.77(-3)$  &  $3.01(-1)\pm 5.55(-3)$  \\
88601  & $1.03(+1)\pm1.16(-1)$ & $2.27(+0)\pm3.59(-2)$ & &  $3.38(+1)\pm 6.22(-1)$  &  $5.96(+0)\pm 4.39(-2)$  &  $1.51(+0)\pm 2.08(-2)$  \\
88972  & $1.14(+0)\pm1.32(-2)$ & $3.15(-1)\pm3.48(-2)$ & &  $6.54(+0)\pm 5.25(-1)$  &  $5.94(-1)\pm 7.65(-3)$  &  $1.73(-1)\pm 3.67(-3)$  \\
89042  & $1.37(+0)\pm1.15(-2)$ & $3.22(-1)\pm3.43(-2)$ & &  $7.88(+0)\pm 6.90(-1)$  &  $7.60(-1)\pm 9.80(-3)$  &  $2.06(-1)\pm 4.36(-3)$  \\
91009  & $7.07(-1)\pm1.12(-2)$ & $1.32(-1)\pm1.34(-2)$ & &  $3.58(+0)\pm 2.47(-1)$  &  $3.63(-1)\pm 4.68(-3)$  &  $1.14(-1)\pm 2.30(-3)$  \\
92043  & $3.34(+0)\pm8.22(-3)$ & $7.70(-1)\pm4.36(-2)$ & &  $1.69(+1)\pm 1.92(+0)$  &  $1.84(+0)\pm 1.69(-2)$  &  $5.15(-1)\pm 7.12(-3)$  \\
95995  & $8.83(-1)\pm8.39(-3)$ & $2.28(-1)\pm9.66(-3)$ & &  $5.07(+0)\pm 4.16(-1)$  &  $4.72(-1)\pm 6.09(-3)$  &  $1.37(-1)\pm 2.78(-3)$  \\
96100  &                       &                       & &  $2.36(+1)\pm 2.81(+0)$  &  $2.23(+0)\pm 1.64(-2)$  &  $6.63(-1)\pm 9.17(-3)$  \\
96441  & $2.20(+0)\pm1.53(-2)$ & $5.44(-1)\pm1.30(-2)$ & &  $1.24(+1)\pm 1.14(+0)$  &  $1.19(+0)\pm 1.53(-2)$  &  $3.62(-1)\pm 7.33(-3)$  \\
97944  &                       & $3.45(-1)\pm2.32(-2)$ & &  $8.44(+0)\pm 9.03(-1)$  &  $7.98(-1)\pm 1.18(-2)$  &  $2.35(-1)\pm 5.63(-3)$  \\
98959  & $8.98(-1)\pm1.48(-2)$ &                       & &  $4.79(+0)\pm 4.02(-1)$  &  $4.60(-1)\pm 6.35(-3)$  &  $1.34(-1)\pm 2.72(-3)$  \\
99240  &                       &                       & &  $4.31(+1)\pm 1.26(+1)$  &  $5.54(+0)\pm 3.57(-2)$  &  $1.50(+0)\pm 1.80(-2)$  \\
99461  &                       &                       & &  $1.72(+1)\pm 1.36(+0)$  &  $2.17(+0)\pm 1.80(-2)$  &  $5.42(-1)\pm 9.49(-3)$  \\
101955 & $7.86(-1)\pm1.26(-2)$ & $2.29(-1)\pm1.69(-2)$ & &  $4.16(+0)\pm 3.22(-1)$  &  $4.08(-1)\pm 5.64(-3)$  &  $1.21(-1)\pm 3.46(-3)$  \\
101997 & $8.39(-1)\pm5.28(-3)$ &                       & &  $4.13(+0)\pm 3.16(-1)$  &  $4.13(-1)\pm 5.70(-3)$  &  $1.21(-1)\pm 3.22(-3)$  \\
103389 & $9.16(-1)\pm7.03(-3)$ & $2.09(-1)\pm2.90(-2)$ & &  $5.15(+0)\pm 3.65(-1)$  &  $4.66(-1)\pm 5.58(-3)$  &  $1.43(-1)\pm 3.41(-3)$  \\
104214 &                       &                       & &                          &                          &                          \\
104217 &                       &                       & &                          &                          &                          \\
105312 & $8.93(-1)\pm1.05(-2)$ & $2.01(-1)\pm3.55(-2)$ & &  $4.90(+0)\pm 4.02(-1)$  &  $4.44(-1)\pm 6.13(-3)$  &  $1.34(-1)\pm 3.08(-3)$  \\
106696 & $5.88(-1)\pm1.25(-2)$ & $1.90(-1)\pm2.43(-2)$ & &  $3.25(+0)\pm 2.04(-1)$  &  $2.91(-1)\pm 3.75(-3)$  &  $8.57(-2)\pm 2.37(-3)$  \\
107350 & $8.70(-1)\pm1.01(-2)$ & $1.93(-1)\pm2.91(-2)$ & &  $4.62(+0)\pm 3.32(-1)$  &  $4.37(-1)\pm 6.04(-3)$  &  $1.33(-1)\pm 3.30(-3)$  \\
107649 & $1.24(+0)\pm1.72(-2)$ & $2.62(-1)\pm3.06(-2)$ & &  $7.07(+0)\pm 5.93(-1)$  &  $6.35(-1)\pm 8.19(-3)$  &  $1.96(-1)\pm 4.50(-3)$  \\
108870 &                       &                       & &                          &                          &                          \\
109378 & $6.55(-1)\pm3.44(-3)$ &                       & &  $3.65(+0)\pm 2.42(-1)$  &  $3.35(-1)\pm 4.94(-3)$  &  $9.83(-2)\pm 2.45(-3)$  \\
109422 & $1.69(+0)\pm3.42(-2)$ & $3.95(-1)\pm1.84(-3)$ & &  $7.65(+0)\pm 3.17(-1)$  &  $1.17(+0)\pm 1.83(-2)$  &  $2.63(-1)\pm 5.81(-3)$  \\
110109 & $1.74(+0)\pm1.42(-2)$ & $4.17(-1)\pm2.61(-2)$ & &  $9.88(+0)\pm 9.20(-1)$  &  $8.93(-1)\pm 1.23(-2)$  &  $2.66(-1)\pm 5.39(-3)$  \\
113357 & $1.49(+0)\pm4.61(-3)$ & $3.23(-1)\pm1.52(-2)$ & &  $8.60(+0)\pm 8.56(-1)$  &  $7.80(-1)\pm 1.08(-2)$  &  $2.25(-1)\pm 4.56(-3)$  \\
113576 & $1.02(+0)\pm7.00(-3)$ & $2.09(-1)\pm6.30(-2)$ & &  $5.21(+0)\pm 4.18(-1)$  &  $5.20(-1)\pm 7.19(-3)$  &  $1.59(-1)\pm 3.81(-3)$  \\
114948 & $9.52(-1)\pm7.55(-3)$ & $2.27(-1)\pm2.70(-2)$ & &  $5.19(+0)\pm 4.21(-1)$  &  $5.00(-1)\pm 6.45(-3)$  &  $1.64(-1)\pm 3.47(-3)$  \\
116745 & $8.22(-1)\pm1.24(-2)$ & $2.56(-1)\pm3.42(-2)$ & &  $4.54(+0)\pm 2.01(-1)$  &  $4.16(-1)\pm 6.51(-3)$  &  $1.30(-1)\pm 2.87(-3)$  \\
120005 &                       &                       & &  $9.30(+0)\pm 6.68(-1)$  &                          &                          \\ 
\hline


\end{longtable}

\onecolumn

\addtocounter{table}{-1}
\addtocounter{subtable}{1}

\begin{longtable}{ll@{ }cl@{ }cl@{ }cc@{}cc}
\multicolumn{10}{c}{\tablename\ \thetable{}. IRAS and {\it Spitzer}/MIPS fluxes.}
\\
\hline
HIP   &\multicolumn{6}{c}{IRAS} & & \multicolumn{2}{c}{MIPS}  \\\cline{2-7}\cline{9-10}
      & 12 $\mu$m& \%   & 25 $\mu$m & \% & 60 $\mu$m & \%    & & \multicolumn{1}{c}{24 $\mu$m}  & \multicolumn{1}{c}{70 $\mu$m} \\
      & \multicolumn{1}{c}{(Jy)} & &  \multicolumn{1}{c}{(Jy)} & & \multicolumn{1}{c}{(Jy)}& & & \multicolumn{1}{c}{(Jy)}  & \multicolumn{1}{c}{(Jy)}  \\
\hline
\endfirsthead

\multicolumn{10}{c}{\tablename\ \thetable{}. Continued.} \\\hline
HIP   &\multicolumn{6}{c}{IRAS} & & \multicolumn{2}{c}{MIPS}  \\\cline{2-7}\cline{9-10}
      & 12 $\mu$m& \%   & 25 $\mu$m & \% & 60 $\mu$m & \%    & & \multicolumn{1}{c}{24 $\mu$m}  & \multicolumn{1}{c}{70 $\mu$m} \\
      & \multicolumn{1}{c}{(Jy)} & &  \multicolumn{1}{c}{(Jy)} & & \multicolumn{1}{c}{(Jy)}& & & \multicolumn{1}{c}{(Jy)}  & \multicolumn{1}{c}{(Jy)}  \\
\hline
\endhead

\hline
\multicolumn{10}{l}{Continued\dots}\\
\endfoot

\\
\endlastfoot
171    & 1.25(+0)       &  7       &  2.96(--1)    &  17     &                &       & &  $2.12(-1) \pm 4.32(-3)$  &$4.41(-2) \pm 1.36(-2)$ \\
544    & 7.67(--1)      &  6       &  2.27(--1)    &  12     &                &       & &  $1.55(-1) \pm 3.16(-3)$  &$1.03(-1) \pm 7.83(-3)$ \\
910    & 1.50(+0)       &  7       &  2.43(--1)    &  20     &                &       & &  $2.49(-1) \pm 5.08(-3)$  &$3.75(-2) \pm 4.48(-3)$ \\
2941   & 1.27(+0)       &  7       &  2.57(--1)    &  15     &                &       & &  $1.99(-1) \pm 4.05(-3)$  &$2.58(-2) \pm 1.19(-2)$ \\
3093   & 1.03(+0)       &  8       &  2.57(--1)    &  14     &                &       & &  $1.92(-1) \pm 3.92(-3)$  &$1.49(-2) \pm 5.69(-3)$ \\
3497   & 4.22(--1)      &  6       &  9.48(-2)     &  18     &                &       & &  $7.54(-2) \pm 1.54(-3)$  &$5.20(-3) \pm 4.41(-3)$ \\
3821   & {\it 7.61(+0)} & {\it 5}  & {\it 1.75(+0)}& {\it 6} &                &       & &  $1.11(+0) \pm 2.27(-2)$  &$1.22(-1) \pm 1.07(-2)$ \\
3909   & 1.12(+0)       &  7       &  3.09(--1)    &  25     &                &       & &  $1.89(-1) \pm 3.85(-3)$  &$2.43(-2) \pm 3.50(-3)$ \\
4148   & 3.79(--1)      &  13      &  1.10(--1)    &  33     &                &       & &  $7.83(-2) \pm 1.60(-3)$  &$3.71(-2) \pm 6.59(-3)$ \\
7513   & 3.12(+0)       &  5       &  7.27(--1)    &  7      &                &       & &  $5.01(-1) \pm 1.02(-2)$  &$5.57(-2) \pm 6.33(-3)$ \\
7978   & 8.14(--1)      &  6       &  2.82(--1)    &  13     & 8.15(--1)      & 13    & &  $1.85(-1) \pm 3.77(-3)$  &$8.63(-1) \pm 5.87(-2)$ \\
8768   & 3.49(--1)      &  9       &               &         &                &       & &  $6.90(-2) \pm 1.41(-3)$  &$2.47(-2) \pm 7.97(-3)$ \\
10138  & 9.15(--1)      &  5       &  1.84(--1)    &  13     &                &       & &  $1.54(-1) \pm 3.14(-3)$  &$6.90(-3) \pm 6.82(-3)$ \\
10798  & 6.74(--1)      &  7       &  1.50(--1)    &  21     &                &       & &  $1.06(-1) \pm 2.17(-3)$  &$1.45(-2) \pm 2.78(-3)$ \\
11452  & 3.85(--1)      &  9       &               &         &                &       & &  $6.18(-2) \pm 1.26(-3)$  &                        \\
11964  & 4.94(--1)      &  5       &  1.55(--1)    &  13     &                &       & &  $9.93(-2) \pm 2.03(-3)$  &$9.60(-3) \pm 8.82(-3)$ \\
12777  & {\it 2.54(+0)} & {\it 6}  &{\it 5.52(--1)}&{\it 14} &                &       & &  $4.06(-1) \pm 8.28(-3)$  &$5.40(-2) \pm 6.43(-3)$ \\
13402  & 1.01(+0)       &  5       &  2.65(--1)    &  16     &                &       & &  $1.84(-1) \pm 3.76(-3)$  &$6.77(-2) \pm 7.06(-3)$ \\
14954  & 1.39(+0)       &  6       &  3.16(--1)    &  19     &                &       & &  $2.18(-1) \pm 4.45(-3)$  &$4.25(-2) \pm 4.76(-3)$ \\
15330  & 1.07(+0)       &  4       &  2.83(--1)    &  9      &                &       & &  $1.85(-1) \pm 3.77(-3)$  &$3.00(-2) \pm 4.66(-3)$ \\
15371  & 1.18(+0)       &  5       &  2.92(--1)    &  10     &                &       & &  $2.22(-1) \pm 4.52(-3)$  &$4.54(-2) \pm 4.95(-3)$ \\
15799  & 5.40(--1)      &  9       &  1.26(--1)    &  21     &                &       & &  $9.74(-2) \pm 1.99(-3)$  &$1.27(-2) \pm 4.68(-3)$ \\
16134  & 3.87(--1)      &  8       &  1.34(--1)    &  20     &                &       & &  $7.14(-2) \pm 1.46(-3)$  &$1.05(-2) \pm 4.46(-3)$ \\
17420  & 4.94(--1)      &  7       &  1.23(--1)    &  22     &                &       & &  $8.22(-2) \pm 1.68(-3)$  &$2.36(-2) \pm 5.44(-3)$ \\
17439  & 4.28(--1)      &  6       &  1.18(--1)    &  16     & 1.36(--1)      & 24    & &  $7.59(-2) \pm 1.55(-3)$  &$8.85(-2) \pm 7.46(-3)$ \\
19849  & 4.97(+0)       &  6       &  1.14(+0)     &  9      &                &       & &  $7.20(-1) \pm 1.47(-2)$  &$8.26(-2) \pm 6.68(-3)$ \\
19884  & 4.58(--1)      &  6       &  1.15(--1)    &  15     &                &       & &  $7.84(-2) \pm 1.60(-3)$  &$1.21(-2) \pm 3.30(-3)$ \\
22263  & 1.12(+0)       &  5       &  2.24(--1)    &  12     &                &       & &  $1.86(-1) \pm 3.80(-3)$  &$1.14(-1) \pm 8.53(-3)$ \\
23311  & 1.31(+0)       &  5       &  2.79(--1)    &  12     &                &       & &  $2.14(-1) \pm 4.37(-3)$  &$2.57(-2) \pm 5.20(-3)$ \\
25110  & 1.11(+0)       &  5       &  2.45(--1)    &  9      &                &       & &  $1.82(-1) \pm 3.71(-3)$  &$3.26(-2) \pm 5.46(-3)$ \\
27887  & 5.03(--1)      &  5       &  1.03(--1)    &  14     &                &       & &  $8.85(-2) \pm 1.81(-3)$  &$1.72(-2) \pm 4.94(-3)$ \\
28103  & 2.88(+0)       &  4       &  7.26(--1)    &  5      & 2.12(--1)      & 22    & &  $5.36(-1) \pm 1.09(-2)$  &$9.39(-2) \pm 7.76(-3)$ \\
28442  & 4.72(--1)      &  6       &  1.16(--1)    &  15     &                &       & &  $7.95(-2) \pm 1.62(-3)$  &$7.50(-3) \pm 3.93(-3)$ \\
29271  & 1.74(+0)       &  3       &  4.01(--1)    &  5      &                &       & &  $3.03(-1) \pm 6.17(-3)$  &$4.26(-2) \pm 1.05(-2)$ \\
29568  & 5.67(--1)      &  5       &  1.42(--1)    &  12     &                &       & &  $9.11(-2) \pm 1.86(-3)$  &$1.53(-2) \pm 2.34(-3)$ \\
32439  & 8.87(--1)      &  5       &  2.18(--1)    &  13     &                &       & &  $1.47(-1) \pm 3.01(-3)$  &$1.58(-2) \pm 2.72(-3)$ \\
32480  & 1.46(+0)       &  6       &  4.12(--1)    &  10     & 3.81(--1)      & 13    & &  $1.93(-1) \pm 3.94(-3)$  &$2.63(-1) \pm 1.83(-2)$ \\
33277  & 7.90(--1)      &  6       &               &         &                &       & &  $1.30(-1) \pm 2.66(-3)$  &$1.31(-2) \pm 1.01(-2)$ \\
34017  & 6.90(--1)      &  7       &               &         &                &       & &  $1.12(-1) \pm 2.29(-3)$  &$1.44(-2) \pm 4.02(-3)$ \\
34065  & 1.42(+0)       &  4       &  3.33(--1)    &  7      &                &       & &  $1.69(-1) \pm 3.44(-3)$  &$2.05(-2) \pm 2.77(-3)$ \\
35136  & 9.62(--1)      &  5       &  2.52(--1)    &  12     &                &       & &  $1.64(-1) \pm 3.34(-3)$  &$2.89(-2) \pm 4.63(-3)$ \\
36439  & 8.86(--1)      &  5       &  2.14(--1)    &  14     &                &       & &  $1.46(-1) \pm 2.98(-3)$  &$1.44(-2) \pm 2.96(-3)$ \\
38382  & {\it1.28(+0)}  & {\it 8}  & {\it 2.84(-1)}& {\it 19}&                &       & &  $2.46(-1) \pm 5.02(-3)$  &$1.34(-2) \pm 4.30(-3)$ \\
38784  & 4.61(--1)      &  6       &  9.72(-2)     &  23     &                &       & &  $8.09(-2) \pm 1.65(-3)$  &$1.06(-2) \pm 2.99(-3)$ \\
40693  & 9.67(--1)      &  6       &  3.41(--1)    &  11     &                &       & &  $2.30(-1) \pm 4.69(-3)$  &$1.51(-2) \pm 2.42(-3)$ \\
40843  & 1.14(+0)       &  18      &  2.74(--1)    &  22     &                &       & &  $1.87(-1) \pm 3.82(-3)$  &$3.25(-2) \pm 5.27(-3)$ \\
42430  & 1.78(+0)       &  4       &  3.80(--1)    &  7      & 1.52(--1)      & 25    & &  $2.93(-1) \pm 5.98(-3)$  &$3.11(-2) \pm 6.07(-3)$ \\
42438  & 8.85(--1)      &  7       &  2.08(--1)    &  16     &                &       & &  $1.56(-1) \pm 3.19(-3)$  &$4.12(-2) \pm 4.16(-3)$ \\
43587  & 9.81(--1)      &  5       &  2.41(--1)    &  26     &                &       & &  $1.69(-1) \pm 3.45(-3)$  &$1.98(-2) \pm 4.41(-3)$ \\
43726  & 6.68(--1)      &  5       &  1.64(--1)    &  17     &                &       & &  $1.19(-1) \pm 2.43(-3)$  &$3.25(-2) \pm 4.13(-3)$ \\
44897  & 5.91(--1)      &  8       &               &         &                &       & &  $1.06(-1) \pm 2.17(-3)$  &$1.87(-2) \pm 4.00(-3)$ \\
45333  & 1.45(+0)       &  5       &  3.36(--1)    &  8      &                &       & &  $2.39(-1) \pm 4.88(-3)$  &$2.39(-2) \pm 5.25(-3)$ \\
45617  & 5.23(--1)      &  8       &               &         &                &       & &  $8.62(-2) \pm 1.76(-3)$  &$2.50(-3) \pm 1.04(-2)$ \\
46580  & 4.74(--1)      &  7       &  1.74(--1)    &  27     &                &       & &  $8.51(-2) \pm 1.74(-3)$  &                        \\
47592  & 1.48(+0)       &  6       &  3.57(--1)    &  10     &                &       & &  $2.42(-1) \pm 4.93(-3)$  &$1.46(-2) \pm 1.14(-2)$ \\
49081  & 1.22(+0)       &  6       &  2.47(--1)    &  18     &                &       & &  $1.96(-1) \pm 4.00(-3)$  &                        \\
49908  & 2.30(+0)       &  6       &  5.85(--1)    &  7      &                &       & &  $4.19(-1) \pm 8.55(-3)$  &$3.87(-2) \pm 4.69(-3)$ \\
51459  & 1.68(+0)       &  4       &  3.91(--1)    &  7      &                &       & &  $2.67(-1) \pm 5.45(-3)$  &$3.38(-2) \pm 4.43(-3)$ \\
51502  & 8.15(--1)      &  5       &  2.01(--1)    &  10     &                &       & &  $1.41(-1) \pm 2.88(-3)$  &$3.96(-2) \pm 3.79(-3)$ \\
53721  & 1.59(+0)       &  5       &  3.10(--1)    &  9      &                &       & &  $2.54(-1) \pm 5.18(-3)$  &$3.14(-2) \pm 4.17(-3)$ \\
54646  & 6.44(--1)      &  8       &               &         &                &       & &  $7.55(-2) \pm 1.54(-3)$  &$1.80(-2) \pm 1.04(-2)$ \\
56452  & 1.13(+0)       &  7       &  2.74(--1)    &  18     &                &       & &  $1.79(-1) \pm 3.66(-3)$  &$2.32(-2) \pm 2.95(-3)$ \\
57507  & 4.82(--1)      &  10      &               &         &                &       & &  $8.28(-2) \pm 1.69(-3)$  &$1.19(-2) \pm 3.59(-3)$ \\
57939  &                &          &               &         &                &       & &  $1.26(-1) \pm 2.57(-3)$  &$1.07(-2) \pm 2.31(-3)$ \\
58345  & 8.48(--1)      &  7       &  1.95(--1)    &  29     &                &       & &  $1.40(-1) \pm 2.86(-3)$  &$9.00(-3) \pm 8.02(-3)$ \\
62145  &                &          &               &         &                &       & &  $8.95(-2) \pm 1.83(-3)$  &$2.89(-2) \pm 2.01(-2)$ \\
62207  & 6.59(--1)      &  8       &  1.59(--1)    &  28     &                &       & &  $1.11(-1) \pm 2.25(-3)$  &$5.57(-2) \pm 5.20(-3)$ \\
62523  & 5.60(--1)      &  9       &  1.53(--1)    &  28     &                &       & &  $9.55(-2) \pm 1.95(-3)$  &$1.35(-2) \pm 3.62(-3)$ \\
64792  & 1.29(+0)       &  6       &  2.67(--1)    &  14     &                &       & &  $2.16(-1) \pm 4.40(-3)$  &$1.91(-2) \pm 5.75(-3)$ \\
64797  & 1.04(+0)       &  8       &  2.77(--1)    &  21     &                &       & &  $1.53(-1) \pm 3.12(-3)$  &$4.68(-2) \pm 1.65(-2)$ \\
65026  & 6.73(--1)      &  8       &  2.01(--1)    &  17     &                &       & &  $1.51(-1) \pm 3.08(-3)$  &$2.51(-2) \pm 1.01(-2)$ \\
65721  & 2.28(+0)       &  7       &  5.37(--1)    &  10     &                &       & &  $3.62(-1) \pm 7.38(-3)$  &$7.90(-2) \pm 8.09(-3)$ \\
67275  & 1.97(+0)       &  7       &  4.85(--1)    &  15     &                &       & &  $3.28(-1) \pm 6.69(-3)$  &$3.32(-2) \pm 6.78(-3)$ \\
67422  & 8.36(--1)      &  6       &  1.78(--1)    &  15     &                &       & &  $7.67(-2) \pm 1.56(-3)$  &$3.11(-2) \pm 1.04(-2)$ \\
67620  & 6.11(--1)      &  7       &  1.84(--1)    &  24     &                &       & &  $9.70(-2) \pm 1.98(-3)$  &$8.60(-3) \pm 3.45(-3)$ \\
68184  & 9.48(--1)      &  5       &  2.10(--1)    &  11     &                &       & &  $1.69(-1) \pm 3.44(-3)$  &$1.81(-2) \pm 6.71(-3)$ \\
68682  & 7.43(--1)      &  6       &  1.91(--1)    &  21     &                &       & &  $1.20(-1) \pm 2.45(-3)$  &$1.61(-2) \pm 1.40(-2)$ \\
69965  & 6.03(--1)      &  8       &  2.43(--1)    &  20     &                &       & &  $1.11(-1) \pm 2.26(-3)$  &$1.95(-2) \pm 9.29(-3)$ \\
70319  & 5.27(--1)      &  7       &               &         &                &       & &  $9.57(-2) \pm 1.95(-3)$  &$1.16(-2) \pm 6.35(-3)$ \\
70857  & 4.40(--1)      &  5       &  1.11(--1)    &  14     &                &       & &  $7.81(-2) \pm 1.59(-3)$  &$6.50(-3) \pm 4.62(-3)$ \\
71181  & 4.88(--1)      &  6       &  1.20(--1)    &  14     &                &       & &  $8.32(-2) \pm 1.70(-3)$  &$2.92(-2) \pm 8.04(-3)$ \\
71681  &                &          &               &         &                &       & &  $1.55(+1) \pm 3.16(-1)$  &$1.01(+0) \pm 6.68(-1)$ \\
71683  &                &          &               &         &                &       & &  $2.85(+1) \pm 5.82(-1)$  &$3.39(+0) \pm 7.02(-1)$ \\
71908  & {\it 4.07(+0)} & {\it 15} &{\it 8.21(--1)}& {\it 15}&                &       & &  $6.70(-1) \pm 1.37(-2)$  &                        \\
72567  & 6.63(--1)      &  6       &  1.07(--1)    &  21     &                &       & &  $1.16(-1) \pm 2.36(-3)$  &$1.08(-2) \pm 2.89(-3)$ \\
72603  & 1.07(+0)       &  7       &               &         &                &       & &                           &                        \\
72848  & 1.13(+0)       &  11      &  2.81(--1)    &  17     &                &       & &  $1.93(-1) \pm 3.93(-3)$  &$3.35(-2) \pm 6.41(-3)$ \\
73100  & 7.30(--1)      &  4       &  1.75(--1)    &  8      &                &       & &  $1.27(-1) \pm 2.59(-3)$  &$2.47(-2) \pm 3.17(-3)$ \\
73182  &                &          &               &         &                &       & &  $2.63(-1) \pm 5.36(-3)$  &$1.60(-2) \pm 3.19(-3)$ \\
73184  &                &          &               &         &                &       & &  $3.96(-1) \pm 8.08(-3)$  &$4.73(-2) \pm 4.37(-3)$ \\
73996  & 1.12(+0)       &  5       &  2.47(--1)    &  11     &                &       & &  $1.96(-1) \pm 4.00(-3)$  &$3.32(-2) \pm 5.84(-3)$ \\
77052  & 8.47(--1)      &  6       &  2.32(--1)    &  15     &                &       & &  $1.50(-1) \pm 3.07(-3)$  &$1.11(-2) \pm 5.25(-3)$ \\
78459  & 1.10(+0)       &  5       &  2.65(--1)    &  8      &                &       & &  $1.97(-1) \pm 4.01(-3)$  &$2.96(-2) \pm 5.76(-3)$ \\
78775  & 4.86(--1)      &  6       &  1.25(--1)    &  15     &                &       & &  $7.88(-2) \pm 1.61(-3)$  &$8.70(-3) \pm 3.55(-3)$ \\
79248  & 5.53(--1)      &  6       &  1.21(--1)    &  13     &                &       & &  $9.05(-2) \pm 1.85(-3)$  &$1.06(-2) \pm 2.60(-3)$ \\
80725  & 4.49(--1)      &  7       &               &         &                &       & &  $7.75(-2) \pm 1.58(-3)$  &$3.73(-2) \pm 1.76(-2)$ \\
82860  & 1.51(+0)       &  4       &  3.58(--1)    &  5      &                &       & &  $2.69(-1) \pm 5.49(-3)$  &$4.47(-2) \pm 6.01(-3)$ \\
83389  & 3.89(--1)      &  7       &  1.08(--1)    &  22     &                &       & &  $6.82(-2) \pm 1.39(-3)$  &$1.10(-2) \pm 9.03(-3)$ \\
84862  & 1.23(+0)       &  4       &  2.75(--1)    &  8      &                &       & &  $2.10(-1) \pm 4.29(-3)$  &$2.46(-2) \pm 4.33(-3)$ \\
85235  & 6.15(--1)      &  5       &  1.52(--1)    &  11     &                &       & &  $1.08(-1) \pm 2.20(-3)$  &$5.80(-2) \pm 4.92(-3)$ \\
85295  & 1.02(+0)       &  7       &  1.57(--1)    &  20     &                &       & &  $1.70(-1) \pm 3.46(-3)$  &$1.38(-2) \pm 4.59(-3)$ \\
86036  & 1.37(+0)       &  3       &  3.02(--1)    &  5      &                &       & &  $2.50(-1) \pm 5.09(-3)$  &$2.74(-2) \pm 4.77(-3)$ \\
86796  & 1.54(+0)       &  7       &  4.49(--1)    &  14     &                &       & &  $2.60(-1) \pm 5.30(-3)$  &$3.10(-2) \pm 8.07(-3)$ \\
88601  & 7.64(+0)       &  5       &  1.80(+0)     &  6      &                &       & &  $1.24(+0) \pm 2.54(-2)$  &$1.44(-1) \pm 1.68(-2)$ \\
88972  & 8.53(--1)      &  3       &  1.86(--1)    &  9      &                &       & &  $1.43(-1) \pm 2.91(-3)$  &$8.60(-3) \pm 3.65(-3)$ \\
89042  & 9.96(--1)      &  5       &  2.20(--1)    &  12     &                &       & &  $1.68(-1) \pm 3.42(-3)$  &$1.71(-2) \pm 3.59(-3)$ \\
91009  & 5.52(--1)      &  6       &  1.33(--1)    &  13     &                &       & &  $9.39(-2) \pm 1.92(-3)$  &$6.40(-3) \pm 5.32(-3)$ \\
92043  & 2.47(+0)       &  4       &  6.02(--1)    &  5      &                &       & &  $4.35(-1) \pm 8.87(-3)$  &$6.98(-2) \pm 8.93(-3)$ \\
95995  & 6.46(--1)      &  3       &  1.56(--1)    &  8      &                &       & &  $1.14(-1) \pm 2.33(-3)$  &$1.27(-2) \pm 5.37(-3)$ \\
96100  & 3.39(+0)       &  3       &  7.74(--1)    &  4      &                &       & &  $5.40(-1) \pm 1.10(-2)$  &$7.32(-2) \pm 7.23(-3)$ \\
96441  & 1.63(+0)       &  6       &  3.10(--1)    &  14     &                &       & &  $2.78(-1) \pm 5.67(-3)$  &$3.68(-2) \pm 4.05(-3)$ \\
97944  & 1.19(+0)       &  6       &  2.94(--1)    &  18     &                &       & &  $1.98(-1) \pm 4.03(-3)$  &                        \\
98959  & 6.83(--1)      &  7       &  1.46(--1)    &  15     &                &       & &  $1.02(-1) \pm 2.07(-3)$  &$1.90(-2) \pm 3.63(-3)$ \\
99240  & 7.00(+0)       &  5       &  1.67(+0)     &  6      & 2.28(--1)      & 19    & &  $1.19(+0) \pm 2.43(-2)$  &$1.39(-1) \pm 1.07(-2)$ \\
99461  & 2.71(+0)       &  5       &  6.30(--1)    &  9      &                &       & &  $4.42(-1) \pm 9.02(-3)$  &$4.85(-2) \pm 1.08(-2)$ \\
101955 & 6.02(--1)      &  6       &  1.98(--1)    &  24     &                &       & &  $1.02(-1) \pm 2.08(-3)$  &$1.15(-2) \pm 7.04(-3)$ \\
101997 & 5.05(--1)      &  8       &               &         &                &       & &  $1.01(-1) \pm 2.06(-3)$  &$7.40(-3) \pm 3.73(-3)$ \\
103389 & 7.12(--1)      &  9       &               &         &                &       & &  $1.16(-1) \pm 2.37(-3)$  &$4.66(-2) \pm 4.93(-3)$ \\
104214 &                &          &               &         &                &       & &  $9.58(-1) \pm 1.95(-2)$  &$6.42(-2) \pm 1.27(-2)$ \\
104217 &                &          &               &         &                &       & &  $7.42(-1) \pm 1.51(-2)$  &$8.33(-2) \pm 1.32(-2)$ \\
105312 & 5.35(--1)      &  9       &  1.51(--1)    &  21     &                &       & &  $1.10(-1) \pm 2.25(-3)$  &$1.36(-2) \pm 5.67(-3)$ \\
106696 & 4.79(--1)      &  9       &  1.25(--1)    &  25     &                &       & &  $7.27(-2) \pm 1.48(-3)$  &$1.92(-2) \pm 5.26(-3)$ \\
107350 & 6.46(--1)      &  7       &  1.30(--1)    &  20     &                &       & &  $1.07(-1) \pm 2.18(-3)$  &$2.84(-2) \pm 3.15(-3)$ \\
107649 & 8.97(--1)      &  6       &  2.00(--1)    &  12     & 3.01(--1)      & 14    & &  $1.55(-1) \pm 3.16(-3)$  &$2.78(-1) \pm 2.15(-2)$ \\
108870 & 5.68(+0)       &  5       &  1.37(+0)     &  5      & 1.79(--1)      & 23    & &  $9.76(-1) \pm 1.99(-2)$  &$1.11(-1) \pm 9.47(-3)$ \\
109378 & {\it 5.28(--1)}& {\it 9}  &               &         &                &       & &  $8.19(-2) \pm 1.67(-3)$  &$7.90(-3) \pm 2.26(-3)$ \\
109422 & 1.20(+0)       &  5       &  2.97(--1)    &  12     &                &       & &  $2.13(-1) \pm 4.35(-3)$  &$1.06(-2) \pm 5.65(-3)$ \\
110109 & 1.31(+0)       &  5       &  3.15(--1)    &  10     &                &       & &  $2.23(-1) \pm 4.55(-3)$  &$2.81(-2) \pm 5.72(-3)$ \\
113357 & 1.08(+0)       &  6       &  2.45(--1)    &  13     &                &       & &  $1.86(-1) \pm 3.79(-3)$  &$2.81(-2) \pm 4.88(-3)$ \\
113576 & 6.71(--1)      &  9       &  2.63(--1)    &  23     &                &       & &  $1.27(-1) \pm 2.60(-3)$  &$1.91(-2) \pm 2.90(-3)$ \\
114948 & 7.39(--1)      &  6       &  1.81(--1)    &  14     &                &       & &  $1.33(-1) \pm 2.72(-3)$  &$6.87(-2) \pm 5.51(-3)$ \\
116745 & 5.89(--1)      &  6       &  1.22(--1)    &  18     &                &       & &  $1.02(-1) \pm 2.09(-3)$  &$1.72(-2) \pm 2.49(-3)$ \\
120005 &                &          &               &         &                &       & &  $1.87(-1) \pm 3.81(-3)$  &$2.13(-2) \pm 2.71(-3)$ \\ 
\hline\noalign{\smallskip}
\multicolumn{10}{l}{Notes: IRAS data are from the Faint Source Catalogue, except those in italics that are from the Point}\\
\multicolumn{10}{l}{\hspace*{0.95cm} Source catalogue.} 

\end{longtable}

\end{center}








%
%
\renewcommand{\thetable}{\arabic{table}}
}

%
%

\onllongtabL{4}{                                                                                 
\begin{center}                                                                                  
\begin{longtable}{llrrrrrrrrr}                                                               
\multicolumn{11}{c}{\tablename\ \thetable{}. Stellar parameters of the DUNES sources.}
\\                                                                                        
\hline
HIP     &  SpT    &T$_{\rm eff}$ & log g &[Fe/H]&  $v \sin i$  &Lbol & Lx/Lbol    & AgeX   &$\log R'_{\rm HK}$  &  Age (Ca {\sc ii})    \\
        &         &(K)          &(cm/s$^2$)& (dex)&(km/s)  &($L_\odot$)& (log)     & (Gyr)   &                   &(Gyr)                  \\
\hline                                                                                    
\endfirsthead                                                                            

\multicolumn{11}{c}{\tablename\ \thetable{}. Continued.}\\ \hline                                                                                    
HIP     &  SpT    &T$_{\rm eff}$ & log g &[Fe/H]&  $v \sin i$  &Lbol &  Lx/Lbol   & AgeX  &$\log R'_{\rm HK}$  &  Age (Ca {\sc ii})    \\
        &         &(K)          &(cm/s$^2$)& (dex)&(km/s)  &$L_\odot$& (log)     & (Gyr)  &                   & (Gyr)                 \\
\hline                                                                                          
\endhead
                 
\hline  
\multicolumn{11}{l}{Continued\dots}\\                                                         
\endfoot
         
\\                                                                                
\endlastfoot 
171   &G3V     &5681    & 4.86  & -0.52 & 1.8  & 0.614&  -5.9  &   3.12    &-4.851 & 3.96  \\     
544   &K0V     &5577    & 4.58  &  0.12 & 3.4  & 0.616&  -4.4  &   0.32    &-4.328 & 0.17  \\     
910   &F5V     &6160    & 4.01  & -0.38 & 3.8  & 3.151&  -7.6  &  12.53    &-4.788 & 3.04  \\     
2941  &K1V+... &5509    & 4.23  & -0.14 & 1.6: & 1.258&        &           &-4.903 & 4.83  \\     
3093  &K0V     &5204    & 4.45  &  0.16 & 1.15 & 0.529&  -6.0  &   4.53    &-4.991 & 6.43  \\     
3497  &G3V     &5670    & 4.48  & -0.24 & 3.0: & 1.012&  -7.6  &  $<$15    &-4.853 & 4.00  \\     
3821  &G0V SB  &5932    & 4.42  & -0.23 & 3.3  & 1.209&  -6.2  &   3.62    &-4.991 & 6.43  \\     
3909  &F7IV-V  &6271    & 4.47  & -0.05 & 3.0  & 1.697&  -5.9  &   1.55    &-4.78  & 2.93  \\     
4148  &K2V     &4940    & 4.70  & -0.15 & 2.0: & 0.297&  -5.2  &   1.79    &-4.83  & 3.64  \\     
7513  &F8V     &6155    & 4.13  &  0.10 & 9.0  & 3.363&  -6.6  &   2.90    &-5.035 & 7.26  \\     
7978  &F8V     &6155    & 4.48  & -0.04 & 5.61 & 1.523&  -5.4  &   0.96    &-4.675 & 1.74  \\     
8768  &K5/M0V  &3865    &       &       &      & 0.077&  -4.6  &   2.03    &-4.59  &       \\     
10138 &K0V     &5165    & 4.56  & -0.22 &      & 0.406&  -5.7  &   3.57    &-4.68  & 1.79  \\     
10798 &G8V     &5371    & 4.62  & -0.46 & 3.0: & 0.435&        &           &-4.997 & 6.54  \\     
11452 &K7V:    &3921    &       &       &      & 0.210&        &           &       &        \\    
11964 &K7V SB  &3790    &       &       &      & 0.092&  -2.8  &   0.18    &-3.997 &        \\    
12777 &F7V     &6314    & 4.31  &  0.03 & 9.0  & 2.250&  -6.0  &   1.44    &-5.071 & 7.92   \\    
13402 &K1V     &5217    & 4.57  &  0.10 & 6.3  & 0.392&  -4.3  &   0.40    &-4.3   & 0.14   \\    
14954 &F8V     &6187    & 4.24  &  0.21 & 8.5  & 3.848&  -6.0  &   1.17    &-4.95  & 5.67   \\    
15330 &G2V     &5716    & 4.57  & -0.22 & 2.7  & 0.761&  -5.4  &   1.32    &-4.86  & 4.11   \\    
15371 &G1V     &5851    & 4.51  & -0.23 & 0.3  & 0.972&  -6.7  &   7.83    &-4.827 & 3.59   \\    
15799 &K0V     &5087    & 4.43  &  0.12 & 1.2  & 0.535&  -4.3  &   0.36    &-4.773 & 2.84   \\    
16134 &K5V     &3995    & *     & -0.13 & 2.0  & 0.136&  -4.6  &   1.44    &-4.35  &        \\    
17420 &K2V     &4990    & 4.52  & -0.12 & 9.9  & 0.294&  -5.0  &   1.35    &-5.04  &        \\    
17439 &K1V     &5166    & 4.44  &  0.05 & 2.9  & 0.402&  -4.9  &   0.93    &-4.534 & 0.76   \\    
19849 &K1V     &5150    & 4.52  & -0.24 & 1.0  & 0.416&  -5.6  &   2.69    &-4.872 & 4.30   \\    
19884 &K5V     &4473    & *     &  0.06 & 1.0  & 0.203&  -5.2  &   2.54    &-4.905 &        \\    
22263 &G3V     &5814    & 4.47  &  0.01 & 2.9  & 0.951&  -4.8  &   0.44    &-4.61  & 1.21   \\    
23311 &K3V     &4827    & 4.69  &  0.3  & 1.1  & 0.276&  -5.9  &   5.09    &-5.29  &         \\   
25110 &F6V     &6307    & 4.17  &  0.10 & 7.7: & 3.224&  -5.9  &   1.21    &-5.216 &         \\   
27887 &K3V     &4870    & 4.49  & -0.29 & 3.0: & 0.252&  -5.9  &   6.65    &-5.037 & 2.60   \\   
28103 &F1V     &7000    & 4.12  & -0.11 & 15.7 & 5.562&  -5.8  &   0.64    &-4.78  &         \\   
28442 &K3/K4V  &4330    &       &       &      & 0.290&        &           &-4.932 &         \\   
29271 &G5V     &5591    & 4.46  &  0.08 &      & 0.847&  -6.1  &   3.37    &-4.94  & 5.49    \\   
29568 &G5V     &5633    & 4.48  & -0.01 & 6.0: & 0.692&  -4.5  &   0.36    &-4.39  & 0.28    \\   
32439 &F8V     &6191    & 4.36  & -0.26 & 5.0: & 1.710&  -5.9  &   1.65    &-5.024 & 7.05    \\   
32480 &G0V     &6086    & 4.35  &  0.09 & 5.0: & 1.752&  -5.8  &   1.38    &-4.985 & 6.32    \\   
33277 &G0V     &5891    & 4.36  & -0.13 & 4.0: & 1.251&        &           &-4.938 & 5.45    \\   
34017 &G4V     &5887    & 4.34  & -0.12 & 4.0: & 1.298&        &           &-4.943 & 5.54    \\   
34065 &G3V...  &5826    & 4.41  & -0.17 & 4.0: & 1.372&        &           &-4.981 & 6.24    \\   
35136 &G0V     &5917    & 4.26  & -0.35 & 4.0: & 1.444&  -8.2  &  $>$15    &-4.927 & 5.25    \\   
36439 &F6V     &6309    & 4.16  & -0.32 & 10.0 & 2.357&        &           &-5.306 &         \\   
38382 &G2V     &5860    & 4.25  & -0.13 & 4.8  & 1.973&        &           &-4.883 & 4.49    \\   
38784 &G8V     &5502    & 4.52  & -0.08 & 2.0: & 0.638&  -5.8  &   3.03    &-4.837 & 3.75    \\   
40693 &K0V     &5405    & 4.46  & -0.04 & 3.1  & 0.595&  -5.9  &   3.27    &-4.991 & 6.43    \\   
40843 &F6V     &6241    & 4.16  & -0.23 & 3.9  & 2.371&        &           &-5.187 &         \\   
42430 &G3/G5V  &5526    & 3.93  &  0.31 & 3.8  & 3.208&  -6.1  &   1.55    &-5.031 & 7.18    \\   
42438 &G1.5Vb  &5807    & 4.42  & -0.11 & 9.8  & 0.970&  -4.5  &   0.31    &-4.38  & 0.26    \\   
43587 &G8V     &5295    & 4.47  &  0.36 & 2.3  & 0.602&  -6.7  &  11.19    &-5.099 & 8.43    \\   
43726 &G3V     &5789    & 4.51  &  0.09 & 1.6  & 1.009&  -5.3  &   0.89    &-4.659 & 1.60    \\   
44897 &F9V     &5980    & 4.45  &  0.03 & 4.0: & 1.258&  -4.7  &   0.37    &-4.608 & 1.20    \\   
45333 &F9V     &5907    & 4.01  &  0.00 & 4.0: & 2.705&        &           &-5.073 & 7.96    \\   
45617 &K3V     &4639    & 4.39  & -0.22 & 6.1  & 0.485&  -5.3  &   1.56    &-4.51  &         \\   
46580 &K3V     &4868    & 4.8   & -0.04 & 1.7  & 0.241&  -4.6  &   0.91    &-4.27  &         \\   
47592 &G0V     &6160    & 4.33  & -0.06 & 6.0: & 1.934&  -6.6  &   4.75    &-4.862 & 4.14    \\   
49081 &G1V     &5759    & 4.34  &  0.19 & 2.2  & 1.375&        &           &-4.969 & 6.02    \\   
49908 &K8V     &4081    & 4.71  & -0.16 & 2.6  & 0.125&  -5.0  &   3.21    &-5     &          \\  
51459 &F8V     &6137    & 4.33  & -0.10 & 2.0: & 1.554&  -5.6  &   1.03    &-4.783 & 2.97     \\  
51502 &F2V     &6710    & 4.32  & -0.26 & 56.2 & 2.840&  -4.9  &   0.30    &-4.55  &          \\  
53721 &G0V     &5908    & 4.37  &  0.03 & 3.2  & 1.600&  -8.3  &  $>$15    &-4.909 & 4.93     \\  
54646 &K8V     &3783    & 5.0   & -1.5  & 2.7  & 0.166&  -4.9  &   2.40    &-4.86  &          \\  
56452 &K0V     &5189    & 4.68  & -0.32 & 4.0  & 0.368&  -6.2  &   7.78    &-4.907 & 4.90     \\  
57507 &G5V     &5559    & 4.47  & -0.26 & 3.0: & 0.690&        &           &-4.924 & 5.20     \\  
57939 &G8Vp    &5052    & 4.70  & -1.27 & 4.0: & 0.228&  -6.4  &  13.30    &-4.896 & 4.71     \\  
58345 &K4V     &4510    & 4.57  &  0.16 & 4.0: & 0.218&  -5.2  &   2.30    &-4.868 &          \\  
62145 &K3V     &5018    & 4.40  &  0.08 & 4.8  & 0.359&  -4.7  &   0.80    &-4.79  &          \\  
62207 &G0V     &5860    & 4.33  & -0.53 & 3.0: & 1.055&        &           &-4.981 & 6.24     \\  
62523 &G7V     &5643    & 4.52  &  0.08 & 2.9  & 0.758&  -5.1  &   0.82    &-4.43  & 0.37     \\  
64792 &G0Vs    &6118    & 4.32  &  0.16 & 7.0  & 2.092&  -4.4  &   0.16    &-4.443 & 0.41     \\  
64797 &K2V     &4952    & 4.55  & -0.15 &      & 0.330&  -5.0  &   1.25    &-4.63  & 1.36     \\  
65026 &K0      &3752    &       &       &      & 0.120&  -4.9  &   2.74    &-4.5226&6        \\   
65721 &G5V     &5513    & 3.94  & -0.07 & 2.7  & 2.989&  -7.0  &   5.57    &-5.069 & 7.89    \\   
67275 &F7V     &6376    & 4.21  &  0.26 & 15.0 & 3.062&  -5.1  &   0.37    &-4.731 & 2.32    \\   
67422 &K2      &4729    & 5.09  & -0.35 & 5.2  & 0.320&  -4.9  &   1.06    &-4.771 &         \\   
67620 &G5V     &5751    & 4.52  &  0.05 & 3.0: & 0.869&  -5.4  &   1.26    &-4.703 & 2.02    \\   
68184 &K3V     &4757    & 4.58  &  0.10 &      & 0.296&  -6.1  &   7.18    &-5.12  &         \\   
68682 &G8V     &5509    & 4.39  &  0.00 & 2.0: & 0.801&        &           &-4.869 & 4.25    \\   
69965 &F7Vw    &6120    & 4.39  & -0.64 & 2.8  & 1.181&        &           &-4.641 & 1.45    \\   
70319 &G1V     &5666    & 4.49  & -0.32 & 2.0: & 0.812&        &           &-4.957 & 5.80    \\   
70857 &G5      &5371    & 4.53  & -0.39 & 1.0: & 0.646&  -5.4  &   1.59    &-4.91  & 4.95    \\   
71181 &K3V     &4809    & 4.67  & -0.09 & 4.3  & 0.252&  -5.2  &   2.25    &-4.97  &         \\   
71681 &K1V     &5178    & 4.56  &  0.15 & 1.1  & 0.444&  -6.0  &   4.27    &-4.923 & 5.18    \\   
71683 &G2V     &5801    & 4.33  &  0.19 & 2.7  & 1.483&  -7.1  &  10.40    &-5.059 & 7.70    \\   
71908 &F1Vp    &7645    & 4.24  &       & 14.0 &11.263&        &           &       &         \\   
72567 &G2V     &6059    & 4.58  &  0.00 & 7.0: & 1.223&  -5.1  &   0.65    &-4.34  & 0.19    \\   
72603 &F3V     &6598    & 4.18  & -0.10 & 3.2  & 3.595&  -5.1  &   0.29    &-4.58  & 1.01    \\   
72848 &K2V     &5313    & 4.57  &  0.10 & 4.0  & 0.498&  -4.8  &   0.70    &-4.52  & 0.69    \\   
73100 &F7V     &6220    & 4.15  & -0.03 & 6.0: & 2.831&  -5.9  &   1.15    &-5.03  & 7.16    \\   
73182 &K5V     &4683    & 3.70  & -0.24 & 1.0: & 0.027&  -5.4  &   5.40    &-5.19  &         \\   
73184 &K4V     &4744    & 4.76  &  0.10 & 1.0  & 0.204&  -5.7  &   4.74    &-4.63  &         \\   
73996 &F5V     &6435    & 4.19  &  0.05 & 44.0 & 3.222&  -5.4  &   0.49    &-4.85  & 3.95    \\   
77052 &G5V     &5659    & 4.45  &  0.06 & 2.0: & 0.844&  -5.3  &   1.10    &-4.8   & 3.20    \\   
78459 &G2V     &5833    & 4.29  & -0.20 & 1.6  & 1.742&        &           &-5.039 & 7.33    \\   
78775 &G8V     &5214    & 4.71  & -0.49 & 2.0: & 0.443&        &           &-4.97  & 6.04    \\   
79248 &K0V     &5336    & 4.48  &  0.43 & 1.6  & 0.653&  -6.5  &   7.45    &-5.175 &         \\   
80725 &K2V     &5040    & 4.53  &       &      & 0.636&  -5.2  &   1.18    &-4.69  & 1.89    \\   
82860 &F6Vvar  &6306    & 4.39  & -0.11 & 9.0  & 2.072&  -5.1  &   0.44    &-4.737 & 2.39    \\   
83389 &G8V     &5488    & 4.56  & -0.07 & 2.0: & 0.617&  -6.2  &   5.57    &-4.843 & 3.84    \\   
84862 &G0V     &5712    & 4.32  & -0.29 & 3.0: & 1.241&  -7.4  &           &-4.963 & 5.91     \\  
85235 &K0V     &5290    & 4.63  & -0.39 & 3.0: & 0.413&        &           &-4.93  & 5.31     \\  
85295 &K7V     &4059    & 4.68  & -0.03 & 3.0: & 0.106&  -4.9  &   2.30    &-4.72  &          \\  
86036 &G0V     &5980    & 4.40  &  0.00 & 4.0: & 1.342&  -5.0  &   0.56    &-4.69  & 1.89     \\  
86796 &G5V     &5787    & 4.29  &  0.29 & 3.4  & 1.821&  -7.8  &  $>$15    &-5.101 &          \\  
88601 &K0V SB  &5312    & 4.54  &  0.05 & 13.0 & 0.594&  -5.2  &   1.01    &-4.586 & 1.05     \\  
88972 &K2V     &5000    & 4.47  & -0.05 & 1.5  & 0.355&  -6.0  &   5.79    &-4.955 & 5.76     \\  
89042 &G0V     &5921    & 4.27  & -0.17 & 3.5  & 1.672&        &           &-4.935 & 5.40     \\  
91009 &K7Vvar  &4200    & 4.50  &  0.05 & 8.5  & 0.232&  -3.2  &   0.12    &-3.66  &          \\  
92043 &F6V     &6431    & 4.08  &  0.04 & 18.0 & 6.141&  -5.3  &   0.30    &-4.898 & 4.74     \\  
95995 &K1V     &5027    & 4.42  & -0.22 & 3.0  & 0.682&        &           &-5.047 & 7.48     \\  
96100 &K0V     &5276    & 4.56  & -0.18 & 8.0: & 0.427&  -5.6  &   2.69    &-4.832 & 3.67     \\  
96441 &F4V     &6954    & 4.04  & -0.03 & 4.3  & 4.147&  -5.4  &   0.39    &-5.234 &          \\  
97944 &K3/K4V  &4687    & 4.40  & -0.20 & 4.0: & 0.770&  -4.7  &   0.49    &-4.582 &          \\  
98959 &G2V     &5735    & 4.46  & -0.22 & 2.0: & 1.000&        &           &-4.857 & 4.06     \\  
99240 &G5IV-V  &5597    & 4.29  &  0.30 & 3.2  & 1.246&  -6.4  &   4.53    &-5.092 & 8.30     \\  
99461 &K2V     &4964    & 4.48  & -0.44 & 1.8  & 0.286&  -6.1  &   8.04    &-4.988 & 6.37     \\  
101955&K5V     &4181    &       &       & 5.6  & 0.331&  -5.0  &   1.20    &-4.972 &          \\  
101997&G8/K0V  &5427    & 4.51  & -0.27 & 2.0: & 0.540&        &           &-4.948 & 5.63     \\  
103389&F7V     &6257    & 4.36  & -0.14 & 13.4 & 2.025&  -4.7  &   0.26    &-4.402 & 0.30     \\  
104214&K5V     &4394    & 4.59  & -0.25 & 2.0: & 0.144&  -5.6  &   6.18    &-4.764 &          \\  
104217&K7V     &4002    & 4.57  & -0.39 & 1.9  & 0.092&  -5.6  &   8.45    &-4.891 &          \\  
105312&G5V     &5479    & 4.45  & -0.36 & 1.6: & 0.690&        &           &-4.988 & 6.37     \\  
106696&K2V     &5053    & 4.62  & -0.18 & 1.9  & 0.306&  -5.1  &   1.56    &-4.702 & 2.01     \\  
107350&G0V     &5952    & 4.44  & -0.07 & 9.7  & 1.090&  -4.4  &   0.24    &-4.48  & 0.53     \\  
107649&G2V     &5912    & 4.44  & -0.01 & 3.0: & 1.258&  -5.7  &   1.40    &-5.02  & 6.98     \\  
108870&K5V     &4629    & 4.36  & -0.06 & 2.0: & 0.215&  -5.5  &   3.66    &-4.851 &          \\  
109378&G0      &5540    & 4.39  &  0.22 & 1.8  & 1.002&        &           &-5.101 &          \\  
109422&F6V     &6339    & 4.15  &  0.08 & 13.6 & 2.801&        &           &-4.823 & 3.53  \\     
110109&G1V     &5870    & 4.45  & -0.19 & 2.2  & 1.144&  -5.6  &   1.32    &-4.918 & 5.09      \\ 
113357&G5V     &5791    & 4.36  &  0.20 & 2.6  & 1.368&  -7.1  &  13.58    &-5.074 & 7.98  \\     
113576&K5/M0V  &3746    & 5.0   & -1.5  & 3.0: & 0.124&        &           &-4.63  &       \\     
114948&F7V     &6249    & 4.31  & -0.21 & 9.0: & 1.867&  -4.4  &   0.18    &-4.434 & 0.38  \\     
116745&K3V     &4750    & 4.51  & -0.30 &      & 0.221&  -5.5  &   3.44    &-4.959 &       \\     
120005&K2      &3769    & 4.71  & -0.40 & 2.9  & 0.086&  -4.1  &   1.01    &-4.42  &            \\
\hline
\label{stellar_parameters}
\end{longtable}
\end{center}
}

\onllongtabL{5}{                                                                                                                       
\begin{center}                                                                                                                         
\label{pacs_aors}                                                                                                                   
\begin{longtable}{lrrrr}
\multicolumn{5}{c}{\tablename\ \thetable{}. PACS AORs.}\\                                                                    
\hline
HIP & PACS & Scan & X-Scan & Time [s]\\                                                                                                   
\hline                                                                                                                                 
\endfirsthead                                                                                                                          
\multicolumn{5}{c}{\tablename\ \thetable{}. Continued.}\\                                                                           
\hline
HIP & PACS & Scan & X-Scan & On-source time [s]\\                                                                                                       
\hline                                                                                                   
\endhead                                                                                                                               
\hline 
\multicolumn{5}{l}{Continued\dots}\\                                                                                                                                
\endfoot                                                                                                                               
\endlastfoot  
171    & 100/160 & 1342212800 & 1342212801 &      900 \\
544    & 100/160 & 1342213512 & 1342213513 &     1440 \\
910    & 100/160 & 1342199875 & 1342199876 &      360 \\
2941   & 100/160 & 1342212844 & 1342212845 &      540 \\
3093   &  70/160 & 1342213242 & 1342213243 &      180 \\
3093   & 100/160 & 1342213244 & 1342213245 &      540 \\
3497   &  70/160 & 1342212704 & 1342212705 &      540 \\
3497   & 100/160 & 1342212706 & 1342212707 &     1440 \\
3821   & 100/160 & 1342213227 & 1342213228 &     1080 \\
3909   & 100/160 & 1342211150 & 1342211151 &      540 \\
4148   &  70/160 & 1342212840 & 1342212841 &      360 \\
4148   & 100/160 & 1342212842 & 1342212843 &     1440 \\
7513   & 100/160 & 1342223326 & 1342223327 &      360 \\
7978   &  70/160 & 1342212838 & 1342212839 &      360 \\
7978   & 100/160 & 1342187141 &            &     4714 \\
7978   & 100/160 & 1342187139 & 1342187140 &      720 \\
8768   &  70/160 & 1342213645 & 1342213646 &      180 \\
8768   & 100/160 & 1342213647 & 1342213648 &     1440 \\
10138  &  70/160 & 1342214175 & 1342214176 &      180 \\
10138  & 100/160 & 1342214177 & 1342214178 &      720 \\
10798  & 100/160 & 1342214185 & 1342214186 &     1440 \\
11452  &  70/160 & 1342223328 & 1342223329 &      180 \\
11452  & 100/160 & 1342223330 & 1342223331 &     1440 \\
11964  &  70/160 & 1342221280 & 1342221281 &      180 \\
11964  & 100/160 & 1342221282 & 1342221283 &     1440 \\
12777  & 100/160 & 1342204215 & 1342204216 &      360 \\
13402  & 100/160 & 1342215731 & 1342215732 &     1440 \\
14954  & 100/160 & 1342216129 & 1342216130 &     1440 \\
15330  & 100/160 & 1342204268 & 1342204269 &      720 \\
15371  &  70/160 & 1342191104 & 1342191105 &     1440 \\
15371  & 100/160 & 1342191102 & 1342191103 &     1440 \\
15799  &  70/160 & 1342216117 & 1342216118 &      180 \\
15799  & 100/160 & 1342216119 & 1342216120 &     1440 \\
16134  & 100/160 & 1342216121 & 1342216122 &     1440 \\
17420  &  70/160 & 1342223592 & 1342223593 &      180 \\
17420  & 100/160 & 1342223594 & 1342223595 &     1440 \\
17439  &  70/160 & 1342222499 & 1342222500 &      540 \\
17439  & 100/160 & 1342222501 & 1342222502 &     1440 \\
19849  & 100/160 & 1342204321 & 1342204322 &      900 \\
19884  &  70/160 & 1342204264 & 1342204265 &      180 \\
19884  & 100/160 & 1342204266 & 1342204267 &     1440 \\
22263  & 100/160 & 1342193112 & 1342193113 &     1440 \\
23311  & 100/160 & 1342193117 & 1342193118 &      360 \\
25110  & 100/160 & 1342219019 & 1342219020 &      720 \\
27887  &  70/160 & 1342203666 & 1342203667 &      180 \\
27887  & 100/160 & 1342203668 & 1342203669 &     1440 \\
28103  & 100/160 & 1342205198 & 1342205199 &      360 \\
28442  &  70/160 & 1342196115 & 1342196116 &      180 \\
28442  & 100/160 & 1342196117 & 1342196118 &     1440 \\
29271  & 100/160 & 1342216043 & 1342216044 &      720 \\
29568  & 100/160 & 1342204431 & 1342204432 &     1440 \\
32439  & 100/160 & 1342196732 & 1342196733 &     1080 \\
32480  &  70/160 & 1342219021 & 1342219022 &      180 \\
32480  & 100/160 & 1342206334 & 1342206335 &     1440 \\
33277  &  70/160 & 1342206326 & 1342206327 &      180 \\
33277  & 100/160 & 1342206328 & 1342206329 &     1260 \\
34017  & 100/160 & 1342205194 & 1342205195 &     1080 \\
34065  & 100/160 & 1342203710 & 1342203711 &      720 \\
35136  & 100/160 & 1342219422 & 1342219423 &      720 \\
36439  & 100/160 & 1342195614 & 1342195615 &     1080 \\
38382  & 100/160 & 1342196123 & 1342196124 &      360 \\
38784  &  70/160 & 1342196753 & 1342196754 &      180 \\
38784  & 100/160 & 1342196755 & 1342196756 &     1440 \\
40693  & 100/160 & 1342196125 & 1342196126 &      900 \\
40843  & 100/160 & 1342196743 & 1342196744 &      540 \\
42430  & 100/160 & 1342198541 & 1342198542 &      360 \\
42438  & 100/160 & 1342196747 & 1342196748 &     1440 \\
43587  & 100/160 & 1342208504 & 1342208505 &      720 \\
43726  & 100/160 & 1342208478 & 1342208479 &     1440 \\
44897  &  70/160 & 1342209375 & 1342209376 &      180 \\
44897  & 100/160 & 1342209377 & 1342209378 &     1440 \\
45333  & 100/160 & 1342206669 & 1342206670 &      540 \\
45617  &  70/160 & 1342206315 & 1342206316 &      180 \\
45617  & 100/160 & 1342206317 & 1342206318 &     1440 \\
46580  &  70/160 & 1342209466 & 1342209467 &      180 \\
46580  & 100/160 & 1342209465 & 1342209468 &     1440 \\
47592  &  70/160 & 1342198545 & 1342198546 &      180 \\
47592  & 100/160 & 1342198547 & 1342198548 &      360 \\
49081  & 100/160 & 1342210618 & 1342210619 &      360 \\
49908  & 100/160 & 1342210610 & 1342210611 &     1440 \\
51459  & 100/160 & 1342210608 & 1342210609 &      360 \\
51502  & 100/160 & 1342197013 & 1342197014 &     1440 \\
53721  & 100/160 & 1342198845 & 1342198846 &      360 \\
54646  &  70/160 & 1342212025 & 1342212026 &      180 \\
54646  & 100/160 & 1342212027 & 1342212028 &     1440 \\
56452  & 100/160 & 1342201166 & 1342201167 &      720 \\
57507  &  70/160 & 1342212828 & 1342212829 &      180 \\
57507  & 100/160 & 1342212830 & 1342212831 &     1440 \\
57939  & 100/160 & 1342212048 & 1342212049 &     1440 \\
58345  & 100/160 & 1342212826 & 1342212827 &     1080 \\
62145  &  70/160 & 1342221293 & 1342221294 &      180 \\
62145  & 100/160 & 1342221295 & 1342221296 &     1440 \\
62207  & 100/160 & 1342212391 & 1342212392 &     1440 \\
62523  & 100/160 & 1342212485 & 1342212486 &     1440 \\
64792  & 100/160 & 1342212812 & 1342212813 &      720 \\
64797  &  70/160 & 1342212808 & 1342212809 &      180 \\
64797  & 100/160 & 1342212810 & 1342212811 &     1440 \\
65026  &  70/160 & 1342198907 & 1342198908 &      180 \\
65026  & 100/160 & 1342198909 & 1342198910 &      720 \\
65721  & 100/160 & 1342213093 & 1342213094 &     1440 \\
67275  & 100/160 & 1342213081 & 1342213082 &      360 \\
67422  &  70/160 & 1342213085 & 1342213086 &      180 \\
67422  & 100/160 & 1342213087 & 1342213088 &     1440 \\
67620  &  70/160 & 1342213844 & 1342213845 &      180 \\
67620  & 100/160 & 1342213846 & 1342213847 &     1080 \\
68184  & 100/160 & 1342213225 & 1342213226 &     1440 \\
68682  &  70/160 & 1342213594 & 1342213595 &      180 \\
68682  & 100/160 & 1342213596 & 1342213597 &     1440 \\
69965  &  70/160 & 1342224194 & 1342224195 &      180 \\
69965  & 100/160 & 1342224196 & 1342224197 &     1080 \\
70319  &  70/160 & 1342213806 & 1342213807 &      180 \\
70319  & 100/160 & 1342213808 & 1342213809 &     1440 \\
70857  &  70/160 & 1342197701 & 1342197702 &      180 \\
70857  & 100/160 & 1342197703 & 1342197704 &     1440 \\
71181  &  70/160 & 1342209500 & 1342209501 &      180 \\
71181  & 100/160 & 1342209502 & 1342209503 &     1440 \\
71681  & 100/160 & 1342224848 & 1342224849 &      360 \\
71908  & 100/160 & 1342205980 & 1342205981 &      360 \\
72567  & 100/160 & 1342213798 & 1342213799 &     1440 \\
72603  &  70/160 & 1342214582 & 1342214583 &      180 \\
72603  & 100/160 & 1342214584 & 1342214585 &     1440 \\
72848  & 100/160 & 1342212768 & 1342212769 &     1440 \\
73100  & 100/160 & 1342209640 & 1342209641 &     1440 \\
73184  & 100/160 & 1342204172 & 1342204173 &      720 \\
73996  & 100/160 & 1342213796 & 1342213797 &      540 \\
77052  & 100/160 & 1342204162 & 1342204163 &      900 \\
78459  & 100/160 & 1342215376 & 1342215377 &      540 \\
78775  &  70/160 & 1342205163 & 1342205164 &      180 \\
78775  & 100/160 & 1342205165 & 1342205166 &     1440 \\
79248  & 100/160 & 1342205996 & 1342205997 &     1440 \\
80725  &  70/160 & 1342205157 & 1342205158 &      180 \\
80725  & 100/160 & 1342205159 & 1342205160 &     1440 \\
82860  & 100/160 & 1342220101 & 1342220102 &     1440 \\
83389  &  70/160 & 1342193509 & 1342193510 &      540 \\
83389  & 100/160 & 1342193511 & 1342193512 &     1440 \\
84862  & 100/160 & 1342193515 & 1342193516 &     1080 \\
85235  & 100/160 & 1342220099 & 1342220100 &     1440 \\
85295  & 100/160 & 1342193057 & 1342193058 &      360 \\
86036  & 100/160 & 1342195413 & 1342195414 &      540 \\
86796  & 100/160 & 1342215572 & 1342215573 &      360 \\
88601  & 100/160 & 1342207023 & 1342207024 &      360 \\
88972  &  70/160 & 1342195696 & 1342195697 &      180 \\
88972  & 100/160 & 1342195698 & 1342195699 &     1080 \\
89042  & 100/160 & 1342207067 & 1342207068 &      720 \\
91009  &  70/160 & 1342196781 & 1342196782 &      180 \\
91009  & 100/160 & 1342196783 & 1342196784 &     1440 \\
92043  &  70/160 & 1342216399 & 1342216400 &      180 \\
92043  & 100/160 & 1342192775 & 1342192776 &      360 \\
95995  &  70/160 & 1342197677 & 1342197678 &      180 \\
95995  & 100/160 & 1342197679 & 1342197680 &     1260 \\
96100  & 100/160 & 1342197771 & 1342197772 &      360 \\
96441  & 100/160 & 1342196779 & 1342196780 &      360 \\
97944  & 100/160 & 1342209045 & 1342209046 &      720 \\
98959  &  70/160 & 1342208851 & 1342208852 &      180 \\
98959  & 100/160 & 1342208853 & 1342208854 &     1440 \\
99240  &  70/160 & 1342187075 & 1342187076 &      288 \\
99240  & 100/160 & 1342195466 & 1342195467 &     1152 \\
99461  & 100/160 & 1342196779 & 1342196780 &      360 \\
101955 &  70/160 & 1342196787 & 1342196788 &      180 \\
101955 & 100/160 & 1342196789 & 1342196790 &     1080 \\
101997 &  70/160 & 1342193526 & 1342193527 &      180 \\
101997 & 100/160 & 1342193528 & 1342193529 &     1440 \\
103389 &  70/160 & 1342193157 & 1342193158 &      180 \\
103389 & 100/160 & 1342193159 & 1342193160 &     1440 \\
104214 & 100/160 & 1342195781 & 1342195782 &      360 \\
105312 &  70/160 & 1342194063 & 1342194064 &      180 \\
105312 & 100/160 & 1342194065 & 1342194066 &     1260 \\
106696 &  70/160 & 1342196105 & 1342196106 &      180 \\
106696 & 100/160 & 1342196107 & 1342196108 &     1260 \\
107350 & 100/160 & 1342195779 & 1342195780 &     1440 \\
107649 &  70/160 & 1342193163 & 1342193164 &      180 \\
107649 & 100/160 & 1342193165 & 1342193166 &     1440 \\
108870 & 100/160 & 1342192760 & 1342192761 &     1440 \\
109378 & 100/160 & 1342211126 & 1342211127 &     1440 \\
109422 & 100/160 & 1342196797 & 1342196798 &      360 \\
110109 & 100/160 & 1342187145 & 1342187146 &     1364 \\
113357 & 100/160 & 1342187255 &            &     1364 \\
113576 & 100/160 & 1342198523 & 1342198524 &      900 \\
114948 & 100/160 & 1342196803 & 1342196804 &     1440 \\
116745 & 100/160 & 1342196805 & 1342196806 &     1440 \\
120005 & 100/160 & 1342209360 & 1342209361 &     1440 \\
\noalign{\smallskip}\hline\noalign{\smallskip}
\end{longtable}
\end{center}
}

\onllongtabL{10}{                                                                                                                       
\begin{center}                                                                                                                         
\label{longtable:positions}                                                                                                                   
\begin{longtable}{rccc}                                                                                                         
\multicolumn{4}{c}{\tablename\ \thetable{}. Optical and PACS100 equatorial positions (J2000) and the positional offset.}\\
\hline                                                                                                                                 
HIP &  ICRS(2000)   &PACS100        &  Offset(arcsec)  \\  
\hline                                                                                                                                 
\endfirsthead                                                                                                                          
\multicolumn{4}{c}{\tablename\ \thetable{}. Continued.}\\                                                                           
\hline                                                                                                                                 
HIP &  ICRS(2000)   &PACS100                                &  Offset \\         
\hline                                                                                                   
\endhead                                                                                                                               
\hline 
\multicolumn{4}{l}{Continued\dots}\\                                                                                                                                
\endfoot                                                                                                                               
\endlastfoot                                                                                                                           
171    & 00 02 10.16   +27 04 56.1 &00 02 10.57   +27 04 56.0 & 5.5  \\ 
544    & 00 06 36.78   +29 01 17.4 &00 06 36.79   +29 01 15.8 & 1.6  \\ 
910    & 00 11 15.86   -15 28 04.7 &00 11 15.88   -15 28 03.4 & 1.3  \\ 
2941   & 00 37 20.70   -24 46 02.2 &00 37 20.54   -24 46 03.9 & 2.8  \\ 
3093   & 00 39 21.81   +21 15 01.7 &00 39 21.84   +21 14 58.9 & 2.8  \\ 
3497   & 00 44 39.27   -65 38 58.3 &00 44 39.18   -65 38 58.8 & 0.7  \\ 
3821   & 00 49 06.29   +57 48 54.7 &00 49 06.51   +57 48 54.3 & 1.8  \\ 
3909   & 00 50 07.59   -10 38 39.6 &00 50 07.60   -10 38 40.8 & 1.2  \\ 
4148   & 00 53 01.14   -30 21 24.9 &00 53 01.33   -30 21 27.7 & 3.7  \\
7513   & 01 36 47.84   +41 24 19.7 &01 36 47.76   +41 24 19.4 & 0.9  \\ 
7978   & 01 42 29.32   -53 44 27.0 &01 42 29.52   -53 44 26.2 & 1.9  \\ 
8768   & 01 52 49.17   -22 26 05.5 &01 52 49.23   -22 26 06.7 & 1.5  \\ 
10138  & 02 10 25.93   -50 49 25.4 &02 10 26.00   -50 49 25.0 & 0.8  \\ 
10798  & 02 18 58.50   -25 56 44.5 &02 18 58.48   -25 56 41.7 & 2.8  \\ 
11452  & 02 27 45.86   +04 25 55.7 &02 27 45.85   +04 25 54.7 & 1.0  \\ 
11964  & 02 34 22.57   -43 47 46.9 &02 34 22.48   -43 47 44.6 & 2.5  \\ 
12777  & 02 44 11.99   +49 13 42.4 &02 44 12.05   +49 13 45.1 & 2.8  \\ 
13402  & 02 52 32.13   -12 46 11.0 &02 52 32.18   -12 46 10.2 & 1.1  \\ 
14954  & 03 12 46.44   -01 11 46.0 &03 12 46.44   -01 11 50.1 & 4.1  \\ 
15330  & 03 17 46.16   -62 34 31.2 &03 17 45.79   -62 34 29.3 & 3.2  \\ 
15371  & 03 18 12.82   -62 30 22.9 &03 18 13.12   -62 30 24.4 & 2.6  \\ 
15799  & 03 23 35.26   -40 04 35.0 &03 23 35.31   -40 04 35.4 & 0.7  \\ 
16134  & 03 27 52.41   -19 48 16.1 &03 27 52.31   -19 48 17.8 & 2.2  \\
17420  & 03 43 55.34   -19 06 39.2 &03 43 55.40   -19 06 39.6 & 0.9  \\ 
17439  & 03 44 09.17   -38 16 54.4 &03 44 09.26   -38 16 54.6 & 1.1  \\ 
19849  & 04 15 16.32   -07 39 10.3 &04 15 16.21   -07 39 11.1 & 1.8  \\ 
19884  & 04 15 56.90   -53 18 35.3 &04 15 57.00   -53 18 35.7 & 1.0  \\ 
22263  & 04 47 36.29   -16 56 04.0 &04 47 36.42   -16 56 04.9 & 2.1  \\ 
23311  & 05 00 49.00   -05 45 13.2 &05 00 48.73   -05 45 13.7 & 4.1  \\ 
25110  & 05 22 33.53   +79 13 52.1 &05 22 34.22    79 13 52.2 & 1.9  \\ 
27887  & 05 54 04.24   -60 01 24.5 &05 54 03.72   -60 01 24.0 & 3.9  \\ 
28103  & 05 56 24.29   -14 10 03.7 &05 56 24.31   -14 10 01.7 & 2.0  \\ 
28442  & 06 00 19.52   -31 01 43.4 &06 00 19.40   -31 01 53.8 &10.5  \\ 
29271  & 06 10 14.47   -74 45 11.0 &06 10 14.53   -74 45 11.1 & 0.3  \\ 
29568  & 06 13 45.29   -23 51 43.0 &06 13 45.38   -23 51 40.2 & 3.1  \\ 
32439  & 06 46 14.15   +79 33 53.3 &06 46 14.44   +79 33 51.0 & 2.4  \\ 
32480  & 06 46 44.34   +43 34 38.7 &06 46 44.23   +43 34 38.8 & 1.2  \\ 
33277  & 06 55 18.67   +25 22 32.5 &06 55 18.68   +25 22 32.7 & 0.2  \\ 
34017  & 07 03 30.46   +29 20 13.5 &07 03 30.49   +29 20 15.6 & 2.1  \\ 
34065  & 07 03 57.32   -43 36 28.9 &07 03 56.77   -43 36 28.4 & 6.0  \\ 
35136  & 07 15 50.14   +47 14 23.9 &07 15 50.18   +47 14 24.5 & 0.7  \\ 
36439  & 07 29 55.96   +49 40 20.9 &07 29 55.88   +49 40 19.6 & 1.5  \\ 
38382  & 07 51 46.30   -13 53 52.9 &07 51 46.30   -13 53 54.9 & 2.0  \\ 
38784  & 07 56 17.23   +80 15 56.0 &07 56 17.50   +80 15 56.4 & 0.8  \\ 
40693  & 08 18 23.95   -12 37 55.8 &08 18 24.10   -12 37 57.9 & 3.0  \\ 
40843  & 08 20 03.86   +27 13 03.8 &08 20 03.88   +27 12 56.6 & 7.1  \\ 
42430  & 08 39 07.90   -22 39 42.8 &08 39 08.10   -22 39 45.3 & 3.7  \\ 
42438  & 08 39 11.70   +65 01 15.3 &08 39 11.74   +65 01 15.3 & 0.3  \\ 
43587  & 08 52 35.81   +28 19 50.9 &08 52 35.89   +28 19 52.0 & 1.5  \\ 
43726  & 08 54 17.95   -05 26 04.1 &08 54 17.91   -05 26 04.3 & 0.6  \\ 
44897  & 09 08 51.07   +33 52 56.0 &09 08 50.84   +33 52 55.7 & 2.9  \\ 
45333  & 09 14 20.54   +61 25 23.9 &09 14 20.92   +61 25 22.8 & 2.9  \\ 
45617  & 09 17 53.46   +28 33 37.9 &09 17 53.35   +28 33 38.6 & 1.6  \\ 
46580  & 09 29 54.82   +05 39 18.5 &09 29 54.82   +05 39 19.9 & 1.4  \\ 
47592  & 09 42 14.42   -23 54 56.0 &09 42 14.26   -23 54 54.8 & 2.5  \\ 
49081  & 10 01 00.66   +31 55 25.2 &10 01 00.68   +31 55 23.7 & 1.5  \\ 
49908  & 10 11 22.14   +49 27 15.3 &10 11 21.87   +49 27 17.2 & 3.2  \\ 
51459  & 10 30 37.58   +55 58 49.9 &10 30 37.24   +55 58 50.9 & 3.0  \\ 
51502  & 10 31 04.66   +82 33 30.9 &10 31 05.98   +82 33 32.5 & 3.0  \\ 
53721  & 10 59 27.97   +40 25 48.9 &10 59 28.01   +40 25 48.9 & 0.5  \\ 
54646  & 11 11 05.17   +30 26 45.7 &11 11 05.58   +30 26 47.6 & 5.6  \\ 
56452  & 11 34 29.49   -32 49 52.8 &11 34 29.45   -32 49 55.1 & 2.4  \\ 
57507  & 11 47 15.81   -30 17 11.4 &11 47 16.04   -30 17 11.2 & 3.0  \\ 
57939  & 11 52 58.77   +37 43 07.2 &11 52 58.36   +37 43 08.9 & 5.2  \\ 
58345  & 11 57 56.21   -27 42 25.4 &11 57 56.37   -27 42 24.2 & 2.4  \\ 
62145  & 12 44 14.55   +51 45 33.5 &12 44 14.46   +51 45 34.1 & 1.0  \\ 
62207  & 12 44 59.41   +39 16 44.1 &12 44 59.33   +39 16 44.3 & 1.0  \\ 
62523  & 12 48 47.05   +24 50 24.8 &12 48 47.17   +24 50 26.2 & 2.2  \\ 
64792  & 13 16 46.52   +09 25 27.0 &13 16 46.63   +09 25 28.6 & 2.3  \\ 
64797  & 13 16 51.05   +17 01 01.9 &13 16 51.11   +17 01 02.5 & 1.0  \\ 
65026  & 13 19 45.66   +47 46 40.9 &13 19 45.66   +47 46 38.3 & 2.6  \\ 
65721  & 13 28 25.81   +13 46 43.6 &13 28 25.72   +13 46 45.3 & 2.1  \\ 
67275  & 13 47 15.74   +17 27 24.9 &13 47 15.95   +17 27 24.2 & 3.1  \\ 
67422  & 13 49 04.00   +26 58 47.7 &13 49 04.07   +26 58 45.9 & 2.0  \\ 
67620  & 13 51 20.33   -24 23 25.3 &13 51 20.10   -24 23 22.8 & 4.0  \\ 
68184  & 13 57 32.06   +61 29 34.3 &13 57 32.19   +61 29 33.7 & 1.1  \\ 
68682  & 14 03 32.35   +10 47 12.4 &14 03 32.37   +10 47 12.3 & 0.3  \\ 
69965  & 14 19 00.90   -25 48 55.5 &14 19 00.88   -25 48 56.1 & 0.7  \\ 
70319  & 14 23 15.28   +01 14 29.6 &14 23 15.29   +01 14 31.5 & 1.9  \\ 
70857  & 14 29 22.30   +80 48 35.5 &14 29 23.25   +80 48 33.6 & 3.0  \\ 
71181  & 14 33 28.87   +52 54 31.6 &14 33 28.81   +52 54 32.9 & 1.4  \\ 
71681  & 14 39 35.08   -60 50 13.8 &14 39 35.54   -60 50 09.7 & 5.3  \\ 
71683  & 14 39 36.50   -60 50 02.3 &14 39 36.05   -60 50 04.9 & 4.2  \\ 
71908  & 14 42 30.42   -64 58 30.5 &14 42 30.22   -64 58 27.3 & 3.4  \\ 
72567  & 14 50 15.81   +23 54 42.6 &14 50 15.79   +23 54 44.4 & 1.8  \\ 
72603  & 14 50 41.18   -15 50 50.1 &14 50 41.06   -15 50 49.9 & 1.7  \\ 
72848  & 14 53 23.77   +19 09 10.1 &14 53 23.94   +19 09 09.0 & 2.6  \\ 
73100  & 14 56 23.04   +49 37 42.4 &14 56 23.14   +49 37 42.0 & 1.1  \\ 
73182  & 14 57 26.54   -21 24 41.5 &14 57 26.42   -21 24 37.0 & 4.8  \\ 
73184  & 14 57 28.00   -21 24 55.7 &14 57 27.86   -21 24 53.0 & 3.3  \\ 
73996  & 15 07 18.07   +24 52 09.1 &15 07 18.20   +24 52 10.3 & 2.1  \\ 
77052  & 15 44 01.82   +02 30 54.6 &15 44 01.98   +02 30 54.8 & 2.4  \\ 
78459  & 16 01 02.66   +33 18 12.6 &16 01 02.60   +33 18 14.3 & 1.9  \\ 
78775  & 16 04 56.79   +39 09 23.4 &16 04 56.75   +39 09 22.0 & 1.5  \\ 
79248  & 16 10 24.31   +43 49 03.5 &16 10 24.34   +43 49 03.8 & 0.4  \\ 
80725  & 16 28 52.67   +18 24 50.6 &16 28 52.58   +18 24 47.6 & 3.3  \\ 
82860  & 16 56 01.69   +65 08 05.3 &16 56 02.03   +65 08 07.6 & 3.1  \\ 
83389  & 17 02 36.41   +47 04 54.8 &17 02 36.38   +47 04 55.3 & 0.6  \\ 
84862  & 17 20 39.57   +32 28 03.9 &17 20 39.42   +32 28 04.1 & 1.9  \\ 
85235  & 17 25 00.10   +67 18 24.1 &17 25 00.20   +67 18 26.0 & 2.0  \\ 
85295  & 17 25 45.23   +02 06 41.1 &17 25 45.23   +02 06 41.5 & 0.4  \\ 
86036  & 17 34 59.59   +61 52 28.4 &17 34 59.30   +61 52 28.5 & 2.1  \\ 
86796  & 17 44 08.70   -51 50 02.6 &17 44 08.66   -51 50 01.9 & 0.8  \\ 
88601  & 18 05 27.29   +02 30 00.4 &18 05 27.27   +02 29 59.4 & 1.0  \\ 
88972  & 18 09 37.42   +38 27 28.0 &18 09 37.36   +38 27 27.4 & 0.9  \\ 
89042  & 18 10 26.16   -62 00 07.9 &18 10 26.10   -62 00 08.5 & 0.7  \\ 
91009  & 18 33 55.77   +51 43 08.9 &18 33 55.87   +51 43 09.3 & 1.0  \\ 
92043  & 18 45 39.73   +20 32 46.7 &18 45 39.62   +20 32 48.3 & 2.2  \\ 
95995  & 19 31 07.97   +58 35 09.6 &19 31 07.92   +58 35 12.2 & 2.6  \\ 
96100  & 19 32 21.59   +69 39 40.2 &19 32 21.01   +69 39 40.2 & 3.0  \\ 
96441  & 19 36 26.53   +50 13 16.0 &19 36 26.58   +50 13 16.4 & 0.6  \\ 
97944  & 19 54 17.75   -23 56 27.9 &19 54 17.96   -23 56 27.5 & 2.9  \\ 
98959  & 20 05 32.76   -67 19 15.2 &20 05 32.60   -67 19 16.5 & 1.6  \\ 
99240  & 20 08 43.61   -66 10 55.4 &20 08 43.53   -66 10 58.1 & 2.7  \\ 
99461  & 20 11 11.94   -36 06 04.4 &20 11 12.28   -36 06 03.5 & 4.2  \\
101955 & 20 39 37.71   +04 58 19.3 &20 39 37.69   +04 58 20.1 & 0.9  \\ 
101997 & 20 40 11.76   -23 46 25.9 &20 40 11.53   -23 46 24.9 & 3.3  \\ 
103389 & 20 56 47.33   -26 17 47.0 &20 56 47.46   -26 17 45.5 & 2.3  \\ 
104214 & 21 06 53.94   +38 44 57.9 &21 06 54.11   +38 44 58.2 & 2.0  \\ 
104217 & 21 06 55.26   +38 44 31.4 &21 06 55.44   +38 44 31.5 & 2.1  \\ 
105312 & 21 19 45.62   -26 21 10.4 &21 19 45.52   -26 21 07.1 & 3.6  \\ 
106696 & 21 36 41.24   -50 50 43.4 &21 36 41.41   -50 50 43.5 & 1.6  \\ 
107350 & 21 44 31.33   +14 46 19.0 &21 44 31.27   +14 46 19.4 & 1.0  \\ 
107649 & 21 48 15.75   -47 18 13.0 &21 48 15.88   -47 18 11.4 & 2.1  \\ 
108870 & 22 03 21.66   -56 47 09.5 &22 03 21.72   -56 47 09.9 & 0.6  \\ 
109378 & 22 09 29.87   -07 32 55.2 &22 09 30.07   -07 32 54.1 & 3.2  \\ 
109422 & 22 10 08.78   -32 32 54.3 &22 10 08.87   -32 32 53.3 & 1.5  \\ 
110109 & 22 18 15.62   -53 37 37.5 &22 18 15.94   -53 37 35.8 & 3.3  \\ 
113357 & 22 57 27.98   +20 46 07.8 &22 57 28.27   +20 46 05.4 & 4.7  \\ 
113576 & 23 00 16.12   -22 31 27.6 &23 00 16.21   -22 31 26.8 & 1.5  \\ 
114948 & 23 16 57.69   -62 00 04.3 &23 16 57.44   -62 00 05.8 & 2.3  \\ 
116745 & 23 39 37.39   -72 43 19.8 &23 39 37.23   -72 43 19.5 & 0.8  \\ 
120005 & 09 14 24.70   +52 41 11.0 &09 14 24.64   +52 41 11.5 & 0.7  \\ 
\noalign{\smallskip}\hline\noalign{\smallskip}
\end{longtable}
\end{center}
}

\onllongtabL{12}{                                                                                                                       
\begin{center}                                                                                                                         
\label{pacs_fluxes}                                                                                                                   
\begin{scriptsize}                                                                                                                     
\begin{longtable}{llrrrrrrrrrrr}                                                                                                         
\multicolumn{13}{c}{\tablename\ \thetable{}. Non excess sources. PACS fluxes, photospheric 
predictions, significance, upper limits 
of the fractional luminosity, MIPS 70 $\mu$m fluxes.}\\    
\hline                                                                                                                                 
HIP     &  SpT    &PACS70     &  S70 &$\chi_{70}$& PACS100         & S100 &$\chi_{100}$& PACS160    &S160  & $\chi_{160}$& L$_{\rm d}$/L$_\star$& MIPS70    \\  
        &         &(mJy)      &(mJy) &          & (mJy)           & (mJy)&           & (mJy)      &(mJy) &            &                   &(mJy)     \\
\hline                                                                                                                                 
\endfirsthead                                                                                                                          
\multicolumn{13}{c}{\tablename\ \thetable{}. Continued.}                                                                              \\         
\hline                                                                                                        
HIP     &  SpT    &PACS70     &  S70 &$\chi_{70}$& PACS100         & S100 &$\chi_{100}$ & PACS160   &S160  & $\chi_{160}$& L$_{\rm d}$/L$_\star$& MIPS70    \\ 
        &         &(mJy)      &(mJy) &          & (mJy)           & (mJy)&           & (mJy)      &(mJy) &            &                   &(mJy)     \\ 
\hline                                                                                                   
\endhead                                                                                                                               
\hline  
\multicolumn{13}{l}{Continued\dots}\\                                                                                                                               
\endfoot                                                                                                                               
\endlastfoot 
   910 & F5V       &                  &  29.51$\pm$0.19&        & 17.66$\pm$ 1.38&  14.46$\pm$0.09&  2.32 & $<$7.5          &  5.61$\pm$ 0.04&     & 8.9e-07&37.50$\pm$4.48  \\
  2941 & K1V+...   &                  &  24.09$\pm$0.43&        & 11.20$\pm$ 1.82&  11.80$\pm$0.21& -0.33 & $<$5.7          &  4.61$\pm$ 0.08&     & 2.0e-06&25.80$\pm$11.93 \\
  3093 & K0V       &   21.93 $\pm$1.69&  22.50$\pm$0.34& -0.33  &  8.21$\pm$ 1.23&  11.02$\pm$0.17& -2.26 & 3.76$\pm$ 3.43&  4.31$\pm$ 0.07&-0.16& 1.7e-06&14.90$\pm$5.69  \\
  3497 & G3V       &    7.97 $\pm$0.95&   8.72$\pm$0.10& -0.78  &  5.23$\pm$ 1.01&   4.27$\pm$0.05&  0.95 & $<$4.8          &  1.67$\pm$ 0.02&     & 2.8e-06& 5.20$\pm$4.41  \\
  3821 & G0V SB    &                  & 131.05$\pm$2.06&        & 60.80$\pm$ 2.04&  64.21$\pm$1.01& -1.50 &15.75$\pm$ 2.66& 25.08$\pm$ 0.39&-3.47& 3.3e-07&122.3$\pm$10.74 \\
  3909 & F7IV-V    &                  &  21.73$\pm$0.41&        & 11.86$\pm$ 1.28&  10.65$\pm$0.20&  0.93 & $<$6.3          &  4.16$\pm$ 0.08&     & 1.1e-06&24.30$\pm$3.50  \\
  7513 & F8V       &                  &  60.22$\pm$0.80&        & 32.78$\pm$ 1.90&  29.51$\pm$0.39&  1.69 &16.49$\pm$ 2.73& 11.53$\pm$ 0.15& 1.82& 6.0e-07&55.70$\pm$6.33  \\
  8768 & K5/M0V    &    4.33 $\pm$1.58&   8.73$\pm$0.12& -2.77  &  4.21$\pm$ 1.07&   4.28$\pm$0.06& -0.07 & 3.50$\pm$ 1.47&  1.67$\pm$ 0.02& 1.25& 9.4e-06&24.70$\pm$7.98  \\
 10138 & K0V       &   14.28 $\pm$1.70&  18.86$\pm$0.16& -2.68  &  7.07$\pm$ 1.31&   9.24$\pm$0.08& -1.65 & 3.78$\pm$ 2.49&  3.61$\pm$ 0.03& 0.07& 2.2e-06&6.900$\pm$6.82  \\
 10798 & G8V       &                  &  13.07$\pm$0.20&        &  6.22$\pm$ 0.89&   6.40$\pm$0.10& -0.20 & $<$4.2          &  2.50$\pm$ 0.04&     & 1.9e-06&14.50$\pm$2.78  \\
 11452 & K7V:      &    5.26 $\pm$1.54&   8.97$\pm$0.17& -2.40  &  3.96$\pm$ 0.78&   4.40$\pm$0.09& -0.56 & $<$3.9          &  1.72$\pm$ 0.03&     & 6.4e-06&                \\
 11964 & K7V SB    &    9.61 $\pm$1.53&  11.60$\pm$0.32& -1.28  &  3.76$\pm$ 0.93&   5.68$\pm$0.16& -2.04 & $<$5.0        &  2.22$\pm$ 0.06&     & 6.5e-06& 9.60$\pm$8.82  \\
 12777 & F7V       &                  &  54.17$\pm$1.34&        & 24.90$\pm$ 2.00&  26.54$\pm$0.66& -0.78 &13.00$\pm$ 3.04& 10.37$\pm$ 0.26& 0.86& 6.5e-07&54.00$\pm$6.43  \\
 15330 & G2V       &                  &  21.56$\pm$0.25&        &  9.18$\pm$ 1.53&  10.57$\pm$0.12& -0.91 & $<$6.3        &  4.13$\pm$ 0.05&     & 1.7e-06&30.00$\pm$4.66  \\
 15799 & K0V       &    9.96 $\pm$1.79&  10.33$\pm$0.31& -0.20  &  4.87$\pm$ 0.96&   5.06$\pm$0.15& -0.20 & $<$3.8        &  1.98$\pm$ 0.06&     & 3.1e-06&12.70$\pm$4.68  \\
 16134 & K5V       &                  &   9.22$\pm$0.18&        &  4.40$\pm$ 0.66&   4.52$\pm$0.09& -0.18 & $<$5.1        &  1.77$\pm$ 0.03&     & 4.9e-06&10.50$\pm$4.46  \\
 19849 & K1V       &                  &  89.30$\pm$0.97&        & 43.12$\pm$ 1.54&  43.76$\pm$0.47& -0.40 & 9.89$\pm$ 2.04& 17.09$\pm$ 0.19&-3.51& 5.6e-07&82.60$\pm$6.68  \\
 19884 & K5V       &    4.60 $\pm$1.60&   9.14$\pm$0.17& -2.81  &  4.35$\pm$ 0.80&   4.48$\pm$0.08& -0.16 & $<$5.1        &  1.75$\pm$ 0.03&     & 4.3e-06&12.10$\pm$3.30  \\
 23311 & K3V       &                  &  25.67$\pm$0.36&        & 12.43$\pm$ 1.54&  12.58$\pm$0.17& -0.10 & 4.86$\pm$ 2.12&  4.91$\pm$ 0.07&-0.02& 2.4e-06&25.70$\pm$5.20  \\
 25110 & F6V       &                  &  22.26$\pm$0.21&        & 12.02$\pm$ 2.26&  10.91$\pm$0.10&  0.49 & 7.30$\pm$ 2.72&  4.26$\pm$ 0.04& 1.12& 1.6e-06&32.60$\pm$5.46  \\
 28442 & K3/K4V    &                  &   9.41$\pm$0.15&        &  3.45$\pm$ 0.61&   4.61$\pm$0.07& -1.90 & $<$12.6       &  1.80$\pm$ 0.03&     & 3.5e-06& 7.50$\pm$3.93  \\
 32439 & F8V       &                  &  17.75$\pm$0.20&        &  7.65$\pm$ 0.94&   8.70$\pm$0.10& -1.11 & 4.67$\pm$ 1.90&  3.40$\pm$ 0.04& 0.67& 9.9e-07&15.80$\pm$2.72  \\
 33277 & G0V       &   15.96 $\pm$1.72&  15.88$\pm$0.29&  0.05  &  7.11$\pm$ 1.00&   7.78$\pm$0.14& -0.66 & 4.85$\pm$ 1.78&  3.04$\pm$ 0.06& 1.02& 1.4e-06&13.10$\pm$10.14 \\
 34017 & G4V       &                  &  13.34$\pm$0.20&        &  7.90$\pm$ 0.68&   6.53$\pm$0.10&  2.01 & 1.79$\pm$ 1.78&  2.55$\pm$ 0.04&-0.43& 1.1e-06&14.40$\pm$4.018 \\
 34065 & G3V...    &                  &  20.25$\pm$0.49&        &  8.35$\pm$ 1.25&   9.92$\pm$0.24& -1.23 & 6.29$\pm$ 2.52&  3.88$\pm$ 0.09& 0.95& 1.4e-06&20.50$\pm$2.77  \\
 35136 & G0V       &                  &  19.16$\pm$0.21&        & 10.60$\pm$ 1.32&   9.39$\pm$0.10&  0.91 & $<$5.9        &  3.67$\pm$ 0.04&     & 1.5e-06&28.90$\pm$4.63  \\
 36439 & F6V       &                  &  17.49$\pm$0.24&        &  8.75$\pm$ 1.42&   8.57$\pm$0.12&  0.13 & $<$5.3        &  3.35$\pm$ 0.05&     & 1.4e-06&14.40$\pm$2.96  \\
 38382 & G2V       &                  &  28.89$\pm$0.45&        & 15.01$\pm$ 1.95&  14.15$\pm$0.22&  0.44 & $<$7.2        &  5.53$\pm$ 0.09&     & 1.5e-06&13.40$\pm$4.30  \\
 40693 & K0V       &                  &  19.84$\pm$0.14&        &  9.33$\pm$ 0.76&   8.95$\pm$0.07&  0.50 & $<$6.0        &  3.50$\pm$ 0.03&     & 1.1e-06&15.10$\pm$2.42  \\
 42430 & G3/G5V    &                  &  35.55$\pm$1.36&        & 15.49$\pm$ 1.28&  17.42$\pm$0.67& -1.33 & $<$10.2       &  6.80$\pm$ 0.26&     & 9.5e-07&31.10$\pm$6.07  \\
 43587 & G8V       &                  &  19.66$\pm$0.22&        &  8.59$\pm$ 0.90&   9.64$\pm$0.11& -1.16 & 4.29$\pm$ 1.93&  3.76$\pm$ 0.04& 0.27& 1.4e-06&19.80$\pm$4.41  \\
 44897 & F9V       &   10.20 $\pm$1.60&  12.36$\pm$0.15& -1.34  &  5.84$\pm$ 0.85&   6.06$\pm$0.08& -0.26 & $<$3.6        &  2.37$\pm$ 0.03&     & 1.4e-06&18.70$\pm$4.00  \\
 45333 & F9V       &                  &  27.09$\pm$0.27&        & 11.49$\pm$ 1.36&  13.27$\pm$0.13& -1.31 & 6.92$\pm$ 2.21&  5.18$\pm$ 0.05& 0.79& 1.1e-06&23.90$\pm$5.25  \\
 45617 & K3V       &    9.81 $\pm$1.67&  10.45$\pm$0.25& -0.38  &  3.96$\pm$ 0.68&   5.12$\pm$0.12& -1.68 & $<$4.8        &  2.00$\pm$ 0.05&     & 2.9e-06&25.00$\pm$10.40 \\
 46580 & K3V       &   11.97 $\pm$1.76&   9.73$\pm$0.22&  1.26  &  6.90$\pm$ 0.83&   4.77$\pm$0.11&  2.54 & $<$5.7        &  1.86$\pm$ 0.04&     & 3.3e-06&                \\
 47592 & G0V       &   29.61 $\pm$1.82&  29.02$\pm$0.63&  0.31  & 11.02$\pm$ 1.86&  14.22$\pm$0.31& -1.69 & $<$8.1        &  5.55$\pm$ 0.12&     & 1.2e-06&14.60$\pm$11.44 \\
 49081 & G1V       &                  &  24.19$\pm$0.36&        &  8.68$\pm$ 1.26&  11.85$\pm$0.18& -2.49 & $<$9.0        &  4.63$\pm$ 0.07&     & 1.2e-06&                \\
 53721 & G0V       &                  &  30.26$\pm$0.50&        & 13.52$\pm$ 1.22&  14.83$\pm$0.25& -1.05 & $<$8.4        &  5.79$\pm$ 0.10&     & 8.7e-07&31.40$\pm$4.17  \\
 54646 & K8V       &   10.54 $\pm$1.53&  11.83$\pm$0.63& -0.78  &  5.03$\pm$ 0.88&   5.80$\pm$0.31& -0.82 & $<$4.2        &  2.26$\pm$ 0.12&     & 6.0e-06&18.00$\pm$10.37 \\
 56452 & K0V       &                  &  21.47$\pm$0.33&        & 11.00$\pm$ 1.11&  10.52$\pm$0.16&  0.43 & 7.21$\pm$ 1.97&  4.11$\pm$ 0.06& 1.57& 1.6e-06&23.20$\pm$2.95  \\
 57507 & G5V       &   12.77 $\pm$1.73&   9.92$\pm$0.17&  1.64  &  5.72$\pm$ 0.82&   4.86$\pm$0.08&  1.05 & $<$4.2        &  1.90$\pm$ 0.03&     & 2.1e-06&11.90$\pm$3.59  \\
 57939 & G8Vp      &                  &  15.66$\pm$0.15&        &  6.71$\pm$ 0.63&   7.67$\pm$0.07& -1.52 & $<$4.5        &  3.00$\pm$ 0.03&     & 1.4e-06&10.70$\pm$2.31  \\
 58345 & K4V       &                  &  16.90$\pm$0.32&        &  7.31$\pm$ 1.12&   8.28$\pm$0.16& -0.86 & $<$8.4        &  3.23$\pm$ 0.06&     & 3.2e-06& 9.00$\pm$8.02  \\
 62145 & K3V       &    7.90 $\pm$1.61&  10.17$\pm$0.19& -1.40  &  3.73$\pm$ 0.69&   4.98$\pm$0.09& -1.80 & $<$5.4        &  1.95$\pm$ 0.04&     & 2.4e-06&28.90$\pm$20.09 \\
 62523 & G7V       &                  &  11.28$\pm$0.12&        &  5.26$\pm$ 0.76&   5.53$\pm$0.06& -0.35 & 4.24$\pm$ 1.91&  2.16$\pm$ 0.02& 1.09& 1.7e-06&13.50$\pm$3.62  \\
 64792 & G0Vs      &                  &  24.05$\pm$0.64&        & 12.28$\pm$ 1.38&  11.79$\pm$0.31&  0.35 & $<$7.5        &  4.60$\pm$ 0.12&     & 1.1e-06&19.10$\pm$5.75  \\
 64797 & K2V       &   14.68 $\pm$1.79&  15.73$\pm$0.54& -0.56  &  7.77$\pm$ 0.95&   7.71$\pm$0.26&  0.06 & $<$3.9        &  3.01$\pm$ 0.10&     & 2.2e-06&46.80$\pm$16.50 \\
 65026 & K0        &   18.68 $\pm$1.66&  17.82$\pm$0.40&  0.50  &  9.02$\pm$ 0.96&   8.73$\pm$0.20&  0.29 & $<$5.4        &  3.41$\pm$ 0.08&     & 4.5e-06&25.10$\pm$1.01  \\
 67275 & F7V       &                  &  38.07$\pm$0.80&        & 14.95$\pm$ 1.08&  18.15$\pm$0.39& -2.78 & 6.24$\pm$ 2.51&  7.09$\pm$ 0.15&-0.34& 4.9e-07&33.20$\pm$6.78  \\
 67422 & K2        &   11.78 $\pm$1.61&  13.53$\pm$0.45& -1.05  &  5.68$\pm$ 0.62&   6.63$\pm$0.22& -1.44 & $<$6.0        &  2.59$\pm$ 0.09&     & 1.9e-06&31.10$\pm$10.41 \\
 67620 & G5V       &    9.61 $\pm$1.82&  10.60$\pm$0.44& -0.53  &  6.88$\pm$ 1.03&   5.19$\pm$0.22&  1.61 & 4.41$\pm$ 1.81&  2.03$\pm$ 0.08& 1.31& 2.3e-06& 8.60$\pm$3.45  \\
 68184 & K3V       &                  &  19.58$\pm$0.20&        &  8.75$\pm$ 1.25&   9.60$\pm$0.10& -0.68 & $<$4.5         &  3.75$\pm$ 0.04&     & 2.6e-06&18.10$\pm$6.71  \\
 68682 & G8V       &   14.90 $\pm$2.03&  13.69$\pm$0.38&  0.59  &  6.63$\pm$ 0.68&   6.71$\pm$0.19& -0.11 & 3.99$\pm$ 1.37&  2.62$\pm$ 0.07& 1.00& 1.3e-06&16.10$\pm$14.04 \\
 69965 & F7Vw      &   11.72 $\pm$1.78&  13.45$\pm$0.32& -0.96  &  5.36$\pm$ 0.98&   6.59$\pm$0.16& -1.24 & $<$5.1        &  2.57$\pm$ 0.06&     & 1.4e-06&19.50$\pm$9.29  \\
 70319 & G1V       &   13.35 $\pm$1.77&  11.40$\pm$0.08&  1.10  &  3.80$\pm$ 1.09&   5.59$\pm$0.04& -1.63 & $<$5.1        &  2.18$\pm$ 0.02&     & 2.3e-06&11.60$\pm$6.35  \\
 70857 & G5        &    8.87 $\pm$2.34&   8.68$\pm$0.25&  0.08  &  4.90$\pm$ 0.63&   4.25$\pm$0.12&  1.01 & $<$6.9        &  1.66$\pm$ 0.05&     & 2.1e-06& 6.50$\pm$4.62  \\
 71681 & K1V       &                  &1510.00$\pm$8.00&        &   670$\pm$41.33& 740.00$\pm$9.00& -1.65 &  211$\pm$61.63&289.00$\pm$ 3.00&-1.26& 8.7e-07&1012.$\pm$667.78\\
 71683 & G2V       &                  &3393.00$\pm$8.00&        &  1339$\pm$78.21&1663.00$\pm$9.00& -4.12 &  479$\pm$67.02&649.00$\pm$ 3.00&-2.53& 5.2e-07&3395.$\pm$702.46\\
 72567 & G2V       &                  &  13.40$\pm$0.17&        &  5.59$\pm$ 0.95&   6.56$\pm$0.08& -1.01 & $<$3.9        &  2.56$\pm$ 0.03&     & 1.4e-06&10.80$\pm$2.89  \\
 72603 & F3V       &   18.81 $\pm$1.00&  18.27$\pm$0.16&  0.53  &  7.49$\pm$ 0.69&   8.95$\pm$0.08& -2.10 & 3.72$\pm$ 0.84&  3.50$\pm$ 0.03& 0.26& 5.8e-07&                \\ 
 73182 & K5V       &                  &  41.28$\pm$1.38&        & 15.08$\pm$ 1.43&  20.23$\pm$0.68& -1.70 & 6.18$\pm$ 2.20&  7.90$\pm$ 0.26&-0.78& 1.5e-06&16.00$\pm$3.19  \\
 73184 & K4V       &                  &  46.53$\pm$1.16&        & 19.82$\pm$ 1.45&  22.80$\pm$0.57& -0.22 & 8.77$\pm$ 1.65&  8.91$\pm$ 0.22&-0.08& 1.3e-06&47.30$\pm$4.37  \\
 73996 & F5V       &                  &  23.23$\pm$0.25&        & 10.01$\pm$ 1.47&  11.38$\pm$0.12& -0.93 &               &  4.45$\pm$ 0.05&     & 1.1e-06&33.20$\pm$5.84  \\
 77052 & G5V       &                  &  17.27$\pm$0.44&        &  8.25$\pm$ 1.20&   8.46$\pm$0.22& -0.17 & 4.53$\pm$ 1.70&  3.31$\pm$ 0.08& 0.72& 1.7e-06&11.10$\pm$5.25  \\
 78459 & G2V       &                  &  23.06$\pm$0.23&        & 10.59$\pm$ 1.18&  11.30$\pm$0.11& -0.60 & $<$5.4        &  4.41$\pm$ 0.04&     & 1.1e-06&29.60$\pm$5.76  \\
 78775 & G8V       &   10.95 $\pm$1.37&  10.56$\pm$0.13&  0.28  &  4.95$\pm$ 0.97&   5.18$\pm$0.06& -0.24 & $<$4.8        &  2.02$\pm$ 0.02&     & 2.9e-06& 8.70$\pm$3.55  \\
 79248 & K0V       &                  &  10.76$\pm$0.11&        &  3.91$\pm$ 0.78&   5.27$\pm$0.06& -1.74 & $<$4.8        &  2.06$\pm$ 0.02&     & 2.1e-06&10.60$\pm$2.60  \\
 80725 & K2V       &    5.90 $\pm$2.43&   9.36$\pm$0.09& -1.43  &  3.76$\pm$ 0.77&   4.59$\pm$0.04& -1.08 & $<$3.3        &  1.79$\pm$ 0.02&     & 2.8e-06&37.30$\pm$17.58 \\
 82860 & F6Vvar    &                  &  29.48$\pm$0.52&        & 16.61$\pm$ 1.00&  14.44$\pm$0.25&  2.10 &               &  5.64$\pm$ 0.10&     & 6.1e-07&44.70$\pm$6.01  \\
 83389 & G8V       &    6.78 $\pm$1.33&   8.16$\pm$0.09& -1.03  &  3.94$\pm$ 0.67&   4.00$\pm$0.04& -0.09 & $<$5.7        &  1.56$\pm$ 0.02&     & 2.2e-06&11.00$\pm$9.03  \\
 84862 & G0V       &                  &  24.83$\pm$0.27&        & 13.00$\pm$ 1.66&  12.17$\pm$0.13&  0.50 & 8.55$\pm$ 2.50&  4.75$\pm$ 0.05& 1.52& 1.6e-06&24.60$\pm$4.33  \\
 86036 & G0V       &                  &  26.26$\pm$1.50&        & 12.54$\pm$ 1.09&  12.87$\pm$0.73& -0.25 & 3.80$\pm$ 2.31&  5.03$\pm$ 0.29&-0.53& 8.6e-07&27.40$\pm$4.77  \\
 86796 & G5V       &                  &  29.08$\pm$0.70&        & 15.46$\pm$ 1.45&  14.25$\pm$0.34&  0.81 & 4.44$\pm$ 2.48&  5.57$\pm$ 0.13&-0.46& 1.1e-06&31.00$\pm$8.07  \\
 88601 & K0V SB    &                  & 154.66$\pm$5.77&        & 73.94$\pm$ 3.05&  75.78$\pm$2.83& -0.44 &33.31$\pm$ 3.19& 29.60$\pm$ 1.11& 1.10& 6.0e-07& 143.7$\pm$16.85\\
 88972 & K2V       &   18.23 $\pm$1.65&  17.41$\pm$0.17&  0.49  &  9.04$\pm$ 0.90&   8.53$\pm$0.08&  0.56 & $<$5.4        &  3.33$\pm$ 0.03&     & 1.8e-06& 8.60$\pm$3.65  \\
 89042 & G0V       &                  &  20.47$\pm$0.21&        &  9.61$\pm$ 1.23&  10.03$\pm$0.10& -0.34 & $<$5.7        &  3.92$\pm$ 0.04&     & 1.3e-06&17.10$\pm$3.59  \\
 91009 & K7Vvar    &    9.78 $\pm$1.61&  10.20$\pm$0.31& -0.26  &  6.19$\pm$ 1.00&   5.00$\pm$0.15&  1.17 & $<$6.3        &  1.95$\pm$ 0.06&     & 5.8e-06& 6.40$\pm$5.32  \\
 95995 & K1V       &   11.66 $\pm$1.71&  13.36$\pm$0.17& -0.99  &  4.98$\pm$ 0.85&   6.55$\pm$0.08& -1.84 & $<$5.7        &  2.56$\pm$ 0.03&     & 2.2e-06&12.70$\pm$5.37  \\
 96100 & K0V       &                  &  64.91$\pm$0.90&        & 30.32$\pm$ 2.08&  31.81$\pm$0.44& -0.70 &14.76$\pm$ 2.41& 12.42$\pm$ 0.17& 0.97& 9.7e-07&77.20$\pm$7.23  \\
 96441 & F4V       &                  &  31.65$\pm$1.03&        & 14.92$\pm$ 1.22&  15.51$\pm$0.50& -0.45 & 9.62$\pm$ 2.88&  6.06$\pm$ 0.20& 1.23& 5.1e-07&36.80$\pm$4.05  \\
 97944 & K3/K4V    &                  &  24.45$\pm$1.02&        &  8.65$\pm$ 1.05&  11.98$\pm$0.50& -2.87 & 8.13$\pm$ 2.38&  4.68$\pm$ 0.20& 1.44& 1.8e-06&                \\
 98959 & G2V       &   10.44 $\pm$2.22&  12.88$\pm$0.12& -1.10  &  5.71$\pm$ 1.01&   6.31$\pm$0.06& -0.59 & $<$5.4        &  2.46$\pm$ 0.02&     & 1.8e-06&19.00$\pm$3.63  \\
 99240 & G5IV-Vvar &  132.20 $\pm$3.83& 144.18$\pm$1.01& -3.02  & 73.53$\pm$ 2.32&  70.65$\pm$0.50&  1.21 &27.23$\pm$ 3.14& 27.60$\pm$ 0.19&-0.12& 4.1e-07&138.5$\pm$10.72 \\
 99461 & K2V       &                  &  48.93$\pm$1.91&        & 23.52$\pm$ 2.26&  23.98$\pm$0.94& -0.19 & $<$9.0        &  9.37$\pm$ 0.37&     & 1.7e-06&48.50$\pm$10.80 \\
101955 & K5V       &    9.22 $\pm$2.68&  12.00$\pm$0.17& -1.03  &  5.79$\pm$ 0.87&   5.88$\pm$0.08& -0.10 & $<$5.4        &  2.30$\pm$ 0.03&     & 4.4e-06&11.50$\pm$7.043 \\
101997 & G8/K0V    &   10.41 $\pm$1.66&  11.62$\pm$0.28& -0.72  &  5.59$\pm$ 0.95&   5.69$\pm$0.14& -0.10 & $<$4.2        &  2.22$\pm$ 0.05&     & 2.3e-06& 7.40$\pm$3.73  \\
104214 & K5V       &                  & 101.17$\pm$2.17&        & 48.31$\pm$ 2.68&  49.57$\pm$1.06& -0.44 &21.63$\pm$ 3.70& 19.36$\pm$ 0.42& 0.61& 1.4e-06&64.20$\pm$12.66 \\
104217 & K7V       &                  &  83.70$\pm$4.38&        & 38.62$\pm$ 2.43&  41.01$\pm$2.15& -0.74 &18.16$\pm$ 3.81& 16.02$\pm$ 0.84& 0.55& 2.0e-06&83.30$\pm$13.15 \\
106696 & K2V       &   10.59 $\pm$1.93&   8.35$\pm$0.14&  1.16  &  3.81$\pm$ 0.79&   4.09$\pm$0.07& -0.35 & 5.55$\pm$ 1.70&  1.60$\pm$ 0.03& 2.33& 3.2e-06&19.20$\pm$5.26  \\
108870 & K5V       &                  & 125.77$\pm$2.91&        & 58.92$\pm$ 1.75&  61.63$\pm$1.42& -1.20 &23.42$\pm$ 1.95& 24.07$\pm$ 0.56&-0.32& 6.2e-07&111.3$\pm$9.47  \\
109422 & F6V       &                  &  24.01$\pm$0.72&        & 10.57$\pm$ 1.80&  11.77$\pm$0.35& -0.65 & $<$8.4        &  4.60$\pm$ 0.14&     & 1.3e-06&10.60$\pm$5.65  \\
110109 & G1V       &                  &  24.71$\pm$0.83&        & 10.14$\pm$ 1.55&  12.11$\pm$0.41& -1.23 & $<$8.4        &  4.73$\pm$ 0.16&     & 1.4e-06&28.10$\pm$5.72  \\
113357 & G5V       &                  &  21.88$\pm$0.24&        & 11.49$\pm$ 2.67&  10.72$\pm$0.12&  0.29 & $<$15.0       &  4.19$\pm$ 0.05&     & 2.8e-06&28.10$\pm$4.88  \\
116745 & K3V       &                  &  12.22$\pm$0.23&        &  5.21$\pm$ 0.86&   5.99$\pm$0.11& -0.90 & $<$4.2        &  2.34$\pm$ 0.04&     & 2.9e-06&17.20$\pm$2.46  \\
120005 & K2        &                  &  27.63$\pm$0.48&        & 11.20$\pm$ 1.19&  13.54$\pm$0.24& -1.93 & 5.05$\pm$ 1.76&  5.29$\pm$ 0.09&-0.14& 3.5e-06&21.30$\pm$2.71  \\
\noalign{\smallskip}\hline\noalign{\smallskip}
\end{longtable}
\end{scriptsize}
\end{center}
}

\end{document}